\documentclass[titlepage,a4paper,11pt,twoside]{report}
\usepackage[ansinew]{inputenc}
\usepackage[]{graphicx}
\usepackage[tbtags]{amsmath}
\usepackage{amssymb}
\usepackage{natbib}
\usepackage{lscape}
\usepackage{ccaption}
\captionnamefont{\small} \captiontitlefont{\small}
\usepackage{fancyhdr}
\usepackage{wasysym}
\usepackage{textcomp}
\usepackage{color}
\usepackage[top=3cm]{geometry}
\usepackage[ngerman,english]{babel}
\usepackage{keyval}
\bibpunct{(}{)};{a}{,}{,}

\def\sun{\hbox{$_\odot$}}			
\def\Earth{\hbox{$_\oplus$}}			

\def\degr{\hbox{$^\circ$}}
\def\arcmin{\hbox{$^\prime$}}
\def\arcsec{\hbox{$^{\prime\prime}$}}
\def\utw{\smash{\rlap{\lower5pt\hbox{$\sim$}}}}
\def\udtw{\smash{\rlap{\lower6pt\hbox{$\approx$}}}}

\def\fdg{\hbox{$.\!\!^\circ$}}

\def\diameter{{\ifmmode\mathchoice
{\ooalign{\hfil\hbox{$\displaystyle/$}\hfil\crcr
{\hbox{$\displaystyle\mathchar"20D$}}}}
{\ooalign{\hfil\hbox{$\textstyle/$}\hfil\crcr
{\hbox{$\textstyle\mathchar"20D$}}}}
{\ooalign{\hfil\hbox{$\scriptstyle/$}\hfil\crcr
{\hbox{$\scriptstyle\mathchar"20D$}}}}
{\ooalign{\hfil\hbox{$\scriptscriptstyle/$}\hfil\crcr
{\hbox{$\scriptscriptstyle\mathchar"20D$}}}}
\else{\ooalign{\hfil/\hfil\crcr\mathhexbox20D}}%
\fi}}
\def\sqr#1#2{{\vcenter{\vbox{\hrule height.#2pt\hbox{\vrule width.#2pt
        height#1pt \kern#1pt\vrule width.#2pt}\hrule height.#2pt}}}}

\captionnamefont{\bfseries} \hangcaption
\begin{document}

    {\pagestyle{empty}
    \begin{titlepage}
        \begin{center}
            {\large \bf
        {\Huge Dissertation}\\ [2ex]
        submitted to the\\[2ex]
        Combined Faculties of the Natural Sciences and Mathematics\\[2ex]
        of the Ruperto-Carola-University of Heidelberg, Germany\\[2ex]
        for the degree of\\[2ex]
        Doctor of Natural Science \\ } 
            \vspace*{14cm}
            {\large \bf
                Put forward by} \\*[2ex]
                {\Large \bf Diplom-Physikerin Eva Meyer} \\*[2ex]
            {\large \bf born in: Essen, Germany \\ [2ex]
                Oral Examination: 21st July 2010 \\}

            \cleardoublepage
	    	\vspace*{2cm}
	    \hrule height 2pt \vspace{0.1cm} \hrule  \vspace{0.7cm}
            {\Huge \bf High Precision Astrometry with Adaptive Optics aided Imaging  \\ \vspace{1.1cm}}
	    \hrule \vspace{0.1cm} \hrule height 2pt \vfill

       \end{center}

             Referees: \hspace{1.5cm}Prof. Dr. Hans-Walter Rix\\ [3ex]
               \hspace*{3.75cm}Prof. Dr. Joachim Wambsgan\ss

    \end{titlepage}

    \cleardoublepage
    \pagestyle{plain}
    \pagenumbering{roman}
    \setcounter{page}{1}
    \section*{Abstract}
Currently more than 450 exoplanets are known and this number increases nearly every day. Only a few constraints on their orbital parameters and physical characteristics can be determined, as most exoplanets are detected indirectly and one should therefore refer to them as exoplanet \textit{candidates}. Measuring the astrometric signal of a planet or low mass companion by means of measuring the wobble of the host star yields the full set of orbital parameters. With this information the true masses of the planet candidates can be determined, making it possible to establish the candidates as real exoplanets, brown dwarfs or low mass stars. In the context of this thesis, an M-dwarf star with a brown dwarf candidate companion, discovered by radial velocity measurements, was observed within an astrometric monitoring program to detect the astrometric signal. Ground based adaptive optics aided imaging with the ESO/NACO instrument was used with the aim to establish its true nature (brown dwarf vs.\ star) and to investigate the prospects of this technique for exoplanet detection. The astrometric corrections necessary to perform high precision astrometry are described and their contribution to the overall precision is investigated. Due to large uncertainties in the pixel-scale and the orientation of the detector, no detection of the astrometric orbit signal was possible.\\
\noindent The image quality of ground-based telescopes is limited by the turbulence in Earth's atmosphere. The induced distortions of the light can be measured and corrected with the adaptive optics technique and nearly diffraction limited performance can be achieved. However, the correction is only useful within a small angle around the guide star in single guide star measurements. The novel correction technique of multi conjugated adaptive optics uses several guide stars to correct a larger field of view. The VLT/MAD instrument was built to demonstrate this technique. Observations with MAD are analyzed in terms of astrometric precision in this work. Two sets of data are compared, which were obtained in different correction modes: pure ground layer correction and full multi conjugated correction.

\newpage
\section*{Zusammenfassung}
Mehr als 450 extrasolare Planets sind zurzeit bekannt und diese Zahl wird fast t\"aglich gr\"osser. Da die meisten Exoplaneten indirekt entdeckt werden, k\"onnen nur wenige Einschr\"ankungen bez\"uglich ihrer Bahnparameter und physikalischen Eigenschaften gemacht werden und sie sollten daher vorl\"aufig als Exoplanet-\textit{Kandidaten} bezeichnet werden. Misst man das astrometrische Signal eines planetaren oder massearmen Begleiters, indem man die Reflexbewegung des Hauptsterns vermisst, so erh\"alt man den vollen Satz an orbitalen Parametern. Mit dieser Information kann die genaue Masse der Kandidaten bestimmt werden und es ist somit m\"oglich, die Planetenkandidaten als wahre Exoplaneten, Braune Zwerge oder massearme Sterne einzustufen. Im Rahmen der vorliegenden Doktorarbeit wurde ein Zwergstern der Spektralklasse M, der einen mittels Radialgeschwindigkeitsmessungen entdeckten wahrscheinlichen Braunen Zwerg als Begleiter hat, innerhalb eines fortlaufenden Beobachtungsprogramms zur Detektion des astrometrischen Signals beobachtet. Bodengebundene Beobachtungen mit dem Adaptiven Optik (AO) Instrument ESO/NACO wurden durchgef\"uhrt, um die wahre Natur des Begleiters zu bestimmen (Brauner Zwerg oder massearmer Stern) und die Aussichten dieser Technik im Bereich der Planetenendeckung zu untersuchen. Die astrometrischen Korrekturen, notwendig um hochpr\"azise Astrometrie zu betreiben, werden in diesem Zusammenhang beschrieben und ihr Beitrag zur Gesamtmessgenauigkeit untersucht. Die gro\{ss}en Unsicherheiten in der Messgenauigkeit der \"Anderung der Pixel-Skala und der Ausrichtung des Detektors verhinderten jedoch, das Signal des astrometrischen Orbits zu messen.\\
\noindent Die Abbildungsqualit\"at eines bodengebundenen Teleskopes ist begrenzt durch die Turbulenz in der Atmosph\"are der Erde. Die dadurch hervorgerufenenen Verformungen der Lichtwellen k\"onnen mit Hilfe der Technik der Adaptiven Optik vermessen und korrigiert werden und somit beinahe beugungsbegrenzte Abbildungen erzeugt werden. Im Fall der klassischen AO mit nur einem Referenzstern ist die Korrektur jedoch nur in einem engen Bereich um den Referenzstern m\"oglich. Multikonjugierte Adaptive Optik verwendet mehrere Referenzsterne, um ein gr\"osseres Gesichtfeld zu korrigieren. Das MAD Instrument wurde gebaut und am Very Large Telescope installiert, um diese neue Technik zu demonstrieren. Beobachtungen mit MAD wurden im Rahmen dieser Arbeit auf ihre astrometrische Genauigkeit hin ausgewertet. Dabei wurden zwei Datens\"atze verglichen, die in unterschiedlichen Korrektur-Modi aufgenommen wurden: zum einen wurde nur die Turbulenzschicht nahe am Boden wurde korrigiert, zum anderen die volle multikonjugierte Konfiguration des Instrumentes genutzt.

    \newpage
    \vspace*{\stretch{1}}
\begin{center}
for my father\\
in loving memory
\end{center}
\vspace*{\stretch{2}}

    \cleardoublepage

    \tableofcontents
    \newpage
    \listoffigures
    \listoftables \cleardoublepage
    \pagestyle{fancy}
    \chapter{Introduction}
    \pagenumbering{arabic}
    \setcounter{page}{1}
    With a few 100 billion stars in our own Milky Way Galaxy and just as many
galaxies in the universe, it would be very ignorant to
believe mankind is alone in this tremendous and beautiful
universe. Even though detecting other life-forms is still far far
away, detecting planets orbiting other stars is already reality.
More than 450 so-called extrasolar planets around more than 380
different stars have already been discovered, with more and more to
come every day.\\
Ever since mankind can remember, the heaven - sprinkled with stars, galaxies and planets - has fascinated humanity and led to the desire to explore, understand and explain what can be seen in the endless space. Since the first use of a telescope for astronomical observations by Galileo Galilei over 400 years age, ever better telescopes and instruments were, are and will be developed. Information achieved of an object on the sky is brought to the observer on Earth via light coming from this object. After travelling through space for thousands and millions of years, this information is altered on the last milli-seconds when passing through Earth's atmosphere and part of the information is lost. One of the most sophisticated methods for telescopes on Earth to retrieve this lost information and to \textquotesingle turn off\textquotesingle ~the twinkling of the stars is a method called \textit{adaptive optics}. Images of celestial objects, blurred out by the atmosphere to the resolution of a backyard telescope, are sharpend to unveil the tiny but great mysteries of the universe.

\noindent The main goal of this work aims at investigating a
technique to detect those planets and low mass companions orbiting other stars than our Sun. High precision
astrometry, supported  by adaptive optics, is used to go for the detection of the tiny motion of a star due to its
unseen companion. Astrometry alone has not been successful to find new worlds yet, and only few known exoplanets could be further characterized by astrometric measurements, but the future is promising with new space missions and instruments to come.\\ 
\noindent Furthermore the technique of adaptive optics is investigated for future astrometric high precision measurements.
Data obtained with an instrument based on a novel concept of adaptive optics instrumentation is analyzed and the astrometric precision achievable is determined.

\subsubsection{Thesis Outline}
This thesis consists of two main parts. In this chapter, an
introduction to the orbital elements of the motion of a
planet/companion around a star (or vice versa) and the various
planet detection methods is given. Brown dwarfs and the brown
dwarf desert are presented at the end of this chapter. In chapter
3 the adaptive optics technique and the NACO instrument, which is used in
this work, are introduced. Chapters 4 and 5 describe the observed target star and its known companion
together with the observing strategy, as well as the necessary
astrometric corrections applied to the data to obtain high precision astrometry. In chapters 6 and 7
the final orbit fit, the resulting companion mass and its significance are
summarized and discussed.\\
\noindent The second part of the thesis deals with a new instrument
and observing technique in the adaptive optics field with respect to astrometric precision. In
chapter 8 multi conjugated adaptive optics and the MAD instrument, which is used in this context, are introduced, followed by the analysis of
the stability and precision of astrometric measurements with MAD
in Chapter 9. The results and a discussion are given in chapter
10.

\newpage
\section{Orbital Elements}
\label{sec:orbitalelements} The orbit of a planet or every other
companion round a star is defined by six orbital elements ($a, e,
P, \omega, \Omega, i$). Fig.~\ref{fig:orbitelements} shows the
definition of the orbital elements:
\begin{figure}
 \begin{center}
 \includegraphics[width=14.5cm]{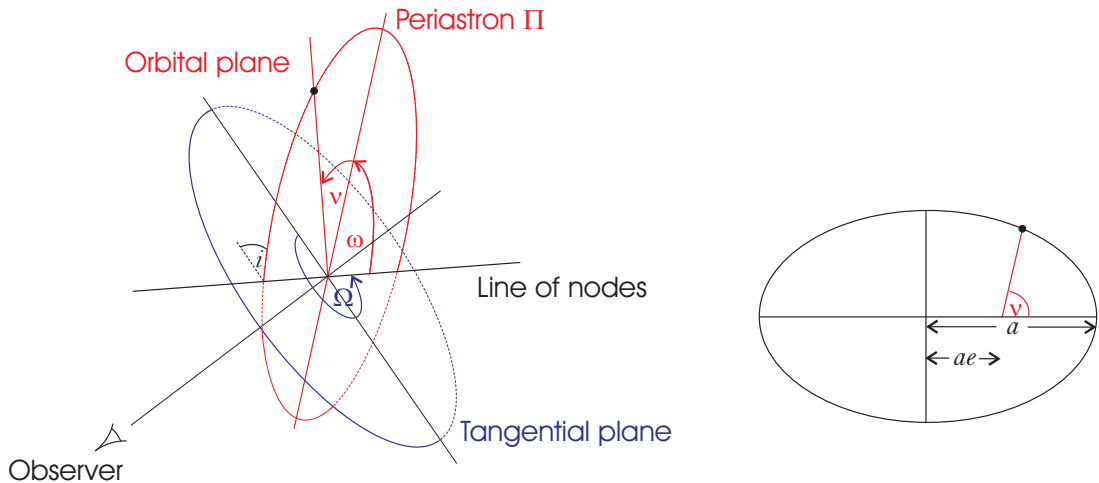}
 \caption{Definition of the orbital elements.}
 \label{fig:orbitelements}
 \end{center}
\end{figure}
The dynamical elements characterizing the size and shape of the
orbital ellipse are the \textit{semi-major axis} $a$, the
\textit{eccentricity} $e$ and the \textit{period} $P$. Often the
\textit{time of periastron} $T_{p}$ is also used to specify the
timing of the orbit.\\
\noindent The position of the orbit in space with respect to the
local coordinate system is characterized by three solid angles.
The intersection of the orbital plane with the plane perpendicular
to our line of sight is called the line of nodes. It connects the
two points where the orbit intersects the tangential plane. These
points are called ascending node and descending node, depending on
whether the companion passes the tangential plane from South to
North or North to South, respectively. The angle from the
coordinate zero point of the reference plane, the projection of
the North celestial pole, to the ascending node is the
\textit{longitude of the
ascending node} $\Omega$. It is measured Eastward.\\
\noindent The angle between the reference plane and the orbital
plane is called the \textit{inclination} $i$. If the orbit of the
companion is direct, i.e. the position angle increases with time,
then $i < 90\degr$, in the case of an retrograde orbit $90\degr <
i < 180\degr$. For an inclination of $0\degr$ or $180\degr$ the
orbit is seen face-on, for $i = 90\degr$ the orbit is seen
edge-on.\\
\noindent The third angle is the \textit{longitude of periastron}
$\omega$. It specifies the orientation of the orbit in the orbital
plane and defines the angle of the direction to the periastron
$\Pi$ from the line of nodes.

\section{Detection Methods}
\subsection{Pulsar Timing}
By surprise the first planetary-mass objects detected were
orbiting a pulsar and had masses close to the terrestrial mass.
Two planetary objects with $2.8~M\Earth$ and $3.4~M\Earth$ with
98.88 days and 66.54 days period, respectively, were found to
orbit the millisecond pulsar PSR B1257~+~12
\citep{Wolszczan1994,Wolszczan1992}. Later also a third component
was found in this system. Pulsars are extremely rapidly rotating
neutron stars which emit mostly radio emission in a very narrow
light-cone. If the alignment with the observer is favorable, a
pulse effect can be observed similar to a lighthouse. These pulses
are very precise and stable in time, which makes it possible to
detect small variations in the periodicity. Such a variation can
occur when a companion is orbiting the pulsar, causing a
positional shift of the pulsar around the barycenter of the
pulsar-companion system. The motion of the pulsar around the
barycenter leads to a change in light travel time of the incoming
pulses, which becomes manifest in a delay or early arrival of the
pulse signals $\tau$.
\begin{equation}
 \tau = \sin{i}\left (\frac{a_{p}}{c}\right )\left( \frac{m_{p}}{M_{*}}\right )
\end{equation}
Here $a_{p}$ is the semi-major axis of the planet's orbit, i the
inclination of the orbit, $M_{p}$ the planetary and $M_{*}$ the
pulsar mass and $c$ stands for the speed of light. With this
method one can measure the period of the planet, its eccentricity
and the projected planet to star mass ratio. Because of the
projection of the true motion of the pulsar onto the radial
direction between the observer and the pulsar, one only measures a
minimum mass for the companion which is still dependent on the
inclination of the orbital plane, assuming a known mass for the
pulsar.

\subsection{Radial Velocity Measurements}
The most successful detection method so far has been the radial
velocity (RV) method. The first extrasolar planet around a solar
type star was discovered this way by Michel Mayor and Didier
Queloz \citep{Mayor1995}. The radial velocity method measures, as
the pulsar timing method, the movement of the star due to an
unseen planet in the direction of the line of sight. In this
process the Doppler-shift of the spectral lines is measured. High
precision spectral line measurements can be performed by comparing
the stellar spectrum with a set of reference lines. This reference
lines are superimposed on the stellar spectrum and can be produced
for example by an iodine cell in the light path of the
spectrograph. If the target star has a planet, it will exhibit a
Doppler shift $\Delta\lambda/\lambda = v/c$, with the same period
as the planetary orbit. The spectral lines will move redward when
the star is moving away from the observer and bluewards when it is
approaching. These variations only measure the component of the
motion projected onto the line of sight of the observer and hence
only a minimum mass, $m_{p}\sin{i}$ of the planet orbiting the
star can be measured. The semi-amplitude of the radial velocity
variation is given by:
\begin{equation}
 K = \left (\frac{2\pi G}{P}\right )^{1/3} ~\frac{m_{p}\sin{i}}{(M_{*} +
 m_{p})^{2/3}}\frac{1}{\sqrt{1-e^{2}}}
\label{equ:RVAmplitude}
\end{equation}
where $P$ is the planetary orbital period, $e$ the eccentricity
and $G$ the gravitational constant. $K, P$ and $e$ can be derived
from the shape of the Doppler curve. Also the argument of
periastron, $\omega$, and the time of periastron, $T_{p}$, can be
derived from the RV curve. Estimating the stellar mass $M_{*}$
from stellar models and assuming $m_{p} \ll M_{*}$ one can
determine $m_{p}\sin{i}$. With Kepler's third law $P[\rm yr] =
(a_{p}[\rm AU])^{3/2}(M_{*}/M\sun)$, where $M\sun$ is one solar
mass, one can
also derive the semi-major axis $a_{p}$ of the planet.\\
\noindent If the mass of the companion cannot be neglected, one
cannot derive $m_{p}\sin{i}$ but has to use the mass function for
the star-planet system instead:
\begin{equation}
 f(m) = \frac{(m_{p}\sin{i})^3}{(M_{*} + m_{p})^2} = \frac{PK^3(1 - e^2)^{3/2}}{2\pi G}
 \label{equ:mass_function}
\end{equation}
and a minimum semi-major axis of the stellar wobble:
\begin{equation}
 a_{*}\sin{i} = K\cdot P\frac{\sqrt{1 - e^2}}{2\pi}
\label{equ:asini}
\end{equation}
\noindent The fact that the RV measurements only yield the
component of the orbital motion in the direction of the observer's
line of sight, can lead to the case that a low-mass star orbiting
another star with a small inclination of the orbit is interpreted
as the signal produced by a planetary companion orbiting at a high
inclination. However, this is very unlikely if one assumes random
orientation of the orbits. The most likely observable inclination
would be close to edge-on, with a median inclination
of $60\degr$ \citep{Kuerster1999}.\\

\noindent The RV technique has the advantage of being mostly
independent of the distance. The only distance related limitation
is that the more distant the stars are the fainter they are,
leading to a lower signal to noise ratio in the spectra. The
precision possible for the RV detection method is about 1 m/s,
limited by intrinsic stellar turbulence and activity in even the most
stable stars. Because of this, spectral types of F, G, and K are
preferred for this technique, as later type stars are often too
faint for adequate signal to noise and early type stars have much
less spectral lines to measure and are limited in the line positioning accuracy
due to the spectral line broadening.  The RV measurements are
strongly biased towards close-in orbits and high masses, as the RV
semi-amplitude $K$ is higher for shorter periods $P$, i.e. smaller
separations from the host star and higher masses of the companion,
explaining the high number of
Hot-Jupiter detections.\\

\noindent For this reason most of the large radial velocity
surveys target non-active main sequence stars
\citep[e.g.][]{Tinney2001, Queloz2000}, but also M-stars
\citep[e.g.][]{Kuerster2006, Zechmeister2009, Bonfils2004} and
young stars \citep{Setiawan2008} are being monitored. Low mass
planets are thought to be found more easily around M-stars, as
their stellar mass is smaller and the effect of perturbations of
smaller planets is easier to detect, but on the other hand M-stars
are fainter and therefore the precision obtained in the RV
measurements is not as high as for solar-type stars.
\begin{figure}
 \begin{center}
  \includegraphics[width=11cm]{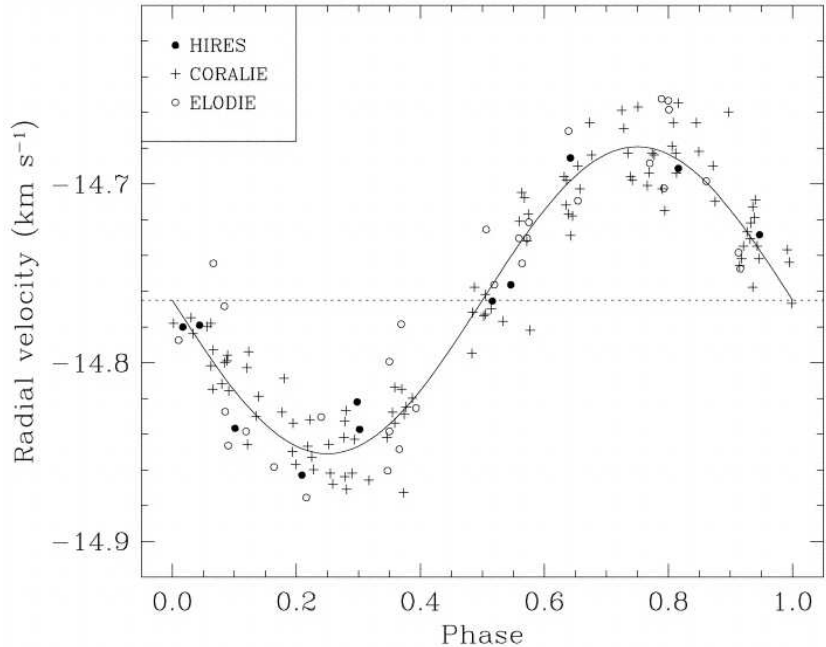}
   \caption[Radial velocity curve of the system HD~209458]
   {Radial velocity curve of the system HD~209458
   \citep[see][]{Mazeh2000}. The planetary
    companion has a mass of $0.685~M_{J}$ and orbits its parent
    star in 3.525 days. The different symbols show the data taken
    with the different instruments and the solid line is the best
    fit Keplerian orbit.}
 \end{center}
\end{figure}

\subsection{Transits}
If a planet passes between its host star's disk and the observer,
the observed flux drops by a small amount. The amount of the
dimming depends on the relative sizes of the planet and star and
its maximum depth is given by $R_{p}/R_{*}^2$, where $R_{p}$ is
the radius of the planet and $R_{*}$ that of the star. So, if one
can estimate $R_{*}$, one has a direct measure for the radius of
the planet, something one can only measure with this method. From
the periodicity of the transit event one gets the orbital period
$P$ and if one can estimate the stellar mass $M_{*}$ one can
derive the semi major axis of the planetary orbit from Kepler's
third law. The shape of the dip in the light curve depends on the
inclination of the system, which has to be close to $90\degr$ to
observe a transit. Due to simple geometric reasons, this is the
case only for a small minority of planets. Additionally  the
probability of a transit is proportional to the ratio of the
diameter of the star and the diameter of the orbit. The longer the
orbital period, the smaller is the chance of a proper alignment.
Also the chance of seeing the transit by measuring at the right
time is decreasing with longer orbital periods. Nevertheless
around 70 planets have already been detected using this method
with likely more to come from the ongoing surveys of the KEPLER
\citep[e.g.][]{Basri2005, Borucki2010} and CoRoT
\citep[e.g.][]{Deleuil2010, Borde2003} missions, which
observe large areas on the sky with thousands of stars.\\
\noindent The transit method for detecting exoplanets is also
biased to close-in orbits, as is the radial velocity method. If
one can combine the two methods and solve the degeneracy of the
orbital inclination, one can compute the true mass of the planet,
and, together with the radius determined from the transit, the
density of the
planet.\\
\begin{figure}
 \begin{center}
  \includegraphics[width=14cm]{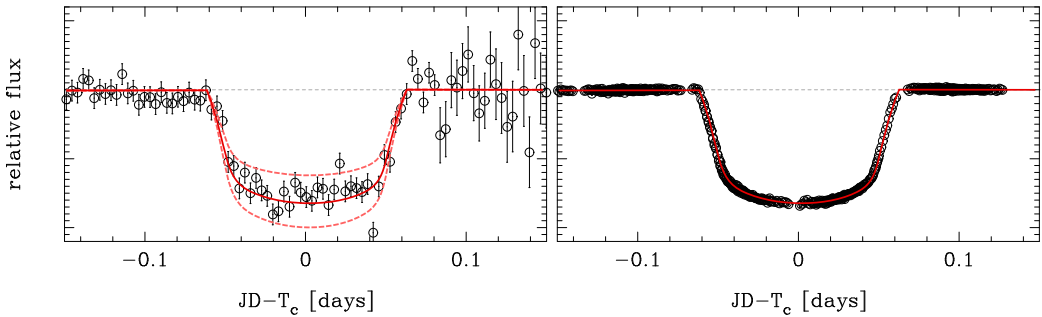}
   \caption[Transit light curves for the planetary system HD~209458]
   {Transit light curves for the planetary system HD~209458. Left the
   first measurements from ground \citep{Charbonneau2000} with the STARE Project Schmidt camera and
   right a light curve obtained with the STIS instrument aboard the Hubble Space Telescope
   \citep{Brown2001}. The solid lines show the transit shape
   for the best fit model and the dashed lines in the left panel
   show additionally the transit curves for a planet with a 10\%
   larger and smaller radius than the one from the best fit, respectively.}
 \end{center}
\end{figure}

\noindent Comparing observations of spectra of the star during
transit and outside of transit can yield spectral features of the
transmission spectrum of the planetary atmosphere if the signal to
noise ratio of the spectra is high enough. Such observations were
conducted for the first transiting planet detected, HD~209458
\citep{Charbonneau2000, Charbonneau2002}.\\

\noindent Likewise one can use this so-called secondary eclipse,
when the planet is behind the star, to measure the thermal flux
emitted from the planet. Hot Jupiters typically have a thermal
flux which is 'only' about $10^{-4}$ times smaller than that of
the star, which makes it possible to indirectly measure it. When
the planet is behind the star, one has the unique opportunity to
measure the true brightness of the star. Subtracting this from the
combined planet + star brightness one can derive the thermal flux,
$F_{p}$ of the planet assuming a known distance of the system.
Under the assumption of blackbody radiation $F_{p}$, is given by:
\begin{equation}
F_{p} = 4\pi R^2_{p}\sigma T^4_{p}
\end{equation}
where $\sigma$ is the Stefan-Boltzmann constant and $T_{p}$ the
effective temperature of the planet. Deducing $R_{p}$ from the
depth of the transit light curve, one can
infer the effective temperature of the planet.\\
\noindent Comparing the spectrum of the star + planet with the one
of the star observed during secondary eclipse, provides the
opportunity to carry out infrared spectroscopy of the planet. The
space telescope Spitzer has been used mainly for this purpose and
the upcoming James Webb Space Telescope (JWST) will provide even
more progress in this field. But also from the ground first
approaches have started to examine the secondary eclipse and its
measurands \citep{Swain2010}.

\subsection{Gravitational Microlensing}
Due to general relativistic effects a light path is bent in the
presence of a gravitational field. In principle any massive object
can act as a lens, bending the light of a background object and
causing a temporary magnification of the brightness of the
background object. Such lensing can be observed on a galactic
scale, where for example a massive cluster in the foreground is
acting as a lens for distant galaxies. In the case of a perfect
alignment the lens would cause the background object to appear as
a ring, the so-called Einstein-Ring, with an angular radius of:
\begin{equation}
 \theta_{E} = \sqrt{\frac{4GM_{*}(D_{S} - D_{L})}{c^2D_{L}D_{S}}}
\end{equation}
where $D_{L}$ is the distance to the lens, $D_{S}$ the distance to
the background source and $M_{*}$ the mass of the lens. In the
case of an imperfect alignment several single images of the
background object are imaged around the lens.\\

\begin{figure}
 \begin{center}
  \includegraphics[width=10cm]{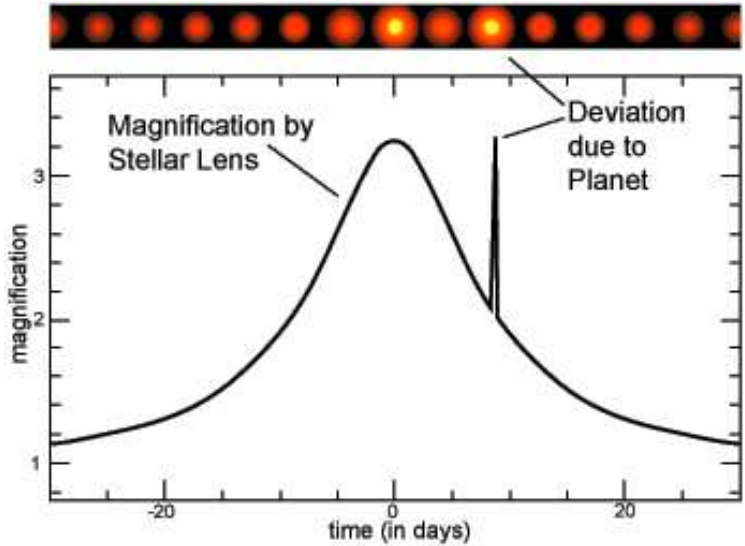}
  \caption[Light curve of a gravitational lensing event of a star with a planet]
  {Light curve of a gravitational lensing event of a star with a
  planet. On top the brightening due to the lensing of the star
  and later the planet is shown. Image taken from the
  \textit{Microlensing Planet Search Project homepage\footnote{http://bustard.phys.nd.edu/MPS/}}.}
 \label{fig:lensing}
 \end{center}
\end{figure}

\noindent In the case relevant for planetary detection with the
gravitational lensing method, a star with a planetary companion
acts as the lens and the background object is a distant star. The
probability of an alignment among two stars is very small, but
increases towards the galactic center. But even there it is only
about one in $10^6$. Contrary to the lensing on a galactic scale
it is not possible with current instrumentation to resolve the
Einstein ring on the stellar scale. Instead one measures the total
magnification which depends on the angular separation between the
lens and the background object $u$ and its change with time. The
magnification factor $Q$ of the event is given by:
\begin{equation}
 Q(t) = \frac{u^2(t) + 2}{u(t)\sqrt{u^2(t) + 4}}
\end{equation}
\noindent As the lens passes the background star, $Q$ changes with
time, and measuring the light curve in a close enough time sample
during the event yields information about the lensing star. If a
planet is in orbit around the lensing star and the already
magnified image of the background star comes close to this planet,
then the planet's own gravitational potential, distorting the
star's potential, becomes also a visible effect. An additional
brightening will occur on top of the brightening due to the
lensing star, causing a sharp peak in the light curve, see
Fig.~\ref{fig:lensing}. This detection method is sensitive down to
very low-mass planets and also to planets orbiting
very distant stars.\\
\noindent As these events do not repeat and two stars need to be
aligned, this approach is challenging. The current approach is to
monitor a large number of planets and alert other collaborating
observatories and institutes as soon as a lensing event is
detected, which then also observe the event if possible. This
provides a good time sampling of the light curve. A very
successful survey is the OGLE survey (Optical Gravitational
Lensing Experiment), which has detected several planets to date
\citep[see e.g.][]{Udalski1993}.

\subsection{Direct Imaging}
Direct imaging of an exoplanet yields a wide range of information
about the planet. One can characterize it spectroscopically,
providing information about the atmosphere and measure its
astrometric motion to derive information about the orbit. But to
directly image a planet next to a bright star is a challenging
task, given the brightness contrast and the small angular
separation typical for exoplanetary systems. Jupiter for example
would only have a 4 arcsecond separation from the sun when viewed
from Alpha Centauri and typical angular distances of known
exoplanets are much smaller.\\

\noindent Exoplanets are cool objects with temperatures in the
range of a few 100~K, which makes their thermal brightness in the
visual negligible. The light observable from the planet in this
wavelength range is reflected light from the primary star. For an
Earth-like planet the flux ratio between planet and star is $\sim
10^{-10}$. This is an almost impossible high contrast for today's
instruments, given the very small separation between the star and
the planet of $0.1-1\arcsec$ for nearby stars, and additional
techniques have to be used. One possibility to nevertheless image
the planet is coronography, where most of the light from the star
is blocked, so the planetary signal becomes visible. Other methods
include spectral differential imaging, the system is imaged
simultaneously in two different filters and the two images are
subtracted afterwards, angular differential imaging, the system is
imaged with two different position angles and the two images are
subtracted afterwards, and nulling interferometry, see below.\\
\noindent In the infrared wavelength regime, the thermal emission
of the planets is higher and peaks in the mid-infrared. The
younger and hotter a planet is, the higher is its IR flux. For an
Earth-like planet the star-planet flux ratio goes down to about
$10^{-7}$. But at the same time, the spatial resolution is getting
worse with longer wavelengths. Using interferometry is one
solution, as it is easier at longer wavelengths and improves the
spatial resolution in addition to the lower flux contrast.
Supplementary nulling interferometry is possible. The light from
two telescopes is brought together with a shift of $\lambda/2$ in
one light path, so the two beams interfere destructively on-axis
where the star is centered. In the ideal case, this cancels out
all light at zero phase, but keeping the flux at other phases and
hence is working like a
coronograph.\\

\begin{figure}
 \begin{center}
  \includegraphics[width=10cm]{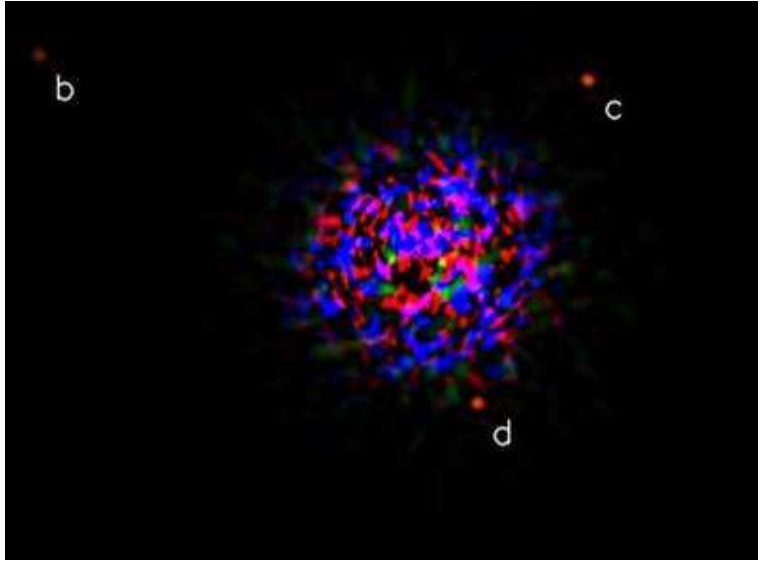}
  \caption[Image of the three planetary companions to the star HR~8799]
  {Image of the three planetary companions to the star HR~8799
  produced by combining J-, H-, and Ks-band images obtained at
  the Keck telescope in July (H) and September (J and Ks) 2008
  \citep{Marois2008}. The three planets b, c, d have masses
  around 7, 10 and $10~M_{J}$ respectively, inferred from
  photometry and fitting evolutionary tracks.}
\label{fig:HR8799}
 \end{center}
 \end{figure}

\noindent Most of the extrasolar planets detected with direct
imaging so far have bigger separations from their host stars than
the ones detected with radial velocity or transits. But with the
already installed and near future instruments for direct imaging
with adaptive optics correction and coronography and/or
interferometry more and
more planets will be detected closer to their stars. \\
One of the first extrasolar planetary systems whose orbital
motions were confirmed via direct imaging is the system HR~8799.
HR~8799 is a young ($\sim 60$ million year old) main sequence star
located 39 parsecs away from the Earth in the constellation of
Pegasus. The system contains three detected massive planets and
also a debris disk. The planets were detected by Christian Marois
with the Keck and Gemini telescopes on Mauna Kea, Hawaii
\citep{Marois2008}. Just recently the first spectrum of an
exoplanet was obtained from the middle one of the three planets
orbiting HR~8799 with the NACO instrument at the VLT
\citep{Janson2010}. This is the first step into a new and amazing
area of extrasolar planet characterization. In
Fig.~\ref{fig:HR8799} the direct image of the three exoplanets is
shown.

\subsection{Astrometry}
\label{subsec:astrometry} Astrometry is the oldest measurement
technique in astronomy. The method consists of precise
measurements of the position of a star on the tangent plane on the
sky relative to a reference frame and has been used for centuries
to measure proper motions, parallaxes and astrometric orbits of
visual binaries. The gravitational influence of an orbiting planet
causes the star to move around their common center of mass in a
small, down-scaled orbital movement. Assuming $M_{*} \gg m_{p}$ a
combination of the law of the lever for the two-body problem with
Kepler's $3^{rd}$ law gives the semi-amplitude $a_{*}$ of the
stellar wobble due to the companion in \textit{radians}:
\begin{equation}
 a_{*} = \frac{m_{p}}{r} ~\left( \frac{G}{4\pi^2} \right)^{1/3} ~\left( \frac{P}{M_{*}} \right)^{2/3}
\label{equ:astrometry}
\end{equation}
Here $r$ is the distance of the system, $M_{*}$ is the
stellar mass , $m_{p}$ the companion's mass, $P$ the orbital
Period and $G$ the gravitational constant. The astrometric signal
becomes stronger the more massive the companion, and/or the less
massive the primary and the bigger the separation between the two
components. This makes this detection method complementary to RV
and transit detection, which are most sensitive to close-in
planets. As one can see from Equ.~\ref{equ:astrometry}, the
astrometric detection of a planetary companion is very sensitive
to the distance of the system, which limits this technique to
applications to nearby stars. The astrometric signals for some of
our own solar systems planets seen from 10~pc distance and
examples for a hot Jupiter and an Earth-like planet orbiting a
solar-type star, as well as several other examples are listed in
Tab.~\ref{tab:astrometry}. The values are calculated for the case
the full major axis $a_{*}$ is measured and are given as the
peak-to-peak astrometric signal, $\alpha = 2 * a_{*}$. If the
orientation of the system is such, that not the full major axis is
measurable, the astrometric signal is even smaller. No matter how
the system is oriented, one can always measure at least the signal
of the minor axis.
\begin{table}[t!]
 \begin{center}
  \begin{tabular}{|l|c|cc|l|} \hline
   Planet &  $a_{p}$ [AU] & \multicolumn{2}{c|}{$\alpha$} & Primary mass \\ \hline
   Jupiter \jupiter & 5 & 0.96 & mas & $1~M\sun$ (sun-like star)  \\
   Jupiter \jupiter & 1 & 0.19 & mas & $1~M\sun$ \\
   hot Jupiter & 0.05 & 9.5 & $\mu \rm as$ & $1~M\sun$ \\
   Neptune \neptune & 1 & 1.3 & $\mu \rm as$ & $1~M\sun$ \\
   Neptune \neptune & 30 & 0.31 & mas & $1~M\sun$ \\
   Earth \earth & 1 & 0.60 & $\mu \rm as$ & $1~M\sun$ \\
   Brown dwarf ($30~M_{J}$) & 0.5 & 2.9 & mas & $1~M\sun$ \\
   Brown dwarf ($30~M_{J}$) & 30 & 171.9 & mas & $1~M\sun$ \\ \hline
   Brown dwarf ($30~M_{J}$) & 1 & 11.5 & mas & $0.5~M\sun$ (M Dwarf)\\
   Brown dwarf ($30~M_{J}$) & 15 & 171.9 & mas & $0.5~M\sun$ \\
   Jupiter \jupiter & 1 & 0.38 & mas & $0.5~M\sun$ \\
   Jupiter \jupiter & 15 & 5.7 & mas & $0.5~M\sun$ \\
   Earth \earth & 1 & 1.2 & $\mu \rm as$ & $0.5~M\sun$ \\
   Earth \earth & 0.1 & 0.12 & $\mu \rm as$ & $0.5~M\sun$ \\ \hline
  \end{tabular}
  \caption[Examples for astrometric signals]{Examples for astrometric signals for planets with different masses
  and semi major axes, orbiting a sun-like
  star or a dwarf star with $0.5~M\sun$. All values are calculated
  for a distance of 10~pc and the case that the full
  major axis is measured.}
  \label{tab:astrometry}
 \end{center}
\end{table}

\noindent  If one has obtained astrometric data points which cover
a sufficient part of the orbit, all orbital parameters can be
determined. Especially with a known distance and stellar mass, the
true mass of the companion can be calculated. Astrometric planet
detection gives therefore more information about the detected
companion than RV does. But to derive the orbital motion of the
star due to the companion, one has to disentangle this motion from
proper motion of the star and the parallactic movement of the
Earth bound observer and the orbital motion. Astrometric position
determination always needs a reference system to which the
position of the target star is referenced. Preferable would be a
fixed system, but this is rather difficult to set up and sometimes
not possible. To a much higher precision, positions can be
determined \textit{relative} to another system. For planet
detection this can be a star asterism in the same field of view
(FoV) as the targeted star. Since the stars used to set up the
reference frame have their own proper motion and parallactic
movement, the proper motion and parallax of the target star can
only be derived relative to this reference frame and do not need
to be the same as the absolute ones. If one is only interested in
the orbital movement due to a companion, one does not need to know
the absolute proper motion and parallax, but of course, it is
always beneficial to know the absolute parameters, e.g. to
calculate the distance of the target.\\

\noindent Astrometric measurements have been used to determine
astrometric binary systems for quite a long time. One of the first
comments about the detection of an unseen companion was made by
William Herschel in the late 18th century, when he claimed an
unseen companion being responsible for the position variations of
the star \textit{70 Ophiuchi}. Other systems were announced in the
coming two centuries, but all of them were later vitiated or are
still under discussion.\\
The only measurements of an astrometric signal due to an unseen
companion were obtained with the Hubble Space Telescope (HST). In
2002 \citeauthor{Benedict2002} succeeded in detecting the
astrometric motion of the previously with RV discovered planet
around the star Gliese 876, and in 2006 the signal of the planet
orbiting $\epsilon$ Eridani \citep{Benedict2006} \citep[see
also][for more examples]{Bean2007, Martioli2010, Benedict2010}.\\
\noindent In 2009 a planet orbiting the ultracool dwarf VB~10 (=
GJ~752B), spectral class M8~V, was discovered from ground with
astrometry using the wide-field seeing limited imager at the
Palomar 200-inch telescope within the Stellar Planet Survey
program (STEPS) \citep{Pravdo2009}. The reflex motion of VB~10
around the system barycenter compared to a grid of reference stars
in the same FoV was monitored over 9 years. The best fit Keplerian
orbit yields a $6.4^{+2.5}_{-3.1}\rm ~M_{J}$ planet in an 0.74
year almost edge-on orbit ($ i = 96.9\degr$). Unfortunately,
lately obtained RV observations with the high precision
spectrograph HARPS at the ESO 3.6~m telescope in La Silla, Chile,
were not able to confirm this planet, but instead ruled out the
astrometric orbit solution \citep{Bean2010}.\\
\begin{figure}
 \begin{center}
  \includegraphics[width=7cm]{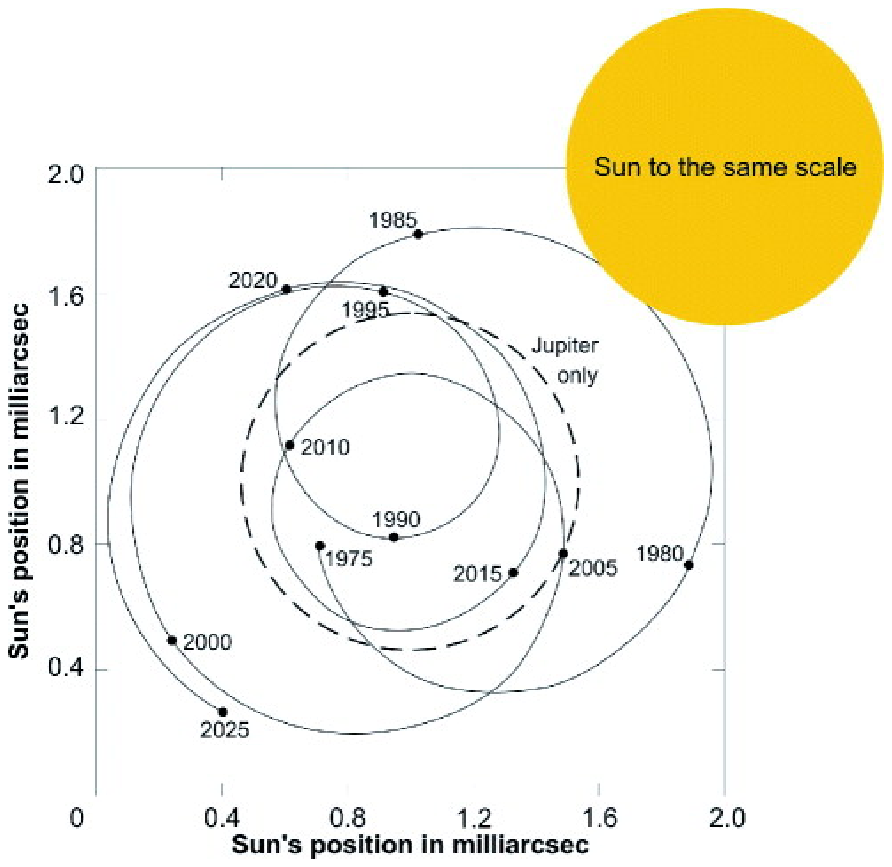}
  \includegraphics[width=7cm]{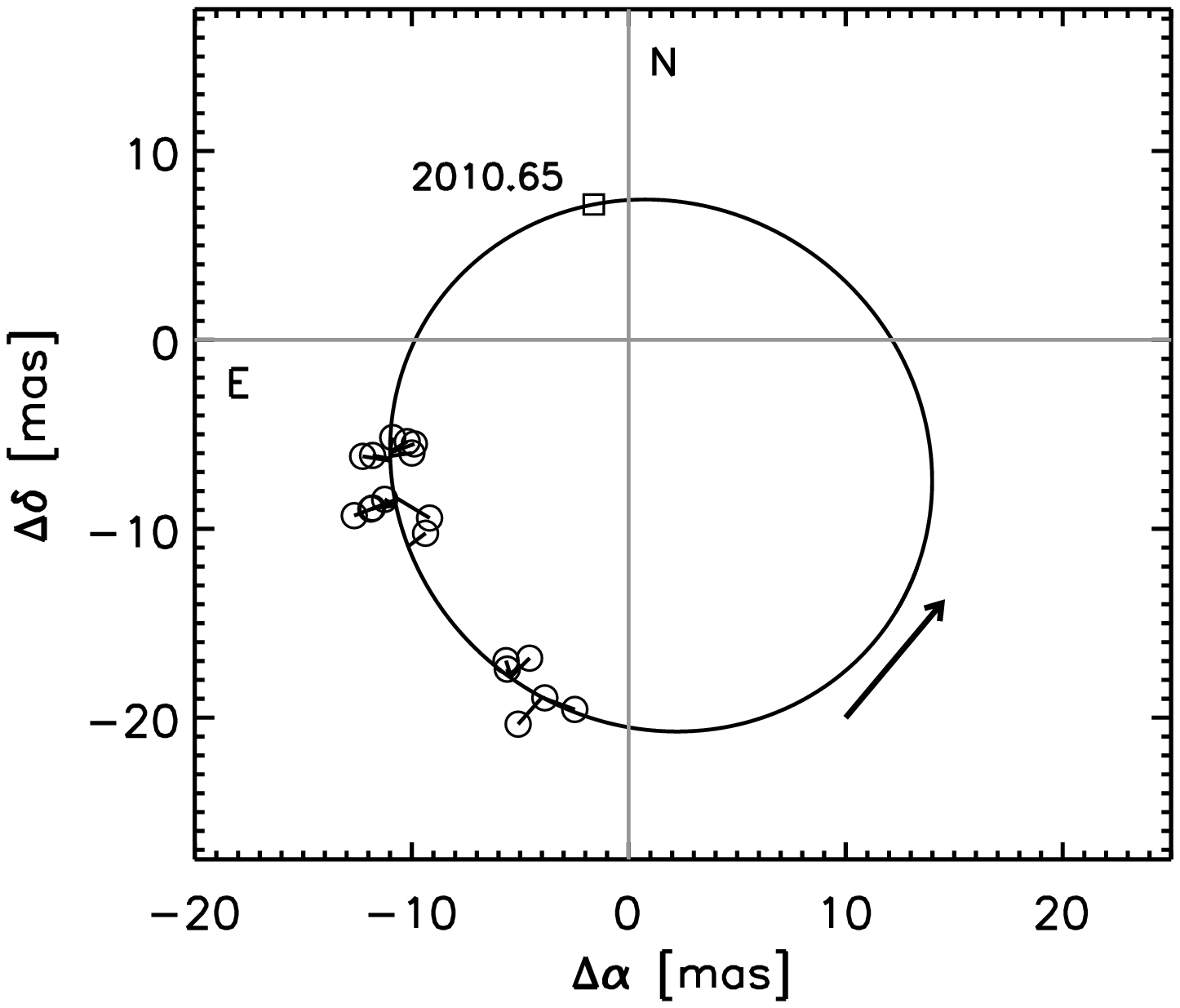}
 \end{center}
  \caption[Examples of astrometric motions of stars due to planets]
  {\textit{Left:} Astrometric motion of the Sun's center around the
  barycenter of the solar system due to all planets over 50 years,
  viewed face-on from a distance of 30 light years \citep{Jones2008}.
  The dashed circular line shows the motion of the Sun due to only Jupiter.
  The size of the disk of the sun is shown in the upper right corner
  for comparison. \textit{Right:} Astrometric orbit of HD~33636,
  a G0 V star at 28.7~pc distance \citep{Bean2007}.
 Open circles are the HST position for each epoch connected with
 a line to the positions calculated by the fit model. The open
 square shows the predicted position of periastron passage.}
\end{figure}

\noindent Measuring an astrometric signal from ground is very
difficult, as the changes of the stellar position are very small
and the atmospheric and systematic distortions, such as
plate-scale variations between different observations, may be
larger. The usage of a correction of the atmospheric distortions
with adaptive optics and determining the plate-solution with great
care, as done in the context of this work, can reduce these
difficulties and enable one to go down to the desired $\sim 1$~mas
precision needed to detect and characterize companions in wider
orbits. However, to detect Earth-like planets orbiting solar-like
stars, one needs a higher precision in the astrometric
measurements. Interferometry makes it possible to achieve
precisions about a few microarcseconds. The already installed and
soon available instrument PRIMA at the VLT will boost the number
of planets detected with astrometry. The spacebound astrometric
mission GAIA, planned launch in August 2011, will also make a huge
contribution in stellar position measurements and with this yield
a huge number of newly detected Jupiter-size planets
\citep{Sozzetti2001}. Special designed missions, such as SIM
PlanetQuest \citep[e.g.][]{Catanzarite2006, Unwin2008}, which are
specialized for astrometric measurements both for parallax
movements and exoplanet detections around stars in the solar
neighborhood, will reach accuracies about one microarcsecond and
will be sensitive down to Earth-like masses. But also the
astrometry obtained from the photometric transit mission KEPLER
may be used for astrometric planet detection, given its stable and
precise pointing.\\

\subsubsection{Combination of Radial Velocity Data and Astrometric Data}
\label{subsec:Astro_RV}
Planet detections with radial velocities only yield a minimum mass
for the companion as one cannot determine the inclination of the
orbital plane with respect to the observer. The measured
semi-amplitude of the velocity change is the projection of the
true motion onto the line of sight from the observer to the star.
Companions detected with radial velocities should therefore be
seen
as exoplanet candidates for the time being.\\
Astrometry yields the whole parameter set necessary to describe
the full orbital motion in space, but typically a large number of
high precision measurements need to be obtained. The complete
astrometric
fit for the motion of a star with a companion in space includes:\\
\begin{tabular}{ll}
$\alpha_{0}, \delta_{0}$ & two coordinate zero points \\
$\mu_{\alpha}, \mu_{\delta}$ & proper motion in right ascension and declination \\
$\pi$ & parallax \\
$a_{*}$ & semi-major axis of the orbit \\
$e$ & eccentricity \\
$P$ & orbital period \\
$\omega$ & longitude of periastron \\
$\Omega$ & longitude of the ascending node \\
$i$ & inclination \\
\end{tabular}

\noindent To solve for these 11 parameters one has to obtain at
least 11 epochs of measurement, preferentially more for reasons of
robustness of the fit and taken the present noise in the data into
account.\\

\noindent Combining radial velocity and astrometric measurements
yields the whole three dimensional orbit as one combines
measurements which are sensitive to the movement in the direction
of the line of sight and in the plane perpendicular to it. Taking
the orbital parameters inferred from radial velocities, $P, e,
\omega, a$, one is left with \textit{only} seven parameters to
solve for with the astrometric fit. If one additionally can infer
the proper motion and parallax independently, only the inclination
and longitude of the ascending node, as well as the coordinate
zero points, need to be determined. However, most of the time this
is not possible, as either the precision of these parameters is
not sufficient, or one has to determine the proper motion and
parallax relative to the local reference frame. Combination of the
two measurement techniques has advantages and disadvantages. The
advantage of this approach is that normally the radial velocity
measurements are obtained with much higher precision than what is
possible for the astrometric measurements. One can therefore hold
the parameters from the RV fit fixed. A disadvantage or rather a
constraint is the circumstance that RV detections are most
sensitive to close-in companions and astrometry to companions in
wider orbits. Therefore the astrometric signal for most of the
planets detected by radial velocities are very small and difficult
to detect. But with the upcoming specialized astrometric
instruments, these will come more and more into range.\\

\noindent The astrometric measurements of radial velocity
exoplanet candidates conducted with the HST Fine Guidance Sensor,
pushed three out of five candidates from the planetary regime into
the brown dwarf or low mass star regime:\\

\hspace{-0.4cm}
\begin{tabular}{|l|c|c|c|l|} \hline 
 Planet cand. & $M\sin{i} [M_{J}]$ & $M [M_{J}]$ & inclination & Reference \\ \hline
 HD~33636~b & 9.3 & $142 \pm 11$ & $4.1\degr \pm 0.1\degr$ & \small{\citealp{Bean2007}} \\
 & & M dwarf star & & \\
 HD~136118~b & 12 & $42^{+11}_{-18}$ & $163.1\degr \pm 3.0\degr$ & \small{\citealp{Martioli2010}} \\
 & & Brown Dwarf & & \\
 HD~38529~c & 13.1 & $17.6^{+1.5}_{-1.2}$ & $48.3\degr \pm 0.4\degr$ & \small{\citealp{Benedict2010}} \\
  & & Brown Dwarf & & \\ \hline
\end{tabular}\\

\noindent This shows how important it is to obtain complementary
measurements to solve for the whole 3D orbit and calculate the
true mass of the exoplanet candidates.\\

\noindent Combination of radial velocity data with astrometry provided by the HIPPARCOS
satellite has also been done \citep{Perryman1996, Mazeh1999,
Zucker2000}. \citet{Han2001} used the HIPPARCOS intermediate
astrometric data to fit the astrometric signal of stars known to
have a possible planetary companion found by radial velocity. Most
of the expected astrometric perturbations were close to or lower
than the precision obtained from the HIPPARCOS data. The
conclusion of this statistical study was that a significant
fraction of the exoplanetary systems are seen edge-on, which would
push their masses to also significantly higher masses.
\citet{Pourbaix2001} later showed that the high inclinations found
by \citeauthor{Han2001} are artifacts of their adopted fitting
procedure. As the reason for that he named the size of the orbit
with respect to the precision of the astrometric measurements,
meaning, the 'measured' astrometric signal was not large enough
compared the astrometric precision. Future astrometric missions,
such as GAIA will lead to much higher precisions in the
astrometric measurements, thus leading to better constraints on
the orbital inclination.


\section{Brown Dwarfs}
\label{sec:BD} In the same year, at the same conference, the first
widely accepted brown dwarf (BD), Gliese 229B
\citep{Nakajima1995}, and the first extrasolar gas giant planet,
51~Peg~b \citep{Mayor1995}, were announced to the astronomical
community. Now there are around 720 BDs known to exist as companions
to nearby stars, in young clusters and most frequently as faint
isolated systems within a few hundred parsecs in the solar
neighborhood. Brown dwarfs are star-like objects with a maximum
mass between 0.07 and 0.08 M\sun, depending on their metallicity.
This mass limit for BDs is defined by the disability to sustain
stable hydrogen fusion reactions in their cores and sets a
division between \textit{stars} and \textit{brown dwarfs}. But BDs
are massive enough to be able to burn deuterium in their cores at
the beginning of their evolution, followed by a steady decline in
their luminosity and effective temperature with time, once their
supply of deuterium is exhausted.\\
\noindent The division between \textit{brown dwarfs} and \textit{giant planets} is yet
not clear and still under debate. Two possible ways of defining
brown dwarfs and giant planets are under discussion. One widely
used definition is based on the mass limit to burn deuterium,
which would define an object with less than $13 \rm ~M_{\rm Jup}$
as a planet \citep{Saumon1996, Chabrier2000}. This is also the
IAU\footnote{www.dtm.ciw.edu/boss/definition.html} definition for
brown dwarfs, which considers objects above the deuterium burning
mass limit as brown dwarfs. The definition can be applied to both
companions and isolated objects and is the reason why very-low
mass objects in clusters are sometimes called free-floating
planets. A drawback of the definition of the border between BDs
and giant planets over the deuterium burning limit is that unlike
the hydrogen burning limit, the ability to fuse deuterium is
insignificant for the physical properties of BDs and therefore
describes no meaningful boundary for the evolution of low mass
objects \citep{Chabrier2007}. In fact there are more differences
in stellar structure and evolution between high and low mass stars
than for low mass brown dwarfs and giant planets. Also the mass
determination with evolutionary models based on the luminosity of
the objects is often uncertain, so a definitive conclusion whether
an object is above or below the deuterium burning mass limit is
difficult. For example the best mass estimate for the object GQ
Lup b is $10-40 \rm ~M_{\rm Jup}$ \citep{McElwain2007,
Seifahrt2007}, hence both definitions, planet and brown dwarf, are
possible. Determining the mass dynamically and therefore
independent of the models puts more constraints on the
evolutionary theories for these low mass objects and help to
better understand their mass distribution and formation processes,
but this is only measurable for brown dwarfs which are bound in a
system.\\
\noindent The other definition to distinguish brown dwarfs and
giant planets is based on their formation processes. Here a planet
is a substellar object formed in a circumstellar disk and a brown
dwarf formed through cloud fragmentation like a star. This
definition also has its drawbacks, as it is obviously difficult to
tell the formation process of a given substellar object. It is for
example most likely for wide massive companions to be formed by
cloud fragmentation rather than in a stellar disk, but for lower
mass companions in intermediate orbits ($8 \rm ~M_{\rm Jup}, @~20
\rm ~AU$) it will be more difficult to determine the formation
mechanism \citep{Luhman2008}. So neither of the definitions
provides a clear distinction between brown dwarfs and giant
planets. The most commonly accepted one at the moment is the one
of the IAU, that distinguishes substellar objects by their mass.
Independent of their formation process, objects below $13 \rm
~M_{\rm Jup}$ which orbit a star or stellar remnant are planets
and objects with masses above this deuterium burning mass limit
and below the hydrogen burning limit ($13 \rm ~M_{\rm Jup} - \sim
80\rm ~M_{\rm Jup}$) are defined as brown dwarfs.

\subsection{Brown Dwarf Formation Processes}
Similar to the difficulty to define a distinctive boundary between
brown dwarfs and giant planets, it is challenging to put
constraints on the formation process of brown dwarfs. Several
scenarios are under discussion, but none of them can explain all of
the available observations. A star-like formation by direct
collapse and fragmentation of a molecular cloud, a scaled version
of the Jeans model, is one possibility for BD formation \citep{Bate2005}.
Another possibility is the ejection of protostellar embryos that
form in a cluster environment, but are ejected due to dynamical
interactions before they can accrete enough material to become a
star \citep{Reipurth2001, Umbreit2005}. This scenario has
difficulties to explain young wide BD binaries, because these
should be ripped off during the ejection, likewise large disks
should not survive an ejection, but are nevertheless observed
\citep[e.g.][]{Scholz2006}. Photoevaporation of their accretion
envelope by the radiation of a nearby massive and hot star
\citep{Whitworth2004}, as well as instabilities in massive disks,
followed by a pulling off of the still substellar companions
through dynamical encounters with other stars can produce isolated
free-floating brown dwarfs \citep{Goodwin2007, Stamatellos2009}.
Turbulence in a molecular cloud can produce local over-densities
that can collapse and form a compact low mass object. This
turbulent fragmentation could explain the formed brown dwarfs as
the low-mass tail of the regular star formation process. Also the
continuity in disk fraction, the ratio of stars or BDs without a disk to those having a disk, from the stellar to the BD regime
\citep{Caballero2007} favors this formation scenario. None of these
scenarios can be seen as the dominant contributor to the brown
dwarf population, neither one can explain all observations. Most
likely the formation processes of brown dwarfs depend on the
stellar environment and its varying initial conditions from case
to case, leading to a changing relevance of the individual scenarios.

\subsection{The Brown Dwarf Desert}
\begin{figure}
 \begin{center}
  \includegraphics[width=14cm]{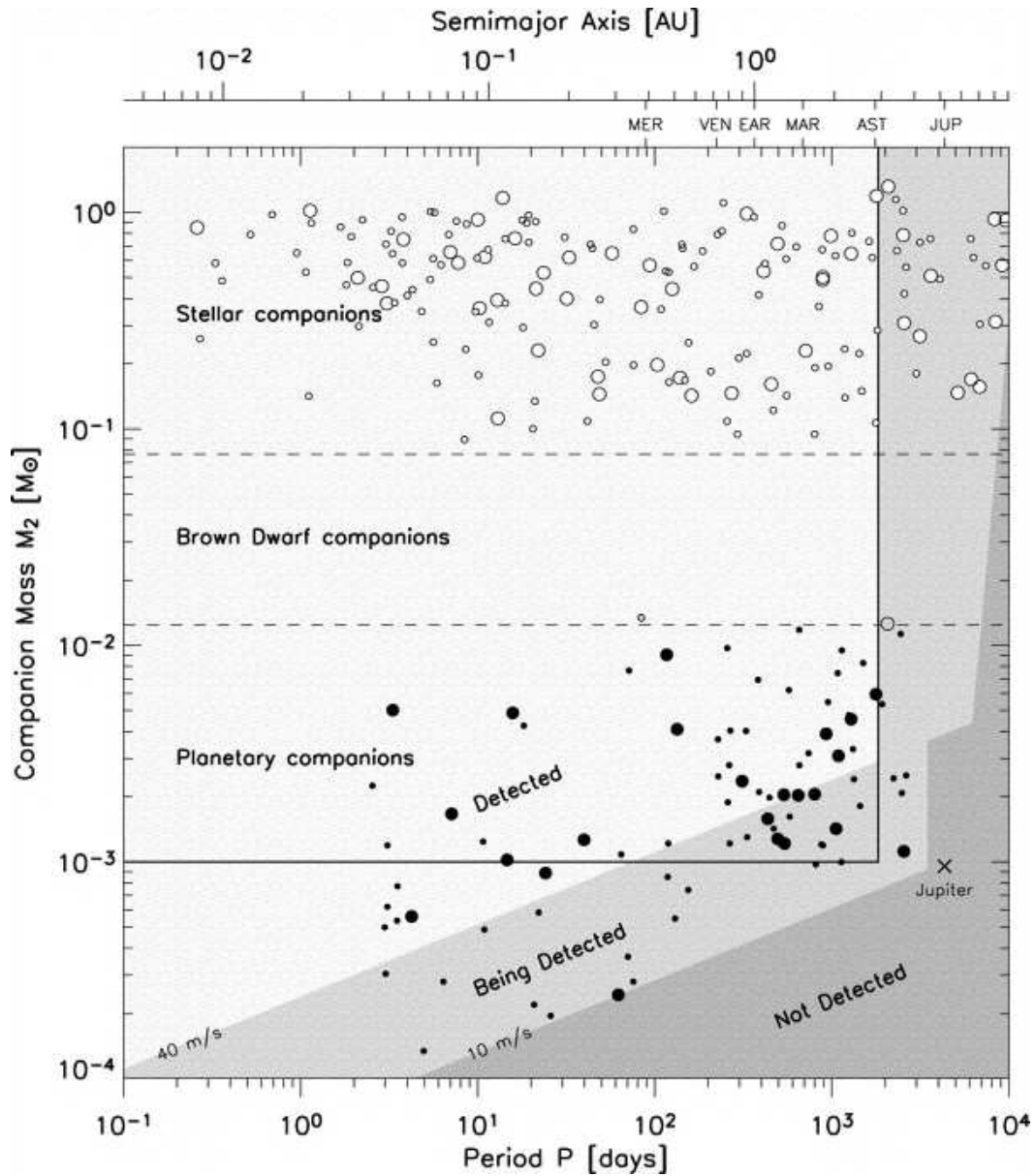}
 \end{center}
\caption[Brown dwarf desert in mass and period]{Brown dwarf desert
in mass and period. Shown are the companion masses
$M_{2}$ versus the period of the companions to sun-like stars. The
solid rectangle defines companions with periods smaller than 5
years and masses between $1 \rm ~M_{\rm Jup}$ and $\sim 1\rm ~M\sun$.
The stellar (open circles), brown dwarfs (blue circles) and
planetary (filled circles) companions are separated by the dashed
lines at the hydrogen and deuterium burning limit. The brown dwarf
desert for companions with orbital periods $P < 5$ years can
clearly be seen. The \textit{detected}, \textit{being detected}
and \textit{not detected} regions of the mass-period space mark
the limit to which high-precision radial velocity measurements
were able to detect companions at the time of the analysis leading
to this plot \citep{Lineweaver2003}. For more details about the
plot and the sample see \cite{Grether2006}.} \label{fig:BDDesert}
\end{figure}

Radial velocity surveys to find substellar companions orbiting
their host stars have led to the detection of a variety of
exoplanets and brown dwarfs. They also led to the definition of
the \textquotedblleft brown dwarf desert\textquotedblright, the
absence of BD companions relative to giant planets and stellar
companions to low-mass stars at separations less than 3~AU
\citep{Marcy2000}. The BD companion frequency at separations
\textgreater ~1000~AU is at least 10 times higher than that of
separations of only a few AU, which is $\approx 0.5~\%$
\citep{Gizis2001, NeuhaeuserGuenter2004}. The frequency of stellar
companions at small separations is $13 \pm ~3\%$
\citep{Duquennoy1991, Mazeh1992}, at least a factor 30 larger than
the BD frequency, whereas at larger separations the ratio of
frequencies of stellar and substellar companions is between $\sim 3$
and 10 \citep{McCarthy2004}. At wider separations to solar-type
stars it appears therefore that the brown dwarf desert is no
longer present \citep{Luhman2007}. Fig.~\ref{fig:BDDesert} depicts
the brown dwarf desert in mass and period. The plot is taken from
\cite{Grether2006} and shows the estimated companion masses versus
the period of the companions. The lack of brown dwarf companions
in short-period orbits can clearly be seen. For more details about the
sample and the companion mass estimates see
 \cite{Grether2006}.\\

\noindent The rareness of
 close-in BD companions is highly significant since the commonly
 employed RV searches for sub-stellar companions to stars are very
 sensitive to such objects as the RV  semi-amplitude \textit{K} is
 higher for shorter periods \textit{P}, e.g. smaller separations
 from the host star (see also Equ.~\ref{equ:RVAmplitude}). Because the RV measurements only result in
 a minimum  companion mass and the HIPPARCOS astrometric precision
 is mostly not sufficient to distinguish BDs from stellar
 companions, the masses of the few known close-in BD candidates are
 often uncertain \citep{Pourbaix2001, PourbaixArenou2001}.
 RV studies of M dwarfs indicate that the BD desert continues also
 into the early M dwarf population\footnote{M. K{\"u}rster, private communication}.

\section{Goal of this Work}

Most planets have been discovered by radial velocity measurements.
But RV measurements only yield a minimum mass for the companions and are
most sensitive to planets in close-in orbits. Transit measurements
can yield the true mass of a companion when combined with RV measurements. But they are also most
sensitive to close-in planets, as the probability of an occurring
transit is higher for those companions.\\
\noindent However, astrometry yields the full set of orbital
parameters and therefore the true mass of a companion. It also
opens a new parameter space of planets with a longer orbital
period.\\

\noindent The goal of this work is to measure the astrometric
signal of the wobble of a star due to its unseen companion. Using adaptive optics aided imaging to detect such
a signal I want to show the
feasibility of this approach for planet and low mass companion detection and characterization. Seeing
limited astrometry has mostly a larger field of view than an adaptive optics imager, but the
achieved accuracy is most times not high enough to measure the
tiny signal of a planet.\\
\noindent To test whether the astrometric signal is
measurable, we chose an object known to harbor a candidate brown
dwarf companion found by RV measurements \citep{Kuerster2008}. When astrometry can be combined with precision RV measurements the number of orbital parameters to be derived from the
astrometric data can be strongly reduced (see Sect.~\ref{subsec:Astro_RV}). Also the astrometric signal is higher for such an
object than for an exoplanet. Additionally, the companion is a very
interesting object as it is a brown dwarf desert candidate. Deriving its true mass, confirming it as a BD or pushing it into
the low mass star regime, would be a very interesting result.

    \cleardoublepage
   \chapter{Introduction to Adaptive Optics}
    \label{chap:AO} 
\begin{figure}
    \begin{center}
        \includegraphics[width=4cm]{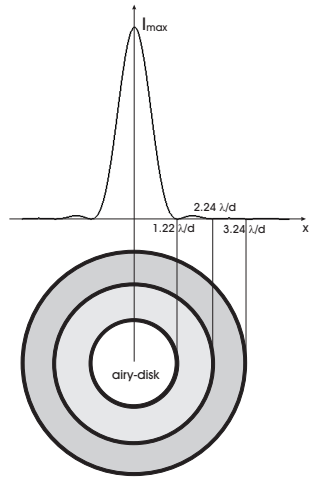}
    \end{center}
    \caption[The Point-Spread-Function (PSF) of a point source]
{The Point-Spread-Function (PSF) of a point source. The diameter
of the main maximum is called Airy-disk and has the radius $\theta
= 1.22\lambda/D$ (from \cite{Stumpf2004}).}
    \label{fig:Airy}
\end{figure}

Stellar wavefronts are assumed to be spherical
waves and are treated as plane waves when they reach Earth because
of the immense distance of the stars. But turbulences in our
atmosphere lead to random deformations of the incoming wave.
Telescope optics collect the light and transform it into an image.
In the ideal case, without any distortions, the angular resolution
of this image is determined by the bending of the light, this is
called the diffraction limit case. For circular apertures the
angular diameter of the image of a point source is given by:
\begin{equation}
 \theta = 1.22 \frac{\lambda}{D}
\end{equation}
where $\lambda$ is the observed wavelength and  $D$ is the
telescope diameter. Such an image of a point source is called
Point-Spread-Function (PSF). The intensity distribution of a PSF
has the form as shown in Fig.~\ref{fig:Airy}.
The inner part of the distribution is called Airy-disc and it has
the radius $1,22\frac{\lambda}{D}$. Two close point sources can
only be imaged separately if the maximum of the one PSF falls onto
the first minimum of the other PSF.\\

\begin{figure}[!t]
    \begin{center}
        \includegraphics[width=10cm]{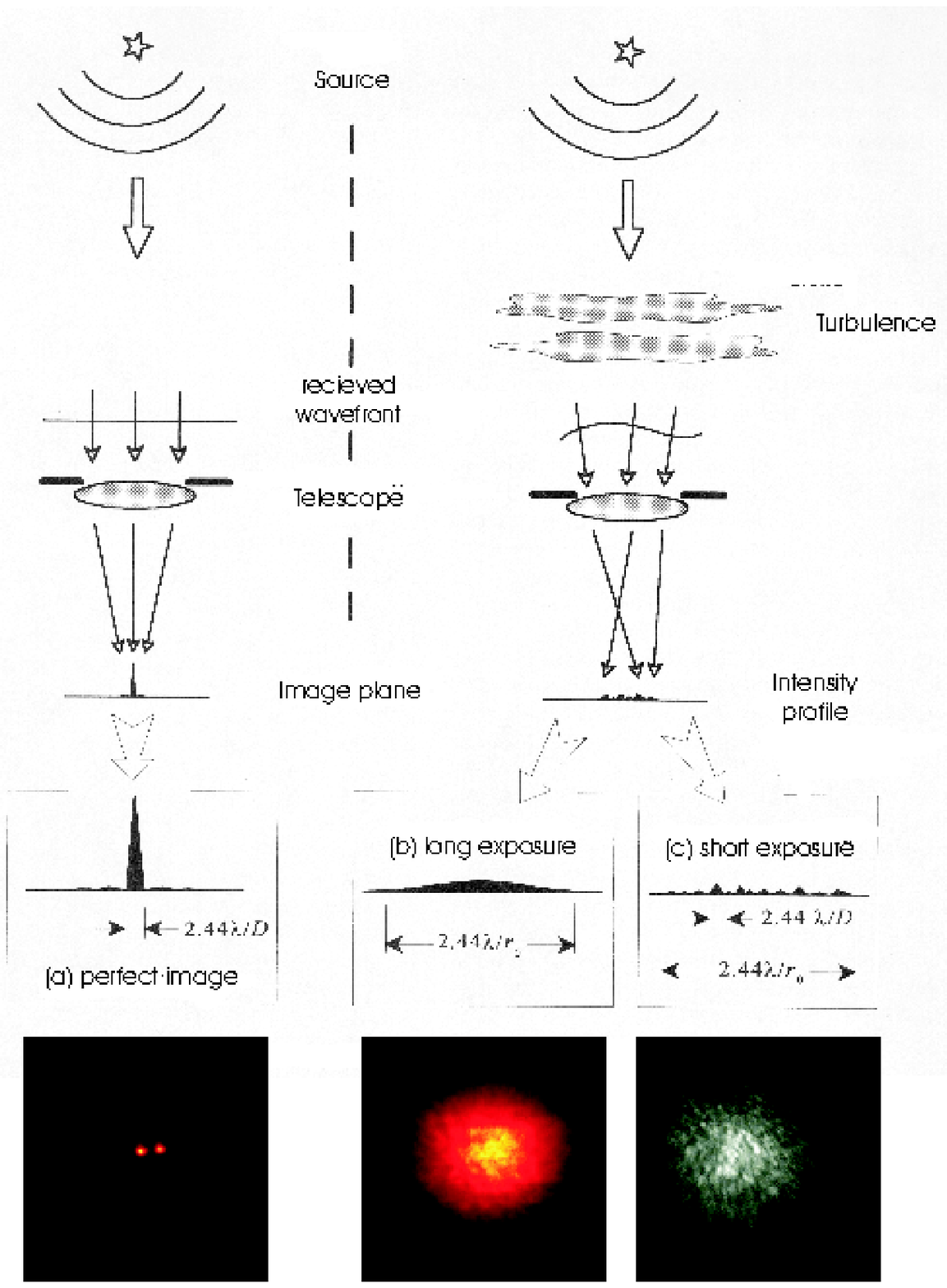}
    \end{center}
    \caption[Effect of atmospheric turbulence on the image of a star]
    {Effect of atmospheric turbulence on the image of a close double star (right).
    Perfect image with no atmosphere (a). Short exposures produce a speckle
    pattern, which shows a snapshot of the momentary distortions (c).
    In longer exposures the pattern is smeared out and results in the
    seeing disk (b). (from \cite{Hardy1998})}
    \label{fig:Exposuretimes}
\end{figure}

\noindent Due to the atmosphere, part of the intensity contained
in the PSF is moved from the maximum to the outer parts of the
light distribution. Details are smeared out and cannot be resolved
anymore. In Fig.~\ref{fig:Exposuretimes} the effects of the
atmosphere on an image of a close binary star are demonstrated. In
the ideal case without any distortions, one obtains a perfect
image for each star with a diameter of $2\theta = 2,44 \lambda/D$
(Fig.~\ref{fig:Exposuretimes}a).
If the exposure time is very short, less than 1/50 second, one
derives an image consisting of many speckles, which all have a
diameter of the diffraction-limited PSF
(Fig.~\ref{fig:Exposuretimes}c). This speckle pattern is changing
randomly in subsequent images and can be seen as a sketch of the
momentary atmospheric distortions. If the exposure time is
increased to a multiple of the timescale of the turbulence change
in the atmosphere, the speckle pattern is smeared out and one
obtains an image of the point source, the so-called seeing disk,
with a diameter which is no longer determined by the ratio
$\lambda/D$ but by $\lambda/r_{0}$
(Fig.~\ref{fig:Exposuretimes}b). The factor $r_{0}$,
called-Fried-Parameter, is an important parameter to describe the
atmospheric turbulence. It will be explained in more detail in
Chapter~\ref{subsec:Fried}. The consequence of the above
circumstance is, that even for the bigger but uncompensated
telescopes the resolution limit is between 0.5-1 arcseconds or
more.\\
\noindent Also the sensitivity of a telescope does not increase in
the same way as in the diffraction limit case, which is again due
to the atmosphere. Generally, the bigger an aperture, the more
light it can collect and the fainter objects it can detect. In the
ideal, diffraction limited case the capability of a telescope to
detect a point source is $\propto D^{4}$. But this capability is
degraded to $\propto D^{2}$ in the seeing limited case, because
the image size is no longer determined by the telescope
diameter.\\
\noindent There are two possibilities to overcome the image
degradations caused by the atmosphere:
\begin{itemize}
 \item One can go to space and leave the atmosphere behind.
 Building space based telescopes has the advantage that one can
 also observe in the wavelength range shorter than $0.3~\mu m$
 (UV) where Earth's atmosphere is opaque, and in the visual and
 infrared between 0.5 and $2.5~\mu m$ where emission lines from
 OH-molecules compromise observations. Disadvantages of satellites
 and space telescopes are their immense costs and technical as
 well as logistic requirements. The transport into orbit limits
 their size and weight. They need to be shock-resistent and
 maintenance or upgrades with new instruments are, if performed at
 all, expensive and dangerous. But there are certain wavelength
 ranges which can only be observed from space.
 \item The other possibility to obtain diffraction-limited images
 is the usage of an adaptive optics (AO) system. This system
 consists of two main components, a wavefront sensor, which
 measures the aberrated wavefront, and a deformable mirror which
 corrects the wavefronts. Such a system is cheaper to build and to
 maintain than a space telescope and can be upgraded or modified
 more easily. In the wavelength range observable from the
 ground, the ground based telescopes equipped with an AO system
 are comparable or superior to space based ones, because their
 primary mirrors can be made considerably larger.\\
 \noindent A more detailed description of an AO system is given in
 Chap.~\ref{sec:AO}.
\end{itemize}

\section{Atmospheric Turbulence}
\label{sec:Turbulence} 
\begin{figure}
    \begin{center}
        \includegraphics[width=12.5cm]{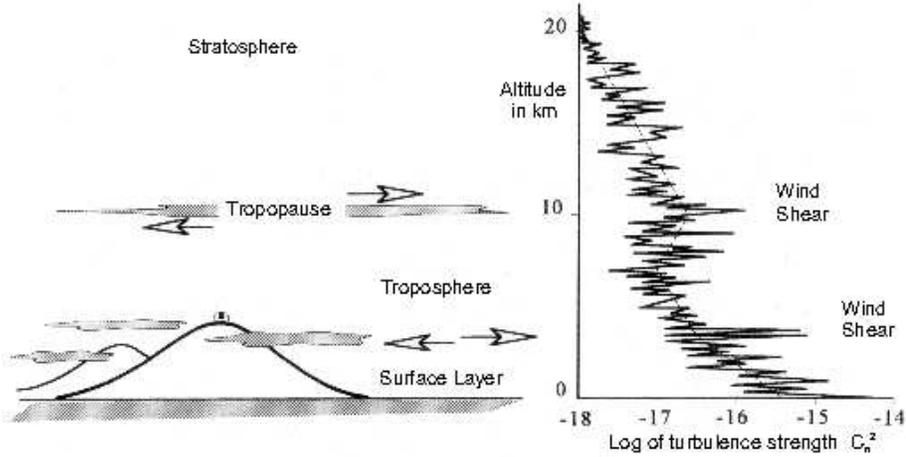}
    \end{center}
    \caption{Structure of the atmosphere with typical turbulence profile. \citep{Hardy1998}}
    \label{fig:Atmosphere}
\end{figure}

The atmosphere is neither static nor
homogeneous. Temperature, density and humidity change continuously
on different scales of time and space. With it comes a permanent
change of the refractive index $n$ of the atmosphere
\citep{Clifford1978}:
\begin{equation}
n(\lambda, P, T) = 1 + 7,76 \cdot 10^{-5} \left( 1 + 7,52 \cdot
10^{-3} \frac{1}{\lambda^{2}} \right) \frac{P}{T}
\end{equation}
where $\lambda$ is the wavelength of the light in $\mu m$, $P$ the
pressure of the air in $mbar$ and $T$ the temperature of the air
in Kelvin. The dependency on the wavelength is small for a broad
range of $\lambda$ and the fluctuations of the pressure balance
with the speed of sound. But the fluctuations of the temperature
are more inertial, and thus dominate the changes of the refractive
index. For two parallel light beams which pass through the
atmosphere, a difference of the refractive index leads to an
adjournment of their relative phases and therefore to
a distortion of the previously plane wavefront.\\
\noindent The air masses in different layers of the atmosphere
move because of different reasons. In Fig.~\ref{fig:Atmosphere}
the structure of the atmosphere and a typical turbulence profile
is shown.
The vertical distribution of the turbulence varies strongly with
height (right panel). $C_{n}^{2}$ is a measure for the vertical
turbulence strength. On average it is strongest near the ground,
the so-called ground layer (GL), because of thermal heating of the
ground from sunshine. In the change-over layer between the
Troposphere and the Tropopause, at an altitude of around 10~km,
the turbulence is dominated by strong winds. A measurement to
describe the movement of the air is the Reynolds-number $Re$. It
describes whether a flow is laminar or turbulent. $Re$ is given
by:
\begin{equation}
Re = \frac{L_{0}\upsilon_{0}}{\eta}
\end{equation}
Here $L_{0}$ is the characteristic size of a turbulence cell,
$\upsilon_{0}$ its characteristic velocity and $\eta$ the
kinematic viscosity of the medium, here the air. If the
Reynolds-number is smaller than the critical value $10^{3}$, the
flow is laminar, otherwise it is turbulent. In air the viscosity
is $\eta = 15 \cdot 10^{-6}m^{2}s^{-1}$ and a typical turbulence
cell has a size of $L_{0} = 15m$ and a velocity of $\upsilon_{0} =
1 m s^{-1}$. This yields an average Reynolds-number for the
atmosphere of $R_{e} = 10^{6}$, which is clearly above the
critical value. So the air flow in the atmosphere is
mostly turbulent.\\
\noindent To describe and analyze a system as complex as the
atmosphere, elaborate models are needed. The most commonly used
one in astronomy is the Kolmogorov model \citep{Kolmogorov1961}.
This model describes the turbulence as originating in energy input
in large air structures, eddies, with a typical size, the
so-called \emph{outer scale}. This large structures transport the
energy loss-free to smaller and smaller structures, the
\emph{inner scale}, till the Reynolds-number is getting smaller
than the critical value. The air flow is then laminar and the
energy dissipates into thermal heating. Typical sizes of the outer
scale are between a few meters at ground level and up to 100~m in
the free atmosphere. The inner scale lies between a few
millimeters at ground and up to a few centimeters close to the
Tropopause.

\subsection{Fried-Parameter}
\label{subsec:Fried} The Fried-Parameter, also called
correlation-length, defines the length over which the mean
divergency of the phase-difference to a plane wavefront does not
exceed the standard deviation of one radiant (1~rad). It was first
introduced and calculated by David L. Fried \citep{Fried1965}:
\begin{equation}
r_{0} = \left[ 0.423 \left( \frac{2 \pi}{\lambda} \right)^{2}\,
(\cos\gamma)^{-1}\, \int\limits_{0}^{\infty} C_{n}^{2}(h)dh
\right]^{-3/5}
\end{equation}
$C_{n}^{2}(h)$ is a measure for the vertical strength of the
turbulence profile and depends on the height $h$. $\gamma$ is the
angle between the line of sight and the zenith. But one can also
interpret $r_{0}$ as the size of an aperture that has the same
resolution as a diffraction-limited aperture without any
turbulence. That means that the resolution of a telescope with a
diameter $D$ larger than $r_{0}$, the Full Width at Half Maximum
(FWHM), is limited to $FWHM \propto \lambda/r_{0}$ and not anymore
$\propto \lambda/D$ as is the case for $D < r_{0}$. The VLT (Very
Large Telescope) has a diffraction-limited resolution of
$\lambda/D = 0.057\arcsec$ at $\lambda = 2.2\mu m$. But the
resolution is lowered by the atmosphere to $\lambda/r_{0} \approx
0.7\arcsec$. One can also see from the equation, that $r_{0}
\propto \lambda^{6/5}$, which means the area over which the
wavefront error is negligible is growing with wavelength. An
$r_{0}$ of typically 10~cm at $0.5~\mu m$ in the visual
corresponds to an $r_{0}$ of 360~cm at $10~\mu m$ in the
infrared and at $2.2\mu m$ $r_{0}$ is typically 60~cm.\\

\subsection{Time Dependent Effects}
In the same way as for the spatial case, a time can be defined
over which the variance of the wavefront changes account to 1~rad.
This time scale is called coherence time $\tau_{0}$. The relation
to $r_{0}$ is given over the wind speed $\mathbf{v}$:
\begin{equation}
\tau_{0} = \frac{r_{0}}{|\mathbf{v}|}
\label{equ:tau}
\end{equation}
With an average $r_{0}$ of 60~cm and $\mathbf{v} = 10$m/s this
yields $\tau_{0}\approx 60$~ms in the infrared at $2.2\mu m$ . To
derive a diffraction-limited image, one needs to correct the
wavefront
errors in this frequency range.\\
\noindent Both $\tau_{0}$ and $r_{0}$ are critical parameters and
the larger they are, the more stable the atmosphere is.

\section{Principles of Adaptive Optics}
\label{sec:AO} A powerful technique in overcoming the degrading
effects of the atmospheric turbulence is real-time compensation of
the deformation of the wavefront by adaptive optics.

\subsection{General Setup of an AO System}
\begin{figure}
    \begin{center}
        \includegraphics[width=14cm]{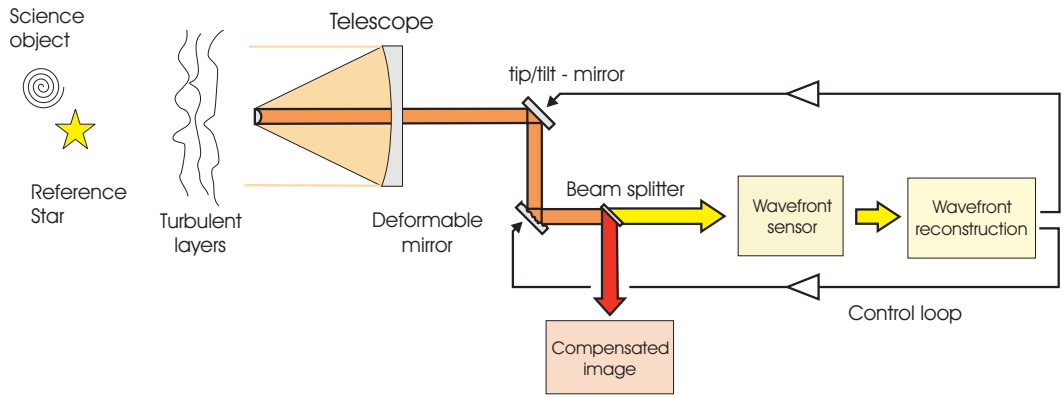}
    \end{center}
    \caption{Principle setup of an AO system.}
    \label{fig:AOSystem}
\end{figure}

The task of an adaptive optics system is the measurement of the
wavefront distortions caused by the atmosphere, the calculation of
a correction and its execution. In Fig.~\ref{fig:AOSystem} the
principle setup of an AO system is depicted.
The light coming from a star is a plane wavefront (WF) until it is
distorted by the Earth's atmosphere. It passes trough the
telescope optics and then reaches the AO system, where it first
passes through the correction unit, so that after the first
correction only the differential wavefront error between the
single measurement cycles has to be corrected. In the ideal case,
the wavefront is plane again after the correction. The correction
unit often consists of two elements, a tip-tilt mirror that is
tiltable in two directions and acts for image stabilization, which
is the largest disturbance generated by the turbulence, and a
deformable mirror (DM) which is deformed inversely to the incoming
WF and compensates for the higher order aberrations.\\
\noindent After the correction unit the light is divided by a beam
splitter into two parts. One part, mostly the infrared part of the
light, is directly led to the science camera. The other part,
mostly the visual light, is led to the wavefront sensor (WFS),
which measures the distortions of the WF. If it is necessary to
observe and sense in the infrared, for very red objects for
example, a intensity-filter is used instead of a dichroic for the
beam splitting. With the WF reconstruction unit, wavefront errors
are translated into control signals for steering the deformable
mirror and the tip/tilt
mirror. With sending these steering signals the loop is closed.\\
\noindent Typical frequencies for one cycle of the loop are 1-2
kHz. The frequency depends on the wavelength $\lambda$ in which
the observation is conducted, because the time dependent change of
the atmospheric disturbance, $\tau_{0}$, is depending on $r_{0}$
and therefore on $\lambda$, see Sec.~\ref{sec:Turbulence}. The
longer the wavelength, the easier it is to correct the wavefront
distortions. Therefore most current AO systems correct for
instruments which work in the infrared or even longer wavelengths.

\subsection{Strehl Ratio}
\label{subsec:Strehl} A good way to describe the quality of a
partially corrected PSF is the Strehl ratio, which basically
corresponds to the amount of light contained in the
diffraction-limited core relative to the total flux,
\citep{Strehl1902}. Due to the wavefront distortions, part of the
intensity of a PSF (Fig.~\ref{fig:Airy}) is moved from the peak to
the halo. Therefore, the maximum intensity of the PSF is
decreased. The Strehl ratio is defined as the ratio between the
actual maximum intensity of a point source, $I_{p}$, and the
intensity which would be reached by a perfect, diffraction-limited
telescope of the same aperture, $I^{*}$. The Strehl ratio can be
calculated analytically, if the aberrated wavefront $\Phi(\rho,
\theta)$ is known, following \citep{Hardy1998}:
\begin{equation}
S = \frac{I_{p}}{I^{*}} = \frac{1}{\pi^{2}} \left|
\int\limits_{0}^{1} \int\limits_{0}^{2\pi} \exp(ik \Phi(\rho,
\theta))\, \rho d\rho\, d\theta \right|^{2} \label{equ:Strehl}
\end{equation}
where, $k$ is the wave number and $\rho, \theta$ are polar
coordinates. But if the mean quadratic error of the wavefront only
amounts up to 2~rad, the Mar\'{e}chal-Approximation
\citep{Marechal1947} is used commonly in astronomy:
\begin{equation}
S \approx \exp^{-\sigma^{2}}
\end{equation}
Here $\sigma$ stands for the mean-square wavefront error.
Typically the Strehl-ratio is less than one percent for a
uncorrected system. In case of good conditions and a bright,
nearby reference source, the correction is good and the resulting
PSF is close to the diffraction limit. A good correction in the
K-Band typically corresponds to a Strehl ratio larger than $30\%$.

\subsection{Anisoplanatism}
\label{sec:Isoplanatism} 
\begin{figure}
 \begin{center}
   \includegraphics{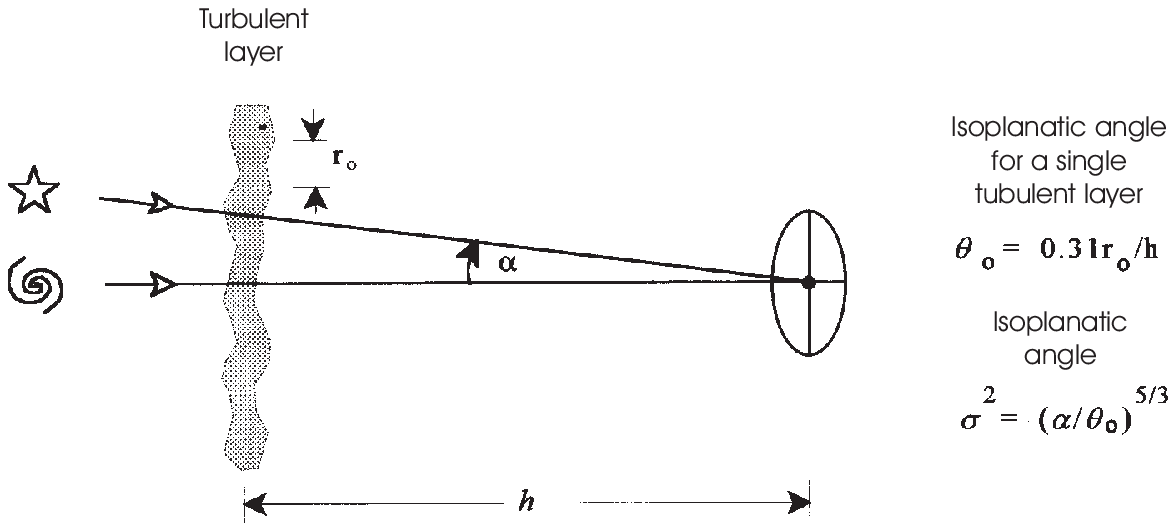}
   \end{center}
   \caption[Angular Anisoplanatism]{Angular anisoplanatism \citep{Hardy1998}.}
    \label{fig:Anisoplanatism}
\end{figure}

All adaptive optics systems need a point
source to measure the wavefront distortions caused by the
atmosphere. Generally this is a suitable reference star, whose
wavefront 'records' on its way through the atmosphere an image of
the actual wavefront distortions. Suitable means that the source
needs to have a certain magnitude to serve as a guide star. This
magnitude is dependent on the AO system and the wavelength in
which one observes. Additionally the reference source should be
close to the target, so the wavefronts of both objects pass
through the same turbulences. If possible the distance of the
target and the reference star should not exceed the isoplanatic
angle $\theta_{0}$, which is defined as the angle over which the
mean quadratic wavefront error $\sigma^{2}$ amounts $1~rad^{2}$
\citep{Fried1982}. The resulting error varies as a function of the
angle $\alpha$ between two light rays
(Fig.~\ref{fig:Anisoplanatism}).
Assuming a single turbulence layer at height $h$, $\theta_{0} =
0.31 r_{0}/h$, with $r_{0}$ being the correlation-length. The
isoplanatic angle is very small, on average $\sim 2\arcsec$ in the
visual and up to $20\arcsec$ in the near infrared.\\
\noindent The correction of the target degrades with the distance
to the guide star and the PSF of the target star is elongated in
the direction to the guide star.\\
\noindent In some cases the observed target can be used as a guide
star itself. In astronomy though, most of the scientific
interesting objects are faint or extended objects, like
protoplanetary disks, star clusters or galaxies. In the case of
protoplanetary disks as well as with observing exoplanets, the
host star can mostly be used as guide star. Also in clusters one
normally finds a suitable bright star. In the cases where no
suitable guide star is close enough, artificial laser guide stars
can possibly be used if implemented. This artificial sources are
replacing the natural guide star as reference objects for the AO.
So the sky-coverage can be enhanced by a huge factor. But laser
guide stars are also introducing new problems, for example the
cone-effect, which arises because of the finite distance of the
laser focus point, which is produced in a height of 90 - 100 km by
resonant fluorescence of a layer which is enriched with natural
sodium atoms, and the therefore only cone-like sampling of the
atmosphere \citep{Bouchez2004}. This effect is even more prominent
in the case of the so-called Rayleigh laser beacons. These are
produced in a height of 5 - 10 km by Rayleigh scattering of the
laser radiation on atmospheric molecules \citep{Fugate1991}.

\section{NACO}
\label{sec:NACO} The instrument NACO is the adaptive optics system
at the VLT and it is mounted at the UT4, Yepun, at the Nasmyth B
focus \citep{NACO1, NACO2}. It consists of two parts, the Nasmyth Adaptive Optics
System NAOS and the High-Resolution Near IR Camera CONICA.\\

\noindent The overall tip and tilt of the wavefront is compensated
by the tip-tilt plane mirror of NAOS. The higher order
aberrations, including static aberrations of NAOS and CONICA, are
compensated by the 185 actuators deformable mirror. A dichroic
acts as the beam splitter to separate the light between CONICA and
the WFS. There are several different dichroics, depending on
whether the WF correction is done in the visible or the IR, how
bright the guide
star is and so on.\\
\noindent NAOS has two WF sensors, one operating in the visible
and one in the near-IR. This kind of sensor consists of a lenslet
array that samples the incoming WF. Each lens forms an image of
the object and the displacement of this image from a reference
position gives an estimate of the slope of the local wavefront at
that lenslet. This WFS works with white light and also with
extended sources and faint stars, although with a lower
performance. For the visible sensor, two Shack-Hartmann sensors,
one with a 14x14 lenslet array, with 144 valid sub-apertures and
one with a 7x7 lenslet array and 36 valid sub-apertures are
available and for the infrared sensor three Shack-Hartman sensors
are available.\\
\noindent Anywhere within a $110\arcsec$ diameter field of view an
off-axis natural guide star can be selected, leading to a maximum
distance between guide star and target of $55\arcsec$. But due to
the anisoplanatism the Strehl ratio is going down to $\sim 9~\%$
for a $V = 10$~mag target with a separation to the guide star of
$30\arcsec$, observed in $\lambda = 2.2~\mu m$ and a seeing of
$0.8\arcsec$, compared to a Strehl ratio of $\sim 47~\%$ for an
on-axis target observed in the same wavelength and under the same
seeing
conditions.\\
\noindent If no suitable guide star is close enough
to the target, a laser guide star (LGS) can be used. The PARSEC
instrument is based on a 4W sodium laser guide star. The laser
which is tuned on the $D_{2}$-line ($\lambda = 589$~nm) of sodium,
is focused at 90~km altitude where a thin, natural layer of atomic
sodium exists. The backscattered light produces a $V\approx
11$~mag
artificial guide star which is used for the AO correction.\\

\noindent The CONICA instrument is an IR imager and spectrograph
in the wavelength range $1-5~\mu m$ and it is fed by NAOS. It is
capable of imaging, long slit spectroscopy and coronographic and
polarimetric observations. Several different plate scales are
available, too.\\
\noindent For imaging, a variety of filters and
pixel scales are offered. In principle three different plate
scales and field of views (FoV) are available, plus one small FoV
for spectral differential imaging (SDI). Table~\ref{tab:cameras} lists
the available cameras with their FoV, pixel scale and spectral
range.
\begin{table}
\begin{center}
\begin{tabular}{|c|c|c|c|} \hline
 Camera & px-scale & FoV & Spectral \\
    & [mas/px] & [arcsec] & range \\ \hline \hline
 S13 & 13.27 & 14 x 14 & $1.0 - 2.5~\mu m$ \\
 S27 & 27.15 & 28 x 28 & $1.0 - 2.5~\mu m$ \\
 S54 & 54.6 & 56 x 56 & $1.0 - 2.5~\mu m$ \\
 SDI & 17.32 & 5 x 5 & $1.6~\mu m$ \\ \hline
 L27 & 27.19 & 28 x 28 & $2.5 - 5.0~\mu m$ \\
 L54 & 54.6 & 56 x 56 & $2.5 - 5.0~\mu m$ \\ \hline
\end{tabular}
\label{tab:cameras}
\caption{List of available cameras for CONICA
with plate scales, field of view and spectral range}
\end{center}
\end{table}
Five broad band filters ($J, H, K_{s}, L', M'$) and 29 narrow and
intermediate band filters are available, plus two neutral density
filters which can be combined with the other filters to reduce the
flux of very bright sources.\\
\noindent The detector of CONICA is a Santa Barbara Research
Center InSb Aladdin 3 array, with a net 1024 x 1024 pixels and a
$27~\mu m$ pixel size. The wavelength range is from $0.8 - 5.5~\mu
m$ and it has a quantum efficiency of 0.8 - 0.9. For bright
objects, a number of ghosts become apparent (see
Chap.~\ref{sec:Obs}).\\
\noindent A single integration with the detector corresponds to
the Detector Integration Time (DIT) and the pre-processor averages
$N$ of these exposures, NDIT, before the result is transferred to
the disk. The number of counts in the final images always
corresponds to DIT and not to the total integration time NDIT x
DIT.\\
\noindent Three readout modes are offered. In case of the
\textit{Uncorr} mode the detector is reset and then read once.
This mode is used when the background is high. The
\textit{Double\_RdRstRd} mode is used when the background is
intermediate between high and low and the detector array is read,
reset and read again. If the background is low, the
\textit{FowlerNsamp} mode is used. Here the array is reset, read
four times at the beginning of the integration ramp and four times
at the end of the integration ramp.\\

\noindent A detailed description of the above mentioned characteristics of
NACO and more information can be found in the NACO User's
Manual:\\
\noindent
http://www.eso.org/sci/facilities/paranal/instruments/naco/doc/

\subsection{Our Observation Configuration}
\label{subsec:OurObs} In our observations, I used the visible
dichroic with the 7x7 optical wavefront sensor and near-IR imaging
with the narrow-band filter $NB\_2.12$ and the S27 camera. Due to
the special arrangement of the stars in our FoV I calculated an
own jitter pattern. For readout I chose the Double\_RdRstRd mode
with an integration time of 0.9 seconds for the target field and
0.4 seconds for the reference field. For more details see
Chap.~\ref{sec:Obs}.

    \cleardoublepage
    \chapter{Observations and Data Reduction}
    \label{chap:observations}
\section{The Target Field}
\begin{figure}
\begin{center}
  \includegraphics[width=12cm]{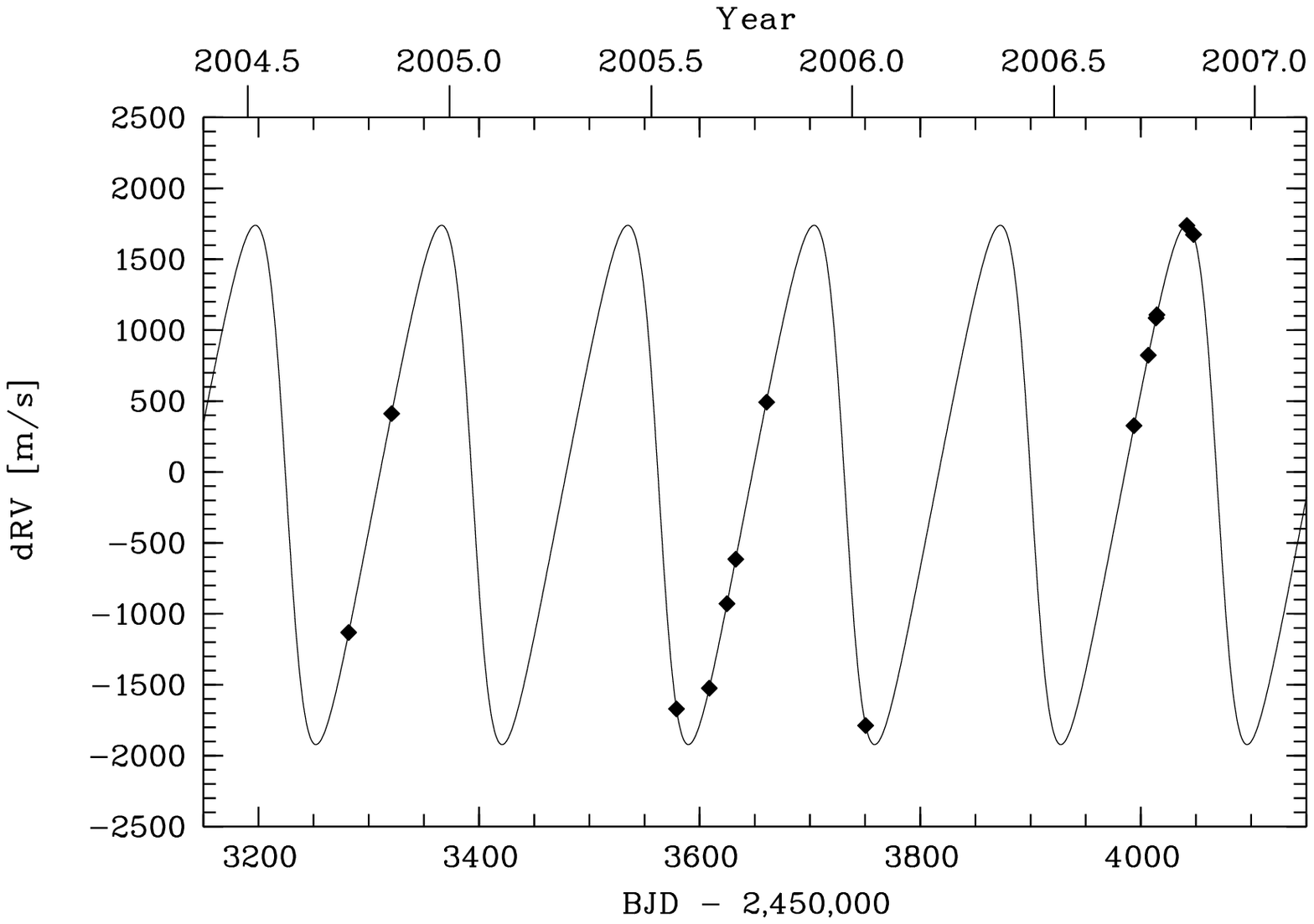}
  \caption[RV time series of GJ~1046]{RV time series of GJ~1046 from \citet{Kuerster2008}.
The solid line corresponds to the Keplerian solution with a period
of 169~d. Note that the average measurement
error is $3.63~{\rm ms}^{-1}$, much smaller than the plot symbols. Additional RV measurements with ESO/FEROS from the last years\footnote{Work done by M. Zechmeister, private communication} fit very well with this orbit, showing that the signal is indeeed due to a companion orbiting the star.}
\label{fig:rv}
\end{center}
\end{figure}

The target star I observed is an M2.5 V dwarf star in the solar
neighborhood. GJ~1046 has an apparent magnitude of V~=~11.61~mag
and K~=~7.03~mag and a stellar mass of $0.398\pm 0.007~\rm M\sun$.
It is a high proper motion star with $\mu_{\alpha} =  1394.10~\rm
mas$ and $\mu_{\delta} = 550.05~\rm mas$ and a parallax of
71.56~mas (i.e. a distance of 13.97~pc) \citep{Hipparcos1997}.\\
\noindent The companion orbiting GJ~1046 was found by radial
velocity measurements with the UVES/VLT spectrograph within a
search for planets around M dwarfs \citep{Kuerster2008,
Kuerster2003, Zechmeister2009}. In Figure~\ref{fig:rv} \citep[from][]{Kuerster2008} the RV time series and best-fit Keplerian
orbit of GJ~1046 is plotted. Assuming a stellar mass, from K-band
mass-luminosity relationship \citep{Delfosse2000}, of $M =
0.398~{\rm M\sun}$, a minimum companion mass $m_{\rm min} =
26.9~{\rm M}_{\rm Jup}$ can be calculated for an inclination $i =
90\degr$, corresponding to an edge-on view of the orbit and
therefore measuring the maximum $RV$ amplitude, from the mass
function $f(m) = (m\sin{i})^3/(M + m)^2 = 9.5\cdot 10^{-5}~M\sun$.
Together with Equ.~\ref{equ:RVAmplitude} for the $RV$
semi-amplitude $K$ an minimum orbital semi-major axis $a=0.42~{\rm
AU}$ was inferred. Its mass and distance to the host star makes
this companion a promising candidate for the brown dwarf desert.
In Tab.~\ref{tab:rvparameters} the stellar parameters and orbital
characteristics inferred from the $RV$ measurements are listed.
These parameters are kept fixed later in the astrometric orbit
fit.
\begin{table}[t]
\begin{center}
\renewcommand{\thefootnote}{\alph{footnote}}
\caption{\large{Stellar and orbital parameters of GJ~1046}}
\begin{tabular}{|l|c|c|c|} \hline
$RV$-derived parameters & value & uncertainty & units \\ \hline
$RV$ semi-amplitude \it K & 1830.7 & $\pm 2.2$ & $ms^{-1}$ \\
Period \it P & 168.848 & $\pm 0.030$ & days \\
Eccentricity \it e & 0.2792 & $\pm 0.0015$ & \\
Longitude of periastron $\omega$ & 92.70 & $\pm 0.50$ & degree \\
Time of periastron $T_{p}$ & 3225.78 & $\pm 0.32$ & BJD-2~450~000 \\
Mass function \it f(m) &  9.504 & $\pm 0.024$ & $10^{-5}~M\sun$ \\ \hline\hline
Inferred parameters & & &\\ \hline
Stellar Mass \it M & 0.398 & $\pm 0.007$ & $M\sun$ \\
Minimum companion mass $m_{min}$ & 26.85 & $\pm 0.30$ & $M_{Jup}$ \\
Min. semi-major axis of companion orbit \it a & 0.421 & $\pm 0.010$ & AU \\
Critical inclination\footnotemark~ $i_{crit}$ & 20.4 & & degree \\
Probability for $i < i_{crit}$ & 6.3\% & & \\ \hline
\multicolumn{4}{l}{\small{$^{\rm a}$ for $m = 0.08~M\sun$}}\\
\end{tabular}
\label{tab:rvparameters}
\end{center}
\end{table}

\noindent To exceed the upper brown dwarf mass limit of $0.08~\rm
M\sun$  the orbital inclination would have to be smaller than
$20.4.\degr$ (or larger than $159.6\degr$). But the probability that the inclination angle $i$ is
by chance smaller than this value, $p(90\degr \geq i \geq \theta)
= \cos{\theta}$ is only 6.3\%, assuming random orientation of the
orbit in space. Combining the RV data with the HIPPARCOS
astrometry of GJ~1046 \citep{Kuerster2008}; (see also
\citep{ReffertQuirrenbach2006}) a $3~\sigma$ upper limit to the
companion mass of $112~ M_{Jup}$ was determined. The
probability for the companion to exceed the star/BD mass threshold
($0.8~M\sun)$ is just 2.9~\%. These two constraints
make it very unlikely that the companion is stellar,
but it rather is a true brown dwarf desert companion.\\
\noindent The expected minimum astrometric signal (see
Equ.~\ref{equ:astrometry}) due to the companion is 3.7~mas
peak-to-peak, which corresponds to 0.136 pixel on the NACO S27
detector. This value is calculated using the HIPPARCOS parallax of
71.56~mas and holds for the case that the orientation of the
system is such that one only sees and measures the minor axis,
$2*b_{1}$ of the orbit. But the true effect is possibly much
larger. For an object at the brown dwarf/star border the full
minor axis would extend 11.5 mas or 0.42 px and the full major
axis 12.1 mas, as the orbit is not very eccentric. At the
HIPPARCOS derived upper limit for the mass of $112~M_{Jup}$, it
would be $2*b_{1} = 15.4\rm ~mas$ or 0.57 px. This is one of the
rare cases where a spectroscopic star-substellar compaion system
comes into reach for astrometric observations.

\begin{figure}
 \begin{center}
  \includegraphics[width=14.5cm]{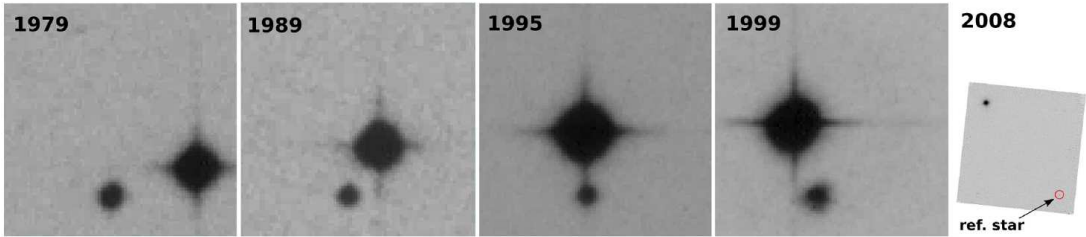}
 \end{center}
\caption[Time series of the movement of GJ~1046 on the sky]{Time series of the movement of GJ~1046 on the sky over nearly 30 years. Due to its high proper motion it passes the star which now is used as the reference star in this work. The first, third and fourth image is taken from the SuperCosmos Sky Survey (SSS), the second one from the ESO archive and the last image is one of my own observations with NACO.}
\label{fig:GJ1046_time}
\end{figure}

\noindent For astrometric measurements a reference star,
preferably close to the observed star is needed. By chance GJ~1046
is located at $\sim 30\arcsec$ separation from a suitable
reference star, see Fig.~\ref{fig:GJ1046} and \ref{fig:GJ1046_time}. This reference star, 2MASS 02190953 -3646596, has
V~=~14.33 and K~=~13.52 (colors taken from the 2MASS/SIMBAD\footnote{2MASS: \citet{Skrutskie2006}},\footnote{SIMBAD: http://simbad.u-strasbg.fr/simbad/} catalogues) which makes it from its color V-K~=~0.81
an F2 star with an effective temperature of \~ 6750 Kelvin \citep{Tokunaga2000}. Assuming the star to be a main sequence star, one can infer an absolute magnitude in the visual of $M_{V}
= 3.7$~mag from theoretical isochrones \citep{Marigo2008} and use the distance modulus $m - M = -5 +
5\log{\big(r[pc]\big)}$ to estimate a distance for the reference
star of $\sim 1337$ pc. At this distance the star would have a
parallax movement of only $\sim 0.75$ mas. Also interstellar
reddening occurs at such distances, which makes the color of the
star redder, so in reality it is even bluer and therefore further
away. So I do not expect a strong influence on the relative
parallax between the two stars compared with the parallax of
GJ~1046 alone in my measurements within the aimed precision. Also
the proper motion of the reference star is not expected to be very high and I therefore
use the HIPPARCOS values for the proper motion and parallax of GJ~1046 as a good first
estimate for the results in the fit later.\\

\section{The Reference Field}
\begin{figure}
 \begin{center}
 \includegraphics[width=6.5cm]{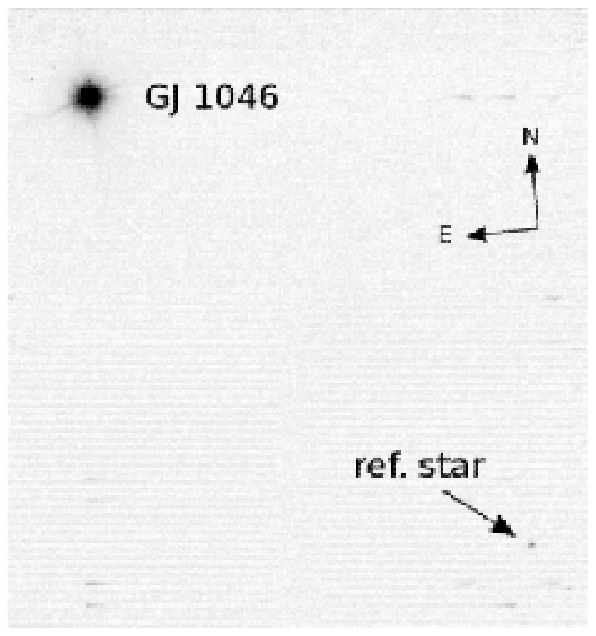} \hspace{0.5cm}
 \includegraphics[width=6.5cm]{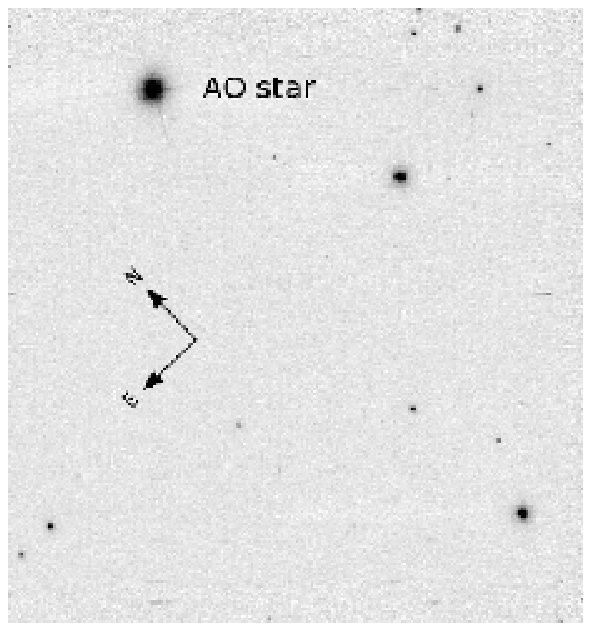}
 \caption[NACO image of GJ~1046 and the reference
  field in the globular cluster 47~Tucanae from July 2008.]
  {NACO image of GJ~1046 (left) and the reference
  field in the globular cluster 47~Tucanae (right) from July 2008.
   GJ~1046 is the bright star in the upper left corner, the reference
    star is located in the lower right corner. Also visible in
    the frames are ghosts produced by the bright star (elongated features, see Chap.~\ref{sec:Obs}).}
 \label{fig:GJ1046}
\end{center}
\end{figure}

Adaptive optics corrections during an observation are not
constant. It is a dynamical process, whose performance depends on
the atmospheric conditions during the observation and changes of
these conditions. Also the telescope focus may have changed
between two observing epochs, inducing a slightly changed
platescale. I rotated the FoV with the derotator, to fit the star
asterism onto the detector, but this rotation only has a finite
accuracy. To check and calibrate for such effects, I observed a
reference field in the rim of the globular cluster 47~Tucanae.
This field contains more stars than our target field and is
observed very close in time to the target field. It is used to
measure the change in platescale between the different epochs. I
do not need to know the absolute platescale, but its change
between the single epochs must be determined to attain subpixel
accuracy. Also the accuracy of the rotation of the detector was monitored with the
reference field, to adopt a reasonable error for the rotation
angle to our data in the target field. To correct for the
uncertainty in the rotation is not possible, because the de-rotator
is turned back to the zero position when moving the telescope to
the reference field and a fixation of the rotated instrument to the angle of the target field was not
possible in service mode, either.\\
\noindent I chose the reference field in the old globular cluster
47~Tucanae, because of its large distance of $4.0 \pm 0.35$~kpc \citep{McLaughlin47Tuc} and accordingly with it
the small intrinsic movement of the single stars in the field. The velocity dispersion of the inner parts of the cluster is 0.609~mas in the plane of the sky and the dispersion in the outer parts being slightly smaller \citep{McLaughlin47Tuc}. The
reference field contains three bright stars and several fainter ones
suitable to check the image scale and the field
rotation, see Fig.~\ref{fig:GJ1046}, right.\\
\noindent As far as possible I checked the cluster membership of the stars
in the field. But for some, especially the faint ones, it was
impossible as no 2MASS magnitudes exist, so that I could not
confirm their membership via the color-magnitude diagram.
McLaughlin et al. observed 47~Tuc with the Hubble Space Telescope
(HST) and calculated proper motions and stellar dynamics for the
stars in the core of the cluster \citep{McLaughlin47Tuc}.
Unfortunately our field lies just outside their radius around the
core, where they obtained their high precision measurements.
However, they also observed the outer part of the cluster where
our field lies and did not mark any of the stars in their tables, meaning they
have not measured an uncommon hich velocity compared to the mean cluster motion for these stars.
Also, comparing the positions of the single stars in the different
epochs in our NACO images did not show any unusual or large motion of one of the stars in one direction. I
therefore assume all the
stars in the observed reference field to be cluster members with common proper motion.\\
The asterism in the reference field was chosen to be similar to
the configuration of the AO and reference star in our target field
plus additional stars for computing the necessary field
distortions between the single epochs. For that I had to rotate
the FoV by $42\degr$ anticlockwise. Altogether I used 11 stars
for the final fit of $x$ and $y$-shift, -scale and rotation between the
epochs.

\section{Adaptive Optics Observations of GJ~1046}
\label{sec:Obs}
\begin{figure}
 \begin{center}
  \includegraphics[width=4.9cm, angle=-90]{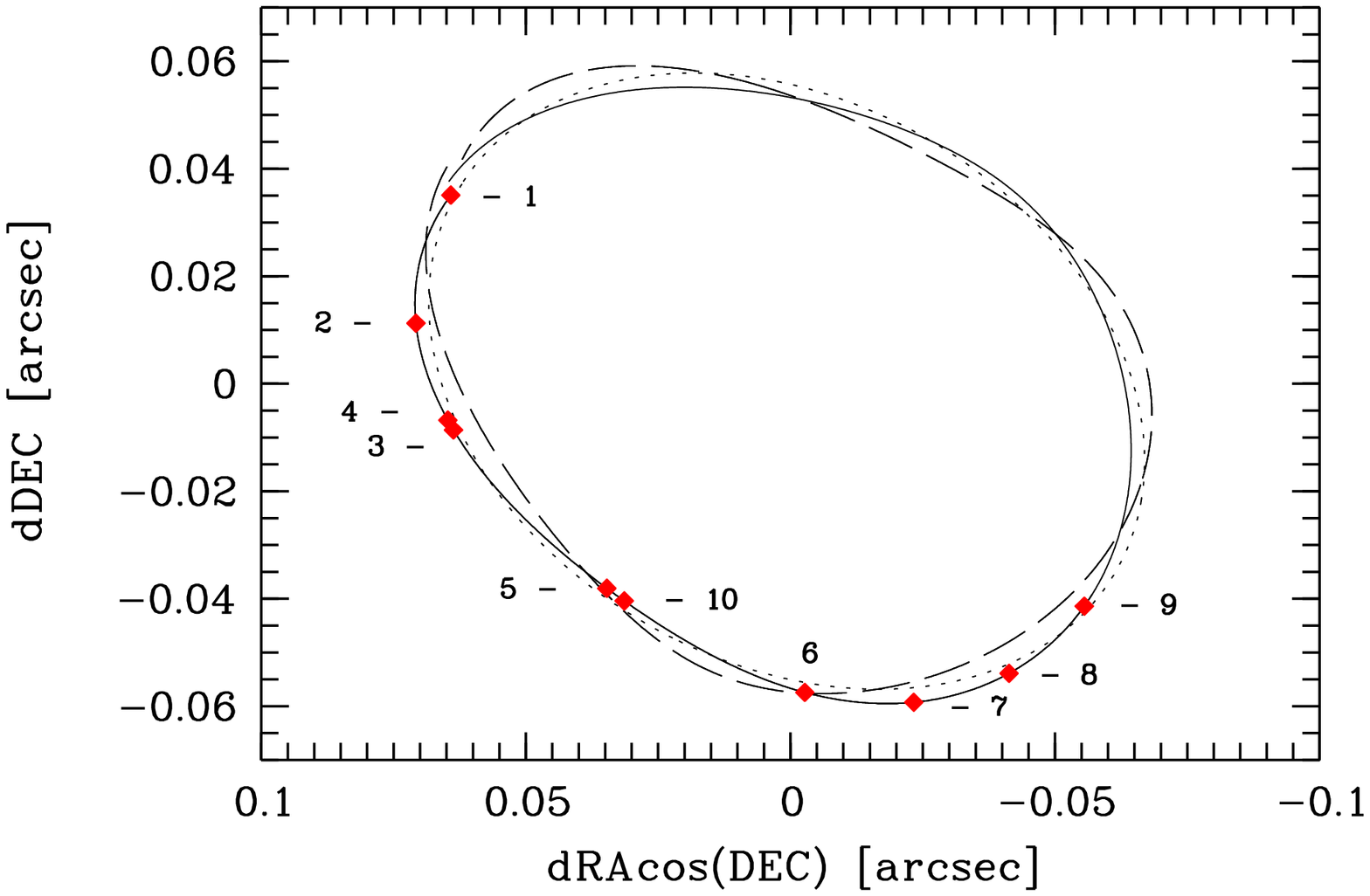} \hspace{0.5cm}
  \includegraphics[width=4.9cm, angle=-90]{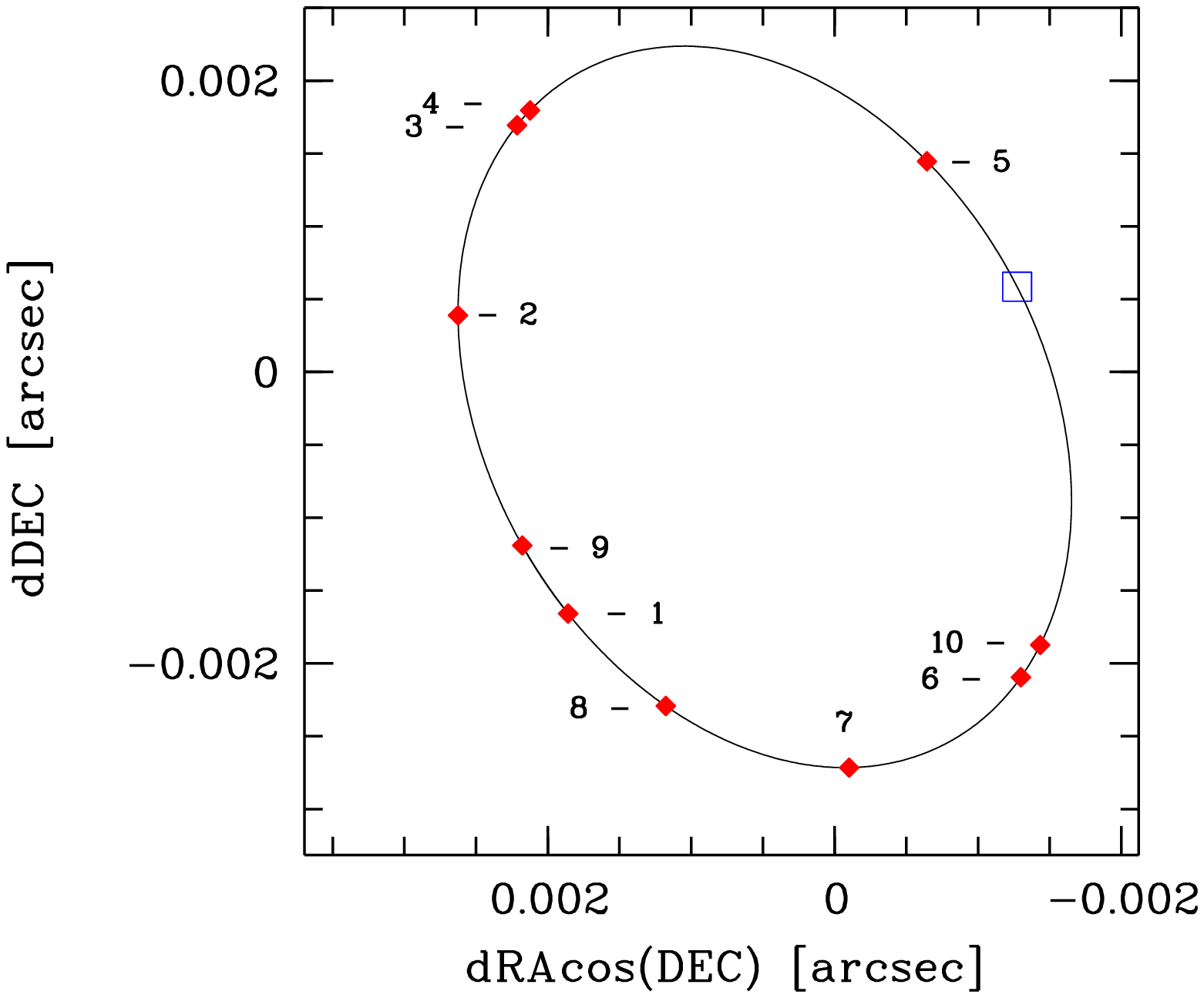}
 \end{center}
\caption[GJ~1046 observations overplotted over a simulated orbit]{GJ~1046 observations distributed over one year in the left panel and over one orbital period in the right panel. In the left panel, the ellipses are simulations with $\varpi = 71.56$ and  $i = 30\degr, ~\Omega = 150\degr$ (dashed line), $i = 45\degr, ~\Omega = 60\degr$ (solid line) and pure parallax movement without any orbital motion (dotted line), proper motion has been subtracted. The right panel shows a simulation of an orbit with $i = 45\degr, ~\Omega = 60\degr$ and parallax and proper motion subtracted. The red squares represent the times of observations overplotted on the simulated orbits, with the number denoting the corresponding epoch. The open blue square marks the periastron passage, $T_{0} = 54745.41$.}
\label{fig:orbit_obs}
\end{figure}

The adaptive optics observations of GJ~1046 analyzed here were
carried out with the adaptive optics instrument NAos COnica (NACO)
at the Very Large Telescope (VLT) of the European Southern
Observatory (ESO) with the UT4 Yepun telescope. All observations
were executed in Service Mode within a monitoring program from
July 2008 till October 2009 to derive the true mass of the
companion via astrometric measurements. Altogether I obtained 10
epochs, where the first nine are in roughly three week intervals
between July and December 2008 and the last epoch was obtained end
of September 2009, see table~\ref{tab:observations} for an overview of the observations. The last row is an indicator for the quality of the data: A = good, B = mostly within specifications, C = outside specifications; with the observing specifications of: seeing better than 0.8\arcsec ~and airmass lower than 1.6. In the target field, GJ~1046 itself was used as
the AO guide star, in the reference field the brightest star in
the upper left corner was used for wavefront sensing. 
\begin{table}[t!]
 \begin{center}
 \begin{tabular}{|c|c|cc|cc|c|c|c|} \hline
  Epoch & Date & \multicolumn{2}{c|}{Target Field} & \multicolumn{2}{c|}{Ref. Field} & \# jitter & \# images per & Quality\\
    &      & DIT & NDIT & DIT & NDIT & position & position &\\ \hline
  1  & 03/07/08 & 0.9 & 110 & 0.4 & 10 & 5 & 5 & A \\
  2  & 02/08/08 & 0.9 & 110 & 0.4 & 10 & 5 & 5 & B\\
  3  & 22/08/08 & 0.9 & 110 & 0.4 & 10 & 5 & 5 & C\\
  4  & 24/08/08 & 0.9 & 110 & 0.4 & 10 & 5 & 5 & A\\
  5  & 27/09/08 & 0.9 & 110 & 0.4 & 10 & 5 & 5 & B\\
  6  & 30/10/08 & 0.9 & 122 & 0.4 & 10 & 5 & 5 & A\\
  7  & 18/11/08 & 0.9 & 122 & 0.4 & 10 & 5 & 5 & B\\
  8  & 07/12/08 & 0.9 & 122 & 0.4 & 10 & 5 & 5 & B\\
  9  & 27/12/08 & 0.9 & 122 & 0.4 & 10 & 5 & 5 & B\\
  10~ & 30/09/09 & 0.9 & 120 & 0.4 & 15 & 5 & 5 & B\\ \hline
 \end{tabular}
 \caption{Overview over the observations, with exposure time, number of jitter positions and quality indicator of the obtained data.}
 \label{tab:observations}
 \end{center}
\end{table}
The target
star is not visible from the Paranal Observatory between February
and June, or only at very high airmass. But as the orbital period
of the companion is 169 days, I covered a full orbit between
July and December 2008 (epochs 1-9). This was very important to better
distinguish the parallactic motion from the half year orbital
motion. It was planed to obtain two more epochs in the period
between July and September 2009 to better constrain the orbit, but
I only got the last epoch in October 2009, which was still
important to extend the time baseline for constraining the relative proper
motion between GJ~1046 and the reference star. In Fig.~\ref{fig:orbit_obs} the observations are shown distributed over one year in the left panel and over one orbital period in the right panel. The plotted ellipses are simulations of the motion of GJ~1046 with parallactic motion, but without proper motion: in the left panel for inclination $i = 30\degr$ and ascending node $\Omega = 150\degr$ (dashed line), $i = 45\degr, ~\Omega = 60\degr$ (solid line) and pure parallax movement without any orbital motion (dotted line). The red squares represent the times of observations overplotted over the simulated motion with $i = 45\degr, ~\Omega = 60\degr$, with the number denoting the corresponding epoch. This plot shows how important it is to have a proper sampling of observations to distinguish the orbital motion from the parallax movement. The right panel shows the times of observations overplotted over a simulated orbit with $i = 45\degr, ~\Omega = 60\degr$ and proper motion and parallax subtracted. As one can see, the full orbit is coverd by the observations. The open blue square marks the periastron passage, $T_{0} = 54745.41$.
(Fig.~\ref{fig:GJ1046}).

\noindent The DIT values were chosen, such that the peak counts
for the bright star in each field are $\leq 80\%$ of the detector
linearity limit of $110,000~e^{-}$ for a small seeing of
$0.4\arcsec$. For the star in the field with the highest
separation from the AO guide star I estimated a signal-to-noise
(S/N) value integrated over the stellar disk and a formal
astrometric precision using photon statistics $(\sigma/(S/N))$ for
a seeing of $0.8\arcsec$. All the performance estimates were made
with the NAOS preparation software and the NACO ETC (Exposure Time
Calculator). In Tab.~\ref{tab:Performance} the estimated
performance for an observation in the middle of the epochs is
listed.

\begin{figure}
\begin{center}
\includegraphics{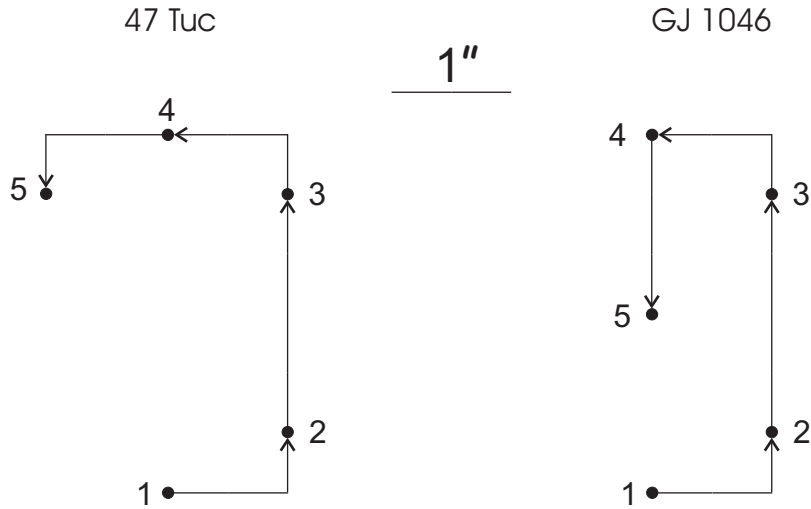}
\end{center}
\label{fig:jitter} \caption[Jitter pattern for the target and
reference field]{Jitter pattern for the target and reference
field.}
\end{figure}

\noindent The observations were made with a narrow-band filter
centered on $2.12~\mu m$ and with a FWHM of $0.022~\mu m$ to minimize the effects of differential
atmospheric dispersion. The reference field in 47~Tuc was observed
each epoch immediately before the science field. I observed the
reference field before the target field and not after it, for reasons
of visibility of the two fields. 47~Tuc culminates roughly half an
hour before GJ~1046 at the location of the Paranal Observatory,
hence the conditions were best for both fields when observed in
this order.\\
\noindent As one can see in Fig.~\ref{fig:GJ1046} left, the target
star GJ~1046 and the reference star just fit in the FoV of the S27
camera of the NACO instrument. GJ~1046 lies in the upper left
corner of the detector and the astrometric reference star in the
lower right corner when the detector is rotated by $6\degr$
anticlockwise. I chose this camera because of its, for the
purpose of this work, suitable pixel-scale of 27 mas/px. I aimed for a
precision of 0.5 mas and with the presumption of an accuracy of
1/50 pixel this camera was the one best suited of the three
offered ones, as the FoV of the S13 camera is too small.\\
\noindent To avoid a badpixel coincident at the pixel positions of the stars
and to better calculate the sky-background, a jitter pattern is
necessary for observations in the infrared. Preferable is a
pattern with an offset of several arcseconds in different
directions,
which is difficult with this arrangement of the stars on the detector.\\
\begin{table}
 \begin{center}
\small{
\renewcommand{\thefootnote}{\alph{footnote}}
\begin{tabular}{|l|ccc|ccc|ccc|c|c|}
\hline
Field         & \multicolumn{3}{|c|}{Strehl$[\% ]$\footnotemark[1] }
              & \multicolumn{3}{|c|}{FWHM$[$mas$]$\footnotemark[1]}
              & \multicolumn{3}{|c|}{Encircled energy$[\% ]$\footnotemark[1]}
          & S/N & $\Delta$ \\
                 & $0.4\arcsec $ & $0.6\arcsec $ & $0.8\arcsec $
                 & $0.4\arcsec $ & $0.6\arcsec $ & $0.8\arcsec $
                 & $0.4\arcsec $ & $0.6\arcsec $ & $0.8\arcsec $
         & @ $0.8\arcsec$ & [mas] \\
\hline
GJ1046 & 53.1 & 47.8 & 41.1 & 72 & 72 & 73  & 59.4 & 54.6 & 48.6 & - & - \\
ref & 39.7 & 21.3 & 7.6  &  76  & 87  & 132  & 48.6 & 31.7 & 16.0 & 142 & 0.39 \\
\hline
47 Tuc & 54.4 & 49.9 & 44.4 & 71 & 72 & 73  & 60.6 & 56.5 & 51.7 & - & -  \\
field  & 45.5 & 30.1 & 14.6 &  74  & 79  & 94 & 53.3 & 39.6 & 23.7 & 351 & 0.11  \\
\hline
\multicolumn{5}{l}{$^{a}$ @ indicated seeing}
\end{tabular}}
\caption[Performance estimates for the middle epoch observation]
{Performance estimates for the middle epoch observation.
Calculated for the AO star and the star furthest away from it. The last column lists the astrometric precision $\Delta$ of the fainter star which is $30.14\arcsec$ separated from GJ~1046 in the target field and $21.28\arcsec$ separated from the AO star in the reference field 47~Tuc. All
the performance estimates were made with the NAOS preparation
software and the NACO ETC.} \label{tab:Performance}
\end{center}
\end{table}
Another problem occurred, because of the brightness of the target
star. Bright stars are known to produce a number of electronic and
optical ghost features on the NACO detector, depending on their
position on the detector. When the position of a bright star is
$(xs, ys)$, the electronic ghosts\footnote{http://www.eso.org/sci/facilities/paranal/instruments/naco/doc/} appear approximately at the
positions $(1024 - xs, ys)$, $(xs, 1024 - ys)$ and $(1024 -xs,
1024 - ys)$. The last one of these would appear very close to our
faint reference star and could therefore influence the positional
measurement. An optical ghost which looks like a set of concentric
rings may also appear, but I did not see that in the
data.\\
Special care has been taken to calculate a jitter pattern which
made sure both stars are always within the FoV and at least 5
sigma afar from the detector edges or any ghost.
 Because the whole configuration plus the additional
jitter pattern brings the two stars close to the edges of the
detector, I had to make sure that the reference star was always
located on the very same pixel at the beginning of the
observations, so the calculated jitter pattern assures a
successful observation. In Table~\ref{tab:jitter} the position of
GJ~1046, the reference star and the closest ghost are listed
together with the final jitter pattern, which is shown in
Fig.~\ref{fig:jitter}. I had to differ the last jitter point in
the observations of the reference field due to the different
positions of the stars and the ghosts of the bright star in this
field.
\begin{table}[b!]
 \begin{center}
 \small{
  \begin{tabular}{|c|cc|cc|cc|c|c|} \hline
   rel. jitter offset & \multicolumn{2}{c|}{GJ~1046} & \multicolumn{2}{c|}{Ghosts} & \multicolumn{2}{c|}{Ref. star} & Dist. to & Dist. to \\
   x [$\arcsec$]~~ ~~y [$\arcsec$] & x [px] & y [px]& x [px] & y [px] & x [px] & y [px] & ref star [px] & ref star [$\sigma$]\\ \hline
    ~0~~  ~~~~~0~~ & 156 & 854 & 868 & 170 & 917 & ~65 & 115.8 & ~9.3 \\
    ~1~~  ~~~~~0.5 & 193 & 873 & 831 & 151 & 954 & ~83 & 140.1 & 11.2 \\
    ~0~~~ ~~~~2~~ & 193 & 946 & 831 & ~78 & 954 & 157 & 145.6 & 11.7 \\
   -1~~  ~~~~~0.5 & 156 & 965 & 868 & ~59 & 917 & 175 & 125.5 & 10.0 \\
    ~0~~  ~~~~-1.5 & 119 & 946 & 905 & ~78 & 880 & 157 & ~82.4 & ~6.6 \\ \hline
  \end{tabular}}
  \caption{Positions of the bright target star, the ghost it is
  producing closest to the reference star and the distance
  of the ghost to the reference star.}
 \label{tab:jitter}
 \end{center}
\end{table}

\section{Data Reduction}
\label{sec:Reduction1} I did not use the NACO-pipeline products
for our astrometric measurements. Instead I reduced the data on
our own. The data reduction is the same for the target and the
reference field. So I only describe the principle chain here and
name differences directly when they occur.\\
\noindent First I created a badpixel-mask using the badpixel routine
included in the eclipse data reduction package\footnote{http://www.eso.org/sci/data-processing/software/eclipse/}.
This routine uses the flatfield images to create the badpixel
mask. If possible and available I used the badpixel masks created
by ESO from sky-flats taken in the same night and in the same
filter as our observations. But this was only the case for two
observations. In the other cases the dome-flats were used. The
mask was then combined with the hotpixel mask, created by ESO
during the pipeline data reduction. The so derived
\textit{strange}-pixel mask is then used to correct the flaged
pixels with the median of their 8 nearest neighbors in the science
frames as well as the dark current frames.\\
Then the dark frames were multiplied by the gain factor of the S27
camera detector, which is 11 e/ADU. Masterdarks were created by
median combining darks with the same exposure times as the target
and reference frames
respectively.\\
Also the dome-flats are multiplied by the gain. The masterdark for
the target field and the reference field are then subtracted for
the corresponding data reduction. Flats with lamps on and flats
with lamps off were median combined, respectively, and the
off-flat was subtracted from the on-flat. Finally the obtained
flat field image
is normalized by dividing it through its mean value.\\
Now I have everything to reduce the science data following
the standard technique.
 The single
frames are multiplied by the gain, masterdark subtracted and
divided by the normalized flat field. Finally the frames are
corrected for the strange-pixels.

\subsection{Sky Subtraction}
I investigated the subtraction of the sky-background in different
ways. As I do not have additional sky images I created a sky
frame from median combining all science frames in the first step.
But this left me with some 'holes' close to the stars after
subtracting the sky frame from the science frames. This is due to
the small jitter pattern I used during the observations and the
fact that I only have 5 different jitter positions. A bigger and
therefore better pattern was not possible, see Chap.
\ref{sec:Obs}. Light from the wings of the stars is not averaged
out and produces somewhat higher values in the sky frame,
which then appear as holes in the sky subtracted science frames.\\
The second step was to remove the influence of the stars by using
$\kappa-\sigma$ clipping. If the mean of a single science frame is
significantly higher than the median value of the sky created
before in step one, pixels in the science frame with values higher
or lower than a certain range around the mean, the stars, are substituted by the mean of the pixels left after
the $\kappa-\sigma$ clipping. These frames, cleaned for stars, are
then used to calculate the sky by
median combination of them.\\
I tested this procedure for different ranges around the mean
value, which define the pixels which are substituted. The holes
indeed were less pronounced, but they also were more irregular,
which introduced a strange pattern into the wings of the PSF of
the bright star. This degrades the possibility to measure
accurately the position of the stars and I decided not to create the
sky this way.\\
Another test was made by creating the sky only by median combining
an inner part of the science frames, where there are no stars. But
this did not lead to a huge improvement in the sky subtraction either.\\
I decided finally to not subtract the sky background at all, as
it is not very high and does not show any slope over the frames.
It can be therefore assumed as a local constant background which should not
disturb the astrometric measurements.

\subsection{50 Hz Noise}
The reduced frames showed a strong noise pattern along the rows.
This phenomenon is known as 50 Hz noise and was
found to be caused by the fans in the front end electronics of the
Infrared Array Control Electronics (IRACE). One can decrease the
effect by subtracting the median of each row from the very same
row. I created a frame with the same size as the science frames
in which the values in the rows have the median value of the
corresponding science row. I created the median value not over
the whole row, but detector quadrant wide. With this I took care
of the fact, that the NACO detector of the S27 camera is read out
quadrant wise.

\subsection{Shift and Add}
\label{subsec:ShiftAdd} Now the frames are fully reduced. The
frames obtained on the same jitter position are added. This is
possible because the telescope was not moved between the single
frames of one jitter position. I am left with five frames per
epoch and field. These frames are added by simple shift and add
using the \textit{jitter} routine \citep{Devillard1999}. So I
have one frame per target and reference field for every epoch. I
also tested shifting and adding all single frames, but this gave
no better result than adding the five
already stacked frames from the different jitter positions.\\

    \cleardoublepage
    \chapter{Analysis and Astrometric Corrections}
   \label{chap:Astrometry} 
\begin{figure}
 \begin{center}
  \includegraphics{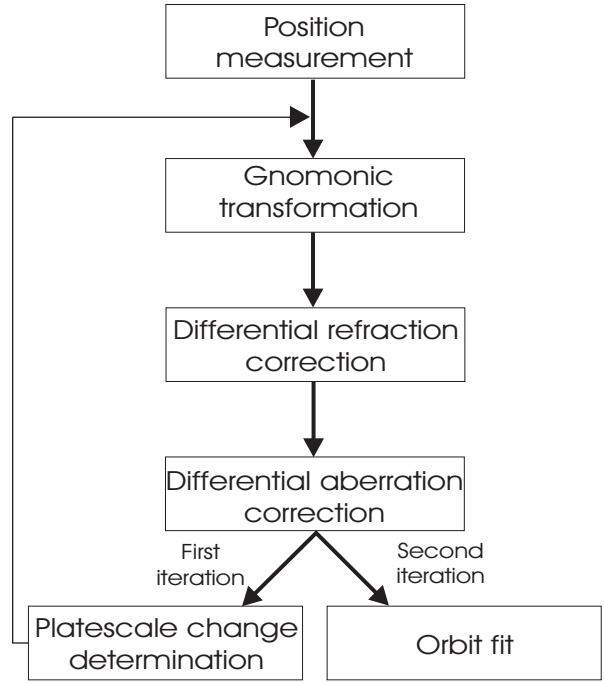}
 \end{center}
 \label{fig:correctionstepss}
 \caption{Stepwise description of the corrections applied
 to the measures pixel coordinates of the stars.}
\end{figure}

Several corrections have to be applied to
the measured positions of the stars before fitting the astrometric
orbit. These include converting pixel coordinates to celestial
coordinates, differential refraction, differential aberration and
change in plate-scale and detector rotation. Logically, these
corrections are executed in the inverse order as they appear. The
detector distortions have to be corrected as first step in
principle, as it is the last effect which results in a
displacement in position of the stars on the detector. But to
avoid cross-talk between this effect and displacements from
aberration and refraction, I corrected for the differential
refraction and aberration before. First the differential
refraction has to be corrected as it deflects the light rays after
the aberration already occurred, then the aberration is corrected.
In an iterative way the correction of the detector distortion
measured as third step, is applied to the originally measured
detector positions of the stars before correction for refraction
and aberration. After that, the correction for differential
refraction and differential aberration is performed again.
Fig.~\ref{fig:correctionstepss} shows the stepwise and iterative
corrections which are described in the following sections in
detail.

\section{Position Measurements}
\label{sec:positions} The positions of the stars were measured by
fitting a Moffat-function to the individual stars using the
non-linear least square fitting package MPFIT2DPEAK, written in
the IDL language and provided by Craig Markwardt
 \citep{Markwardt2009}. The Moffat function is a modified Lorentzian
 with a variable power law index $\beta$ \citep{Moffat1969}. It
better represents the form of a PSF corrected by AO than a simple
 Gaussian-function, but the Gaussian-function is contained in
the Moffat-function as a limiting case with $\beta \rightarrow
\infty$:
\begin{equation}
f(x,y) = c + I_{0}~\left[\left(\frac{x-x_{0}}{\rho_{x}}\right)^2 +
\left(\frac{y - y_{0}}{\rho_{y}}\right)^2 + 1\right]^{-\beta}
\label{equ:Moffat}
\end{equation}
c is a constant, $x_{0}, y_{0}$ are the center of the function and
$\rho_{x}, \rho_{y}$ define the FWHM of the PSF: $FWHM_{x} =
\rho_{x}(2\frac{1}{\beta} - 1)^{1/2}$. The Moffat-function has two
advantages over the Gaussian. It is numerically well behaved in
the treatment of narrow PSFs, because of the use of polynomials
instead of exponential expressions. But it also allows the wings
of the PSF to be fitted. This makes it a better fitting function
for AO corrected PSFs.\\
\noindent For the PSF-fit, the stars in the stacked frames were
marked and a small box of $15 \times 15$~px ($407 \times 407$~mas)
was cut around them. After finding the px with the maximum, a
bigger box with $31 \times 31$~px ($842 \times 842$~mas) was cut
around the star. Then the Moffat was fitted to the PSF. I also
tried to just fit the Center of Weight (CoW), but the fitted CoW
of the distribution is dependent on the box size, whereas the
fitted Moffat peak was very stable for all kinds of box-sizes. I
therefore decided to fit a rotated Moffat-function to determine
the peak position of the
PSFs.\\
A phenomenon in adaptive optics images which occurs for sources
which are not the guide star, is known as radial breathing. The
PSFs of the objects are elongated in the direction to the guide
star, but not in the direction perpendicular to it. The effect
gets stronger the farther away a source is from the guide star, so
the shapes of the PSFs get more and more elliptical. This is
important to know for photometry, but should not influence the
center of the distribution.\\
The fit routine also outputs values for the Full Width at Half
Maximum (FWHM). I could have used this for an estimate of the
positional error, if I assumed photon statistics. But to also
cover the error, which may arise from the fit itself, I adopted
another method for the positional uncertainty.

\subsection{Positional Error Estimate - Bootstrapping}
\label{subsec:Bootstrap} To estimate the positional error
resulting from the fit, I re-sampled the intensity distribution in
the cut boxes, PSF plus background, 100 times with the bootstrap
method and fitted each single realization again with a
Moffat-function. The bootstrap method is a general technique for
estimating for example standard errors for estimators and was
first introduced by Bradley Efron \citep{Efron1979}. I numbered
all photons in each PSF distribution and generated the same amount
of random numbers. The random numbers defined which of the photons
was picked and put at the same position in the re-sampled PSF. But
after that, the photon is \textquotesingle\textit{put
back}\textquotesingle, so it could be picked up again. To do so,
one needs to know the total number of photons, $N_{ph}$, in the
box. To get that number, one has to take into account the
gain-factor of the detector, ($11~e^{-}/ADU$ for NACO), and the
fact that the images are averaged twice. First during the
acquisition, when N frames with the exposure time DIT are
averaged, and then when the frames are stacked to obtain the final
frame (\ref{sec:NACO} and \ref{subsec:ShiftAdd}). Because of the
massive rising computational amount when multiplying the averaged
image by NDIT, the number of photons is a lot larger, and because
I did not want to underestimate the error, I decided to stay
conservative and worked on the averaged image without multiplying
by NDIT. Only the gain and the average factor from adding the
single frames is taken into account. In this way I generated the
re-sampled PSFs by executing the pick and put-back procedure as
many times as there were photons in the original distribution.
This was done 100 times for each star in the target field as well
as in the reference field.\\
The standard deviations in x and y of the 100 fits
from the mean position are used as an estimate for the errors.\\
\noindent To check whether the obtained fitting errors are
reasonable, I also fitted a rotated Gaussian distribution to the
same PSFs in the boxes and calculated the positional error if we
only assume photon statistics, $dx =
\frac{\sigma_{x}}{\sqrt{N_{ph}}}$. The errors are of the same
order as the ones from the bootstrap re-sampling, but due to the
fact that a Gaussian does not represent the form of the AO
corrected PSFs very well and that I wanted to include the error
contribution of the fit, I used the errors from the bootstrap
re-sampling as the positional uncertainties. In
Table~\ref{tab:booterror} the errors from the bootstrapping and
the Gaussian approximation are listed.\\
\noindent Additionally to this error one has to take into account
the errors originating from the differential refraction,
differential aberration and the plate-scale and rotational errors.
These contributions to the error budget are described in the
following sections.
\begin{table}
 \begin{center}
  \begin{tabular}{|l|l|l|ll|ll|} \hline
    \multicolumn{3}{|c|}{ } & \multicolumn{2}{c|}{Bootstrap error} & \multicolumn{2}{c|}{Gauss error} \\
    \multicolumn{3}{|c|}{ } & ~~~~~~x & ~~~~~~y & ~~~~~~x & ~~~~~~y  \\ \hline
    & & [px] & $2.62\cdot 10^{-4}$ & $2.48\cdot 10^{-4}$ & $2.15\cdot 10^{-4}$ & $2.02\cdot 10^{-4}$ \\
    & \raisebox{2.0ex}[-2.0ex]{AO star}~ & [mas]~~ & $7.11\cdot 10^{-3}$ & $6.73\cdot 10^{-3}$ & $5.84\cdot 10^{-3}$ & $5.48\cdot 10^{-3}$ \\ \cline{2-7}
    \raisebox{2.5ex}[-4.0ex]{\textbf{GJ~1046}}~~ & & [px] & $0.70\cdot 10^{-2}$ & $0.73\cdot 10^{-2}$ &  $1.64\cdot 10^{-2}$ & $1.61\cdot 10^{-2}$ \\
    & \raisebox{2.0ex}[-2.0ex]{ref star} & [mas] & 0.19 & 0.70 & 0.45 & 0.44 \\ \hline
    & & [px] & $5.48\cdot 10^{-4}$ & $5.96\cdot 10^{-4}$ &  $3.70\cdot 10^{-4}$ & $3.47\cdot 10^{-4}$ \\
    & \raisebox{2.0ex}[-2.0ex]{AO star} & [mas] & 0.01 & 0.02 & 0.01 & $9.42\cdot 10^{-3}$ \\ \cline{2-7}
    \raisebox{2.5ex}[-4.0ex]{\textbf{47~Tuc}} & & [px] & $1.63\cdot 10^{-2}$ & $1.60\cdot 10^{-2}$ & $1.31\cdot 10^{-2}$ & $1.14\cdot 10^{-2}$ \\
    &\raisebox{2.0ex}[-2.0ex]{ref star} & [mas] & 0.44 & 0.43 & 0.36 & 0.31 \\ \hline
  \end{tabular}
  \caption[Positional error from the PSF fit calculated with bootstrap re-sampling]{Positional error from the PSF fit calculated with bootstrap re-sampling and simple photon statistics. In the 47~Tuc reference field, the
  star furthest away in the opposite corner of the detector is
  used as the reference star (ref star).}
 \label{tab:booterror}
 \end{center}
\end{table}

\section{Astrometry with FITS-Header Keywords}
\label{sec:FITS}
\subsection{World Coordinates in FITS}
World coordinates are coordinates that serve to locate a
measurement, as for example frequency, wavelength or longitude and
latitude, in a multidimensional parameter space. The
representation of world coordinates in the Flexible Image
Transport System (FITS), which is used by all observatories since
the General Assembly of the IAU (resolution R11), was first
introduced by \cite{Wells1981}. This initial description was very
simple and \cite{Greisen2002} later described it in more detail
with more possible extensions. Keywords in the FITS header are
used to describe the parameters necessary to convert the $x$ and
$y$ coordinates from the detector into the world coordinates. Each
axis of the image has a certain coordinate type and a reference
point, for which the coordinate value, the pixel coordinate and an
increment are given. The basic keywords are:\\
\begin{tabular}{ll}
 CRVAL\textit{n} & coordinate value at reference point \\
 CRPIX\textit{n} & pixel coordinate at reference point \\
 CDELT\textit{n} & coordinate increment at reference point \\
 CTYPE\textit{n} & axis type \\
\end{tabular}\\

\begin{figure}[t!]
 \begin{center}
 \includegraphics[height=10cm]{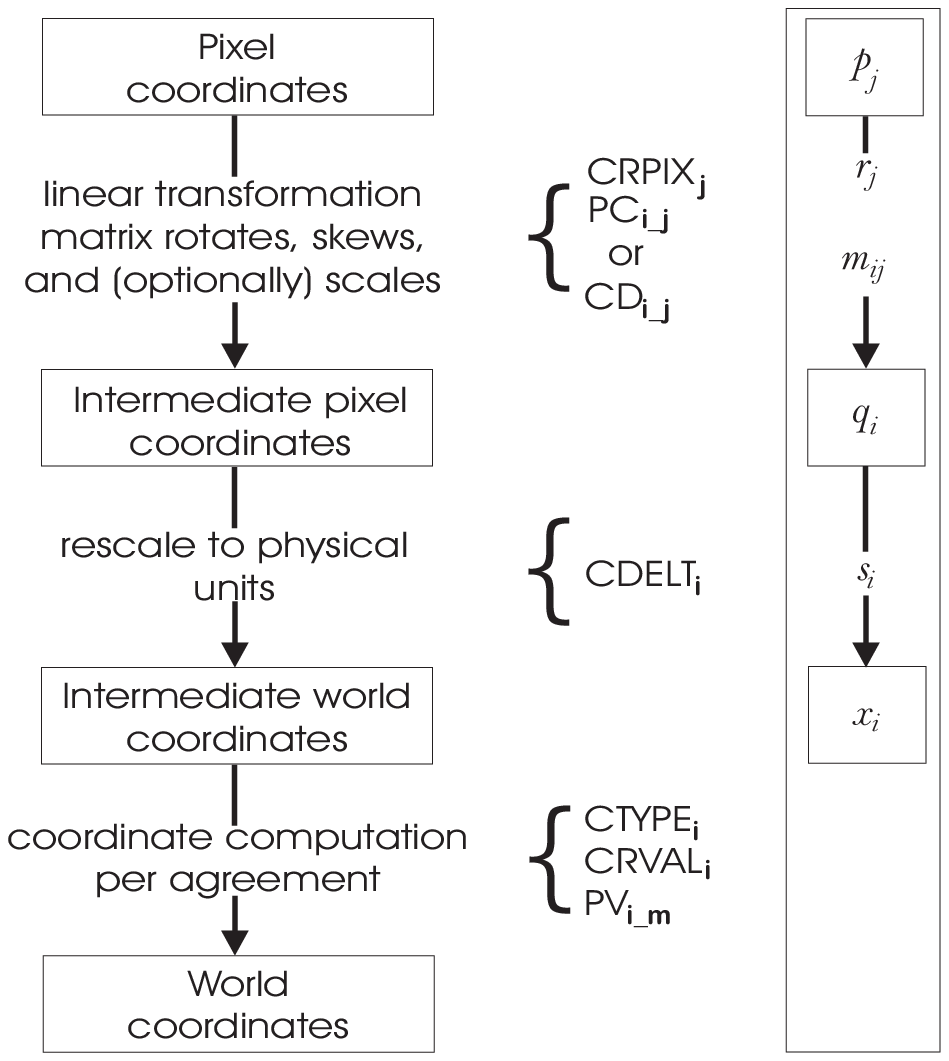}
 \end{center}
 \caption{Conversion of pixel coordinates to world coordinates (after \cite{Calabretta2002}).}
 \label{fig:px2wcs}
\end{figure}

\noindent To convert pixel coordinates to world coordinates a
multi-step process is needed, whose principle steps are shown in
Fig.~\ref{fig:px2wcs}.
The first step is a linear transformation via matrix
multiplication from the pixel coordinates $p_{j}$ to intermediate
pixel coordinates $q_{i}$:
\begin{equation}
 q_{i} = \sum_{j=1}^N m_{ij}(p_{j} - r_{j})
\label{equ:qi}
\end{equation}
$r_{j}$ are the pixel coordinates of the reference point, given by
the CRPIX$_{j}$ elements. In the following notation the index $j$
refers to the pixel axis and $i$ to the world coordinate axis. The
$m_{ij}$ matrix is a non-singular square matrix with the dimension
of $N \times N$. $N$ is given by the keyword value of NAXIS, which
gives the dimension of the data array, but not necessarily that of
the world coordinates. If the dimensions of the World Coordinate
System (WCS) are different, the keyword WCSAXES is used, which
then gives the maximum value of the index of any WCS keyword. The
resulting intermediate pixel coordinates $q_{i}$ are offsets from
the reference point along axes that are coincident with those of
the intermediate world coordinates $x_{i}$. The transformation to
the corresponding $x_{i}$ is then a simple scaling:
\begin{equation}
 x_{i} = s_{i}q_{i}
\label{equ:xi}
\end{equation}
There are two formalisms that describe the transformation matrix
in FITS. One is the PC\textit{i\_j}, where the matrix elements of
$m_{ij}$ are given by the PC\textit{i\_j} header keywords and
$s_{i}$ by CDELT\textit{i}. If PC\textit{i\_j} and CDELT\textit{i}
are not given in the FITS header, the second formalism is used.
This one combines Eqs.~\ref{equ:qi} and \ref{equ:xi} to :
\begin{equation}
 x_{i} = \sum_{j=1}^N(s_{i}m_{ij})(p_{j} - r_{j})
\label{equ:xisum}
\end{equation}
with the keywords CD\textit{i\_j}$ = s_{i}m_{ij}$. This formalism
omits the second step on the transformations.\\ The last step
depicted in Fig.~\ref{fig:px2wcs} from the intermediate world
coordinates to the world coordinates applies a non-linear
transformation, depending on the world coordinate system one is
aiming at.

\subsection{Celestial Coordinates in FITS}
For the transformation to celestial coordinates, the  intermediate
world coordinates are interpreted as Cartesian coordinates in the
plane of the projection. The final step is divided into two steps
which are a spherical projection defined in terms of a convenient
coordinate system, the native spherical coordinates, and a
spherical rotation of these coordinates to finally obtain the
required celestial coordinates \citep{Calabretta2002}. Depending
on the kind of projection, the equations for the transformation to
the native longitude and latitude $(\phi, \theta)$ are specified
in the keyword CTYPE\textit{i}. There is a wide variety of
projecting a sphere onto a plane and vice versa. The one used in
data set of this work is the so-called gnomonic projection. It is
described in more detail in Sec.~\ref{subsec:xyad}. The values of
the keywords
are:\\
\begin{tabular}{ll}
CTYPE1 & = 'RA - - - TAN'\\
CTYPE2 & = 'DEC - - TAN'\\
\end{tabular}\\

\noindent The last three characters define the projection type and
the leftmost four characters are used to identify the celestial
coordinate system. In this case a gnomonic projection
(\ref{subsec:xyad}) and equatorial coordinates.\\
\noindent For the last step, the spherical rotation from native
coordinates to celestial coordinates, one needs three Euler angles
which specify the rotation. The celestial coordinates for the
reference point, $(\alpha_{0},\delta_{0})$, specified in the
header via the CRVAL\textit{i} keyword, are associated with a
native coordinate pair $(\phi_{0}, \theta_{0})$, which are defined
explicitly for each projection. The difference of these native
coordinates for different projections is due to the fact that the
projections diverge at different points. The Mercator projection
diverges for example at the native pole, while the gnomonic one
diverges at the equator. Therefore they cannot have the reference
point at these points, because that would mean infinitive values
for CRVAL\textit{i}. The projection equations are then constructed
in a way so that $(\phi_{0}, \theta_{0})$ transform to the
reference point $(x, y)$.\\
There are several other FITS keywords which define the necessary astrometry:\\
\begin{tabular}{ll}
 RADECSYS & defines the reference frame, e. g. FK5\\
 EQUINOX & defines the coordinates epoch, e. g. 2000.0\\
 MJD-OBS & defines the Modified Julian Date of the observation start\\
\end{tabular}\\

\noindent The combination of CTYPE\textit{i}, RADECSYS and EQUINOX
define the coordinate system of the CRVAL\textit{i} and of the
celestial coordinates resulting from the transformations.

\subsection{Transformation from xy-Coordinates into RA/DEC}
\label{subsec:xyad} To convert the pixel values measured for the
peak positions of the stars in the frames to celestial
coordinates, I used the astrometry in the FITS header. The
projection given by the keyword CTYPE\textit{i} is a special form
of a zenithal projection, the gnomonic projection. The name
deduces from the Greek word Gnomon $(\gamma\nu\acute{o}\mu o\nu)$,
which stands for a shadow-stick used as an astronomical
instrument. The shadow cast by the tip of the stick was already
used to measure the time with a sundial in antiquity. The gnomonic
projection is the oldest map
projection, developed by Thales in the 6th century BC.\\
\noindent The gnomonic projection displays all great circles on a
sphere as straight lines. The surface of the sphere is projected
from its center, hence perpendicular to the surface, onto a
tangential plane and the least distortions occur at the tangent
point. In Astronomy, the Earth's radius is small compared to the
distance to the stars and can be neglected, so the observatory can
be seen as being in the center of the projection. This directly
implies the negligence of the diurnal parallax (see
Chap.~\ref{subsec:parallax}).

\noindent A problem occurred after stacking the images with the
\textit{jitter} routine, as described in
Chap.~\ref{subsec:ShiftAdd}. The image obtained is slightly larger
in $x$- and $y$-direction than the single images before. The new
size of the image is updated in the FITS header by the jitter
routine, but not the resulting coordinate shift of the reference
point. Additionally, no information is given on how much the
zero-point of the array is shifted. To compute the new pixel
coordinates of the reference point in the stacked image, I had to
execute some more
steps.\\
\noindent  As the star I used as a guide star for the AO system is
quite bright and also very well corrected in the single frames I
could measure its pixel coordinates very accurately. I measured
its position in the first stacked image of the first jitter
position in the same way as described in
Chap.~\ref{sec:positions}. Then I calculated the distance to the
given coordinates for the reference point. After measuring all
positions in the final image, I recalculated the pixel coordinates
of the reference point by using the beforehand calculated distance
to the AO guide star and updated the
CRPIX\textit{i} values in the FITS header.\\

\noindent The transformation of the coordinates was then performed
with the IDL routine \textit{xyad.pro} from the IDL
astrolib\footnote{http://idlastro.gsfc.nasa.gov/} which uses the
formalism described in detail in paper II of the series of papers
describing coordinates in the FITS formalism
\citep{Calabretta2002}.

\section{Astrometric Corrections}
The exact position where an object appears on the sky does not
only depend on the coordinates of the observed object, but also on
various effects which are connected to the relative velocity of
the observer, i.e. the \textit{aberration}, and the atmosphere of
the Earth, i.e. the \textit{atmospheric refraction}.

\subsection{Theory of Atmospheric Refraction}
The atmospheric refraction decreases the true zenith distance of
an object. Light, passing a surface that separates two layers with
different refractive indices $n$ and $n + dn$ is refracted in a
way described by Snell's Law. Let $\eta$ be the angle of incident
in a medium with refractive index $n$. The angle of refraction in
a medium with refractive index $n + dn$ is then $\eta + d\eta$
following
\begin{equation}
(n + dn)\sin{(\eta + d\eta)} = n\sin{\eta}
\end{equation}
The total bending of the light can be expressed as
\begin{equation}
R = z_{t} - z_{0}
\end{equation}
where $R$ is the total refraction angle experienced by the light
ray, $z_{t}$ the true zenith distance in degrees and $z_{0}$ the
apparent observed zenith distance. In the infrared wavelength
regime, the refraction can reach tens of arcseconds, where the
effect is
larger in the J-Band than in the K-Band.\\
In theory one needs to know $n$ at all points of the light-path
through the atmosphere to determine $R$. In practice this is
normally not possible and one has to represent the atmosphere by a
model which leaves $R$ only as a function of atmospheric
conditions.\\
Using a spherical model for the atmosphere means assuming
iso-refractive index spheres around the center of the Earth.\\
\begin{figure}
 \begin{center}
  \includegraphics[width=12cm]{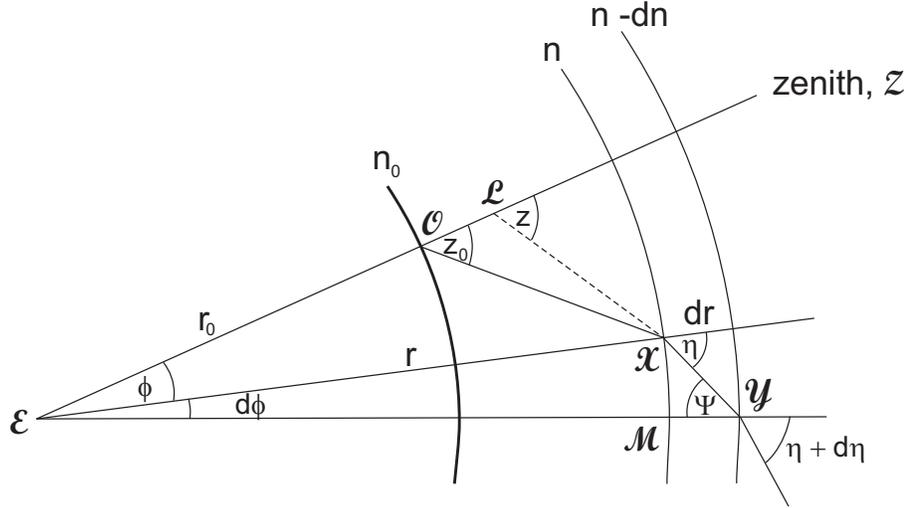}
 \end{center}
 \caption{Refraction in the atmosphere. For more details see text.}
 \label{fig:Refraction}
\end{figure}

\noindent In figure \ref{fig:Refraction}, $r_{0}$ is the geocentric
distance of the observer $\mathcal{O}$ from the center of the
Earth $\mathcal{E}$, $n_{0}$ the refractive index at ground level
and $z_{0}$ the apparent distance  of an celestial object to
the local zenith $\mathcal{Z}$.\\
If one considers a small layer with width $dr$ and the refractive
index $n$, then the range $\mathcal{XY}$ is the linear path of the
light ray coming from the observed object through this layer. At
the point $\mathcal{Y}$ one can apply Snell's law to the incoming
light ray which comes from a medium with the refractive index $n -
dn$. $\eta + d\eta$ is then the incident angle in the point
$\mathcal{Y}$ and $\Psi$ the refractive angle. This leads to:
\begin{equation}
n\sin{\Psi} = (n - dn)\sin{(\eta + d\eta)} \label{equ:nsinpsi}
\end{equation}
Naming the angle between the two zenith directions of the observer
and the one in point $\mathcal{X}$, $\phi$, the increment of this
angle towards the direction $\mathcal{M}$, $d\phi$ and the radius
vector from $\mathcal{E}$ to $\mathcal{X}$, $r$, one can write:
\begin{equation}
r\sin{\eta} = (r + dr)\sin{\Psi} \label{equ:nsineta}
\end{equation}
This equation comes from the sine relation in the triangle $\mathcal{EXY}$.\\
If one now multiplies respectively the right and left sides of
\ref{equ:nsinpsi} and \ref{equ:nsineta} one gets
\begin{equation}
nr\sin{\eta} = (r + dr)(n - dn)\sin{(\eta + d\eta)}
\label{equ:nrsineta}
\end{equation}
Again with Snell's law this proves $nr\sin{\eta}$ to be invariant.
Transformed to the observer's position and together with
Equ.~\ref{equ:nsinpsi} one can write the important relation:
\begin{equation}
nr\sin{\Psi} = n_{0}r_{0}\sin{z_{0}} \label{equ:nrsinpsi}
\end{equation}
If one looks at the triangle $\mathcal{LEX}$ one can see that $z =
\phi + \Psi$ and from the triangle $\mathcal{XMY}$ one gets
$\tan{\Psi} = \frac{rd\Phi}{dr}$. Combining these two equations
one gets
\begin{equation}
dz = d\Psi + \frac{\tan{\Psi}}{r} dr \label{equ:dz}
\end{equation}
Now one first differentiates \ref{equ:nrsinpsi}, which yields
\begin{equation}
nr\cos{\Psi} d\Psi + r\sin{\Psi} dn + n\sin{\Phi} dr = 0
\end{equation}
or written in a different way
\begin{equation}
nr\cos{\Psi}\left(d\Psi + \frac{dr}{r}\tan{\Psi}\right) =
-r\sin{\Psi} dn
\end{equation}
Together with Equ.~\ref{equ:dz} one has:
\begin{equation}
dz = -\frac{dn}{n}\tan{\Psi} \label{equ:dz2}
\end{equation}
Now one can look at Equ.~\ref{equ:nrsinpsi} in a different way and
write it, with $nr\cos{\Psi} = \sqrt{r^2n^2 - r^{2}_{0}n^{2}_{0}\sin^{2}{z_{0}}}$, as:
\begin{equation}
\tan{\Psi} = \frac{r_{0}n_{0}\sin{z_{0}}}{\sqrt{r^2n^2 -
r^{2}_{0}n^{2}_{0}\sin^{2}{z_{0}}}}
\end{equation}
This one substitutes into the previously obtained Equ. \ref{equ:dz2}
and finally derives after integration over dz
\begin{equation}
R = r_{0}n_{0} \sin{z_{0}}~\int_{1}^{n_{0}} \frac{dn}{n(r^{2}n^{2}
- r_{0}^{2}n_{0}^{2} \sin^{2}{z_{0}})^{1/2}} \label{equ:R}
\end{equation}
This general refraction formula requires a detailed knowledge of
the refractive index $n$ and therefore for the whole atmosphere
itself, as $n$ depends on the wavelength $\lambda$ of the incoming
light and the temperature $T$, pressure $P$ and water pressure
$P_{w}$ of the atmosphere. This formula is an exact description of
the refractive index of the air and can be calculated numerically
if the variation of the refractive index with height, $r = r(n)$,
is postulated. However, a simplification is possible when only
modest zenith distances are allowed. One can then expand
Equ.~\ref{equ:R} to \citep{Green1985,Gubler1998}:
\begin{equation}
R = A\tan{z_{0}} + B\tan^{3}{z_{0}} \label{equ:RAB}
\end{equation}
$A$ and $B$ are constants and depend on the wavelength $\lambda$,
$T$, $P$ and the relative humidity $H$\footnote{$H=P_{w}/ P_{sat}$
with $P_{sat}$ the saturation water vapor at temperature $T$}.
This expansion makes the calculation of the differential
refraction a lot easier, see next section.\\

\subsection{Differential Atmospheric Refraction} The differential
atmospheric refraction between two stars, $\Delta R$, is due to
two main effects: The different color of the stars (meaning
different temperature of the stars and therefore different
effective wavelength in the used filter) and the different zenith
distance of the two stars. The color dependent effect is smaller
(around 20 times) than the effect due to the differing zenith
distances for moderate zenith distance differences of $\sim 15\arcsec$ and temperature differences of $\sim 3000$ Kelvin in the K, K' and H band \citep{Gubler1998}. Assuming two stars with refraction
$R_{1}$ and $R_{2}$ respectively, $\Delta R = R_{1} - R_{2}$,  and
an apparent zenith distance difference of $\Delta
z_{0}=z_{1}-z_{2}$, one can write the refraction experienced by
star 2 expressed by $z_{1}$ and $\Delta z_{0}$ and (\ref{equ:RAB})
as:
\begin{equation}
 \begin{split}
 R_{2}(z_{2}) & = R_{2}(z_{1} - \Delta z_{0}) \\
 & = A_{2}\tan{(z_{1} - \Delta z_{0})} + B_{2} \tan^{3}{(z_{1} -
\Delta z_{0})} \label{equ:R2}
 \end{split}
\end{equation}
Expanding this equation around $z_{1}$ to second order in $\Delta
z_{0}$ then yields
\begin{equation}
 \Delta R = \Delta R_{\lambda} + \Delta R_{z}
 \label{equ:Rlz}
\end{equation}
where $\Delta R_{\lambda}$ denotes the color dependent effect and
$\Delta R_{z}$ the differential zenith distance dependent effect.
The two components of $\Delta R$ have the form:
\begin{align}
\Delta R_{\lambda} & = (A_{1} - A_{2})\tan{z_{1}} + (B_{1}
-B_{2})\tan^{3}{z_{1}} \\
\Delta R_{z} & = (1 + \tan^{2}{z_{1}})\bigl((A_{2} +
 3B_{2}\tan^{2}{z_{1}})\Delta z_{0}  \notag \\
 & - \left[A_{2}\tan{z_{1}} + 3B_{2}(\tan{z_{1}} + 2\tan^{3}{z_{1}})\right]\Delta z_{0}^{2}\bigr)
\label{equ:dR}
\end{align}
\noindent Usually the constants $A$ and $B$ are determined
empirically. Gubler and Tytler did this by first changing the
integration variable in Equ.~\ref{equ:R} from $n$ to $z$ which
yields
\begin{equation}
R = -\int_{0}^{z_{0}} \frac{rdn/dr}{n + rdn/dr}~dz \label{equ:Rdz}
\end{equation}
Full integration of this integral for two different zenith
distances yields the refraction indices corresponding to these two
values of $z$. A curve fit to Equ.~\ref{equ:Rdz} then gives the
two constants $A$ and $B$. I am using characteristic but fixed
values for
 the atmospheric parameters hereafter, so the constants $A$ and $B$
are only dependent on the wavelength:
\begin{align*}
T_{0} & = 278~\rm K \\
P_{0} & = 800~\rm mbar \\
H_{0} & = 10~\%
\end{align*}
One can then derive an expression for $A$ and $B$ in seconds of
arc with only a wavelength dependence. This was done by Gubler and
Tytler \citep{Gubler1998} empirically for the K-Band (centered on
$2.2~\mu m$):
\begin{eqnarray}
A(\lambda) & = & 45.95126 + \frac{0.26147}{\lambda^{2}} \\
B(\lambda) & = & -0.05083862 - \frac{0.0002622385}{\lambda^{2}}
\label{equ:AB}
\end{eqnarray}
I used during the observations a narrow-band filter centered at
$2.12 \mu m$. So we could suppress and therefore neglect the color
dependent part of the differential refraction, $\Delta
R_{\lambda}$,  and only corrected for the effect due to the
different zenith distances of the stars, $\Delta R_{z}$. With
$\lambda = 2.12 \mu m$, the two constants
then take the values $A = 46.0094\arcsec$ and $B = -0.050897\arcsec$.\\
Looking at Equ. \ref{equ:dR} one can see that even with the most
extreme values during the observations, $\Delta z_{0} = 1
7.2\arcsec$ and $z_{1} = 48\degr89$, the quadratic term never
exceeds $\sim 1.26~\mu as$. The following expression for $\Delta
R_{z}$ is therefore a good approximation:
\begin{equation}
\Delta R_{z} = \frac{(1 + \tan^{2}{z_{1}})(A_{2} +
 3B_{2}\tan^{2}{z_{1}})}{206265}\Delta z_{0}
 \label{equ:deltaR}
\end{equation}
The factor in the denominator comes from the fact that $\Delta
z_{0}$ is now expressed in arcseconds instead of radians. Also,
$\Delta
R_{z}$ is now expressed in seconds of arc.\\

\noindent As stated above, the values for $A$ and $B$ are
calculated for fixed standard values of atmospheric temperature,
pressure and humidity, $T_{0}, P_{0}, H_{0}$. But in reality, the
actual circumstances during the observations are different.
Depending on how much the actual temperature, pressure and
humidity vary from these standard values, the differential
refraction changes, too. \cite{Gubler1998} investigated this
phenomenon and gave values for the change in $\Delta R$ due to
changes differences of temperature, pressure and humidity form the
standard values. The relation between the parameters is linear,
the differential refraction goes as the inverse of the
ground-level temperature and changing the pressure by a given
fraction changes $\Delta R$ by the same fraction. A change in the
humidity does not change the differential diffraction
significantly. Changing the humidity from 10\% to 100\% only leads
to a change in the differential refraction of a few
micro-arcseconds.\\
\noindent While I took the change in
temperature and pressure into account, I did not correct for any
change in the humidity, as it was close to the standard value of
10\% anyway during all observations. The correction factors were
taken into account as follows. Taking the values given by Gubler
and Tytler (tables 3 and 4) I made a linear least-squares
approximation in one-dimension to the data, yielding values for
the slope $m$ and the interception point with the y-axis $b$ for
the relation between temperature and diff. refraction and pressure
and diff. refraction, respectively:
\begin{align}
 \Delta R_{T} &= m_{T} \cdot \Delta T + b_{T} \\
 \Delta R_{P} &= m_{P} \cdot \Delta P + b_{P}
\label{equ:deltaRTP}
\end{align}
Here $\Delta T$ is the difference to $T_{0}$ in Kelvin, $\Delta P$
the difference to $P_{0}$ in mbar and $\Delta R$ is given in
percent. The derived values for $m$ and $b$ are: $m_{T} = -0.363 \pm 0.049$, $b_{T} = 0.086 \pm 0.409$, $m_{P} = 0.125 \pm 0.018$ and $b_{P} = 0.000 \pm 0.380$. After converting the changes in $\Delta R$ from
percentage to arcseconds, the final differential refraction is
given by:
\begin{equation}
 \Delta R = \Delta R_{z} + \Delta R_{T} + \Delta R_{P}
\label{equ:deltaRztp}
\end{equation}

\subsection{Correction for Differential Refraction}
\label{subsec:DiffRefCorr} With the results from
Equs.~\ref{equ:deltaR} and \ref{equ:deltaRztp}, one can start to
correct the measured positions of the stars for differential
refraction. As this is the last distortion before the light
arrives on the detector, it has to be corrected first and before
the
differential aberration.\\

\noindent The celestial coordinates given for the reference point
in the FITS header are already refraction and aberration
corrected\footnote{Communication with Claudio Mela via ESO USD help}. So one only has to
correct for the differential part of the refraction between the
reference point and the stars. This is done by first undoing the
aberration correction for the reference point (see Chapter
\ref{chap:aberration}), so that one knows at which point the light
rays really entered the atmosphere. With these 'new' coordinates
for the reference point, the celestial coordinates of the stars
are calculated in the way described in Chapter \ref{sec:FITS}.
Then the zenith distances $z_{0}$ of the reference point and the
stars are given by:
\begin{equation}
\cos{z_{0,i}} = \sin{\delta_{i}}\sin{\varphi} +
\cos{\delta_{i}}\cos{\varphi}\cos{HA_{i}} \label{equ:z_dist}
\end{equation}
Here $\varphi$ denotes the local latitude of the observatory,
which is for the VLT/UT4 ~$\varphi = -24\fdg6270$, $\delta$ the
declination of the star and $HA = LST - \alpha$ the local hour
angle of the star. $i$ stands for the different stars. As the
zenith distance changes with time, $HA$ and $z_{0}$ were
calculated over the whole observing time of one epoch. This means
for every single frame of the field in that epoch, which leaves me
with 25 (35 for epoch 1, see Tab.~\ref{tab:observations}) values
for $HA$ and $z_{0}$ for each night. I therefore introduce another
index, $j$, which stands for the different frames at an individual epoch.\\
\begin{figure}
 \begin{center}
  \includegraphics[width=14cm]{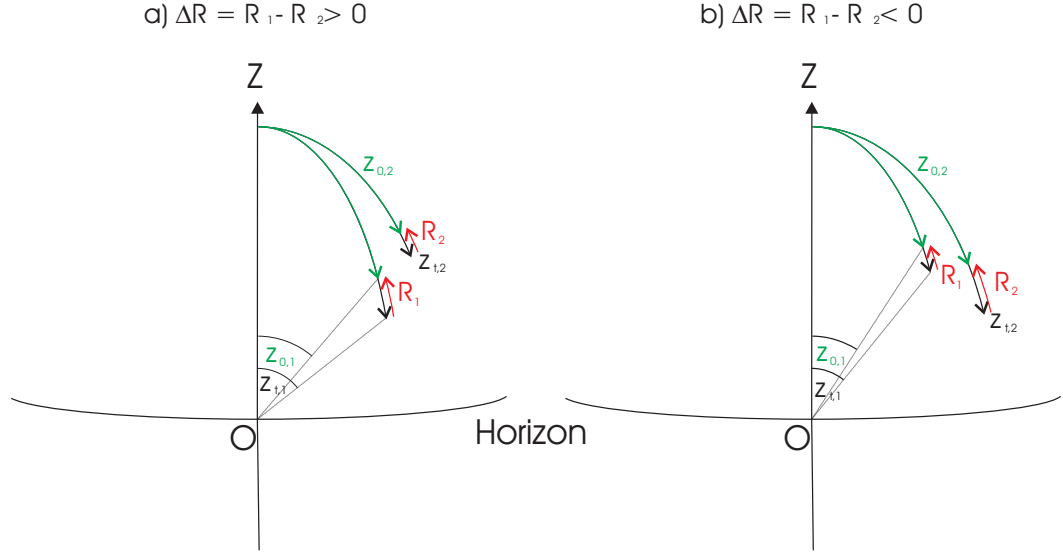}
 \end{center}
\caption{Differential refraction}
\label{Delta_ref}
\end{figure}

\begin{figure}[t!]
 \begin{center} \hspace{0.8cm}
  \includegraphics[width=12cm, angle=90]{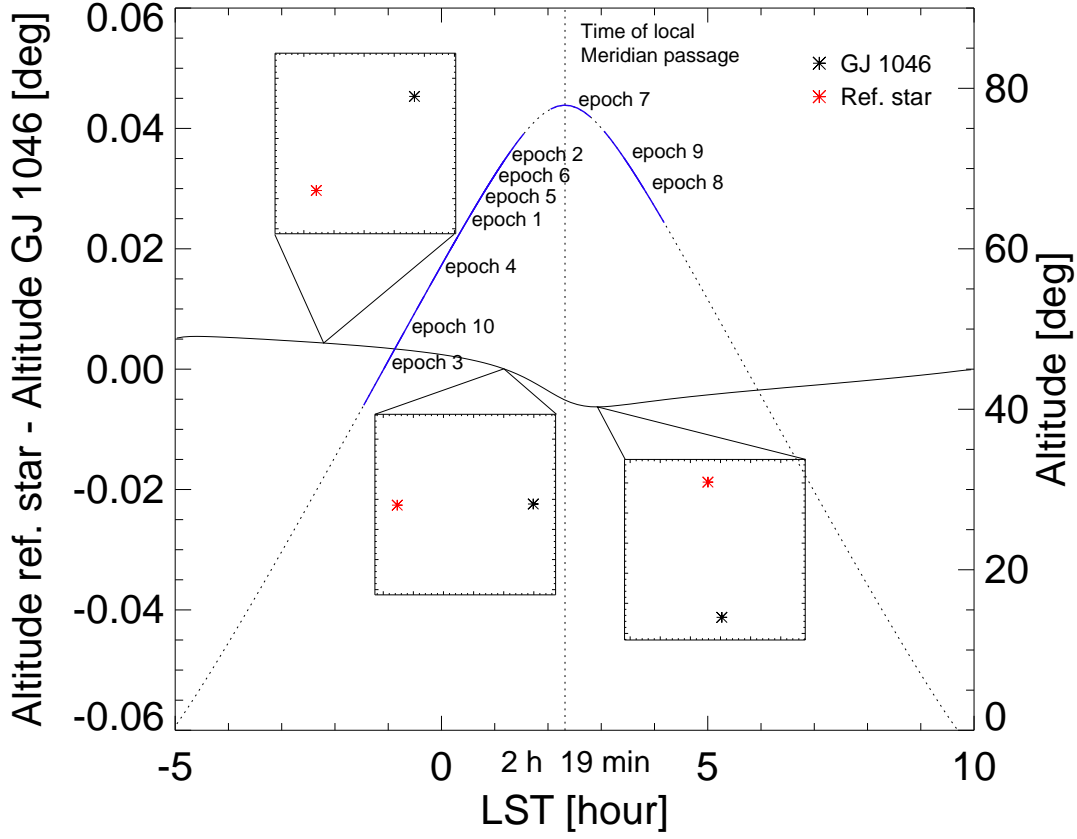}
 \end{center}
 \caption[Sign of the differential refraction correction during
 different times of observations]{Sign of the differential refraction
 correction during different times of observations. Plotted as the
 solid curve is the difference in altitude of the two stars in the
 field: \textit{Alt(reference star) - Alt(GJ 1046)}. The dotted curve
 shows the visibility of the objects above the local horizon at the
 Paranal Observatory in mid of October. The right x-axis gives the
 altitude in degree. The vertical dashed line marks the time of the
 local Meridian Passage of GJ~1046. The small inlets show the
 orientation
 of the two stars at different phases of their visibility; the left
 one, when the reference star is at a lower altitude, the middle one
 when they are nearly at the same height above the horizon and the
 right on when GJ~1046 is at a lower altitude. Also shown in blue
 are the times of observation in LST for all epochs.}
\label{fig:zenithdistance}
\end{figure}

\noindent Now one can calculate $\Delta z_{0_{i,j}} =
z_{0_{ref,j}} - z_{0_{i,j}}$, and after that $\Delta R_{i,j}$ for
each star, using Equ.~\ref{equ:deltaRztp} and $z_{1} =
z_{0_{ref,j}}$ in Equ.~\ref{equ:deltaR}. $\Delta T$ was in the
range between 3.72 and 8.71 K, and $\Delta P$  between -57.3 and
-54.7 mbar for all epochs. I calculated $\Delta R_{T{i,j}}$ and
$\Delta R_{P{i,j}}$ for each frame in each epoch with the
parameters $m_{T}, m_{P}, b_{T}$ and $b_{P}$ derived from the
linear fits. The sign of $\Delta z_{0}$ and therefore $\Delta R$
depends on the time the observation was conducted and therefore
the orientation of the star asterism on the local sky relative to
the reference point. If the zenith distance $z_{0_{ref}}$ of the
reference point is bigger than the zenith distance of any star
$z_{0_{i}}$, then $\Delta z_{0_{i}} ~\textgreater~0$ and also
$\Delta R_{i}~\textgreater~0$ for this star. If
$z_{0_{ref}}~\textless~z_{0_{i}}$ then $\Delta z_{0_{i}},\Delta
R_i~\textless~0$. In Figure \ref{Delta_ref} a and b, the two cases
are shown. Depending on whether the observation was carried out
before, after, close to or even during the passing of the stars
through the local meridian, the time-depending behavior of $\Delta
z_{0_{i}}$ and all the following parameters is different. In
Fig.~\ref{fig:zenithdistance} one can see different cases during
different phases of the visibility of the stars. Plotted as the
solid curve is the difference in altitude of the two stars in the
field: \textit{Alt(reference star) - Alt(GJ 1046)}. The dotted
curve shows the visibility of the objects above the local horizon
at Paranal Observatory in mid-October. The right $y$-axis gives
the altitude in degrees. The vertical dashed line marks the time
of the local Meridian Passage of GJ~1046. Due to the slightly
different coordinates of the two objects, they \textit{overtake}
each other during the night, leading to a change in the sign of
the differential refraction correction. The reference point of the
detector, to which the differential refraction is measured, lies
between the two stars. The small inlets show the configuration of
the two stars at different phases of their visibility; the left
one, when the reference star is at a lower altitude, the middle
one when they are nearly at the same height above the horizon and
the right on when GJ~1046 is at a lower altitude. Also shown are
the times of observation in LST (in blue) for all individual
epochs.

\noindent The true zenith distance of the stars is then:
\begin{equation}
 z_{t_{i,j}} = z_{0_{i,j}} + \Delta R_{i,j}
\end{equation}
Now one needs to recalculate $\alpha$ and $\delta$ of the stars. Here
the fact that the refraction changes the zenith distance, but not
the azimuth of the stars helps. The azimuth $AZ$ of a given frame
in a given observation for a star is:
\begin{equation}
 \cos{AZ} = \frac{(\sin{\delta} - \cos{z_{0}}\sin{\varphi})}{\sin{z_{0}}\cos{\varphi}}
\end{equation}
Rearranging this equation and replacing $z_{0}$ with $z_{t}$ lets
one calculate the declination corrected for differential
refraction:
\begin{equation}
 \delta_{corr_{i,j}} = \sin^{-1}\left({\cos{z_{t_{i,j}}}\sin{\varphi} + \sin{z_{t_{i,j}}}\cos{\varphi}\cos{AZ_{i,j}}}\right)
\end{equation}
To compute the right ascension one needs to recalculate the hour
angle $HA$ first, as $\alpha_{i,j} = LST_{j} - HA_{i,j}$:
\begin{equation}
 HA_{i,j} = \cos^{-1}\left( \frac{\cos{z_{t_{i,j}}} - \sin{\delta_{corr_{i,j}}}\sin{\varphi}}{\cos{\delta_{corr_{i,j}}}\cos{\varphi}}\right)
\end{equation}
From this I calculated offsets from the positions in right
ascension and declination $(\alpha, \delta)$ without the
differential refraction correction:
\begin{equation}
\Delta \alpha_{i,j} = \alpha_{i,j} - \alpha_{corr_{i,j}}
\end{equation}
Again I had 25 (35) values of $\Delta \alpha$ for each star. To
get one value for correction I calculated the mean change in
position, $\Delta\alpha_i = \frac{\sum_{\scriptstyle
j=1}^{\scriptstyle N} \Delta\alpha_{i,j}}{N}$,  N = 25 (35), same
for $\delta$, and the corresponding error of the mean value,
$\sigma_{\alpha_{i}}/\sqrt{N}, ~\sigma_{\delta_{i}}/\sqrt{N}$,
where $\sigma_{\alpha_{i}}$ and $\sigma_{\delta_{i}}$ are the
standard deviations of the $\Delta\alpha_{i,j}$ and
$\Delta\delta_{i,j}$. The coordinates corrected for differential
refraction are then derived by:
\begin{eqnarray}
 \alpha_{corr_{i}} & = & \alpha_{i} - \Delta \alpha_{i}\\
 \delta_{corr_{i}} & = & \delta_{i} - \Delta \delta_{i}
\end{eqnarray}
In Tab.~\ref{tab:deltaRADEC} the corrections in right ascension
and declination are listed for the two stars in the target field
for each epoch. Now one has the coordinates of the stars corrected
for differential  refraction. The next step is then to correct for
differential aberration.
\begin{table}[t!]
 \begin{center}
  \begin{tabular}{|c|cc|cc|} \hline
     & \multicolumn{2}{c|}{GJ~1046} & \multicolumn{2}{c|}{reference star} \\
    epoch & $\Delta\alpha$~[mas] & $\Delta\delta$~[mas] & $\Delta\alpha$~[mas] & $\Delta\delta$~[mas] \\ \hline
    1  & ~~$0.678\pm 0.141$ &  $-0.126\pm 0.023$ & $-1.891\pm 0.149$ & ~~$0.510\pm 0.009$\\
    2  &  $-0.006\pm 0.066$ & ~~$0.043\pm 0.037$ & $-0.441\pm 0.010$ & ~~$0.195\pm 0.046$\\
    3  & ~~$3.553\pm 0.099$ &  $-0.034\pm 0.019$ & $-5.740\pm 0.149$ & ~~$0.057\pm 0.030$\\
    4  & ~~$1.782\pm 0.061$ &  $-0.272\pm 0.004$ & $-3.114\pm 0.092$ & ~~$0.478\pm 0.010$\\
    5  & ~~$1.998\pm 0.059$ &  $-0.275\pm 0.006$ & $-1.898\pm 0.087$ & ~~$0.531\pm 0.005$\\
    6  & ~~$0.349\pm 0.068$ &  $-0.122\pm 0.025$ & $-0.993\pm 0.103$ & ~~$0.393\pm 0.028$\\
    7  & ~~$0.248\pm 0.141$ & ~~$2.123\pm 0.068$ & $-0.358\pm 0.184$ &  $-2.732\pm 0.109$\\
    8  & ~~$2.902\pm 0.021$ & ~~$1.186\pm 0.046$ & $-4.251\pm 0.041$ &  $-1.733\pm 0.063$\\
    9  & ~~$2.780\pm 0.039$ & ~~$1.540\pm 0.057$ & $-3.960\pm 0.066$ &  $-2.186\pm 0.047$\\
    10 & ~~$3.475\pm 0.101$ &  $-0.172\pm 0.017$ & $-4.743\pm 0.132$ & ~~$0.237\pm 0.024$\\ \hline
  \end{tabular}
  \caption[Differential refraction corrections]{Differential
  refraction corrections applied to the two stars in the target
  field. Values are given for correction in right ascension and
  declination plus the error arising from the change in zenith
  distance of the stars during the observation. The total
  differential effect between the two stars is the difference
  between their values.}
\label{tab:deltaRADEC}
 \end{center}
\end{table}

\subsection{Errors from Differential Refraction Correction}
Correcting for differential refraction can introduce several error
terms:
\begin{itemize}
\item The first one, which we already showed, is the error coming
from neglecting the quadratic term in Equ.~\ref{equ:dR}. This
error is even for our most extreme observation configurations only
$\sim 1.26~\mu as$ and well
below our aimed precision.\\
\item Another error source are the not perfectly known celestial
coordinates of the stars to calculate $z_{0}$. Although the given
absolute coordinates may not have a precision better than a few
arcseconds, this is no problem as the error in the differential
refraction is only about $\sim 5~\mu as$ if the zenith distance
$z_{0}$ is only known to a precision of $0\degr01 = 36\arcsec$
\citep{Gubler1998}. This is well within the pointing accuracy of
the VLT. Also, I am not aiming for absolute astrometry to obtain
the orbital motion due to the companion, but instead only need
relative astrometry to the reference star in the
image.\\
\item The biggest error
term comes from the linear fit made to
calculate the correction factors due to different temperature and
pressure during the observations. The deviation of the data from a
linear relation can be used to calculate errors for the derived
slope and intersection values, which then lead to errors of the
correction factor in $\Delta R$ of $\sim 0.5 - 1.0$\%. This
translates into a small error of the calculated $\Delta
\alpha_{ij}$ and $\Delta \delta_{ij}$. The errors for the
coordinate offsets  $\Delta \alpha_{i}$ and $\Delta \delta_{i}$
are then calculated by $\sqrt{1/\sum_{\scriptstyle
j=1}^{\scriptstyle N} \frac{1}{\sigma_{i,j}}}$, where
$\sigma_{i,j}$ stands for the single errors of the $\Delta
\alpha_{ij}$ and $\Delta \delta_{ij}$. The errors arising from this fit are between 0.129 - 1.779 mas for $\Delta
\alpha_{i}$ and between 0.050 - 0.313 mas for $\Delta \delta_{i}$. The large upper limit results from one epoch, where the temperature and pressure values were deviating a lot stronger than at the other observing epochs. The errors of all other epochs are below the milli-arcsecond range. Still the errors resulting from the correction due to differing atmospheric conditions are pretty high and probably need a better model to correct for if one wants to go down to small micro-arcsecond precision.
\item Additional errors arise from the change of the zenith distance during the
observation. The observations took around 45 min for the target
field, which lead to a change in zenith distance depending on the
time of observation, before, close to or after the crossing of the
targets through the local meridian. The result is a variation of
$z_{0}$ between 0.5\degr ~in the best case (epoch 7) and 16.1\degr
~in the worst case (epoch 1), which translates into $0.35 -2.29 \rm ~mas$
peak-to-peak difference in the correction $\Delta \alpha_{ij}$ and
$0.07 - 0.98 \rm ~mas$ in $\Delta \delta_{ij}$ for the different
epochs. The larger the zenith distance the stronger the
refraction, because of the higher airmass the light has to travel
through.
The error from calculating the mean value for $\Delta \alpha_{i}$
and $\Delta \delta_{i}$ varies therefore between $21 - 184~\mu as$
and $4 - 109~\mu as$ for right ascension
and declination, respectively, see also Tab.~\ref{tab:deltaRADEC}.\\
This error and the error from the correction necessary due to the
different temperature and pressure are taken into account by
quadratic addition with the error from the PSF-Fit.
\end{itemize}

\subsection{Theory of Aberration}
\label{chap:aberration}

\begin{figure}
 \begin{center}
  \includegraphics[width=8cm]{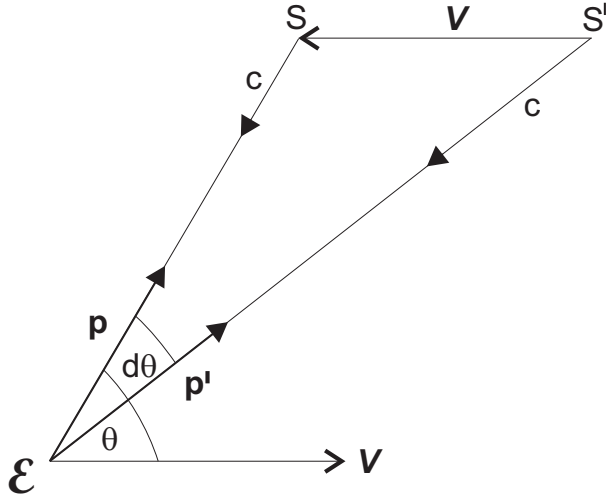}
 \end{center}
 \caption{Stellar aberration}
 \label{fig:Aberration}
\end{figure}

The observer's velocity through space relative to an observed
object is responsible for a phenomenon called \textit{Bradley
aberration} or \textit{Stellar aberration}. It was discovered by
James Bradley in 1727 \citep{Bradley1727}. He was trying to
measure the stellar parallax of g Draconis, in order to confirm
the Copernican theory of the Solar System, as it proves the motion
of the Earth. Instead he measured an annual variation which was
not consistent with the expected parallax. The variation was
strongest for stars in the direction perpendicular to the orbital
plane of the Earth. As $\gamma$ Draconis passes right through the
zenith in Greenwich, where Bradley did his observations, it was by
chance a perfect target to detect stellar aberration. Bradley
concluded rightly that the displacement he saw was not due to
changes in the Earth's position, but rather due to changes in
Earth's velocity. The aberration is caused by the relative
velocity of the source, e.g. star, and the observer and the finite
speed of light during the light travel time. The relative
positions between the observer on Earth and the source change and
the light seems to be coming from a direction different from the
direction from which it was
emitted.
The movement of the observer can be divided into three different
motions leading to the following three effects:
\begin{itemize}
 \item The diurnal aberration caused by the daily rotation of the Earth
 \item The annual aberration, due to the movement of the Earth around
 the barycenter of the Solar System
 \item The secular aberration due to the motion of the Solar System barycenter in space
\end{itemize}
The secular aberration is a displacement due to the relative
motion of the stars and the Solar System barycenter and is equal
to the proper motion of the stars multiplied by the light time.
Since this is rarely well known and the barycentric position of a
star is mostly of marginal
 interest, this effect is normally ignored.\\
The absolute effect of the annual aberration can be approximated
by
\begin{equation}
 A_{year} \approx k_{year} \sin{\theta}
\end{equation}
Here $k = V/c \approx 10~km~s^{-1}/300000~km~s^{-1} \approx 20.5\arcsec$ is
the annual aberration constant and $\theta$ is the angle between
the velocity vector of the Earth and the direction of the light
coming from the star, see Fig. \ref{fig:Aberration}. The
differential effect between two stars at a separation of 30\arcsec
~can then reach up to 3 mas over one year and needs to be
corrected for. The differential effect of the diurnal aberration
is smaller, around $42~\mu as$ per
day.\\
But to derive and correct the effect of the aberration precisely
one needs to take the relativistic addition of velocities into
account, where the distinction between annual and diurnal
aberration is not possible. This is done by applying the Lorentz
transformation between two sets of coordinates, $x, y, z,
(\mathbf{r})$ and time $t$ in a given reference frame $S$, and $x', y', z', (\mathbf{r'})$ and $t'$ in another
frame $S'$ which is moving with the constant velocity $\mathbf{V}$
with respect to $S$. The generalized Lorentz transformation from
the moving frame to the stationary one can then be written as:
\begin{align}
 \mathbf{r} & = \mathbf{r'} + \gamma \mathbf{V}t' + (\gamma - 1)\frac{\mathbf{V}(\mathbf{V}\cdot\mathbf{r'})}{V^2}\\
 t & = \gamma\left( t' + \frac{\mathbf{V}\cdot\mathbf{r'}}{c^2}\right)
 \label{equ:rt}
\end{align}
here $\gamma = (1 - V^2/c^2)^{-1/2}$, is the Lorentz factor.  A
point $\mathbf{r}$ in the stationary system $S$ has the velocity
$\mathbf{U} = \text{d}\mathbf{r}/\text{d}t$. In the moving system
its coordinates are $\mathbf{r'}, t'$ and its velocity would be
$\mathbf{U'} = \text{d}\mathbf{r'}/\text{d}t'$. Now one can
differentiate Equ.~\ref{equ:rt}:
\begin{align}
 \frac{\text{d}\mathbf{r}}{\text{d}t} & = \frac{\text{d}\mathbf{r'}}{\text{d}t'}\frac{\text{d}t'}{\text{d}t} + \gamma~\mathbf{V}\frac{\text{d}t'}{\text{d}t} + (\gamma-1)\left( \mathbf{V}\cdot\frac{\text{d}\mathbf{r'}}{\text{d}t'}\frac{\text{d}t'}{\text{d}t}\right) \frac{\mathbf{V}}{V^2}\\
 \text{d}t & = \gamma\left( \text{d}t' + \left( \mathbf{V}\cdot\frac{\text{d}\mathbf{r'}}{\text{d}t'}\text{d}t'\right) /c^2\right)
\end{align}
And with the new notation
\begin{equation}
 \frac{\text{d}t'}{\text{d}t} = \frac{1}{\gamma(1 + \mathbf{V}\cdot\mathbf{U'}/c^2)}
 \label{equ:dt'/dt}
\end{equation}
one can than write:
\begin{equation}
 \mathbf{U} = \frac{\mathbf{U'} + \gamma\mathbf{V} + (\gamma -1)(\mathbf{V}\cdot\mathbf{U'})\mathbf{V}/V^2}{\gamma(1 + \mathbf{V}\cdot\mathbf{U'}/c^2)}
 \label{equ:U}
\end{equation}
This is the formula one needs to correct for aberration.
Now one can apply this result to the case of the moving Earth. The
velocity of the Earth, $\mathcal{E}$ with respect to the
stationary stars is $\mathbf{V}$. In Fig. \ref{fig:Aberration},
the unit vector $\mathbf{p}$ is in the direction to the geometric,
true position of an object $S$ at time $t$ in the stationary
frame. The velocity vector of the light coming from this object in
the very same frame is then $-c\mathbf{p}$. But the observer on
Earth will see the object at time $t$ at the position $S'$ whose
direction is defined by adding the velocities of the Earth and the
light coming from $S$, $-c\mathbf{p'} = \mathbf{V} + c\mathbf{p}$,
with the unit vector $\mathbf{p'}$. Summarizing the important
notations
\begin{itemize}
 \item velocity of light in the geometric direction: $-c\mathbf{p}$
 \item velocity of light in the apparent direction: $-c\mathbf{p'}$
 \item velocity of the observer on Earth: $\mathbf{V}$
\end{itemize}
and applying them to the previously derived equation \ref{equ:U} one obtains:
\begin{equation}
 c\mathbf{p} = \frac{c\mathbf{p'} - \gamma\mathbf{V} + (\gamma - 1)(\mathbf{V}\cdot c\mathbf{p'})\mathbf{V}/V^2}{\gamma(1 - c\mathbf{p'}\cdot\mathbf{V}/c^2)}
\end{equation}
Dividing by $c$ and expanding by $\gamma/\gamma$ one gets:
\begin{equation}
 \mathbf{p} = \frac{\gamma^{-1}\mathbf{p'} - \mathbf{V}/c + (1 - \gamma^{-1})(\mathbf{V}\cdot \mathbf{p'})\mathbf{V}/V^2}{(1 - \mathbf{p'}\cdot\mathbf{V}/c)}
\end{equation}
Using the definition of $\gamma = (a - V^2/c^2)^{-1/2}$ and writing
\begin{equation}
 (1 - \gamma^{-1})(1 + \gamma^{-1}) = \mathbf{V}^2/c^2
\end{equation}
one finally gets the vector in the geometric direction of the
aberration corrected coordinates:
\begin {equation}
 \mathbf{p} = \frac{\gamma^{-1}\mathbf{p'} - \mathbf{V}/c + (1 + \gamma^{-1})(\mathbf{p'}\cdot\mathbf{V}/c )(\mathbf{V}/c)}{(1 - \mathbf{p'}\cdot\mathbf{V}/c)}
 \label{Equ:p}
\end {equation}
The inverse formula is:
\begin {equation}
 \mathbf{p'} = \frac{\gamma^{-1}\mathbf{p} + \mathbf{V}/c + (1 + \gamma^{-1})(\mathbf{p}\cdot\mathbf{V}/c )(\mathbf{V}/c)}{(1 + \mathbf{p}\cdot\mathbf{V}/c)}
 \label{Equ:p'}
\end {equation}
In the following correction for aberration these last two
equations are used.

\subsection{Correction for Differential Aberration}
\label{subsec:diff_aber_corr} Looking at the coordinates given for
the reference point in the FITS-Header, the celestial coordinates
given by the keyword CRVAL\textit{i} are already corrected for
aberration and refraction. So the only thing I have to take care
of, is the correction for the differential effect between this
reference point and the stars. The correction for differential
refraction was already done in the previous step, see
Sec.~\ref{subsec:DiffRefCorr}. I did not correct for the
differential effect of the aberration directly, as this is not
possible, but calculated for each star the celestial coordinates
without aberration correction and then corrected for each star for
total aberration. This directly corrects also for the differential
effect. This is done in two steps.\\
First I calculated the celestial coordinates of the reference
point before they were corrected for aberration. So in practice,
I undid the correction for aberration. This was done the
following way:\\
A set of three linearly independent unit-vectors, $X, Y, Z$ was
calculated from the celestial coordinates, $\alpha_{1},
\delta_{1}$ of the reference point in the aberration corrected
frame:
\begin{align}
X & = \cos{\delta_{1}} * \cos{\alpha_{1}} \notag \\
Y & = \cos{\delta_{1}} * \sin{\alpha_{1}} \label{equ:unitvec} \\ 
Z & = \sin{\delta_{1}} \notag
\end{align}
$X, Y, Z$ describe the equatorial system, therefore $X \rm ~and
~Y$ lie in the equatorial plane and $X$ points to the vernal
point with $\alpha = 0$ hours and $Y$ points to the direction of $\alpha = 6$ hours. $Z$ points to the north pole. Then the current velocity of
the Earth is extracted from the JPL DE405
ephemerides\footnote{http://ssd.jpl.nasa.gov/,
http://cow.physics.wisc.edu/$\sim$craigm/idl/ephem.html}. This set
of ephemerides is the most recent and precise one. It includes
nutations and librations and is tied to the International Celestial
Reference Frame (ICRS) through VLBI observations of the Magellan
spacecraft in orbit around Venus. The origin of the coordinate
system is the barycenter of the Solar System. \\
\noindent To calculate the current velocity of the observer on
Earth's surface, one needs to take into account the position of
the observer on the Earth. The observer moves eastwards, as the
Earth rotates. For points on the celestial sphere which pass the
local Meridian, the right ascension is equal to the Local Siderial
Time (LST). As the direction to the East, the direction of the
movement of the observer on Earth, is perpendicular to the
Meridian, a factor of six hours has to be added to the LST. The
velocity in x and y direction then amounts to:
\begin{align}
V_{x} & = V_{x} + V_{r} * \cos{(LST + 6)} \notag \\
V_{y} & = V_{y} + V_{r} * \sin{(LST + 6)} \notag
\end{align}
where $V_{r}$ denotes the rotational velocity of the Earth at the
position of the observer. The z direction is not influenced.
Taking Equ.~\ref{Equ:p'} and splitting $\mathbf{p}$ into its
components $X, Y, Z$, we can calculate the new vector $\mathbf{p'}
= (X', Y', Z')$. After dividing $X', Y', Z'$ with the length of
the calculated vector, $\mid{\mathbf{p'}}\mid = X'*X' + Y'*Y' +
Z'*Z'$, one has the now unaberrated linearly independent unit
vectors $X', Y', Z'$ for the reference point. The declination and
right ascension can then simply be computed by:
\begin{align}
\delta_{0} & = \sin^{-1}{Z'} \notag \\
\alpha_{0} & = \tan^{-1}{\frac{Y'}{X'}}
\label{equ:sinZ}
\end{align}
Here the coordinates $\alpha_{0}, \delta_{0}$ are the celestial
coordinates of the reference point before aberration correction.\\
\noindent I updated the CRVAL\textit{i} values in the FITS header
with these coordinates and calculated the celestial coordinates of
the other stars in the field. These coordinates are then taken to
be corrected for differential refraction afterwards as described
in
Sec.~\ref{subsec:DiffRefCorr}.\\
\noindent Now one can proceed with the correction for differential
aberration. Each star's set of celestial coordinates is
transformed into their corresponding set of linearly independent
vectors $X', Y', Z'$. But this time Equ.~\ref{Equ:p} is applied to
calculate the vector $\mathbf{p'} = (X, Y, Z)$ in the geometric
direction of the velocity of light. These Cartesian coordinates
are finally transformed with Equ.~\ref{equ:sinZ} (with only $X',
Y', Z'$ exchanged by $X, Y, Z$) into $\alpha, \delta$ of celestial
coordinates now corrected for differential refraction and
differential aberration.

\subsection{Errors from Differential Aberration Correction}
To calculate the effect of aberration on the coordinates of an
object and then correct for it, one has to know the apparent
position of the object. The pointing accuracy of VLT/NACO is at
least accurate to one second of arc. To estimate the error of the
correction of the differential aberration effect we looked at the
effect of the total aberration. During our observations the total
amount of aberration correction was $d\alpha\sim 10\arcsec$ in
right ascension and $d\delta\sim 12\arcsec$ in declination. We
took these values to estimate the error in the differential effect
originating from the inaccuracy of knowing the exact apparent
celestial coordinates. We changed the reference coordinates from
the reference point by $\alpha_{0}\pm d\alpha$ and $\delta_{0}\pm
d\delta$ and calculated the correction again. The absolute values
of the coordinates for the stars changed by roughly the amount we
added and subtracted, as expected. But the measured distance
between the stars almost did not change. The error in the
differential aberration correction, introduced by an error in the
absolute coordinates is therefore very small. It is in the range
of a few micro arcseconds in both directions of right ascension
and declination.\\

\subsection{Light Time Delay}
\label{subsec:LTD} Light emitted at a time $T_{1}$ takes longer or
shorter to reach the observer than light emitted at an earlier
time $T_{0}$:
\begin{equation}
 \Delta T = (D(T_{1}) - D(T_{0}))/c
\end{equation}
Where $D$ is the distance of the star and $c$ the speed of light.
At the time of the second observation the star has moved a little
bit closer or further away due to its radial motion in space. Assuming two
stars in a field with different radial velocities, then their
distances to us change by a different amount. The light reaching
us at the same time during the observation was emitted at two
different times in the past and also this time difference changes
over time, as the two stars move with their own spatial velocity.
Together with the proper motion, this leads to a steady change in
angular separation observed on the sky. This light time delay
typically is of the order of hours or days, leading to a change in
angular separation of about
$10-100\mu as$ over the course of a few years\footnote{Numbers taken from the lecture
\textit{Modern Astrometry: Methods and Applications} given by S.
Reffert at the University of Heidelberg, Germany in 2008}.\\
\noindent Let us assume a radial velocity difference for my target
star and the reference star of 20 km/s and a maximum epoch
difference of the observations of 1.5 yrs. That would mean that
the two stars are $9.5\cdot 10^8$~km further apart (or closer
together) in the radial direction at the last observation compared
with the first one. This then corresponds to a light travel time
of $\Delta T = 53~min$. With a proper motion difference of $\sim
1500~mas/yr$ one star travels within this 53 min only $150\mu as$
with respect to the other. So this is the changing angular
separation between the two stars over 1.5 yrs and it is way below
the aimed measurement precision, especially as the effect is even
smaller between the single epochs. I can therefore neglect this
effect in my observations and data analysis.

\subsection{Differential Tilt Jitter}
As described previously (Sec.~\ref{sec:AO}) a tip/tilt mirror is
used in adaptive optics observations to compensate for the image
motion of the guide star. The image of the guide star is
stabilized with high accuracy with respect to the imager. But as
the guide star is not necessarily the target star, or like in my
case more than one star in the field is of interest, an effect
known as Differential Tilt Jitter (DTJ) comes into play. The
difference of the tilt component of turbulence along any two lines
of sight in the FoV causes a correlated, stochastic change in
their measured separation. Light from the target and the reference
star passes through different columns of atmospheric turbulence
that are sheared. Arising from this shearing effect, the
decorrelation in the tilt component of the wavefront phase
aberration yields the DTJ. This leads to a fluctuation in the
relative displacement of the two objects, which is random,
achromatic and isotropic \citep{Cameron2009}. Turbulence at higher
altitudes contributes most to the differential tilt jitter as the
light from the two stars traverses more through common parts of
the atmosphere near the ground. The effect is bigger along the
separation axis than perpendicular to it. The longer the exposure
time of the observation, the more the effect of differential tilt
jitter averages out. Cameron et al. showed that in their 1.4
seconds exposure already part of the effect averaged out, as the measured magnitude of the tilt jitter was smaller than the expected one from theoretical models.\\
As I have exposure times of 99 sec for the target field and 4 sec
for the reference field, plus I add all single frames to one
image before measuring the positions and distances between the
stars, I assume the effect to be averaged out in the images.

\subsection{Parallax}
\label{subsec:parallax}
The stars one observes are not at infinity and as a result of
their finite distance, the motion of the observer through space
produces a displacement of the position of the star with respect
to the fixed reference system. This parallactic displacement is
due to several motion components of the observer:
\begin{itemize}
 \item The motion of the geocenter around the barycenter of the
 Earth-Moon system
 \item The motion of the barycenter of the Earth-Moon system
 around the barycenter of Solar system. This combination of
 motions results in a yearly periodic displacement called
 \textit{annual parallax} and a small monthly component and
  is due to the change in viewing angle from different
  positions on Earth's orbit.
 \item The motion of the observer around the center of mass
 of the Earth, caused by the daily rotation of the Earth.
 This motion is called the \textit{diurnal parallax}.
\end{itemize}
\begin{figure}
 \begin{center}
 \includegraphics[width=14.5cm]{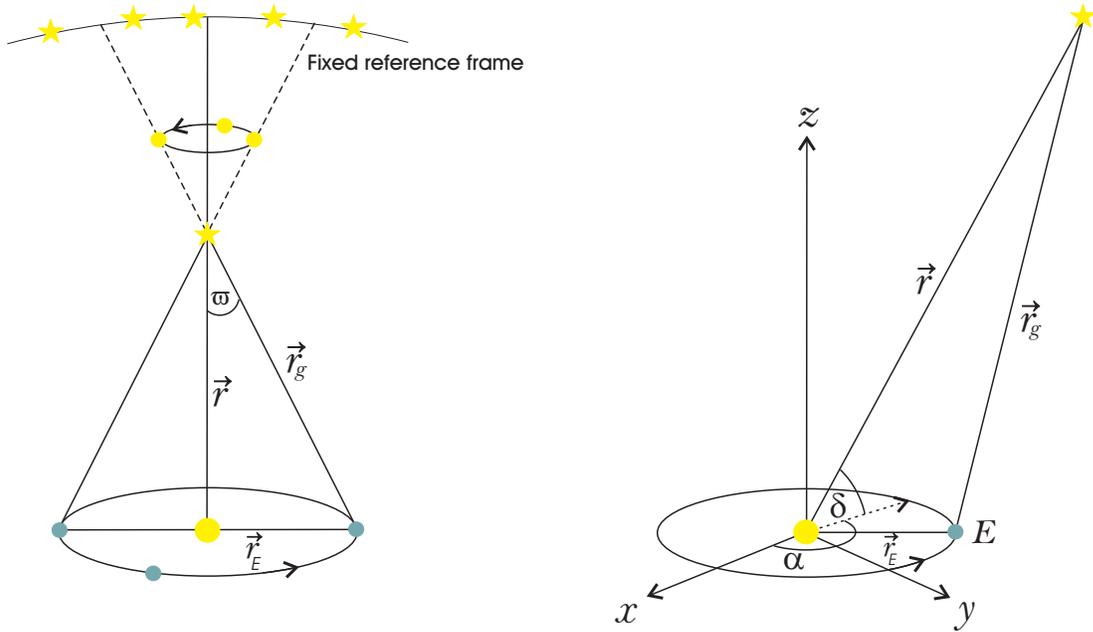}
\end{center}
\caption{Parallax movement and correction of a star's position}
\label{fig:parallax}
\end{figure}
The stellar or trigonometric parallax $\varpi$ is known as the motion of a
star with respect to a fixed reference system, e.g. background
stars, over one year.
\begin{equation}
 \varpi = \frac{k}{r}
\end{equation}
with $k = \pi/648000$ and $r$, the distance of the star, expressed
in astronomical units (AU) and $\varpi$ the angle between the
vectors $\vec{r}_{g}$ and $\vec{r}$ in Fig.~\ref{fig:parallax},
left side. The denominator of k erases from the transformation of
$\varpi$ from radians into arcseconds. More conveniently is the
expression of $r$ in parsec, $1~pc = 648000\rm ~AU/\pi = 206264.81
\rm AU$. The parallax can then be written as ~$\varpi = 1~ AU/r$
and one parsec corresponds to the distance under which an 1~AU
displacement is seen as a $1\arcsec$ angle. The apparent
geocentric motion of the star is an ellipse with its semi-major
axis equal to the parallax if in a first approximation the Earth
orbit is described as an ellipse with its semi-major equal to one
AU. As one can see, the parallax decreases with the distance of
the source. The approximation $\tan{\varpi}\approx \varpi$ made
here is good to about $2\cdot 10^{-17} \text{radians} \approx
10^{-7} \text{mas}$\footnote{Numbers taken from the lecture
\textit{Modern Astrometry: Methods and Applications} given by S.
Reffert at the University of Heidelberg, Germany in 2008}, so in
practice it can be used for all parallax
applications.\\

\noindent The diurnal parallax effect is smaller than the annual
one by a factor equal to the ratio of Earth's radius $(R_{E} =
6378 \rm km)$ and the astronomical unit $(1~AU = 1.496\cdot 10^{8}
\rm km)$. For the parallax motion of my target star (71.56~mas)
this would mean a correction of $3\mu as$, which is well below the
aimed precision in the measurements, so we can use the geocentric
coordinates for our parallax correction.\\

\noindent To correct for all motion effects at once, one should
use precise Earth ephemerides, giving the exact position of the
Earth, $\vec{r} = \big(x_{E}(t), y_{E}(t), z_{E}(t)\big)$ in the
above defined system of unit vectors (Equ.~\ref{equ:unitvec}), at
the time $t$ of observation with respect to the barycenter of the
Solar system. The movement of the Earth should never be
approximated as a circle. The position of a star in barycentric,
$\vec{r}$, and geocentric, $\vec{r}_{g}$, coordinates is given by:
\begin{align}
\vec{r} & = r \hspace{0.2cm}
\begin{cases}
    \cos{\delta}\cos{\alpha} & \hspace{4.5cm}\scriptstyle \alpha:~ \rm right~ ascention \notag \\
    \cos{\delta}\sin{\alpha} & \hspace{4.5cm}\scriptstyle \delta:~ \rm declination \notag \\
    \sin{\delta} & \hspace{4.5cm}\scriptstyle r:~ \rm distance \notag
\end{cases} \\
\vspace{1.0cm}
\vec{r}_{g} & = r_{g}
\begin{cases}
    \cos{\delta_{g}}\cos{\alpha_{g}} \notag \\
    \sin{\alpha_{g}}\cos{\delta_{g}} & \hspace{4.2cm}\scriptstyle \rm g:~geocentric \notag \\
    \sin{\delta_{g}} \notag
\end{cases}
\end{align}
From Fig.~\ref{fig:parallax} one can see that
\begin{equation}
 \vec{r} = \vec{r}_{g} + \vec{r}_{E}
\label{equ:r}
\end{equation}
With the approximation of $r\approx r_{g}$ and $r = 1/\varpi$ and
plugging this into Equ.~\ref{equ:r} one obtains:
\begin{subequations}
\begin{align}
 \cos{\delta}\cos{\alpha} & = \cos{\delta_{g}}\cos{\alpha_{g}} + x_{E}\varpi \\
 \cos{\delta}\sin{\alpha} & = \cos{\delta_{g}}\sin{\alpha_{g}} + y_{E}\varpi \\
 \sin{\delta} & = \sin{\delta_{g}} + z_{E}\varpi
 \end{align}
\label{equ:paralla}
\end{subequations}
Introducing the parallactic corrections, $\Delta\alpha$ and
$\Delta\delta$, which need to be added to the geocentric
coordinates to derive the barycentric coordinates $(\alpha,
\delta)$ we have:
\begin{subequations}
\begin{align}
 \alpha & = \alpha_{g} + \Delta\alpha & \Leftrightarrow & & \Delta\alpha &= p_{\alpha}\varpi \label{equ:pialpha} \\
 \delta & = \delta_{g} + \Delta\delta & \Leftrightarrow & & \Delta\delta &= p_{\delta}\varpi \label{equ:pidelta}
\end{align}
\end{subequations}
with $p_{\alpha}, p_{\delta}$ called the parallax factors. These
factors define the shape of the parallax ellipse, depending on the
position on sky and $\varpi$ defines the size of this ellipse. To
derive them one has to do some transformations and small angle
approximations to Equations~\ref{equ:paralla} which yield
\begin{align}
 \Delta\alpha &= \frac{y_{E}\varpi\cos{\alpha} - x_{E}\varpi\sin{\alpha}}{\cos{\delta_{g}}} \\ \notag
 \Delta\delta &= -(x_{E}\varpi\cos{\alpha} + y_{E}\varpi\sin{\alpha})\sin{\delta} + z_{E}\varpi\cos{\delta}
\end{align}
Comparison with Equations.~\ref{equ:pialpha} and
\ref{equ:pidelta} yield for the parallax factors the relation:
\begin{subequations}
\begin{align}
 p_{\alpha} &= \frac{y_{E}\cos{\alpha} - x_{E}\sin{\alpha}}{\cos{\delta_{g}}} \label{equ:pialphaII}\\
 p_{\delta} &= -(x_{E}\cos{\alpha} - y_{E}\sin{\alpha})\sin{\delta} + z_{E}\cos{\delta}
\label{equ:pideltaII}
\end{align}
\end{subequations}
The parallax factors change, depending on where on the Earth orbit
the observation was taken. One can see this in the dependency of
$p_{\alpha}$ and $p_{\delta}$ on $x_{E}, y_{E}, z_{E}$.

\noindent As one notices, to calculate the parallax factors one
needs the coordinates one wants to compute. So an iterative
approach is necessary, but in practice the difference between the
geocentric and barycentric coordinates is mostly negligible for
the computation of the parallax factors and one iteration is
sufficient.\\

\noindent If one only corrects for the parallax, the position
$(\alpha, \delta)$ of a star at epoch T is given by:
\begin{align}
\alpha &= \alpha_{0} + p_{\alpha}\varpi \notag \\
\delta &= \delta_{0} + p_{\delta}\varpi
\end{align}
where $\alpha_{0}, \delta_{0}$ are the coordinates of the star at
some reference epoch $T_{0}$.

\noindent As we will fit the parallax motion in our orbit fit we
need the parallax factors, which we can compute with our measured
geocentric coordinates for each epoch. We then have to fit only
the size of the parallax, $\varpi$. A detailed description of the
orbit fit will be given in Sec.~\ref{sec:Fit}.

\subsection{Proper Motion}
Each star has its own proper motion $\mu$, leading to a
displacement on the celestial sphere when observed at an epoch $T$
compared with an observation at epoch $T_{0}$. This displacement
can be in any direction. A few stars have a high proper motion of
the order of $1\arcsec/$ per year or more, but mostly the annual
movement is only a small fraction of that. A star's proper motion
depends on its space motion relative to the center of the
celestial sphere, the barycenter, but also on its distance. So in
general the further away a star with a given space velocity, the
smaller its proper motion. The velocity of a star, resulting in
its proper motion is assumed to be uniform. The path of the star
is then an arc of a great circle on the celestial
sphere.\\
Ideally, proper motion is measured by comparing the position of a
star in observations many years apart. The longer the baseline,
the more accurate the proper motion can be derived. But the
positions measured are only relative to the other stars observed
with the target star and one has to assume that their positions
and proper motions are known for each epoch. The best solution is
to measure the positions relative to a fixed reference frame. Most
satisfying would be to use objects outside our Milky Way as
reference, such as other galaxies or quasars (QSO = Quasi-Stellar
Object). QSO are preferable as they are more point like and many
of them are also radio sources, whose position can be measured to
a very high accuracy with radio interferometry. They are galactic
nuclei at enormous distances and are therefore unaffected by any
galactic motion.\\

\noindent Taking equatorial coordinates the proper motion in right
ascension and declination is the time derivative of the
coordinates at epoch $T_{0}$
\begin{align}
\mu_{\alpha} &= \left(\frac{d\alpha}{dt}\right)_{T=T_{0}} \notag\\
\mu_{\delta} &= \left(\frac{d\delta}{dt}\right)_{T=T_{0}}
\end{align}
$\mu_{\delta}$ corresponds to a full (great circle) angle on the
sky, but $\mu_{\alpha}$ is reckoned on the equator. Its actual
component along the local small circle is given by
$\mu_{\alpha}\cos{\delta}$. $\mu_{\alpha}$ and $\mu_{\delta}$ are
normally expressed in arcseconds per year, however if
$\mu_{\alpha}$ is given in seconds of time per year, one has to
multiply it by a factor of 15
to get arcseconds.\\
\noindent A star's motion can be separated into a radial motion
$V_{r}$ (see Sec.~\ref{subsec:LTD}) and a transverse motion on a
tangent plane to the celestial sphere. The transverse motion
equals the proper motion $\mu =
\sqrt{\mu_{\alpha}^2\cos^2{\delta}+ \mu_{\delta}^2}$. With $\phi$
being the position angle of this motion measured from North over
East, the components of the proper motion in right ascension and
declination can be written as:
\begin{align}
\mu_{\alpha}\cos{\delta} &= \mu\sin{\phi} \notag \\
\mu_{\delta} &= \mu\cos{\phi}
\end{align}

\noindent One can now calculate a star's position for any given
epoch, if one only assumes proper motion and parallactic motion.
The position $(\alpha, \delta)$ of a star at epoch T is then given
by:
\begin{align}
\alpha &= \alpha_{0} + (T - T_{0})\mu_{\alpha}^{*} + p_{\alpha}\varpi \notag \\
\delta &= \delta_{0} + (T - T_{0})\mu_{\delta} + p_{\delta}\varpi
\end{align}
with $\mu_{\alpha}^{*}$ being short for
$\mu_{\alpha}\cos{\delta}$.

\subsubsection{Secular Acceleration}
The secular acceleration is a variation in the proper motion and
is a purely geometric effect, due to an object which is
approaching or receding. Writing the proper motion as a function
of the object's space velocity $V$, its angle $\Theta$ with the
direction barycenter-star and its distance $r$
\begin{equation}
\mu = \frac{V\sin{\Theta}}{r}
\label{equ:secularacc}
\end{equation}
and calculating its derivative with respect to time $t$, one gets:
\begin{equation}
\frac{d\mu}{dt} = -\frac{V}{r^2}\sin{\Theta}\frac{dr}{dt} +
\frac{V}{r}\cos{\Theta}\frac{d\Theta}{dt}
\end{equation}
With $\frac{d\Theta}{dt}= -\mu$ and $V\cos{\Theta} = V_{r} =
\frac{dr}{dt}$ and using Equ.~\ref{equ:secularacc} one can write
\begin{equation}
\frac{d\mu}{dt} = -\frac{2\mu}{r} V_{r}
\end{equation}
Expressing $\mu$ in arcseconds per year , the distance $\varpi =
\frac{1}{r}$ in arcseconds and $V_{r}$ in km/s, the secular
acceleration can be calculated by:
\begin{equation}
\frac{d\mu}{dt} = -2.05\cdot 10^{-6} \mu\varpi V_{r}
\end{equation}

\noindent Using the values for proper motion ($\mu =
1.5\arcsec$/year) and parallax ($\varpi = 71.56 \rm ~mas$) from
the HIPPARCOS catalog and a radial velocity $V_{r} = 63 \rm
~km/s$ for our target star\footnote{Mathias Zechmeister (private
communication; based on measurements with the ESO FEROS
spectrograph)}, the secular acceleration amounts to $-13.84\cdot
10^{-6} \rm ~arcsec~ yr^{-1}$. This would result in a change in
proper motion of 0.02 mas/yr over the 1.5 yrs time baseline of our
observations, which we will not be able to measure. We can
therefore stay with the truncation of the Taylor series of the
star's motion after the first term and assume the proper motion of
our target as constant in time. As the reference star is much
further away and its proper motion therefore is way smaller, the
effect for it will be even smaller.

\section{Plate-scale and Detector Rotation Stability}
\label{sec:plate_scale} Doing astrometry with NACO, one has to
take care of two effects. The global pixel- or plate-scale can
change and field distortions can be present, both effects can
change with time. To monitor and correct the possible change of
pixel-scale and also the rotation of the detector, I observed a
reference field in the globular cluster 47~Tuc every time right
before the target field. To minimize these effects, I centered the
stars on the same pixel positions at the beginning of each
observation in both, the target and reference field, and executed
the same jitter pattern each time. As long as the jitter pattern
is the same in each observation, I do not have to take care about
the absolute distortions, but have to monitor, if they change with
time. Experience from other observers using NACO, indicates that
the distortion pattern is stable over time\footnote{Andreas
Seifahrt (private communication)}. If the distortions are constant
with time, they should be the same for each observation, and only
the change in pixel-scale has to be corrected. Unfortunately, it
is not so easy to separate the change in the globular pixel-scale
and the field distortions, which create different local
pixel-scales. Also, the final image, on which I work, is the sum
of the shifted and added single images. The distortions, present
in each single image, are smeared out in the final image, making
it very difficult to model them, as they are dependent on the size
and distribution of the jitter offsets. NACO is at the Nasmyth
focus, but it does not have an optical de-rotator. It rotates as a
whole instrument, instead. Therefore variable flexures exist in
this Nasmyth instrument and the distortions from the reference
field can probably not be used
to correct the target field.\\
\noindent The only thing left is to
model a global pixel-scale, assuming it only changes on a global
scale and has no direct cross-talk with the distortions.

\begin{figure}
 \begin{center}
  \includegraphics[width=10cm]{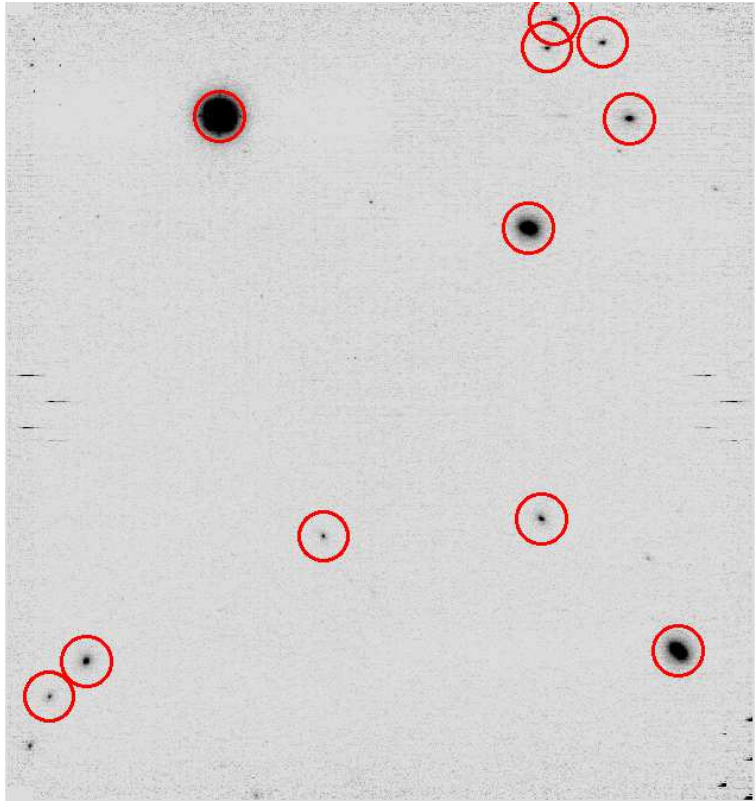}
 \end{center}
\caption{Stars in the reference field in 47~Tuc used to calculate
the change in pixel-scale and rotation between the different
epochs.} \label{fig:platescale_stars}
\end{figure}

\noindent To calculate the changes between the different epochs, I chose the
same 11 stars for each epoch in the observed reference field. In
Fig.~\ref{fig:platescale_stars} the stars used for the distortion
fit are marked.
As I do not need to measure an absolute plate-scale or rotation of
the detector, but only relative values, I chose epoch 9 as the
reference epoch to which all other epochs are mapped. This epoch
was chosen, because it is the middle one in time of all
observations. I then mapped the positions of the stars in each
epoch to the positions measured in epoch 9 with a linear
coordinate transformation, calculating a shift in $x$ and $y$,
scale in $x$ and $y$, and a rotation and skew. The skew is
realized by allowing a different rotation for the $x$ and $y$
axes. The parameters were calculated with the \textit{geomap}
program, which is part of the IRAF reduction and data analysis
software\footnote{http://iraf.noao.edu/}.\\

\begin{figure}[t!]
 \begin{center}
  \includegraphics[width=14.5cm]{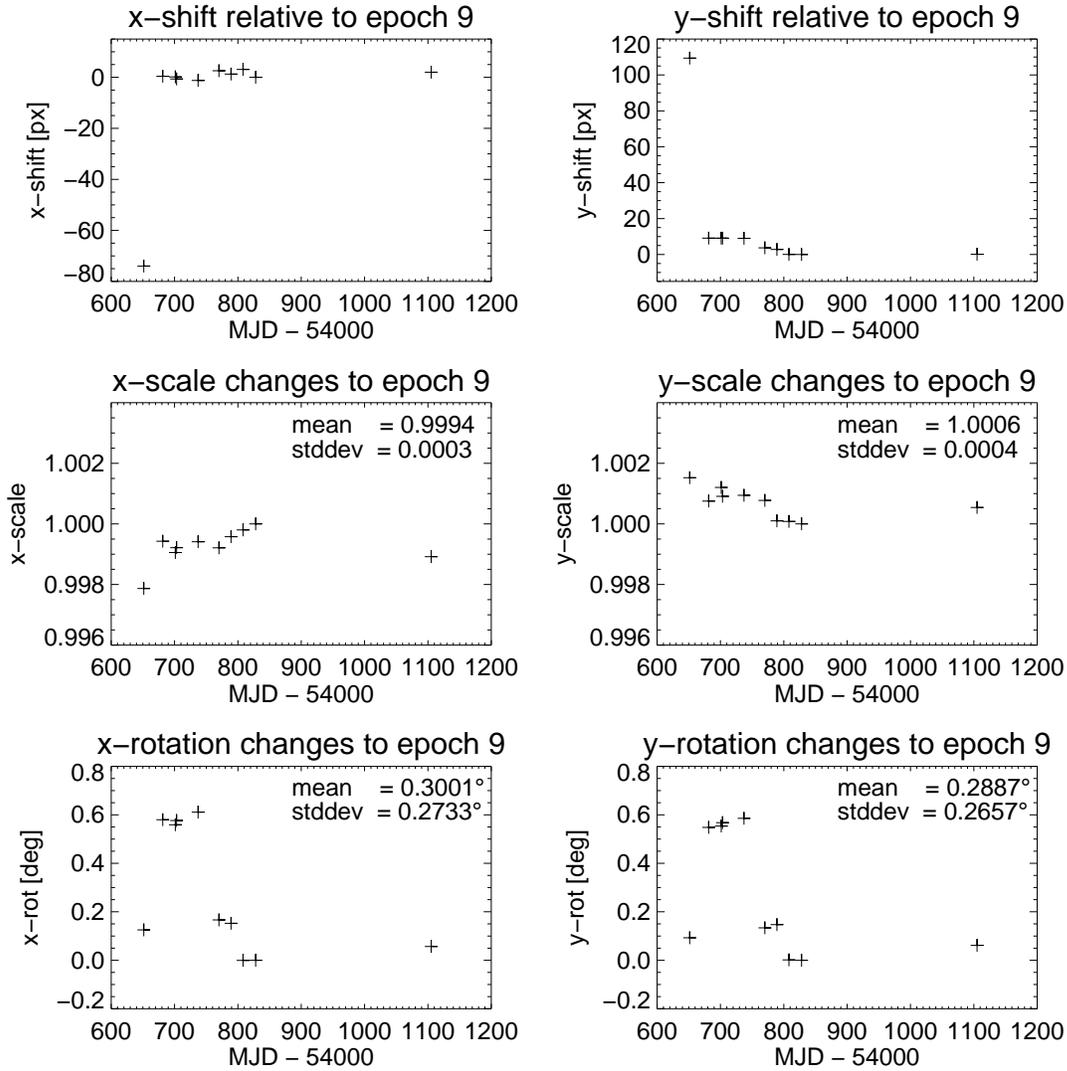}
 \end{center}
\caption[Calculated distortion parameters of all epochs]
{Calculated distortion parameters of all epochs. The parameters
for shift, scale and rotation in $x$ and $y$, respectively, are
shown from top to bottom versus the MJD of the observation. For
the scale and rotation parameters the mean value and the standard
deviation are given, for an impression of the stability of these
parameters over time.}
\label{fig:distortion_params}
\end{figure}

\noindent The calculated values for each epoch are shown in Fig.~\ref{fig:distortion_params}
The panels show from top to bottom the calculated shift, scale and
rotation for the $x$ and $y$ direction versus the Modified Julian
Date (MJD), for each epoch mapped to epoch 9. The much larger
values for the shift of the first epoch are due to a rotation of
the frame about 90\degr, due to an error in the value of the
applied rotation of the detector. I corrected the coordinates of
the stars to this rotation, but their positions on the frames
nevertheless differ by roughly +75 pixel on the $x$-axis and -100
pixel on the $y$-axis. But the shift is not the important
parameter, as it does not change the distance between the stars.
The more interesting values are the scale and rotation. Here the
mean value and the standard deviation are also given in the
respective panels. The standard deviation gives an impression of
the stability of the parameters over time. It seems the larger the
separation in time is for the first nine epochs, the more
different is the plate-scale, but this would have to be explored
in more detail with more data doing a distortion analysis for
NACO. The calculated rotation between the different epochs does
not show a trend with time. For the 47~Tuc observations, the
detector was rotated by $42\degr$ anticlockwise. The scatter of
the measured rotation between the epochs gives the accuracy with
which the applied rotation of the detector can be performed. The
scatter is quite large, 0.26 - 0.27 degree, but there is no large
difference for the two axes, indicating only a very small skew.
Because the detector is again rotated between the observations of
the reference and the target field, and I cannot assume that the
rotation error is the same for both fields in the same night, I
cannot correct for the rotation between the epochs. As there are
only two stars in the target field, an independent estimate of the
differential rotation between the epochs for this field is not
possible. I added the scatter of the rotation parameter over time,
measured as a mean of the $x$- and $y$-rotation, in the reference
field as the uncertainty of the later calculated
position angle between the two stars in the target field.\\

\begin{figure}[t!]
 \begin{center}
  \includegraphics[height=12cm, angle=90]{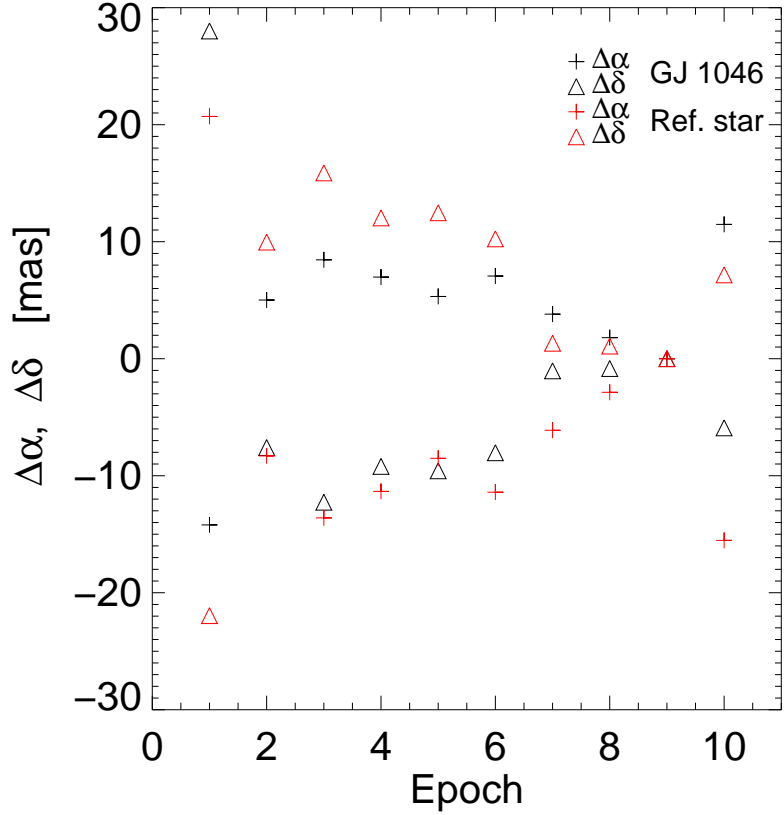}
 \end{center}
\caption[Relative change in positions due to the plate-scale
correction] {Relative change in positions due to the plate-scale
correction. Shown are the values for $\Delta~\alpha = \alpha -
\alpha_{corr}$ and $\Delta~\delta = \delta - \delta_{corr}$ for
both stars in the target field. The opposite directions of the
changes for GJ~1046 and the reference star are expected, due to
the fact the the coordinates are calculated from the reference
point, which is almost in the middle of the detector. The large
displacement i n epoch 1 is due to the larger change in
pixel-scale compared with the other epochs. The visible trends
correspond to ones already visible in
Fig.~\ref{fig:distortion_params}.}
\label{fig:dRA_pltscl}
\end{figure}
\subsection{Plate-scale Correction}
\label{subsec:pltscl_corr} 
The detector distortions are the last
effect changing the true positions of the stars. Therefore one
should correct for them first. I assumed that the change in
pixel-scale at any epoch with respect to the one in epoch 9 is the
same in the reference field and the target field. To correct for
the change in pixel-scale I multiplied the FITS header keywords
CD\textit{i\_j}$ = s_{i}m_{ij}$ (Equ.~\ref{equ:xisum}) with the
scale factors derived from the distortion fit:
\begin{align}
 \rm new~ CD_{11} &=\rm CD_{11} * \text{x-scale} \\ \notag
 \rm new~ CD_{12} &=\rm CD_{12} * \text{x-scale} \\ \notag
 \rm new~ CD_{21} &=\rm CD_{21} * \text{y-scale} \\ \notag
 \rm new~ CD_{22} &=\rm CD_{22} * \text{y-scale}  \notag
\end{align}
This was done for every epoch (for the first epoch the $x$- and
$y$-scale are exchanged due to the 90\degr ~rotation) and
following this, the measured pixel coordinates of the stars were
transformed to celestial coordinates, as described in
Chap.~\ref{subsec:xyad}, using the new plate-scale. After that,
the corrections for differential refraction and aberration were
calculated and applied again (Chap.~\ref{subsec:DiffRefCorr} and
\ref{subsec:diff_aber_corr}). That the celestial positions changed
indeed due to the plate-scale correction can be seen in
Fig.~\ref{fig:dRA_pltscl} where the change, after the
transformation from $x/y$ coordinates to celestial coordinates, in
right ascension and declination is plotted in
milli-arcseconds for each epoch.\\
\noindent A scaling of the image itself before measuring the
positions would involve interpolation, which again can introduce
errors. I therefore chose the applied way of correction, because
here I do not have to work on the images directly to change the
pixel-scale. I therefore corrected directly the coordinates of the
stars, which is more precise.
\noindent The RMS given by the fit for the coordinate mapping
takes all transformations into account. I cannot use this value as
an estimate for the plate-scale correction alone, as it also takes
for example the rotation into account. To estimate the precision
of the plate-scale correction I took star pairs in the reference
field in one epoch and calculated their separation. Taking the
separation of the same star pairs in epoch 9, the reference epoch,
as the separation with a plate-scale of the theoretical value of
27.15 mas/px, I could calculate the pixel-scale of the other
epochs by comparing the separations:
\begin{equation}
 \text{plate-scale\_i} = \frac{\text{separation\_epoch\_9}}{\text{separation\_epoch\_i}} * 27.15~\frac{mas}{pixel}
\end{equation}
I did this for 100 random star pairs for each epoch to have good
statistics. The standard deviation of the distribution of
pixel-scales calculated this way is then taken as the uncertainty
in the pixel-scale correction for the later separation measurement
of the two stars. In Tab.~\ref{tab:deltaRADEC_pxscl} the
uncertainties in the pixel-scale are listed for each epoch.
\begin{table}[h!]
 \begin{center}
  \begin{tabular}{|c|c|c|} \hline
   Epoch & pixel-scale & pixel-scale \\
         & uncertainty [$\frac{mas}{px}$] & uncertainty [\%] \\ \hline
   ~1 & 0.046 & 0.171\\
   ~2 & 0.035 & 0.127\\
   ~3 & 0.027 & 0.100 \\
   ~4 & 0.018 & 0.066\\
   ~5 & 0.016 & 0.058\\
   ~6 & 0.045 & 0.164\\
   ~7 & 0.014 & 0.051\\
   ~8 & 0.032 & 0.117\\
   ~9 & 0.000 & 0.000\\
   10 & 0.029 & 0.105\\ \hline
  \end{tabular}
 \caption[Uncertainties in the pixel-scale]{Uncertainties in the pixel-scale.
  Calculated by randomly taking 100 star pairs in the reference field,
   calculating the changed pixel-scale relative to epoch 9, and taking
    the standard deviation of the distribution of derived pixel-scales
     as uncertainty.}
  \label{tab:deltaRADEC_pxscl}
 \end{center}
\end{table}

\noindent These first results in this chapter show, that in
multi-epoch astrometry the most challenging task is to measure and
correct the pixel-scale very precisely to obtain milli- or even
micro-arcsec astrometric solutions.

    \cleardoublepage
    \chapter{The Orbit Fit}
    \label{chap:orbit}
\section{Preparing the Coordinates for the Orbital Fit}
\label{sec:dist_posang} To fit an orbit to the positions of the
two  stars derived so far, one needs to apply some last
transformations and calculations. In a first approach I
re-transformed the celestial coordinates of the target star
GJ~1046 and the reference star 2MASS 02190953 -3646596 to pixel
coordinates with the IDL routine \textit{adxy.pro} from the IDL
astrolib\footnote{http://idlastro.gsfc.nasa.gov/}. This routine is
the inverse of \textit{xyad.pro}, which I used to transform pixel
coordinates into celestial coordinates in Chap.~\ref{subsec:xyad}.
I then finally corrected the coordinates for the nominal rotation
of 6\degr ~of the detector, relative to the North direction. The $x$
and right ascension axes, as well as the $y$ and declination axes
are now parallel. One can now work with the separations of the two
stars in $x$ and $y$ and their change with time.\\
\noindent
Additionally, I calculated the separation and position angle of the
two stars. Because of the large separation of the stars, one
cannot just calculate the separation in Cartesian coordinates. One
has to calculate the separation $\rho$ of the two stars along a great
circle using the \textit{cosine} formula \citep[][p.
12]{Green1985}:
\begin{equation}
 \rho = \cos^{-1}{\left (\sin{\delta_{1}} \sin{\delta_{2}} + \cos{\delta_{1}} \cos{\delta_{2}} \cos{\Delta \alpha}\right )}
\end{equation}
Where $\delta_{1}$ and $\delta_{2}$ are the declination of the two
stars, star~1 = reference star and star~2 = GJ~1046, and $\Delta
\alpha$ is the difference in their right ascension.\\
\noindent The position angle $\Theta$ of GJ~1046 relative to the
reference star, measured from North through East from the Meridian
containing the reference star is calculated using the
\textit{four-parts} formula \citep[][p. 12]{Green1985}:
\begin{equation}
 \Theta = \tan^{-1}{\left (\frac{\sin{\Delta \alpha}}{\cos{\delta_{1}} \tan{\delta_{2}} - \sin{\delta_{1}} \cos{\Delta \alpha}}\right )}
\end{equation}
The errors of the separation and position angle, resulting from
the positional uncertainties of the stars, are between 0.22 -
1.45~mas for the separation and between 0.00045\degr -
0.00252\degr ~for the position angle. However, a much larger
uncertainty due to the pixel-scale variability and the detector
rotation uncertainty had to be added for each epoch. The resulting uncertainties
in the separation are between 15.30~mas and 50.90~mas and 0.26072\degr ~in the
position angle with only very small differences for
the single epochs. A summary of the calculated separation and
position angle values together with their errors is given in
Tab.~\ref{tab:sep_pa_error} for each epoch. In column two the
separation and position angle are noted, column three lists the
combined errors due to the position measurement and the correction
for differential refraction and aberration. In column four the
error due to the pixel-scale uncertainty and the detector rotation
are given and column five finally lists the combined
errors of all these effects.\\
\begin{figure}[b!]
 \begin{center}
  \includegraphics[width=6.8cm]{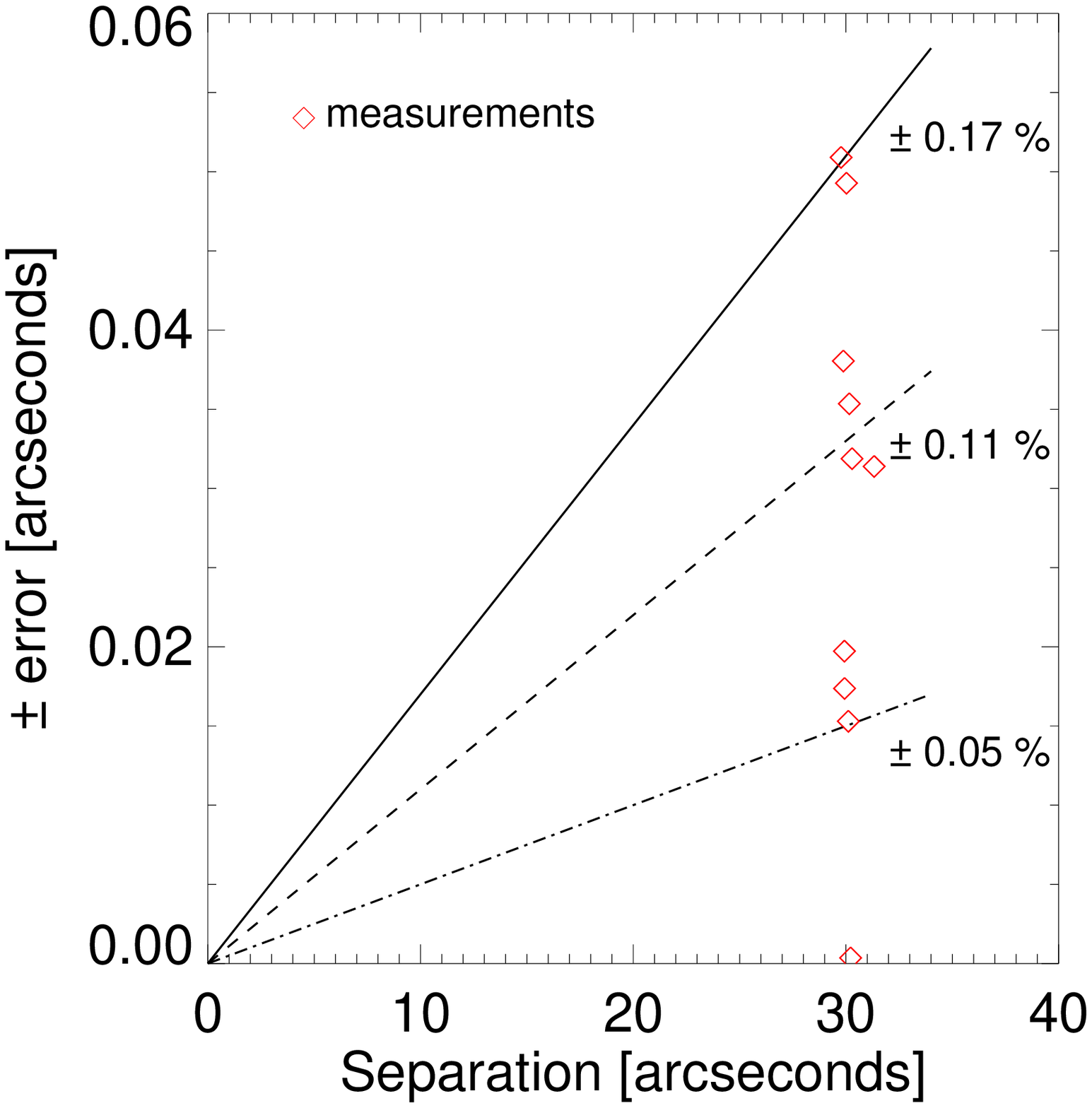}
  \includegraphics[width=7.5cm]{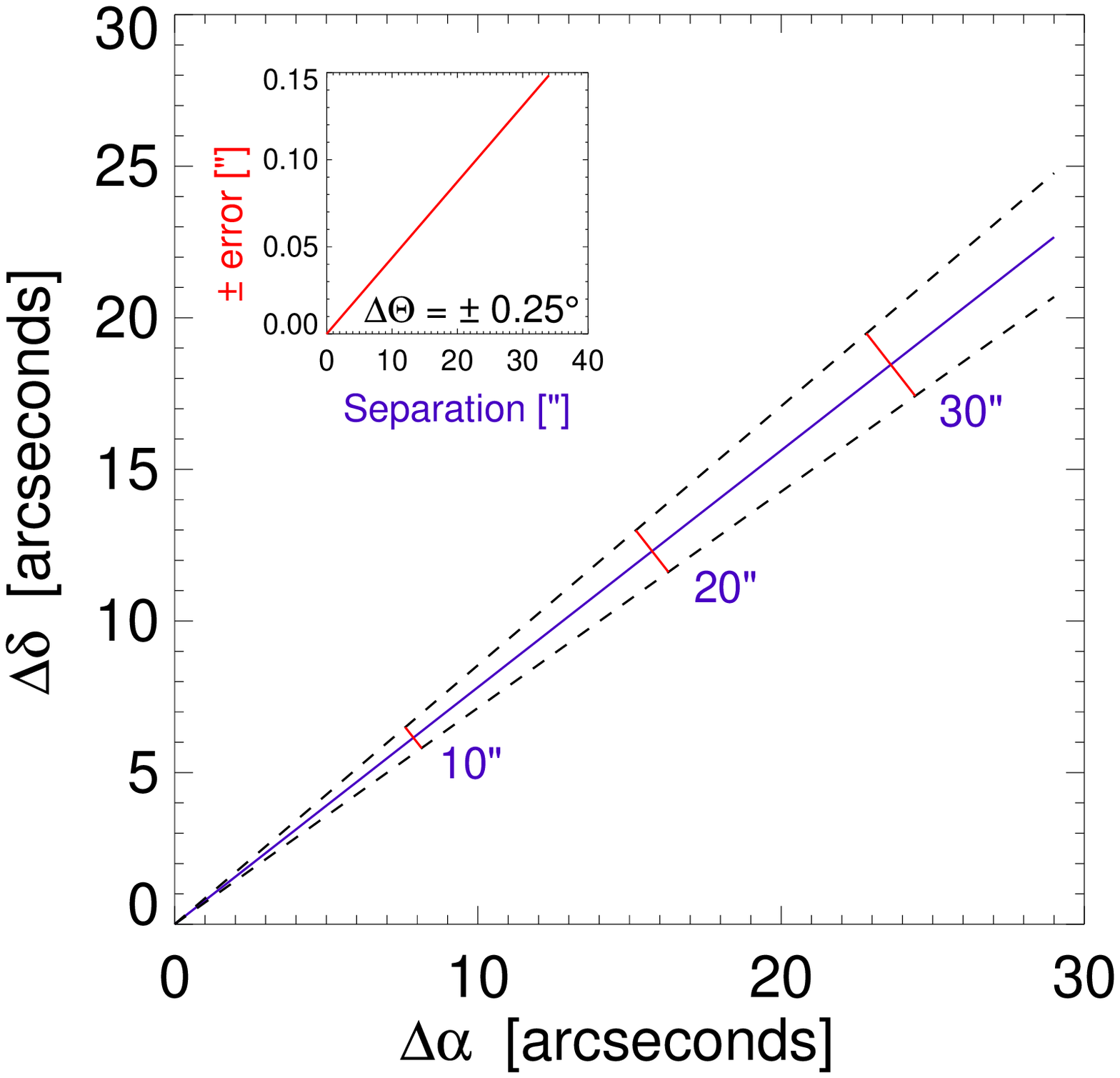}
 \end{center}
 \caption[Effect of separation between the target and reference
 star on the precision obtainable]{Effect of separation between
 the target and reference star on the precision obtainable for
 the separation (left panel) and the position angle (right panel).
  For more details and explanation see text.}
\label{fig:dist_posang}
\end{figure}
\noindent This already shows, how crucial it is to measure the
pixel-scale and the rotation of the detector very accurately. The
large separation of the target and reference star yields this
large separation and position angle uncertainty. In
Fig.~\ref{fig:dist_posang} the effect of the separation between the
target and reference star on the obtainable precision is shown.
The left panel shows the error in the separation in arcseconds as
a function of pixel-scale uncertainty and separation. The lines
represent the errors in the separation between the stars for three
different values of uncertainty ($\pm 0.05\%, ~\pm 0.11\%, ~\pm
0.17\%$) in the pixel-scale. The red diamonds represent my
measured uncertainties. The right panel shows the error in the
separation in right ascension ($\Delta\alpha$) and declination
($\Delta\delta$) due to the uncertainty in the position angle.
Shown is the case for a position angle of 42\degr ~(solid line)
with an uncertainty of $\pm 2.5\degr$ (dashed lines). This 10 times
larger error than measured was taken for reasons of better
depiction of the effect. The separation between the stars is indicated along
the lines by marks at 10\arcsec, 20\arcsec ~and 30\arcsec.
Projecting the error onto the directions of right ascension
and declination leads to large errors in these two values. In the small inlet the size of the uncertainty of the position angle, $\Theta \pm 0.25\degr$ (red in the big panel), along the direction perpendicular to the separation (blue in the big panel) is shown in seconds of arc as function of the separation.\\

\begin{landscape}
\begin{table}
 \begin{center}
  \begin{tabular}{|c|cc|cc|cc|cc|} \hline
   & & & \multicolumn{2}{c|}{\small{Pos., Refrac., Aberr.}} & \multicolumn{2}{c|}{\small{pixel-scale, de-rotator}} & $\sqrt{\sum_{\scriptstyle i} \Delta\rho_{i}^2}$ & $\sqrt{\sum_{\scriptstyle i} \Delta \Theta_{i}^2}$  \\
   Epoch & $\rho ~[\arcsec]$ & $\Theta ~[\degr]$ & $\Delta\rho_{1}$ [mas] & $\Delta\Theta_{1} ~[\degr]$ & $\Delta\rho_{2}$ [mas] & $\Delta\Theta_{2} ~[\degr]$ & [mas] & [\degr] \\ \hline
   1 & 29.7709 & 37.96 & 0.23 & $4.4\cdot 10^{-4}$  & 50.91 & 0.26 & 50.90 & 0.26 \\
   2 & 29.8758 & 38.54 & 0.54 & $9.9\cdot 10^{-4}$  & 37.94 & 0.26 & 38.04 & 0.26 \\
   3 & 29.9273 & 38.57 & 1.06 & $19.0\cdot 10^{-4}$ & 29.93 & 0.26 & 29.98 & 0.26 \\
   4 & 29.9330 & 38.62 & 0.28 & $5.0\cdot 10^{-4}$  & 19.76 & 0.26 & 19.70 & 0.26 \\
   5 & 30.0251 & 38.78 & 0.40 & $7.5\cdot 10^{-4}$  & 17.41 & 0.26 & 17.43 & 0.26 \\
   6 & 30.1134 & 38.38 & 0.47 & $8.6\cdot 10^{-4}$  & 49.39 & 0.26 & 49.42 & 0.26 \\
   7 & 30.1570 & 38.54 & 0.33 & $6.1\cdot 10^{-4}$  & 15.38 & 0.26 & 15.32 & 0.26 \\
   8 & 30.2174 & 38.47 & 0.36 & $6.7\cdot 10^{-4}$  & 35.35 & 0.26 & 35.42 & 0.26 \\
   9 & 30.2864 & 38.53 & 0.40 & $7.1\cdot 10^{-4}$  & ~0.00 & 0.26 & ~0.40 & 0.26 \\
  10 & 31.3265 & 38.62 & 1.45 & $25.2\cdot 10^{-4}$ & 32.89 & 0.26 & 33.00 & 0.26 \\ \hline
  \end{tabular}
  \caption[Summary of the measured separation and position angle together with their uncertainties.]{Summary of the measured separation $\rho$ and position angle $\Theta$ together with their uncertainties. The third column list the combined errors due to the PSF fit and the correction for differential refraction and aberration. In column four the errors due to the pixel-scale uncertainty and the detector rotation
are given and column five finally lists the combined
errors of all these effects.}
  \label{tab:sep_pa_error}
 \end{center}
\end{table}
\end{landscape}

\section{Theory of Deriving the Orbital Elements}

\begin{figure}
 \begin{center}
  \includegraphics{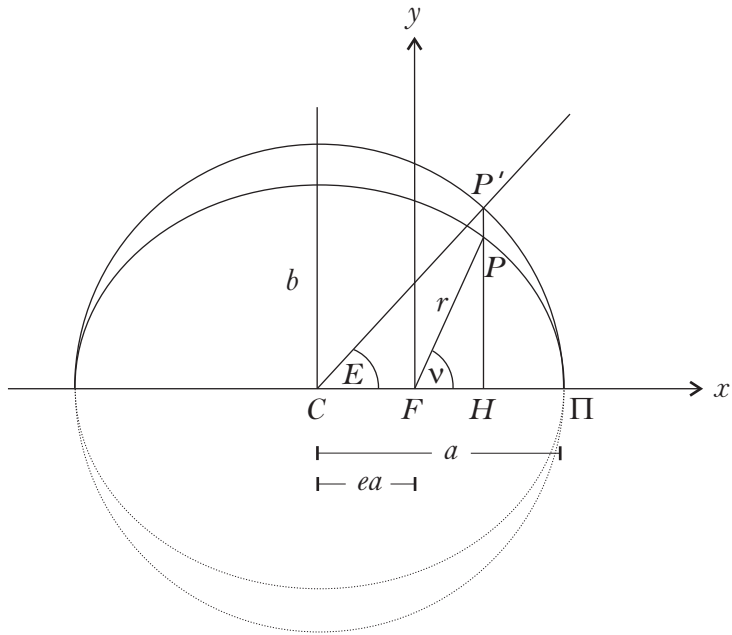}
 \end{center}
\caption{Principle elements of an ellipse with the definition of
the eccentric anomaly $E$ and the true anomaly $\nu$.}
\label{fig:ellipse}
\end{figure}

The orbit of a planet around its host star, as well as the motion
of the star due to a companion, are ellipses with the common center of mass of the
two bodies at one of the focal points of both ellipses. To describe
the orbital motion of the companion or the host star, one needs to
know where the body is at any time $t$. The orbital elements are
the same for the two bodies, except the size of the semi-major
axis. To derive the orbital elements one can regard the ellipse as
a projection of a circle. In Fig.~\ref{fig:ellipse} the principal
components of such an ellipse are depicted.
The denotation is as follows:\\
\begin{tabular}{ll}\vspace{-0.3cm}
 $C$ & center of the ellipse\\ \vspace{-0.3cm}
 $F$ & focal point of the ellipse\\ \vspace{-0.3cm}
 $\Pi$ & periastron\\ \vspace{-0.3cm}
 $a$ & semi-major axis\\ \vspace{-0.3cm}
 $b$ & semi-minor axis\\ \vspace{-0.3cm}
 $e$ & eccentricity\\ \vspace{-0.3cm}
 $ea$ & distance from $C$ to $F$\\ \vspace{-0.3cm}
 $P$ & point on ellipse at time $t$\\ \vspace{-0.3cm}
 $P^{'}$ & point on circle which is projected onto the ellipse\\ \vspace{-0.3cm}
 $r$ & radius vector\\ \vspace{-0.3cm}
 $E$ & eccentric anomaly\\ \vspace{-0.3cm}
 $\nu$ & true anomaly\\
\end{tabular} \vspace{0.2cm}

\noindent The eccentric anomaly $E$ is defined as the angle
between the semi-major axis $a$ and the direction from the center of
the ellipse to the point $P^{'}$ on the circle. It is zero at the
periastron $\Pi$ and increases by $2\pi$ during one orbit. The true
anomaly is the angle between the distance $\overline{F\Pi} = a - ae
= a(1-e)$ and the distance $\overline{FP} = r$, which is the
radius vector at time $t$. Looking at Fig.~\ref{fig:ellipse}, one
can see that for a Cartesian coordinate system with its origin at $F$, the $x$ and $y$ values of the point $P$ in polar
coordinates in the plane of the orbit are given by:
\begin{align}
 X = r\cos{\nu} &= a(\cos{E} - e) \notag \\
 Y = r\sin{\nu} &= a\sqrt{1 - e^2}\sin{E}
\label{equ:rcosnu}
\end{align}
and the distance of $P$ from the focal point $F$ is:
\begin{equation}
 r = a(1 - e\cos{E}) = \frac{a(1 - e^2)}{1 + e\cos{\nu}}
\label{equ:r=ecosE}
\end{equation}

\noindent Defining the mean anomaly $M$ at time $T$ by the mean
motion $n = \frac{2\pi}{P}$, with $P$ being the period and $T_{p}$
the time of periastron:
\begin{equation}
 M = n(T - T_{p}) = E - e\sin{E}
\end{equation}
one can calculate the eccentric anomaly $E$ at any time using an
iterative approach. The last expression $M = E - e\sin{E}$ is
known as Kepler's equation. Once $E$ is determined one can compute
$r$ and $\nu$ from Equ.~\ref{equ:r=ecosE} and from:
\begin{equation}
 \tan{\frac{\nu}{2}} = \sqrt{\frac{1 + e}{1 - e}} \tan{\frac{E}{2}}
\end{equation}

\noindent In the simple case of visual binaries, one centers the
primary in the local coordinate system to describe the orbital
motion of the companion with respect to the primary star, but the
equations are also true for the motion of the primary around the
secondary. The coordinates of the companion on the
tangential plane on the sky with respect to the primary, expressed in seconds of arc, are then
defined by the separation $\rho$ and the position angle $\theta$,
starting from North on eastward:
\begin{align}
 x = \rho\sin{\theta} \notag \\
 y = \rho\cos{\theta}
\end{align}

\subsection{The Thiele-Innes Constants}
Each observation of a visual binary yields a pair of coordinates
at a given time $T$. These coordinates are the separation of the
two stars and the position angle. After a hopefully sufficiently
long time interval of observations one has a series of values
($\rho, \theta, T$) or equivalently $(x, y, T)$. These measured
values give the apparent orbit in the $x/y$ plane on the sky. \\
\noindent The true orbit of the companion about the primary, which
one wants to obtain, is an ellipse with the primary situated at one
focus $F$. Projecting this ellipse onto the plane of the sky yields
again an ellipse, the apparent orbit. This apparent ellipse is not
a Keplerian ellipse anymore. The focus $F$ of the true ellipse doe not coincident with a focus of the apparent ellipse, meaning the primary is not
seen at a focus. The semi-major axis of the apparent orbit does not correspond to the semi-major axis of the true orbit either and
therefore does not lie along the projection of the diameter that
contains the periastron $\Pi$. Only the projection of the center
of the true orbit appears at the center of the apparent orbit and
the areas on the apparent ellipse are projections of the areas of
the true ellipse. Kepler's $2^{nd}$ law of proportionality of the
areas swept by the radius vector still holds. \\

\noindent With the orbital parameters defined in
Chap.~\ref{sec:orbitalelements} one can transform the local
apparent coordinates $(\rho\sin{\Theta},~\rho\cos{\Theta})$ in the coordinate system $OXY$ to the true coordinates in the coordinate system $Oxy$. With the scale of
the true orbit in arcseconds, $a\arcsec = a\varpi$ with the parallax $\varpi$ and the semi-major axis $a$, one can write
the true coordinates in the plane of the orbit, also expressed in arcseconds, as
\begin{equation}
\mathbf{B}
\begin{cases}
 X = \varpi r\cos{\nu} \notag \\
 Y = \varpi r \sin{\nu}
\end{cases}
\end{equation}
To transform $\mathbf{B}$ into $\mathbf{r} = (\rho\sin{\theta},
\rho\cos{\theta})$ one has to perform three transformations.\\
\noindent The first one is a rotation by $-\omega$ around the
$Z$-axis which is perpendicular to the $X/Y$-plane, the true
orbital plane. After this, one has to account for the projection
angle $i$, the inclination, between the true orbital plane and the
plane perpendicular to the line of sight. The intermediate
coordinates can be written as:
\begin{align}
 X_{inter} &= X\cos{\omega} - Y\sin{\omega} \notag \\
 Y_{inter} &= \cos{i}(X\sin{\omega} - Y\cos{\omega})
\label{equ:XYinter}
\end{align}
One then has to rotate these coordinates by $-\Omega$ around the
z-axis, which is in the direction of the line of sight and
perpendicular to the tangential plane. Finally a permutation of
the abscissa and ordinate needs to be performed to obtain the
coordinate system $Oxy$ in the tangential plane
\begin{align}
 x &= X_{inter}\sin{\Omega} + Y_{inter}\cos{\Omega} \notag \\
 y &= X_{inter}\cos{\Omega} + Y_{inter}\sin{\Omega} \notag
\end{align}
Combining this with Equ.~\ref{equ:XYinter} one finally can write
the apparent coordinates expressed by the true coordinates as:
\begin{align}
 x =& X(\cos{\omega}\sin{\Omega} + \sin{\omega}\cos{\Omega}\cos{i}) \notag \\
    & + Y(-\sin{\omega}\sin{\Omega} + \cos{\omega}\cos{\Omega}\cos{i}) \notag \\
 y =& X(\cos{\omega}\cos{\Omega} - \sin{\omega}\sin{\Omega}\cos{i}) \notag \\
    & + Y(-\sin{\omega}\cos{\Omega} - \cos{\omega}\sin{\Omega}\cos{i}) \notag
\end{align}
With the Thiele-Innes constants defined as:
\begin{align}
 X_{TI1} &= a(\cos{\omega}\sin{\Omega} + \sin{\omega}\cos{\Omega}\cos{i}) \notag \\
 Y_{TI1} &= a(-\sin{\omega}\sin{\Omega} + \cos{\omega}\cos{\Omega}\cos{i}) \notag \\
 X_{TI2} &= a(\cos{\omega}\cos{\Omega} - \sin{\omega}\sin{\Omega}\cos{i}) \notag \\
 Y_{TI2} &= a(-\sin{\omega}\cos{\Omega} - \cos{\omega}\sin{\Omega}\cos{i})
\label{equ:T-I-Const}
\end{align}
one can write in short form:
\begin{align}
 x &= \frac{r}{a}\cos{\nu}X_{TI1} + \frac{r}{a}\sin{\nu}X_{TI2} \notag \\
 y &= \frac{r}{a}\cos{\nu}Y_{TI1} + \frac{r}{a}\sin{\nu}Y_{TI2}
\label{equ:xy_orbit}
\end{align}

\section{The Astrometric Orbit Fit}
\label{sec:Fit}

\noindent In the context of this work, I do not measure the
companion's motion around the star directly, but instead the star's
motion about the common center of mass of the system. Therefore
the orbital motion ($x_{orbit}, y_{orbit}$) I see, is related to
the star's motion and semi-major axis. $x_{orbit}$ and $y_{orbit}$
are the expressions derived in Equ.~\ref{equ:xy_orbit} with an index $orbit$ added to point out that this is the orbital motion of the star due to the companion in the following equations.

\noindent In my case I already have some of the
orbital elements from the radial velocity measurements. These are: the Period $P$, the eccentricity $e$, the longitude of
periastron $\omega$ and the time of
periastron $T_{p}$. The semi-major axis $a$ is related to the inclination $i$ over Equ.~\ref{equ:asini}, so the only two orbital parameters one has to solve for, after fixing the spectroscopic parameters, are the inclination of the orbit and the longitude of the ascending node $\Omega$.
These two angles of the orbit of the star are the same for the companion. I could fix the values of $P,~e,~\omega, \rm ~and T_{p}$ to the values derived in the spectroscopic orbit fit, because the accuracy from this fit is much higher than the accuracy with which they could be derived in the astrometric fit.

\noindent One can write the position of
the star on the sky depending on proper motion and parallax of the
center of mass and the reflex motion of the star at any time $T$ as:
\begin{align}
 \alpha &= \alpha_{0} + (T - T_{0})\mu_{\alpha} + p_{\alpha}\varpi + x_{orbit} \notag \\
 \delta &= \delta_{0} + (T - T_{0})\mu_{\delta} + p_{\delta}\varpi + y_{orbit}
\label{equ:fullorbit}
\end{align}
where $\alpha_{0}, ~\delta_{0}$ are the position of the star at epoch $T_{0}$, $\mu_{\alpha}$ and $\mu_{\delta}$ are the proper motion and $p_{\alpha}, ~p_{\delta}$ are the parallax factors in right ascension and declination, respectively. As I measure relative positions and not absolute ones, the positions $\alpha_{0}, ~\delta_{0}$ are replaced by the separation of the two stars in right ascension and declination $\Delta\alpha_{0}, ~\Delta\delta_{0}$ and $\alpha, ~\delta$ then correspond to separations, too.\\

\begin{figure}
 \begin{center}
  \includegraphics[width=14.5cm]{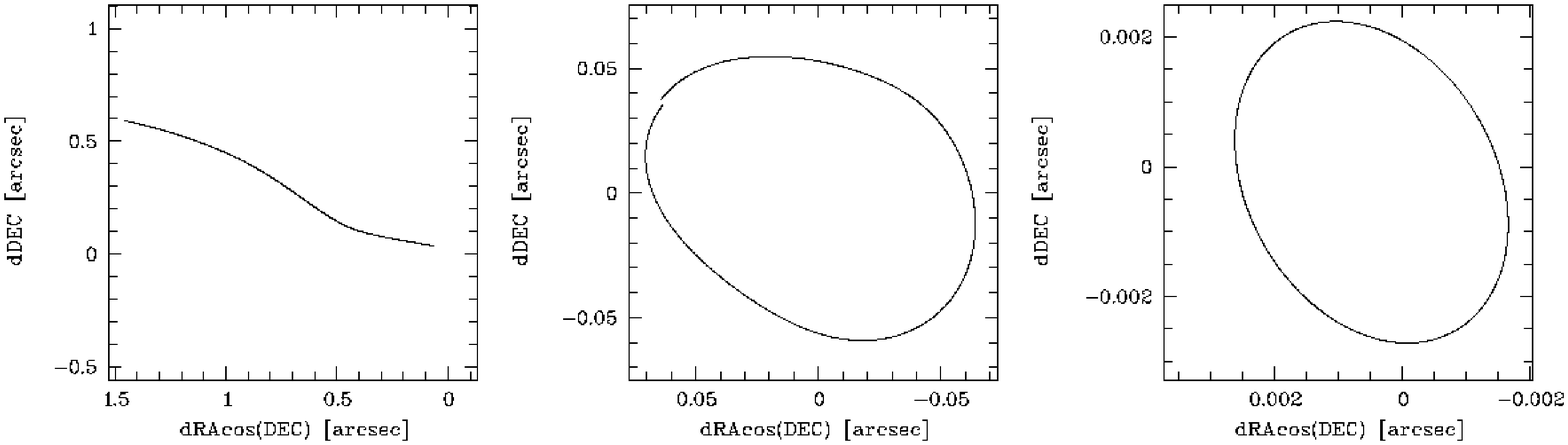}
 \end{center}
 \caption[Simulated change in declination versus right ascension of the orbit of GJ~1046]{Simulated change in declination versus right ascension of the orbit of GJ~1046. The left panel shows a simulated astrometric signal for GJ~1046 (see text for details). The other two panels show the simulated astrometric signal without proper motion and with (middle) and without (right) parallax, respectively. Note the different scales in the changes of right ascension and declination in the three different cases.}
 \label{fig:pm_pi_orbit}
\end{figure}

\noindent To solve this equation one also has to fit for the proper motion and parallax of the target star \textit{relative} to the reference star. As seen in the previous sections, the uncertainty in the pixel-scale and the orientation of the detector lead to large errors in the final separation and position angle, with the error in the separation being still a lot smaller than that of the position angle. The orbital period of the companion of GJ~1046, and therefore of GJ~1046 itself, is 169 days, almost half a year. Special care has to be taken to disentangle the orbital motion from the parallax motion. For this, the observations were timed in such a way that measurements over almost a full orbit were obtained. In Fig.~\ref{fig:pm_pi_orbit} an example of a simulated astrometric signal of GJ~1046 with proper motion, parallax and orbital motion is shown as a change in right ascension and declination over one year. The HIPPARCOS proper motion and parallax are taken and $i = 45\degr$ and $\Omega = 60\degr$ are assumed. As the proper motion of GJ~1046 is very high, it totally dominates its motion and pulls the parallax + orbital ellipse apart into a wave-like motion (left panel). Subtracting the proper motion in the simulation, one can see the orbital motion due to the companion, but still dominated by the parallax (middle panel). The curve is not a perfect ellipse as in the case of pure parallax motion and is not closing after one full orbit of the Earth around the sun. Finally the pure orbital motion without parallax and proper motion is displayed (right panel). As one can see, it is very important to know/measure the parallax motion to distinguish the orbital motion from it. Note the different scales in the changes of right ascension and declination in the three different cases.

\noindent The fit for the orbital parameters was performed by Rainer K\"ohler from the Landessternwarte Heidelberg, who is experienced in deriving orbital solutions, using his orbit fit program. A model is calculated based on Equations~\ref{equ:fullorbit} which is then compared with the actual measurements by minimizing $\chi^2$, using a Levenberg-Marquardt algorithm (LMA) to \citep{Press1992}.\\
\noindent A first fit was performed, accounting only for the measurements of the separations of GJ~1046 from the reference star, because of the relatively smaller error of this parameter. No orbital motion was taken into account at this step, so Equations.~\ref{equ:fullorbit} reduce to:
\begin{align}
 \alpha &= \Delta\alpha_{0} + (T - T_{0})\mu_{\alpha} + p_{\alpha}\varpi \notag \\
 \delta &= \Delta\delta_{0} + (T - T_{0})\mu_{\delta} + p_{\delta}\varpi
\label{equ:pm+pi}
\end{align}
We took the values for the proper motion and parallax from the HIPPARCOS catalog for a first test, $\mu_{\alpha} = 1394.10$ mas/yr, $\mu_{\delta} = 550.05$ mas/yr, $\varpi = 71.56$ mas, so the only free parameters are the separation zero-points $\Delta\alpha_{0} \rm ~and ~\Delta\delta_{0}$. The calculated separations from the model parameters for $\alpha$ and $\delta$ are then compared with the measured ones for each epoch.\\

\noindent The second approach was to include the orbital movement in the fit. This was done by scanning the possible angles for the inclination $i$ and the longitude of the ascending node $\Omega$ in one degree steps and calculating a $\chi^2$ value for each ($i,\Omega$) pair. Again, the only free parameters in Equ.~\ref{equ:fullorbit} for the LMA were $\Delta\alpha_{0} \rm ~and ~\Delta\delta_{0}$, as $i$ and $\Omega$ were given at each step. For each ($i,\Omega$) pair a model was calculated for the separation $\rho$ and this time also for the position angle $\Theta$. Inclination $i$ and ascending node $\Omega$ are needed to calculate the Thiele-Innes constants (Equ.~\ref{equ:T-I-Const}) to derive the orbital influence on the separation and position angle. After scanning $i$ from 0-180\degr ~and $\Omega$ from 0-360\degr ~we have a $\chi^2$ map for the full parameter space of these two angles. We again used the proper motion and parallax from the HIPPARCOS catalog. We can now search the resulting $\chi^2$ map for a minimum.

    \cleardoublepage
    \chapter{Results}
    \label{sec:results1} Due to the large uncertainties in the
detector orientation and pixel-scale correction I was not able to
detect any astrometric motion of the star due to its brown dwarf
companion. Only a formal value for the inclination of the system
and thus for the mass of the companion could be derived, but with
low significance of the results, see
Sect.~\ref{sec:orbit_results}.\\

\begin{figure}[t]
 \begin{center}
  \includegraphics[height=14.5cm, angle=90]{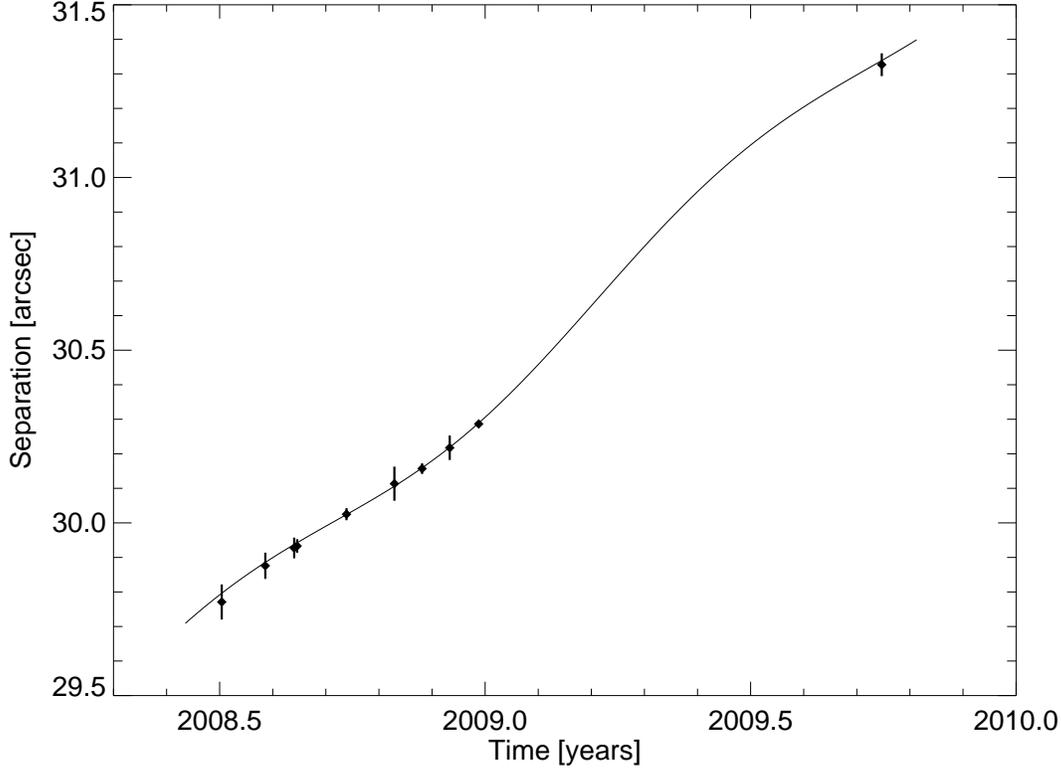}
 \end{center}
\caption[Measured separation between GJ~1046 and the reference
star vs. time for the 10 observed epochs]{Measured separation between GJ~1046
 and the reference star vs. time for the 10 observed epochs. The error bars
 represent the uncertainty in the separation due to the pixel-scale
 uncertainty. Epoch 9 has only a very small error bar as it is the reference epoch
 for the measurement of the pixel-scale changes (see Sect.~\ref{sec:plate_scale}).
 Overplotted is the model for the separation changes calculated with the HIPPARCOS
 values for proper motion and parallax (solid line). No orbital motion is taken into account.
 The good agreement of the measurements with the model shows that the reference star
 has indeed only a very small intrinsic parallax and proper motion.}
\label{fig:Dist_over_time}
\end{figure}

\noindent Using the HIPPARCOS values for proper motion and
parallax and not taking any orbital motion into account for the
model fit, results in a good agreement of the model with the measured
separation changes. In Fig.~\ref{fig:Dist_over_time} the measured
separation between the stars is plotted versus time. Overplotted
as the solid line is the model with the motion calculated from the
HIPPARCOS values for parallax and proper motion. The very good agreement
indicates that the assumption of a small or negligible parallax
and proper motion of the reference star was right and that we only
 see and detect the motion of GJ~1046 itself. 
The $\chi^2$ value of this model fit is $\chi^2 = 7.63$, with 18 degrees of freedom (DoF) the chance probability of this value is $p(\chi^2) = 0.984$. Therefore the confidence for rejecting the model without orbital motion is onlt 1.6\%. In principle one should stop at this point and accept the model, which only takes proper motion and parallax movement of the star into acoount, as precise enough to represent the data. However, from the radial velocity measurements, I know about the presence of a companion, hence also a model including orbital motion is calculated and fit to the data.

\section{The Orbit}
\label{sec:orbit_results} 
\begin{figure}[t]
 \begin{center}
  \includegraphics[height=14.5cm, angle=90]{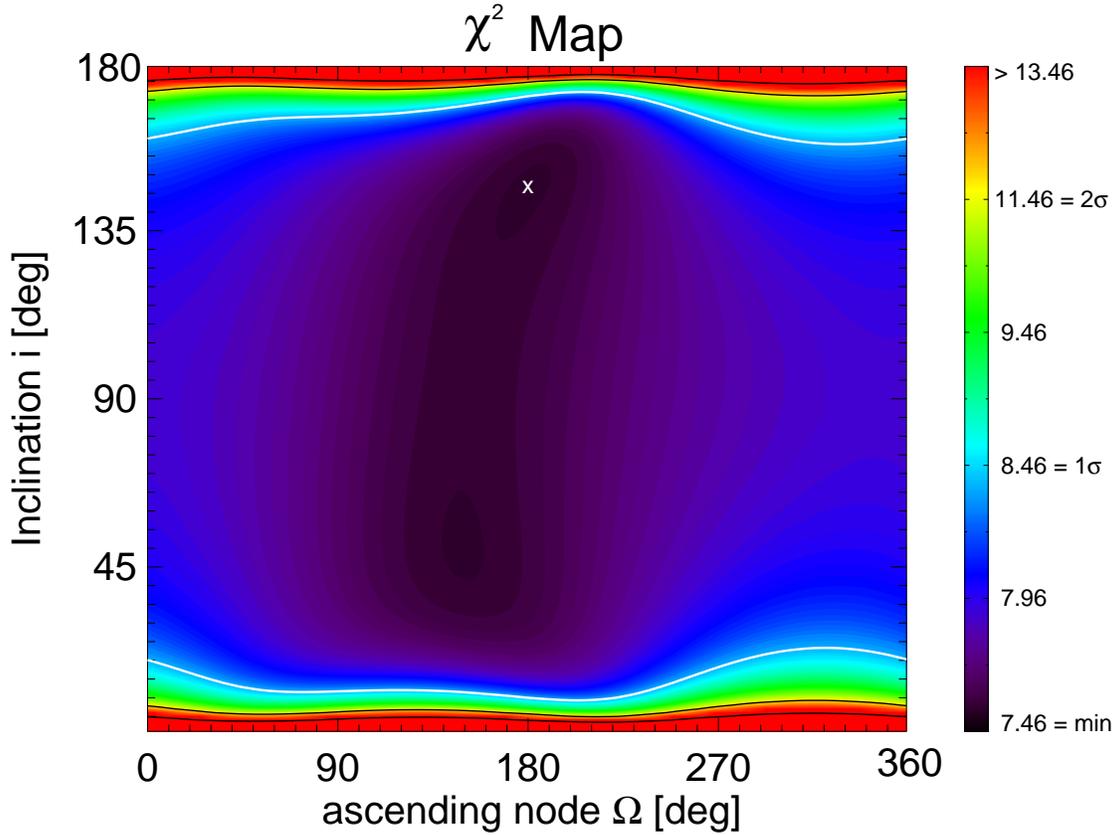}
 \end{center}
\caption[$\chi^2$ contour map for fitting the orbital motion to the
separation and position angle measurements]{$\chi^2$ contour map for
fitting the orbital motion to the separation and position angle measurements.
The spectroscopic parameters were fixed, as well as the parallax and proper
motion (HIPPARCOS values). The inclination $i$ and
the ascending node $\Omega$ were looped through, leaving the zero-point
separation in declination and right ascension as the only free fit
parameters. The white contour represents the $1\sigma$ (68.3\%) confidence level and the black contours the
$2\sigma$ (95.4\%) and $3\sigma$ (99.7\%) confidence levels. The white cross marks the formally best fit solution found at $i = 145\degr$ and $\Omega = 180\degr$.}
\label{fig:Chi2_map}
\end{figure}

\begin{figure}[t]
 \begin{center}
  \includegraphics[height=14.5cm, angle=90]{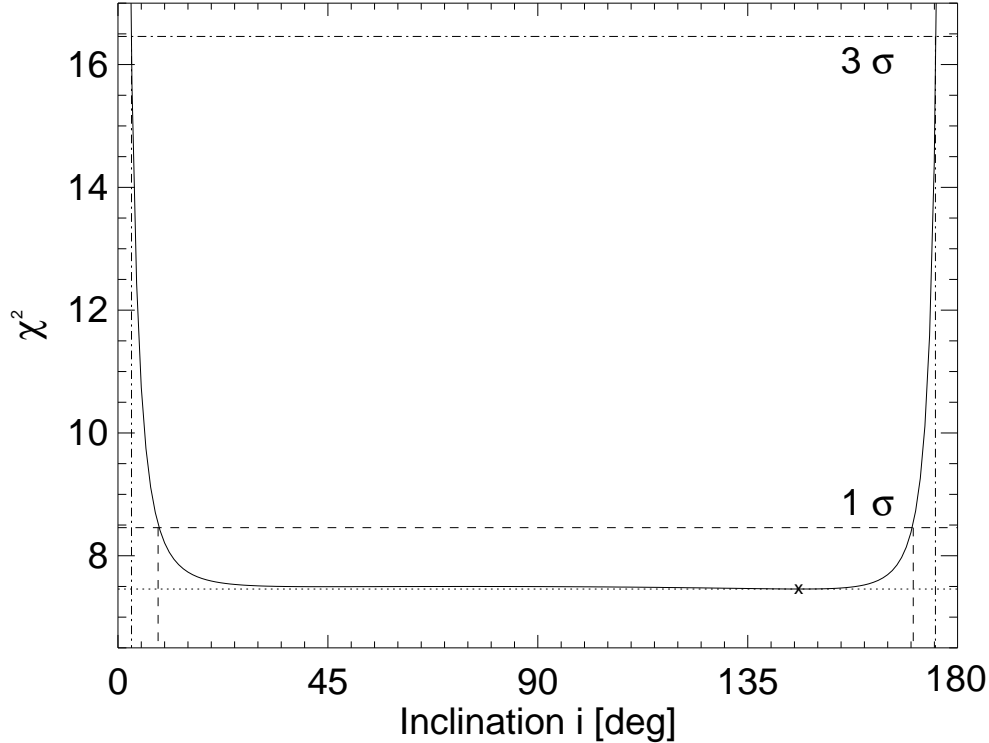}
 \end{center}
\caption[$\chi^2$ of the astrometric fit as a function of only the
inclination]{$\chi^2$ of the astrometric fit as a function of only
the inclination together with the $1\sigma$ and $3\sigma$ confidence
levels, represented as the dashed and dash-dotted horizontal lines,
respectively. $\Omega$ is treated as uninteresting, therefore the
confidence levels correspond only to the single parameter inclination
and are located at the levels $\chi^2 + 1$ and $\chi^2 + 9$, respectively. 
The cross marks the formally best fit solution.}
\label{fig:Chi2_sigma}
\end{figure}

Due to the large uncertainty in the
alignment of the detector a transformation of the separation onto
the right ascension and declination axes results in large error
bars in these two directions. An attempt to fit a model based on
Equ.~\ref{equ:fullorbit} with the separation at epoch zero
$\Delta\alpha_{0}, ~\Delta\delta_{0}$, parallax $\varpi$, proper
motion $\mu_{\alpha}$ and $\mu_{\delta}$, inclination $i$ and
longitude of the ascending node $\Omega$ as free parameters led to
resulting values of the proper motion and parallax strongly
diverging from the HIPPARCOS values. Even though a small deviation
is in principle possible, as only relative motions to the
reference star are measured, a deviation of up to 200~mas, as derived in the fit, is not
possible. Additionally, the results from the fit without orbital
motion using the HIPPARCOS values showed a very good agreement with
these values. Therefore, the proper motion and parallax were fixed
to the HIPPARCOS values in the following approach to fit the
orbital
motion, as we are not able to improve these values.\\
\noindent In Fig.~\ref{fig:Chi2_map} the resulting $\chi^2$
contour map from the model fit with orbital motion to the measured
separation and position angle is depicted. The white contour
represents the $1\sigma$ (68.3\%) and the black contours the
$2\sigma$ (95.4\%) and $3\sigma$ (99.7\%) confidence levels.
The formally best fit is achieved with an inclination $i =
145.0\degr$ ($180\degr - i = 35\degr$) and a longitude of the
ascending node $\Omega = 180\degr$, leading to a formal mass of
the companion of $m_{p} = 48.5 ~M_{Jup}$, calculated with the
mass function (Equ.~\ref{equ:mass_function}). This would mean that
the companion is indeed a brown dwarf, residing in the brown dwarf desert.
However, this result cannot be considered significant, as already the simplier model without orbital motion could not be rejected with sufficient confidence. Likewise the model including orbital motion cannot be rejected. The confidence for rejection of this model is only 3.7\% (DoF = 16, $\chi^2 = 7.46$), but setting better constraints on the orbital parameters is not possible. This is also indicated by the large formal errors derived for the inclination $i$ and the ascending node $\Omega$.
Already within
$1\sigma$, inclinations from $8.5\degr - 170.5\degr$ are possible. The
$1\sigma,~2\sigma$ and $3\sigma$ levels span the entire parameter space for
the ascending node, which is therefore completely undetermined. In
Fig.~\ref{fig:Chi2_sigma} the $\chi^2$ of the fit as a function of
only the inclination is plotted together with the $1\sigma$ and
$3\sigma$ confidence levels. Only inclinations smaller than 3\degr ~and
larger than 175\degr ~can be excluded with a $3\sigma$ confidence. The
formally best mass estimate is at the same time the lower mass limit for
the companion, as both the upper and lower $1\sigma$ limit for the
inclination yield higher masses. The $1\sigma$ upper limit for
the companion mass is $236 ~M_{Jup}$ for an inclination of
$8.5\degr$. Calculating the $3\sigma$  upper limit of the companion mass leads to the very unlikely case that the \textquotesingle companion\textquotesingle ~is more massive than the primary, which is already excluded by the upper limit of $112 \rm ~M_{Jup}$ derived from the combination of the RV data with the HIPPARCOS astrometric data \citep{Kuerster2008}. All masses are calculated with the mass function (see Equ.~\ref{equ:mass_function} and Tab.~\ref{tab:rvparameters}). In Tab.~\ref{tab:fit_results} the parameters derived from the astrometric orbit fit are summarized.
\begin{table}
 \begin{center}
  \caption{\large{Parameters derived from astrometry}}
  \begin{tabular}{|l|c|l|} \hline
   inclination $i$ & 145\degr & formal optimum \\
   longitude of the & & \\
   ascending node $\Omega$ & 180\degr & formal optimum \\
   companion mass $m_{p}$ & $48.52~M_{Jup}$ & formal optimum \\
   minimum $i$ & 3\degr & $3\sigma$ limit \\
   maximum $i$ & 175\degr & $3\sigma$ limit \\ \hline
  \end{tabular}
 \label{tab:fit_results}
 \end{center}
\end{table}

\section{Discussion and Conclusion}
I have observed an M Dwarf with a known Brown Dwarf desert candidate
companion with adaptive optics aided imaging to detect and measure
the astrometric signal of the star due to its unseen companion.
The observations were conducted with the NACO instrument at the
VLT UT4 telescope. A star at 30\arcsec ~separation was used as a
reference for the astrometric measurements. The expected
astrometric peak-to-peak signal is minimum 3.7~mas and up to
15.4~mas for an object at the upper mass limit derived from
HIPPARCOS astrometry (see Chap.~\ref{chap:observations}).\\
\noindent To measure the relative positions of the target star and
the reference star very precisely, one has to correct for several
effects which change the true positions of the stars. After
measuring the positions on the detector by fitting a Moffat
function to the PSFs, I calculated the error due to the fit with
the bootstrap method. The position accuracy is well below
milli-arcsecond accuracy at this step.\\
\noindent To be very precise in the relative positions between
the stars, I corrected for differential refraction, which takes
the different zenith distances of the two stars into account. As I
could not work on the single frames, because of the faintness of
the reference star, I calculated a mean correction for the
positions in right ascension and declination. The main error
contributions in this case come from the long duration of the
observations and the correction for different temperature and pressure in the individual observing epochs. Depending on when the observations were conducted,
close to, before or after the local Meridian passage, the zenith
distance and therefore the differential refraction changes faster
or slower with time. I could not correct for the time dependent
effect, but rather applied the mean change in right ascension and
declination as the correction factor and the standard deviation of
the measured values as the error. Additionally, the error from the correction for the different temperature and pressure is taken into account. The positional precision after
this correction is still below the milli-arcsecond range, except in one case, where the temperature and pressure during the observation deviated strongest from the standard conditions.\\
\noindent  The next correction I applied is the one for
differential aberration. Due to the movement of the Earth through
space the positions observed are different from the true ones.
Depending on the position of the star on the celestial sphere
relative to the observer, a correction for aberration has to be
applied which is different for each star. \\
noindent The by far biggest uncertainty in the relative
separation of the two stars comes from the change in the
plate-scale between the different epochs and the uncertainty of
this change. Because of the large separation of the two stars an
uncertainty of 0.05\% adds a large error to the overall error
budget. Also the error in the applied detector rotation amounts to
large error bars of the position angle. The detector rotation is
stable to about 0.26\degr ~as measured in the reference field 47~Tuc.
Again, the large separation of the target and reference star
leads to large errors.\\

\noindent The formally best fit with fixed spectroscopic
parameters and including orbital motion yields an
inclination $i = 145\degr$ and therefore a companion mass $m_{p} =
48.5 ~M_{Jup}$. The $1\sigma$ upper limit for the mass is
$236 ~M_{Jup}$. Unfortunately the result has no significance,
as already the model without orbital motion could not be rejected. The $1\sigma$ limit almost spans the whole parameter
space for the inclination, only angles smaller than 3\degr ~and
larger than 175\degr ~can be excluded with $3\sigma$ confidence.\\
\noindent The formal inclination value derived in this work is in rough agreement with the value ($i = 125.9\degr$) derived by the combination of the radial velocity
data with the HIPPARCOS astrometric data \citep{Kuerster2008}. Even though the HIPPARCOS
astrometry could also only yield a formal best fit value for the
inclination and the ascending node with low significance,
they could set stronger constraints on the
inclination. A $3\sigma$ upper limit for the companion mass of $112~M_{Jup}$ could
be set with a lower limit for the inclination of 15.6\degr ~and an upper limit of $i = 161\degr$.\\
\noindent I therefore could not detect the astrometric signal of
the companion to GJ~1046, nor could I set stronger constraints on
its true mass.

\noindent Another point one has to keep in mind is the possible light contamination from the companion. When determining the position of GJ~1046, I measured the photocenter of the light distribution. If the companion itself has a certain brightness, the position of the photocenter is shifted from the primary to a position between the two objects. The direction of this shift changes with the orbital motion of the companion and can scale down the observable orbital motion of the primary. The amount of flux the companion contributs to the combined flux distribution depends on the mass and age of the companion. As the energy distribution of very low mass stars and brown dwarfs peaks in the near-IR and my observations are obtained in the K band, the contribution of the companion will be more severe than for example in the V band, where the spectroscopic measurement were conducted by \cite{Kuerster2008}. For a given age one can calculate flux ratios as a function of the mass ratio of the two objects with theoretical models \citep[see e.g.][Fig.~1]{Burrows2001}. But as the companion is probably a brown dwarf which cools and dims with time, its age also plays an important role in terms of its brightness. The age of GJ~1046 is likely \textgreater 1~Gyr, but the brightness of a 1~Gyr and a 5~Gyr old brown dwarf already differs notedly \citep[see e.g. models by][]{Baraffe2003}. This makes it so difficult to conclude masses of isolated brown dwarf from photometry alone. As my measurement are not precise enough to detect any orbital motion, I did not take the possible contamination by the companion into account in my analysis. If, on the other hand, my measurements would have been precise enough to detect the orbital motion, I would have needed to simulate the possible effect on the measurements of the position of GJ~1046 and include the results either by correction of the measured position or by adjusting the error budget.

\noindent Current AO imaging instruments have only small FoVs of a
few tens of arcseconds. This makes it difficult to find targets
for astrometry which have several suitable reference stars close
by. The results from this work have shown, how important it is to
have more than one star in the same FoV which one can use as
astrometric reference points. Even a third star can already help to
constrain the rotation of the detector in the very same field and
therefore offer the ability to correct for a differential rotation
between different observing epochs. Additional reference stars
enhance the precision with which the position of the target star
is measured with respect to the other stars. The so-called
plate-solution can be derived in the target field itself, making
it possible to correct for differential distortions and
plate-scale changes between the epochs.\\
\noindent Also a smaller separation between target and reference
star is preferable. As shown, the uncertainties in the final
fitting parameters scale with the separation between target and
reference source. Measurements of other groups have shown a
similar precision of the plate-scale and detector rotation.
\cite{Neuhaeuser2008} measured a pixel-scale uncertainty of $\pm
0.05$~mas/px ($= 0.38\%$) and a rotation stability of $\pm
0.25\degr$. The calibration was made with a HIPPARCOS astrometric
binary and the smaller camera S13 of NACO, which has a FoV of
$14\arcsec\times 14\arcsec$. \cite{Koehler2008} used images
obtained in the Orion Trapezium cluster to monitor and calibrate
the pixel-scale and detector orientation. The results are a
scatter of the pixel-scale smaller than 1\% and a rotator
precision of a few tenths of a degree. Here the S13 camera was
used, too. In the case of the calibration with the binary, the
change in pixel-scale is only measured in one direction and the
calibration measurements of both groups are obtained on a smaller
FoV than the measurements presented in this work. The scatter of
the pixel-scale is roughly the same in mas/px in all measurements,
but due to the larger FoV and pixel-scale of the S27 camera the
relative scatter is smaller. This shows that a correction for
differential refraction and aberration before measuring the change
in pixel-scale enhances the precision. On the other hand, my
measurements with different scales fitted to the $x$- and
$y$-axes show different values for the two directions (Fig.~\ref{fig:distortion_params} and a
fit with only one scale yields bigger overall fitting errors.
This indicates a differential scale change for different
directions in the field, hence the length of a given distance on
the detector depends on its location on the detector. The
relatively large scatter in the pixel-scale suggests that a global
pixel-scale is probably not the correct ansatz for obtaining high
precision astrometry over larger separations, as there probably are
local pixel-scale changes due to distortions present in the
frames. Knowledge of and correction for these distortions, and
correction of a left over change in pixel-scale will lead to a higher
precision in the separation.\\
\noindent At one epoch, without correction for a change in the
pixel-scale to another epoch, the positional precision achieved in
this work is already promising. The change-over to multi-epoch
astrometry and the involved need for very well known pixel-scales
and distortions is the most challenging task. A very well
characterized distortion analysis, both spatial and temporal, would
enhance the precision additionally.\\
\noindent A group
analyzing the galactic center nicely shows the possible precision,
but also the limitation of astrometry with NACO \citep{Trippe2008,
Gillessen2009, Fritz2010}. They use the inner few arcseconds of
the S13 camera for their analysis and the S27 camera to set up a
reference frame. As they have sufficient reference stars in their
FoV ($\sim100-200$), they are able to calculate and perform a
distortion correction, already including a correction for the
scale. With this method they are able to obtain position
uncertainties for the stars down to 0.6 mas for the S27 camera
\citep{Trippe2008}, for the S13 camera and the smaller FoV they
get even smaller uncertainties. This shows that with a good
distortion correction between the frames a precision sufficient to
detect astrometric signals of large planetary companions is
possible, but it also shows that astrometry on large scales with
NACO will probably not be able to detect or characterize
Earth-like planets.

    \cleardoublepage

    \chapter{Introduction to MCAO and MAD}
      \section{MCAO - The Next Generation of Adaptive Optics}
\label{sec:MCAO} As seen in the introduction to classical adaptive
optics correction with one guide star (Chap.~\ref{sec:AO}), the
FoV is limited by the effect of anisoplanatism, because only the
integrated phase error over the column above the telescope in the
direction to the guide star is measured. Turbulence outside this
column, e.g. in the direction of the target, if it cannot be used
as guide star, is not mapped and cannot be corrected. In the case
of a laser guide star as reference source this problem is even
more severe, due to the low focussing altitude and the resulting
cone-effect (Chap.~\ref{sec:Isoplanatism}).\\

\noindent Multi Conjugated Adaptive Optics (MCAO)
\citep{Beckers1988, Ellerbroeck1994} is an approach to achieve
diffraction limited image quality over bigger FoVs of up to 2
arcminutes and hence overcome anisoplanatism. Moderate
Strehl-ratios, 10-25\%, can be achieved, but with a higher
uniformity of the PSF shape over the FoV. This is desired for
resolving structures of extended sources, such as galaxies or
cores of star clusters. In MCAO the 3-dimensional structure of the
turbulence is reconstructed by means of the information coming
from several guide stars, i.e. natural or laser guide stars.
Instead of correcting the turbulence integrated over the column
above  the telescope, which size is defined by the isoplanatic
patch, at once, turbulence from different layers is corrected with
several deformable mirrors conjugated to these layers. Typically
two layers are being corrected, the ground layer close to the
telescope and a higher layer at around 8 - 10 km height. A full
correction for the higher layer is only guaranteed if this layer
is fully covered by the footprints of the columns of the
beam-paths from the reference stars to the telescope. So the
number of needed guide stars depends on the altitude one wants to
correct.  Two different approaches exist to combine the signals
from the different reference stars, the Star Oriented (SO)
approach and the Layer Oriented (LO) approach.
\begin{figure}
 \begin{center}
  \includegraphics[width=6.0cm]{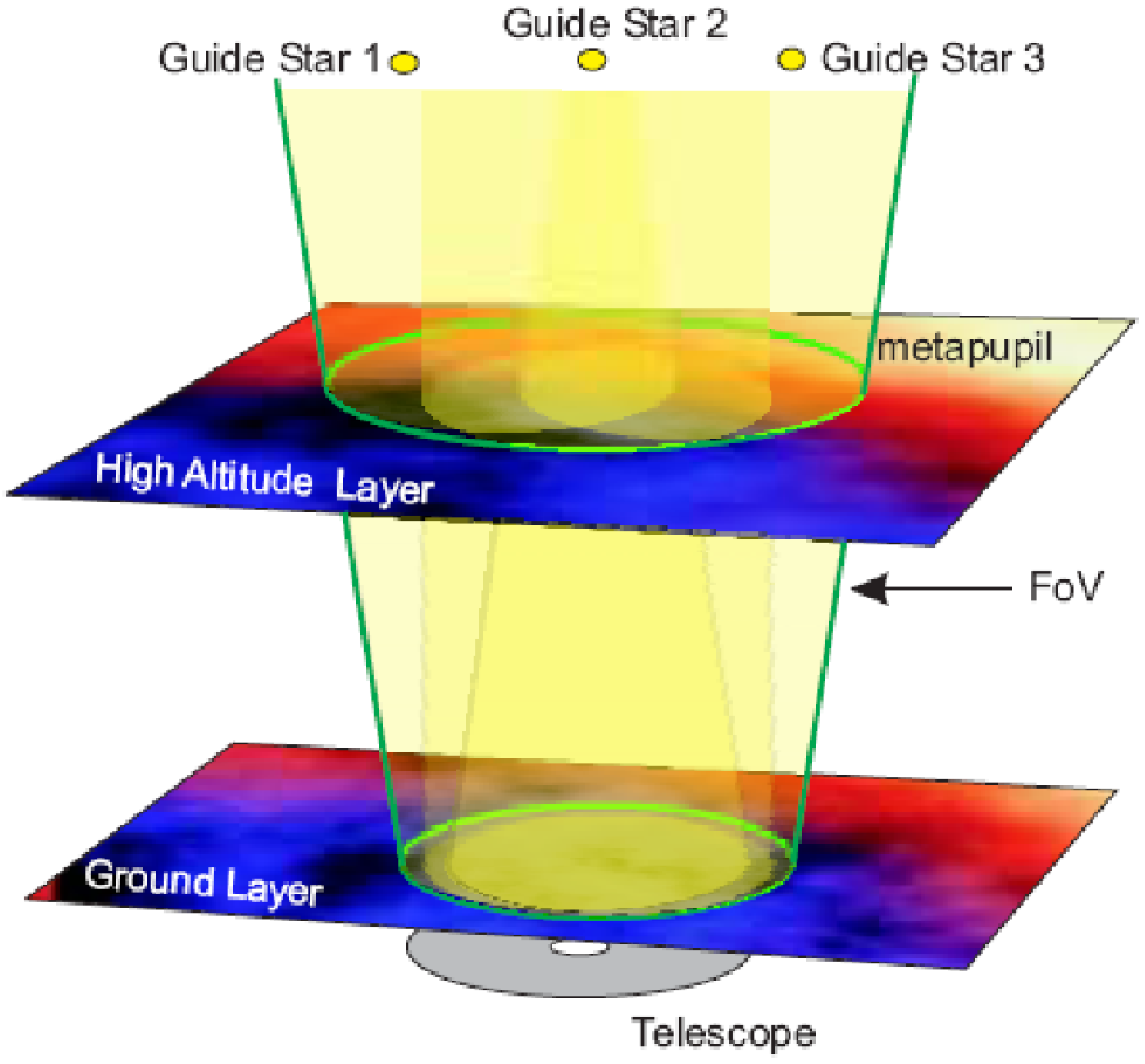} \hspace{1.0cm}
   \includegraphics[width=6.0cm]{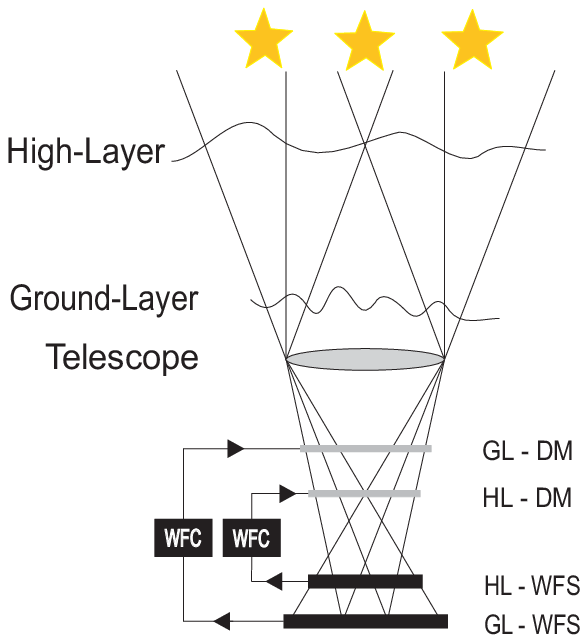}
 \end{center}
 \caption[Principle of Multi Conjugated Adaptive Optics correction]
 {\textit{Left:} Principle of Multi Conjugated Adaptive Optics
 correction. Several guide stars are used to probe the 3-dimensional structure of the atmosphere By illuminating the volume within the FoV and distinctive
 layers are corrected. A full
correction for the higher layer is only guaranteed if this layer
is fully covered by the footprints of the guide star metapupils. (Image taken from \cite{KellnerPHD}).
 \textit{Right:} Schematics of the layer oriented MCAO approach with
 two layers. WFS, DM, and WFC are labels for the wavefront sensor,
 deformable mirror and wavefront computer for the high layer (HL)
 and ground layer (GL), respectively.}
\end{figure}

\subsection{Star Oriented Approach}
In the star oriented mode, each reference star is observed by one
wavefront sensor and typically one detector. The information from the different
directions of the guide stars is combined to generate 3D
information of the atmosphere within the mapped FoV. In this
approach of turbulence tomography \citep{Tallon1990} the influence
of a single layer can be computed and corrected with one
deformable mirror conjugated to this layer. The first verification
of this approach was done in an open loop measurement at the
Telescopio Nazionale di Galileo (TNG) \citep{Ragazzoni2000}.

\subsection{Layer Oriented Approach}
In the layer oriented approach \citep{Ragazzoni2000b}, each WFS
and detector is conjugated to one layer in the atmosphere instead
to a single star. The light of several guide stars is optically
co-added to increase the SNR on the detector, such that also
fainter stars can be used as guide stars. This increases the sky
coverage, the fraction of regions on the sky that can offer a
suitable asterism, substantially for this approach. Also the
number of needed wavefront sensors and detectors is reduced,
reducing the detector read-out-noise and the needed computing
power compared to the SO approach. Only as many detectors are
needed as layers are being corrected and not as many guide stars
are used. Information from one WFS, and therefore one layer, can
directly be fed to the corresponding DM. This results in a
combination of independent control loops for the different layers
and allows to adjust the integration time and bin size on
the detector independently
to the characteristics of the conjugated layer.\\
Distortions introduced by turbulence in layers close to the layer
for which correction is attempted contribute stronger to the
measurements than those from layers further away. The further away
a layer, the more its introduced aberrations are smoothed out
\citep{Diolaiti2001}.

\noindent A limitation of this approach can be the different
brightnesses of the used guide stars. This can lead to an
overestimation of the turbulence from a certain direction, when
the light is co-added \citep{Nicolle2004}. An asterism of stars
with similar brightness is therefore preferable.

\subsection{Ground Layer Adaptive Optics}
As already pointed out in Chap.~\ref{sec:Turbulence} most of the
turbulence in the atmosphere is generated in the ground layer.
Correcting only this layer, one can remove the major contributor
to the phase aberrations of the incoming wavefronts
\citep{Rigaut2002}. Additionally, the correction is valid over a
large FoV, as the light coming from different directions passes
through the same region of turbulence near the ground, because of
its small distance from the telescope pupil. In principle Ground
Layer Adaptive Optics (GLAO) corrections can be operated by any
MCAO system, operating just one single correction loop conjugated
to the ground layer. In this context a Rayleigh Laser Guide Star
(LGS), which is focused to about 5 - 10 km altitude, can be used,
as it automatically illuminates only the ground layer and is not
usable for correction of higher layers but still valid for ground
layer corrections \citep{Morris2004}.\\

\subsection{Current and Future MCAO Systems}
Several MCAO instruments are planned and being built for different
telescopes. The first on-sky tested MCAO system is the MAD
instrument at the ESO/VLT. It will be described in detail in the
next section.\\
\noindent At the Gemini South observatory on Cerro Pachon, Chile,
a laser guide star assisted MCAO System, GeMS, is being installed
with first light expected soon. The system will consist of 5 laser
guide stars and additionally up to three natural guide stars. The
single 50~W laser is split into five beams which are launched from
behind the secondary mirror. The laser beacons will be located in
the middle and the four corners of the $1.2\arcsec \times
1.2\arcsec$ FoV. Three deformable mirrors are going to be used to
correct for turbulence in three layers: at ground level, 4.5~km
and 9~km altitude. Shack-Hartmann wavefront sensors will be used
in this star oriented MCAO approach.\\
\noindent The Fizeau-Interferometer LINC-NIRVANA for the Large
Binocular Telescope (LBT) on Mt. Graham in Arizona, will be
equipped with four layer oriented correction units, two for each
telescope, which will correct the ground layer and a high layer
\citep[e.g.][]{Farinato2008}. In the so-called LINC mode the
instrument will work with a classical single guide star AO system.
In the later implementation phase called NIRVANA, the MCAO system
will be able to use up to 12 NGS per side for the GLAO plus up to
8 NGS for the high layer adaptive optics correction (HLAO). The
stars for GL correction can be located anywhere in a ring with an
inner diameter of $2\arcmin$ and an outer diameter of $6\arcmin$
around the central science field of $10.5\arcsec \times
10.5\arcsec$ and the guide stars for the HLAO inside a circle with
a diameter of 2 arcminutes.

\section[MAD - Multi conjugated Adaptive optics Demonstrator]{MAD - Multi conjugated Adaptive optics Demonstrator}
\label{sec:MAD}
\begin{figure}[b]
 \begin{center}
  \includegraphics{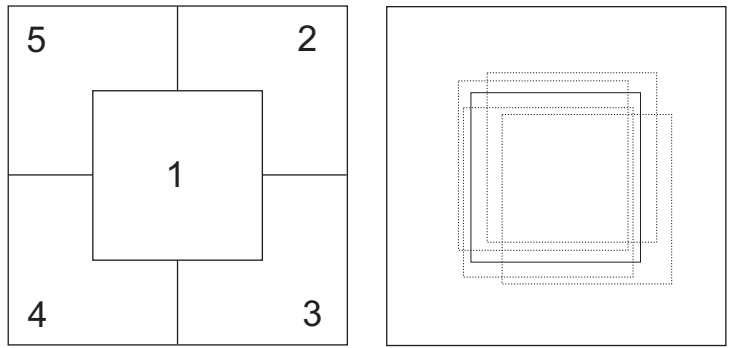}
 \end{center}
 \caption[Examples for jitter offsets with MAD]{Examples
 for jitter offsets with MAD to scan the $2\arcmin \times 2\arcmin$ FoV. \textit{Left:} Mosaic of 5 pointings covering the full
 $2\arcmin \times 2\arcmin$ FoV. \textit{Right:} Jitter offsets
 for 5 pointings around the center of the FoV.}
 \label{fig:MAD_jitter}
\end{figure}
The ESO Multi conjugated Adaptive optics
Demonstrator (MAD) is a prototype MCAO instrument, which was used
to test different MCAO reconstruction techniques in the laboratory
and on sky \citep{Hubin2002, Marchetti2003, Arcidiacono2006}. After extensive
testing in the laboratory it was installed at the Nasmyth-focus
platform of the ESO VLT UT3 telescope Melipal in the beginning of 2007. Because the
instrument bench is fixed to the Nasmyth-platform, the pupil
co-rotates with the field and an optical de-rotator at the
entrance of the adaptive optics system is needed, as well as for
the science camera.\\
\begin{figure}[t]
 \begin{center}
  \includegraphics[height=9.5cm]{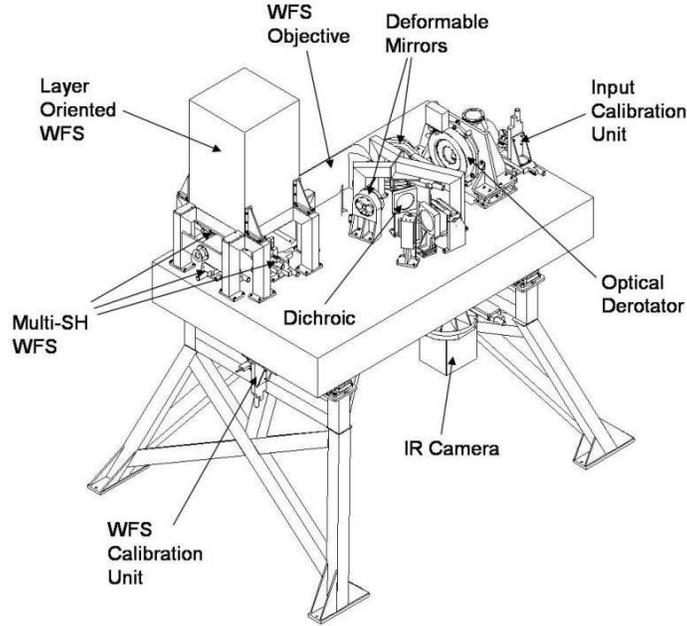}
 \end{center}
  \caption[3-dimensional view of the MAD bench]{3-dimensional view of
 the MAD bench with the major instrument components marked.
 The light beam enters at the de-rotator, passing the two
 deformable mirrors (DM) and is split by the dichroic into the
 IR part which is led to the science camera and the visible part
 which is led to the wavefront sensor objective. The layer
 oriented wavefront sensor is located above the star oriented multi Shack-Hartmann (SH) sensor.}
\label{fig:MADLayout}
\end{figure}

\noindent MAD is designed to characterize the performance of both
MCAO approaches, the star oriented one and the layer oriented one.
It is optimized for corrections in the $\rm K'$ band ($2.2\mu m$)
over a circular $2\arcmin \times 2\arcmin$ FoV using natural guide
stars (NGS). For the SO approach a multi Shack-Hartmann wavefront
sensor, consisting of three movable Shack-Hartmann WFS (SHS) to
look at any star present in the FoV, is implemented. SHS consist
of a lenslet array that samples the incoming WF. Each lens forms
an image of the guide star and the displacement of this image from
a reference position gives an estimate of the slope of the local
wavefront at that lenslet. The guide stars used for sensing the
distortions should not be fainter than $m_{v} = 14$. The layer
oriented approach uses a multi Pyramid wavefront sensor (PWS)
which is capable to sense up to 8 NGS simultaneously. The PWS
consists of a four-sided glass pyramid which tip is located in the
focal point of the telescope and it measures, similar to the SHS,
the local slope of the wavefront. A uniform distribution of these
stars is preferable but they can be everywhere in the $2\arcmin
\times 2\arcmin$ FoV. The two different WFSs cannot be used
simultaneously, a selector folds the light
either to the SO-WFS or the LO-WFS.\\
\noindent Two deformable mirrors conjugated to the ground layer
and a layer at 8.5~km altitude correct for the turbulence induced
phase errors. A dichroic splits the light into an IR part ($1.0 -
2.5~\mu m$), led towards the science detector and a visual part
($0.45 - 0.95~\mu m$) led towards the wavefront sensor path. The
CAMCAO (CAmera for MCAO) IR camera, built by the Faculdade de
Ci\^{e}ncias da Universidade de Lisboa (FCUL), is the science
camera of MAD \citep {Amorim2004}. It consists of a $2k \times 2k$
Hawaii2 IR detector with a pixel scale of 0.028 arcsec/pixel and a
FoV of $57\arcsec \times 57\arcsec$. The CAMCAO optics provide
diffraction limited images down to the J ($1.25~\mu m$) band, and
it is equipped with standard IR band filters for J ($1.25~\mu m$),
H ($1.65~\mu m$), $\rm K_{s}$ ($2.2~\mu m$), Br-gamma ($2.165~\mu m$) and
Br-gamma continuum. In Fig.~\ref{fig:MADLayout} a detailed sketch
of the layout of the instrument is shown. Unlike other
instruments, with MAD the camera and not the telescope is moved to
jitter within the field of view. Jitter movements are small
offsets from the central pointing to avoid bad pixel coincidences
on always the same position in the science field, to better
estimate the sky background and/or to scan a larger FoV. The
detector can cover the full $2\arcmin \times 2\arcmin$ FoV by
moving into the four adjacent $1\arcmin \times 1\arcmin$ quadrants
using linear x-y stages, see Fig.~\ref{fig:MAD_jitter} for
examples of jitter offsets for observations with MAD.\\

\noindent On sky testing of the star oriented mode was started in
February/March 2007 with the first closing of the MCAO loop on
March 25th \citep{Marchetti2007}. This first demonstration run
consisted of 8.5 effective nights spread over 12 nights in total,
with another demonstration run following for the SO mode. During
the third demonstration run in September 2007 the LO mode was
tested \citep{Arcidiacono2008, Ragazzoni2008, Falomo2009}. After
the very successful demonstrations of the performance of MAD,
three public science demonstration runs for the star oriented mode
followed in November 2007, January 2008 and August 2008, where
high resolution galactic as well as extragalactic science was
performed \citep[see e.g.][]{Bouy2009, Wong2009}.

\section{Goal of this Work}
The reason for analyzing MAD data was the uniqueness of this very
first MCAO data. I want to analyze the stability of the multi
conjugated adaptive optics correction regarding the potential
achievable astrometric precision. MCAO observations offer the
advantage of a big field of view with a resolution close to the
diffraction limit. Although most of this advantage is seen in the
enhancement in photometric studies, there are also some very
interesting applications in the field of astrometry. Cluster
dynamics, measuring velocity dispersions and common proper motions
is just one of the many possibilities. Another very interesting
case is the possibility of detecting the astrometric signal of a
planetary companion orbiting its star. Limitations of today's AO
based searches are often the small FoV of classical AO imagers.
Having an $1\arcmin \times 1\arcmin$ FoV with several reference
stars will enable more precise astrometric measurements. But
before starting with such science, one should check and analyze
the performance of these systems, to see which effects this
special kind of AO system has on astrometric measurements. When
the first observations with such a system, the MAD instrument,
were performed I had the chance to analyze some of these data. I
decided to work with the data obtained in the layer oriented mode,
as this will be the approach the future LBT instrument LINC
NIRVANA will work with.\\
\noindent The goal of this work is to see how stable the AO
performance of MAD is over time in terms of astrometric stability
and to measure the achievable positional precision in the first
MCAO layer oriented data.

    \cleardoublepage
     \chapter{Astrometry with MAD}
       \section{Observations}

The observations analyzed here were conducted with the MAD
instrument and the multi pyramid sensor in the LO mode. I
analyzed data from two globular clusters, 47~Tucanae (NGC~104) and
NGC~6388, which were observed during the demonstration run in
September 2007 by our colleagues from the INAF Osservatorio
Astronomico di Padova, Italy \citep{Arcidiacono2008,
Ragazzoni2008, Falomo2009}. Due to bad luck with the observing
conditions, these two sets are the best data obtained during the
only layer oriented run carried out with MAD. In the case of the
data of 47~Tuc I analyzed the data of the central $57\arcsec
\times 57\arcsec$ FoV in the context of this work. In the case of
the NGC~6388 cluster I analyzed a data set observed under good
initial seeing conditions $0.46\arcsec$), which lies at an outer
part of the cluster. Another field in the center of the cluster
was also observed, but with an initial seeing of $1.76\arcsec$,
which I did not analyze in this work. Unfortunately there will be
no more data to compare my results
with, as MAD is no longer offered.\\

\noindent I want to emphasize that the MAD instrument, as its name
already says, was built to \textit{demonstrate} that MCAO
correction over a big field of view is possible. It did this with
great success! But one also has to keep in mind, when interpreting
the results shown in the next chapters, that this data is test
data with all the possible problems during first observations. Not
the full performance of this new AO correction technique can be
expected, for this one has to wait for the next generation of MCAO
instruments, which will be fully optimized. Nevertheless, this is
the first attempt to analyze and characterize a layer oriented
MCAO system with respect to its astrometric performance.

\subsection{GLAO - 47~Tuc}
\begin{figure}
 \begin{center}
  \includegraphics[width=12cm, angle=90]{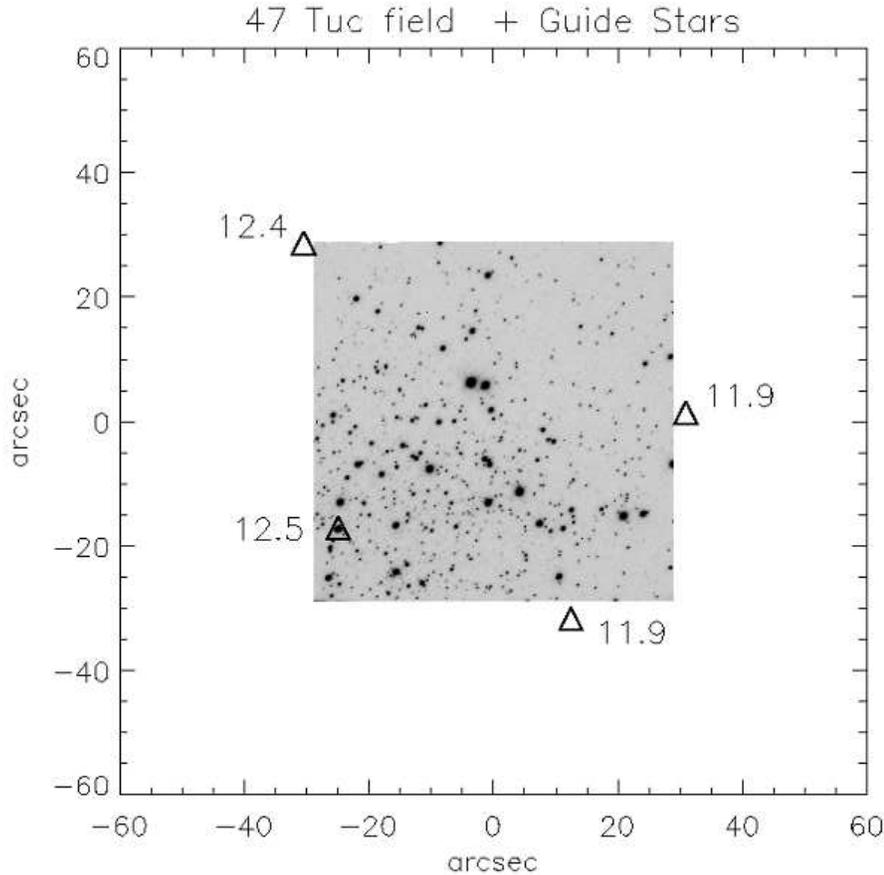}
 \end{center}
 \caption[MAD image of the core of the globular cluster 47~Tuc]
 {MAD image of the core of the globular cluster 47~Tuc.
 The triangles mark the positions of the AO guide stars relative
 to center of the observed FoV. The numbers close to the stars
 correspond to their F606W (visual) magnitude.}
 \label{fig:47Tucguidestars}
\end{figure}

The observations of the globular cluster 47~Tuc were obtained on
September 22nd 2007 using only the Ground Layer Adaptive Optics
(GLAO) approach. The center of the cluster, RA(J2000)=00:24:05.6,
DEC(J2000)=-72:04:49.4, was observed in the Br-$\gamma$ filter.
The camera was not moved during the exposures to scan the full FoV
but instead stayed at the same position, observing a $57\arcsec
\times 57\arcsec$ field. Four guide stars with magnitudes V =
11.9, 11.9, 12.4 and 12.5 mag\footnote{HST F606W photometry data},
corresponding to an integrated magnitude of 10.63
\citep{Arcidiacono2008}, positioned around this field with one
guide star in the lower left corner of this field were used to
sense the wavefront distortions due to the atmosphere, see
Fig.~\ref{fig:47Tucguidestars}. Altogether I have 19 frames with
NDIT = 15 exposures of an integration time of DIT $= 2$ seconds.
Each frame is averaged to correspond to a 2 seconds exposure.
Seven sky frames were obtained before the cluster observation with
the same values for DIT and NDIT and also averaged to frames with
an exposure time of 2 seconds each. In
Table~\ref{tab:observingconditions1} the observation is summarized
together with the atmospheric conditions during the observations,
indicated by the seeing value and the performance of the system
indicated by the value of the fitted FWHM.
\begin{table}[t]
\begin{center}
\small{
\begin{tabular}{|l|c|c|c|c|c|c|c|c|} \hline
Object & frame & DIT & NDIT & \multicolumn{2}{c|}{rel. jitter} & AO correct. & seeing [$\arcsec$] & FWHM [$\arcsec$] \\
 & & & & ~~x~~ & y & & in $V$ & in $Br_{\gamma}$ \\ \hline
 & 1  & 2 & 15 & 0 & 0 & GLAO & 1.09 & 0.178\\
 & 2  & 2 & 15 & 0 & 0 & GLAO & 1.15 & 0.186\\
 & 3  & 2 & 15 & 0 & 0 & GLAO & 1.13 & 0.206\\
 & 4  & 2 & 15 & 0 & 0 & GLAO & 1.08 & 0.169\\
 & 5  & 2 & 15 & 0 & 0 & GLAO & 1.09 & 0.178\\
 & 6  & 2 & 15 & 0 & 0 & GLAO & 1.08 & 0.147\\
 & 7  & 2 & 15 & 0 & 0 & GLAO & 1.15 & 0.157\\
 & 8  & 2 & 15 & 0 & 0 & GLAO & 1.17 & 0.193\\
 & 9  & 2 & 15 & 0 & 0 & GLAO & 1.17 & 0.173\\
47~Tuc & 10 & 2 & 15 & 0 & 0 &  GLAO & 1.15 & 0.178\\
 & 11 & 2 & 15 & 0 & 0 & GLAO & 1.14 & 0.145\\
 & 12 & 2 & 15 & 0 & 0 & GLAO & 1.15 & 0.148\\
 & 13 & 2 & 15 & 0 & 0 & GLAO & 1.11 & 0.144\\
 & 14 & 2 & 15 & 0 & 0 & GLAO & 1.15 & 0.166\\
 & 15 & 2 & 15 & 0 & 0 & GLAO & 1.14 & 0.183\\
 & 16 & 2 & 15 & 0 & 0 & GLAO & 1.15 & 0.183\\
 & 17 & 2 & 15 & 0 & 0 & GLAO & 1.19 & 0.187\\
 & 18 & 2 & 15 & 0 & 0 & GLAO & 1.13 & 0.200\\
 & 19 & 2 & 15 & 0 & 0 & GLAO & 1.11 & 0.174\\ \hline
\end{tabular}
 \label{tab:observingconditions1}
 \caption[Summary of the observations of the cluster 47~Tuc]
 {Summary of the observations of the cluster 47~Tuc. The
 seeing value is measured by the DIMM seeing
monitor in V band and the FWHM value corresponds to the one
measured in the data (see Chap.~\ref{sec:mad_positions})}}
\end{center}
\end{table}

\begin{figure}[t]
 \begin{center}
  \includegraphics[width=12cm, angle=90]{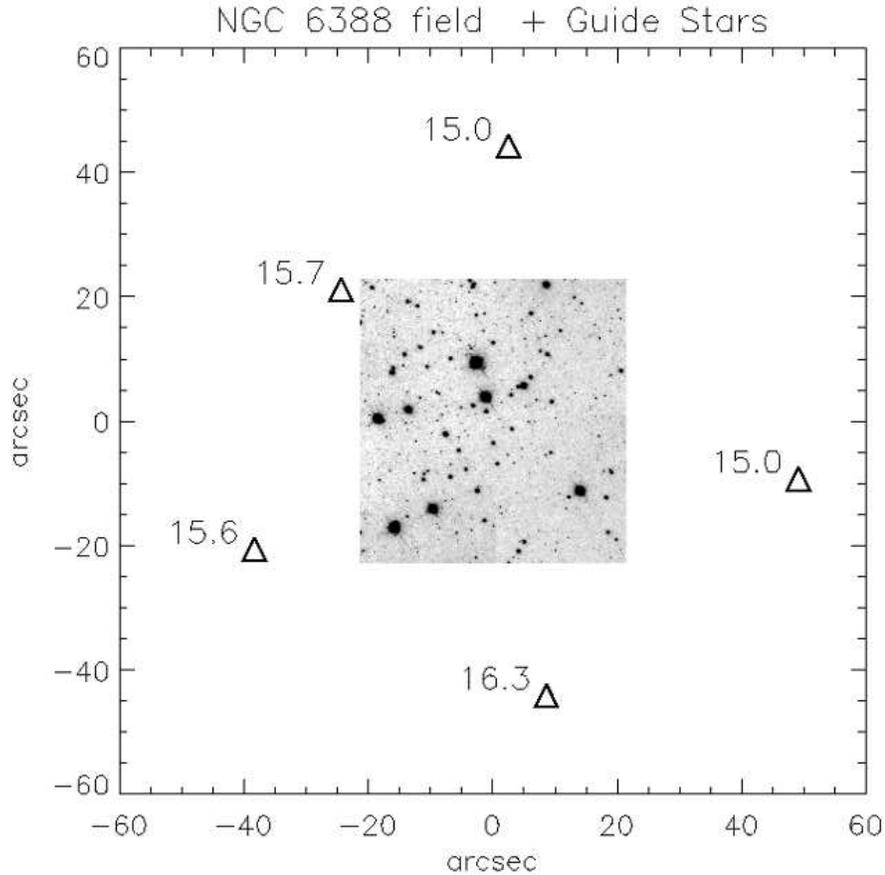}
 \end{center}
 \caption[MAD image of the globular cluster NGC~6388]
 {MAD image of the globular cluster NGC~6388. The triangles mark the
 positions of the AO guide stars relative to center of the observed
 FoV. The numbers close to the stars correspond to their F606W (visual) magnitude.}
 \label{fig:NGC6388guidestars}
\end{figure}
\subsection{MCAO - NGC~6388}
\begin{figure}[t]
 \begin{center}
  \includegraphics[width=12cm]{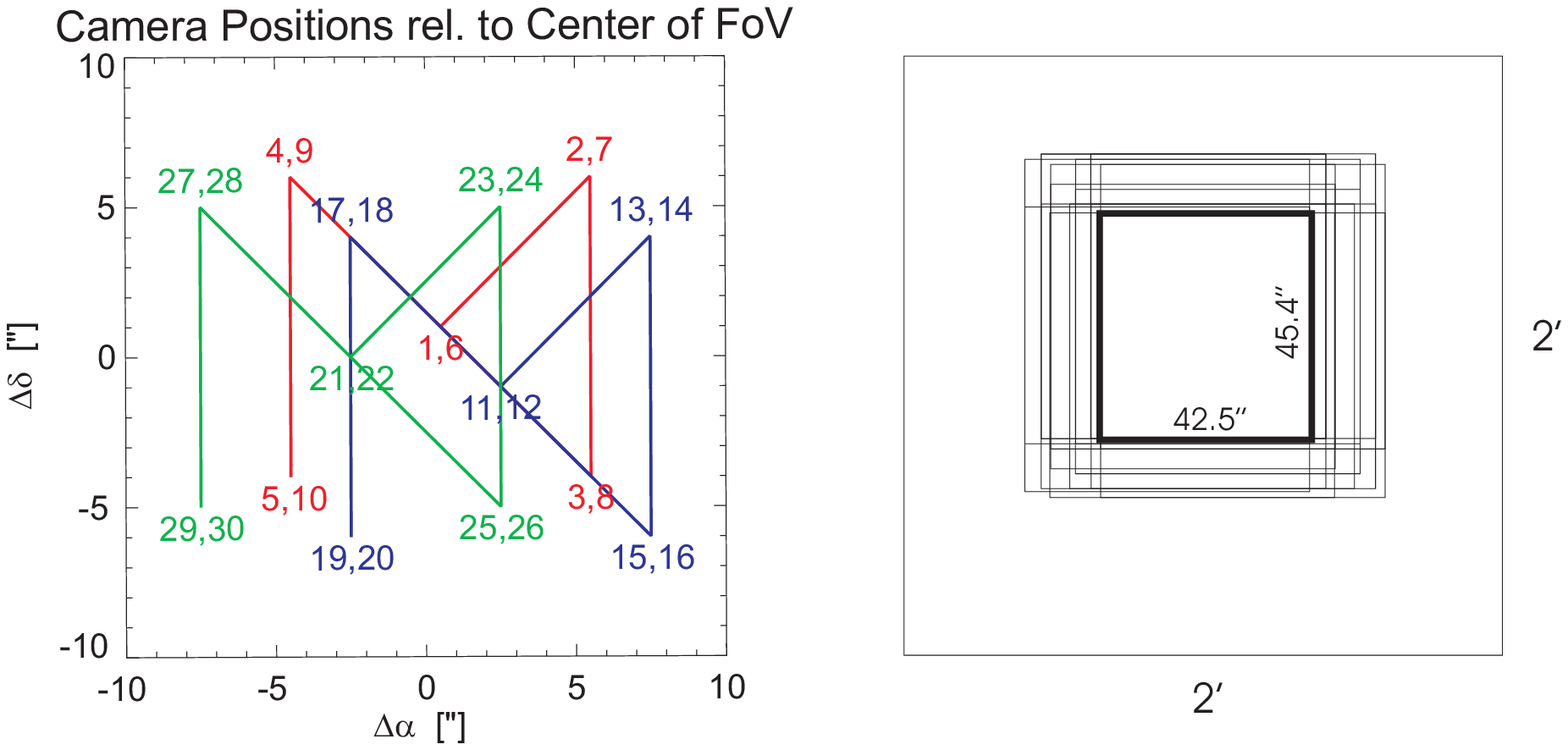} 
 \end{center}
 \caption[Jitter pattern of the observations of NGC~6388]
 {\textit{Left:} Jitter pattern of the observations of NGC~6388. The offsets are
 in seconds of arc and relative to the center of the FoV. The numbers
 indicate the frame which was taken at that position. Note the change
 of taking two images at one position for frames 11-30 and then moving
 to the next jitter position instead of taking one frame per position
 and execute the jitter pattern twice for frames 1-10. \textit{Right:}
 Frames taken in the $2\arcmin \times 2\arcmin$ FoV and the finally
 cut-out area common to all frames drawn with thick lines. This common
 area was used to investigate the astrometric precision.}
 \label{fig:NGC6388jitterpattern}
\end{figure}

The data of the globular cluster NGC~6388 was obtained on
September 27th 2007 using the full MCAO capability of MAD. The
observations are in the $K_{s}$ filter using 5 guide stars with $V
= 15.0, 15.0, 15.6, 15.7$ and 16.3 mag, corresponding to an
integrated  magnitude of 13.67 \citep{Arcidiacono2008}. The field
together with the guide stars is shown in
Fig.~\ref{fig:NGC6388guidestars}. The observed field lies at the
lower left corner of the cluster at RA(J2000)=17:36:22.86,
DEC(J2000)=-44:45:35.53. All together 30 frames were obtained, the
first five in GLAO and the last 25 in full MCAO. A jitter pattern
of five positions was used, repeated six times with three slightly
different central points, to scan part of the $2\arcmin \times
2\arcmin$ FoV (Fig.~\ref{fig:NGC6388jitterpattern}). The first 10 frames were obtained with DIT = 10 seconds and NDIT =
24, and in the last 20 frames the number of exposures was reduced
to NDIT = 12. All frames are averaged to correspond to a 10 second exposure.\\
\begin{table}
\begin{center}
\small{
\begin{tabular}{|l|c|c|c|c|c|c|c|c|c|} \hline
Object & frame & DIT & NDIT & \multicolumn{2}{c|}{rel. jitter} & AO correct. & seeing [$\arcsec$] & FWHM [$\arcsec$]\\
 & & & & ~~x~~ & y & & in $V$ & in $K_{s}$\\ \hline
 & 1  & 10 & 24 & 0 & 7 & GLAO & 0.43 & 0.098\\
 & 2  & 10 & 24 & 5 & 12 & GLAO & 0.41 & 0.094\\
 & 3  & 10 & 24 & 5 & 2 & GLAO & 0.49 & 0.099\\
 & 4  & 10 & 24 & -5 & 12 & GLAO & 0.55 & 0.094\\
 & 5  & 10 & 24 & -5 & 2 & GLAO & 0.51 & 0.090\\
 & 6  & 10 & 24 & 0 & 7 & MCAO & 0.41 & 0.095\\
 & 7  & 10 & 24 & 5 & 12 & MCAO & 0.38 & 0.097\\
 & 8  & 10 & 24 & 5 & 2 & MCAO & 0.37 & 0.098\\
 & 9  & 10 & 24 & -5 & 12 & MCAO & 0.39 & 0.103\\
 & 10 & 10 & 24 & -5 & 2 & MCAO & 0.40 & 0.106\\
 & 11 & 10 & 12 & 2 & 5 & MCAO & 0.45 & 0.126\\
 & 12 & 10 & 12 & 2 & 5 & MCAO & 0.43 & 0.117\\
 & 13 & 10 & 12 & 7 & 10 & MCAO & 0.45 & 0.130\\
 & 14 & 10 & 12 & 7 & 10 & MCAO & 0.50 & 0.158\\
NGC~6388  & 15 & 10 & 12 & 7 & 0 & MCAO & 0.51 & 0.130\\
 & 16 & 10 & 12 & 7 & 0 & MCAO & 0.49 & 0.155\\
 & 17 & 10 & 12 & -3 & 10 & MCAO & 0.48 & 0.170\\
 & 18 & 10 & 12 & -3 & 10 & MCAO & 0.49 & 0.140\\
 & 19 & 10 & 12 & -3 & 0 & MCAO & 0.45 & 0.119\\
 & 20 & 10 & 12 & -3 & 0 & MCAO & 0.41 & 0.133\\
 & 21 & 10 & 12 & -3 & 6 & MCAO & 0.42 & 0.120\\
 & 22 & 10 & 12 & -3 & 6 & MCAO & 0.44 & 0.131\\
 & 23 & 10 & 12 & 2 & 11 & MCAO & 0.54 & 0.139\\
 & 24 & 10 & 12 & 2 & 11 & MCAO & 0.56 & 0.158\\
 & 25 & 10 & 12 & 2 & 1 & MCAO & 0.43 & 0.176\\
 & 26 & 10 & 12 & 2 & 1 & MCAO & 0.50 & 0.153\\
 & 27 & 10 & 12 & -8 & 11 & MCAO & 0.46 & 0.143\\
 & 28 & 10 & 12 & -8 & 11 & MCAO & 0.48 & 0.135\\
 & 29 & 10 & 12 & -8 & 1 & MCAO & 0.47 & 0.120\\
 & 30 & 10 & 12 & -8 & 1 & MCAO & 0.47 & 0.134\\ \hline
\end{tabular}
 \label{tab:observingconditions2}
 \caption[Summary of the observations of the cluster
NGC~6388]{Summary of the observations of the cluster NGC~6388. The
 seeing value is measured by the DIMM seeing
monitor in V band and the FWHM value corresponds to the one
measured in the data (see Chap.~\ref{sec:mad_positions})} }
\end{center}
\end{table}

\noindent After the science frames, five sky frames were taken
outside the cluster using the same jitter pattern as in the
science frame. At each position 24 frames with 10 seconds exposure
time, averaged to one frame of 10 seconds each, were obtained.

\noindent In Table~\ref{tab:observingconditions2} the observations
are summarized together with the atmospheric conditions during the
observations, indicated by the seeing value.

\section{Data Reduction}
\subsubsection{NGC~6388}
Each science frame of the NGC~6388 cluster data was flatfielded by
the flatfield image obtained from sky flats at the beginning of
the night and badpixel corrected, by replacing the marked pixels
with the median of the pixels themselves and their 8 nearest
neighbors, with a badpixel mask obtained from the same flat-field
images. I did not create the flatfield and badpixel mask by my own
in this case, but used those made by our colleagues from Italy. A
description of how these frames were created can be found in
\citep{Moretti2009}. Briefly, the obtained sky flats were median
combined to one flatfield and also used to create the badpixel
mask.\\
\begin{figure}[b!]
 \begin{center}
  \includegraphics[width=14cm]{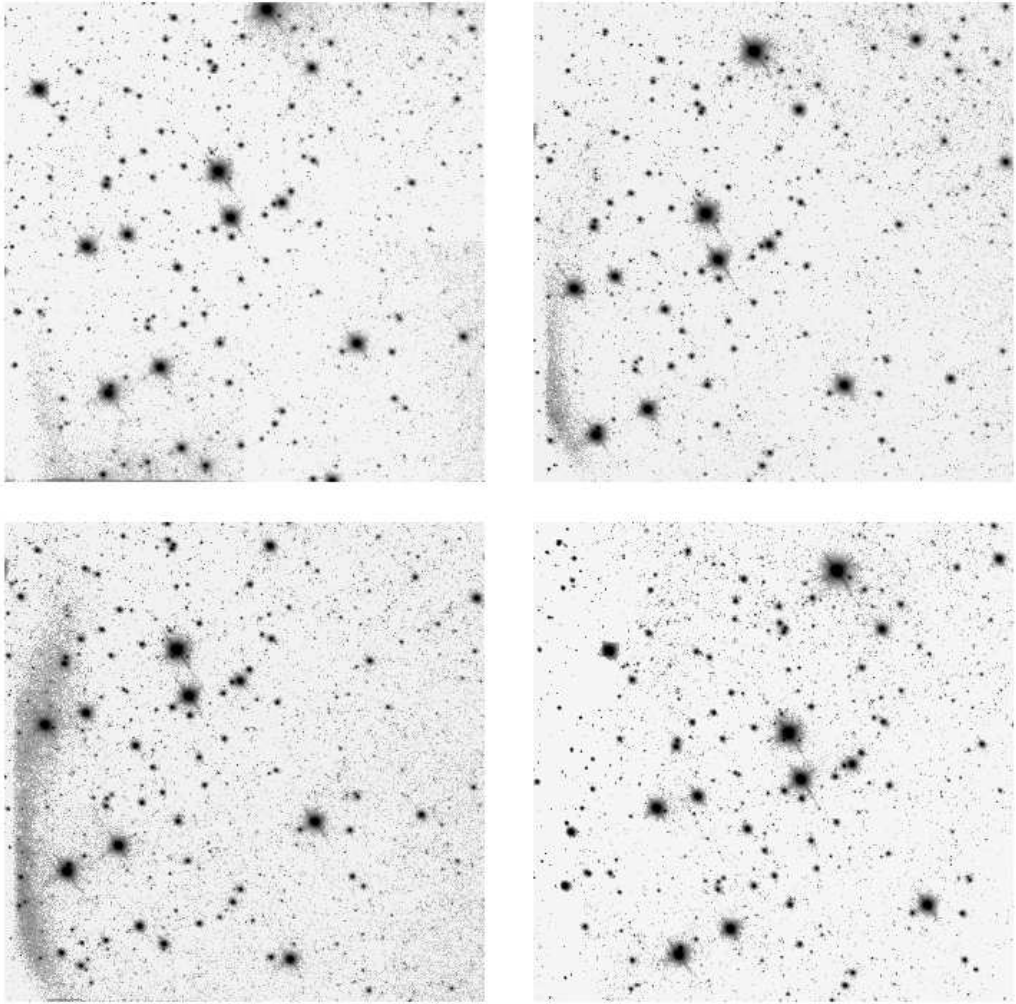}
 \end{center}
 \caption[Arc in the MAD frames produced by unshielded and unfiltered reflected light]
 {Arc in the MAD frames produced by unshielded and unfiltered reflected light.
 Several examples from different jitter positions show the intensity
 dependence of this arc on the jitter position. Some frames have a
 prominent arc, whereas it is not visible in others.}
\label{fig:NGC6388arc}
\end{figure}
\noindent Sky subtraction was done by median combining all
sky and science frames to get one single sky frame. Using only the
sky frames obtained right after the science frames to create the
sky frame, did not work satisfactorily. As only five sky-frames
were available, the remaining stars in the frames did not average
out perfectly, leaving small holes in the science frames after
subtracting the sky frame. Also $\kappa -\sigma$ clipping did not
yield a satisfactory result. Using all science frames together
with the sky frames to create a median image for the sky
estimation yielded the best result. Therefore, this resulting
image was used for sky subtraction. This sky frame was then
normalized to
the median counts of the science frame before subtraction.\\
\noindent Additionally, NaN and infinite values which were still
in the frames after the data reduction, were substituted by the
median of their 8 nearest neighbors, as the routines for the
following analysis had problems with such pixel values.\\
\noindent In the first demonstration runs with MAD a problem
occurred. Unfiltered light was reflected on the instrument bench
and could enter the science camera, producing a banana like arc on
the lower left side of the images. The intensity and position of
this arc depends on the position of the camera while jittering to
scan the FoV. This reflection could be avoided in the later
science demonstration runs by shielding the camera entrance with a
tube. Nevertheless, I see this arc in the images, see
Fig.~\ref{fig:NGC6388arc}. As it is position dependent, the arc
cannot be removed with the flatfield correction or sky
subtraction. The distribution of the arc is always close to the
border of the detector and only affecting a small portion of the
stars. While this extra, unshielded light is for sure a problem
for photometry, its influence on the astrometric positions of the
stars may be small. Nevertheless, I tried to avoid stars,
affected by the arc in my analysis.\\

\noindent In the case of the NGC~6388 data, jittering was used
during the observations and I cut the images to the common FoV
which was part of all images after the data reduction. That left
us with a slightly  smaller field of the size of $1517~\rm~px
\times 1623~\rm~px$ ($42.5\arcsec \times 45.4\arcsec$), see
Fig.~\ref{fig:NGC6388jitterpattern}, right panel. Only the stars in
this common field are taken into account for the
following astrometric analysis.\\

\subsubsection{47~Tuc}
In the case of 47~Tuc no flat-field images were taken in
Br$\gamma$ (central wavelength $= 2.116 \mu m$) in that night. As
I had no other choice I used the same flatfield image for
correction as for the NGC~6388 data observed in $K_{s}$ (central
wavelength $= 2.15 \mu m$), assuming that the pixel to pixel
variations are the same or very similar in these two filters,
whose central wavelength is not too different. As the data I got
from our collaborators for the cluster 47~Tuc was already sky
subtracted and corrected for the reflecting arc, I only corrected
for the flatfield, badpixel and NaN and infinite pixel values in
the same way as for the NGC~6388 data. The data of 47~Tuc was
obtained without moving the camera. All frames contain the same
field and I did not cut the single frames.\\

\section{Strehl Maps}
\label{sec:Strehl} 
\begin{figure}
 \begin{center}
   \includegraphics[width=6.3cm, angle=90]{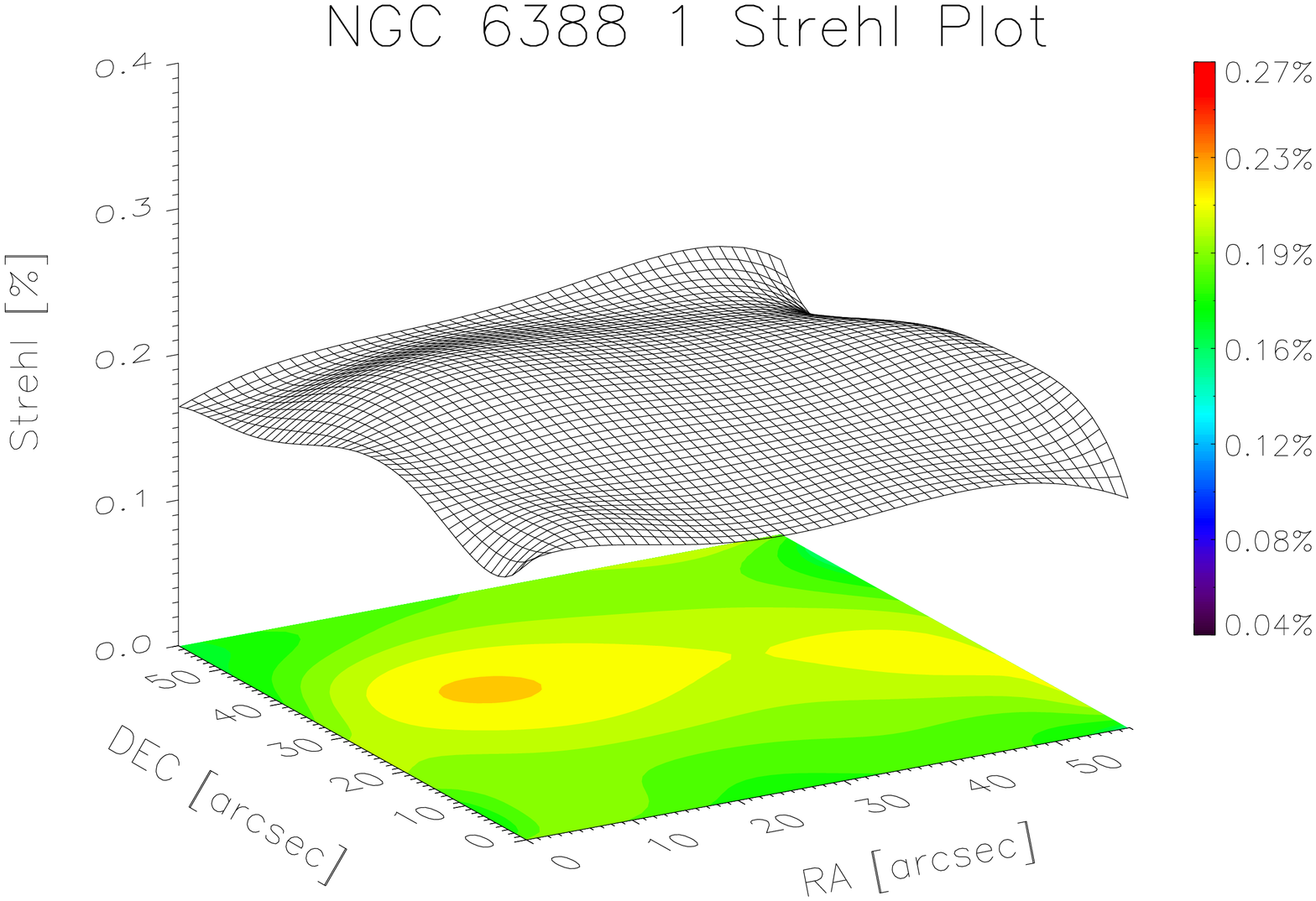}\\ \vspace{0.8cm}
    \includegraphics[width=6.3cm, angle=90]{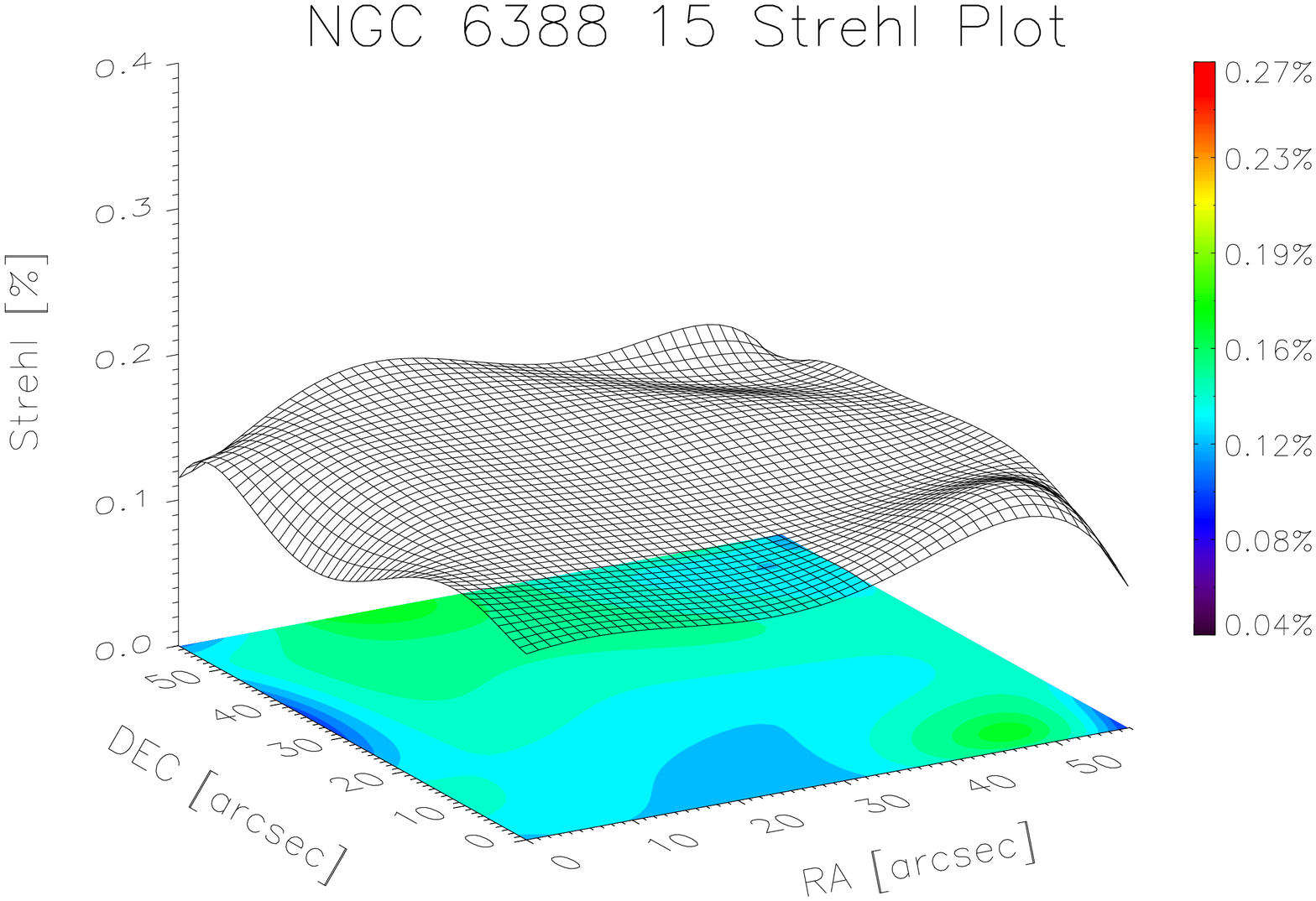}\\ \vspace{0.8cm}
    \includegraphics[width=6.3cm, angle=90]{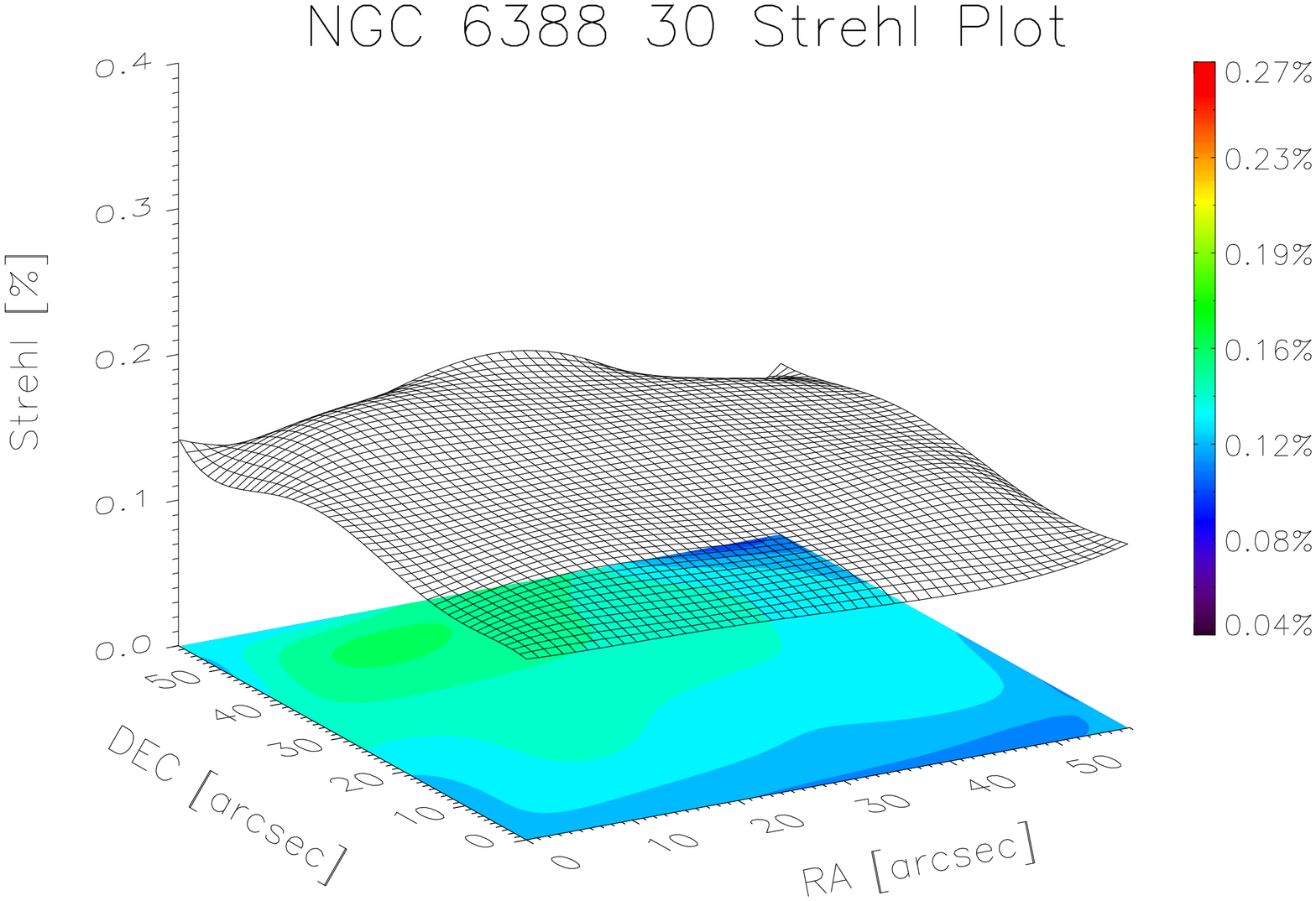}
 \end{center}
 \caption[Strehl-maps for some of the mad frames of the cluster NGC~6388]
 {Strehl-maps for some of the mad frames of the cluster NGC~6388.
 Shown are the first (GLAO), middle and last exposure (both MCAO).
 The degradation in performance with time, as described in the text,
 can be seen.}
\label{fig:mad_NGC6388_strehl}
\end{figure}

\begin{figure}
 \begin{center}
    \includegraphics[width=6.5cm, angle=90]{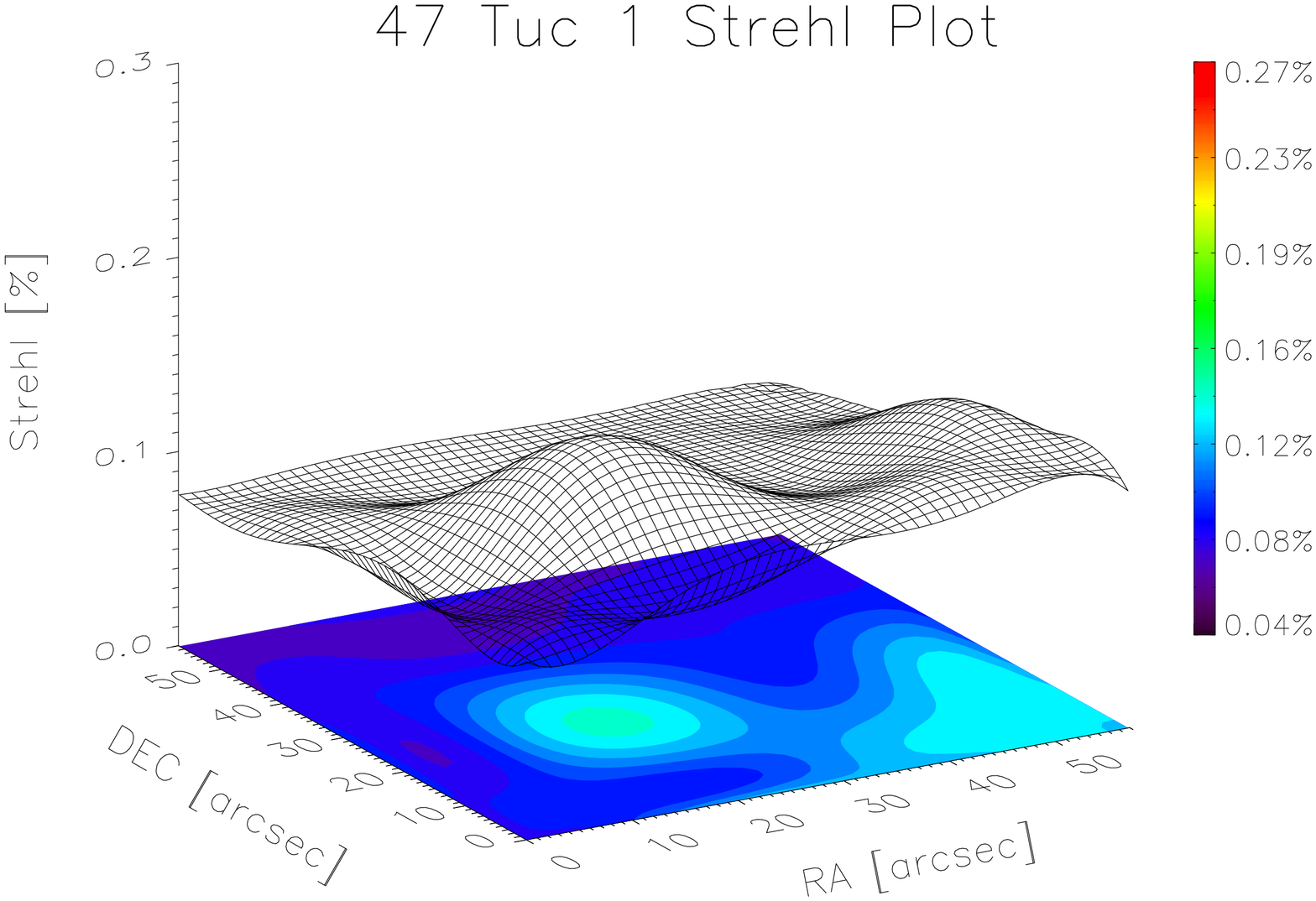}\\ \vspace{0.8cm}
    \includegraphics[width=6.5cm, angle=90]{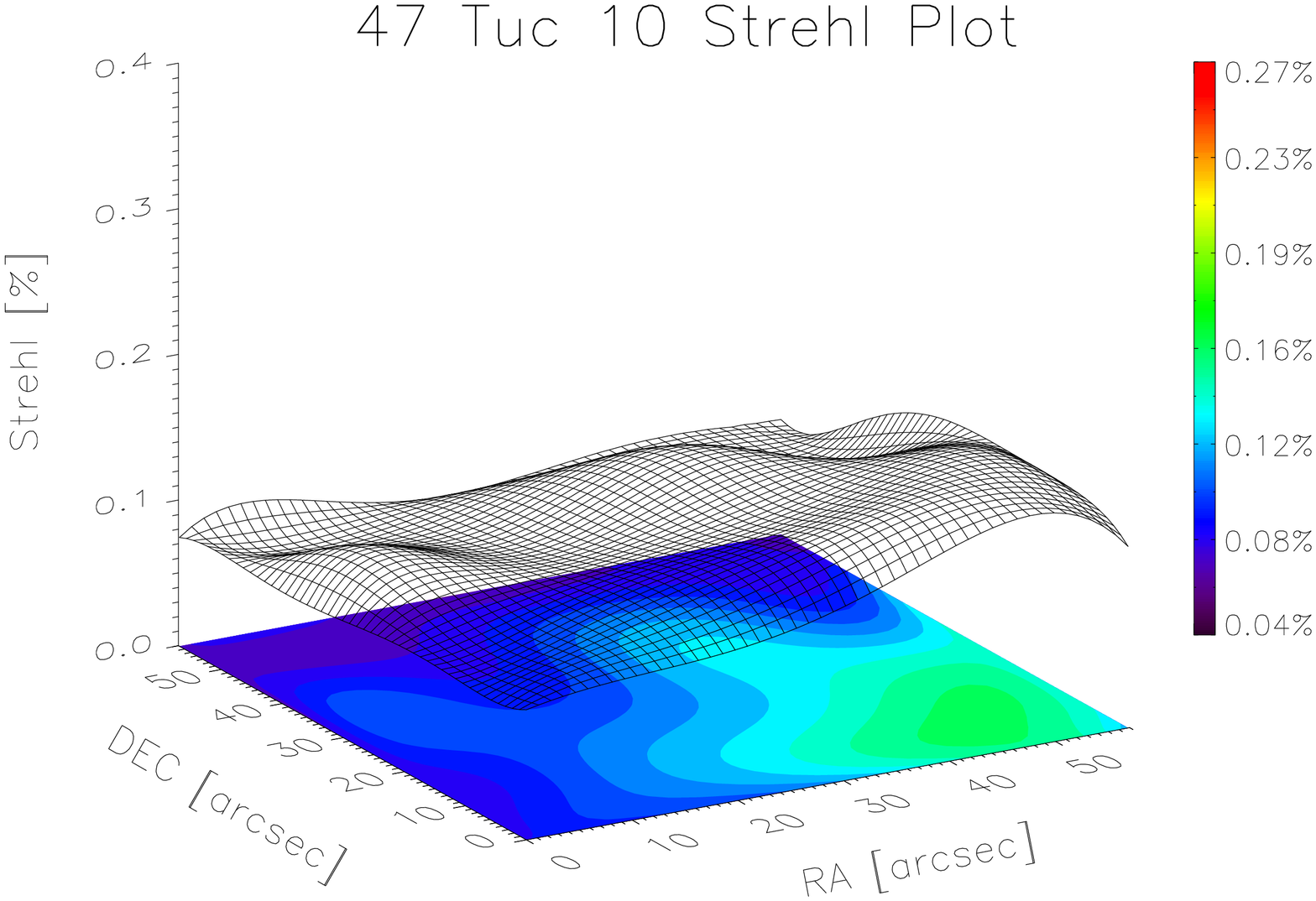}\\ \vspace{0.8cm}
    \includegraphics[width=6.5cm, angle=90]{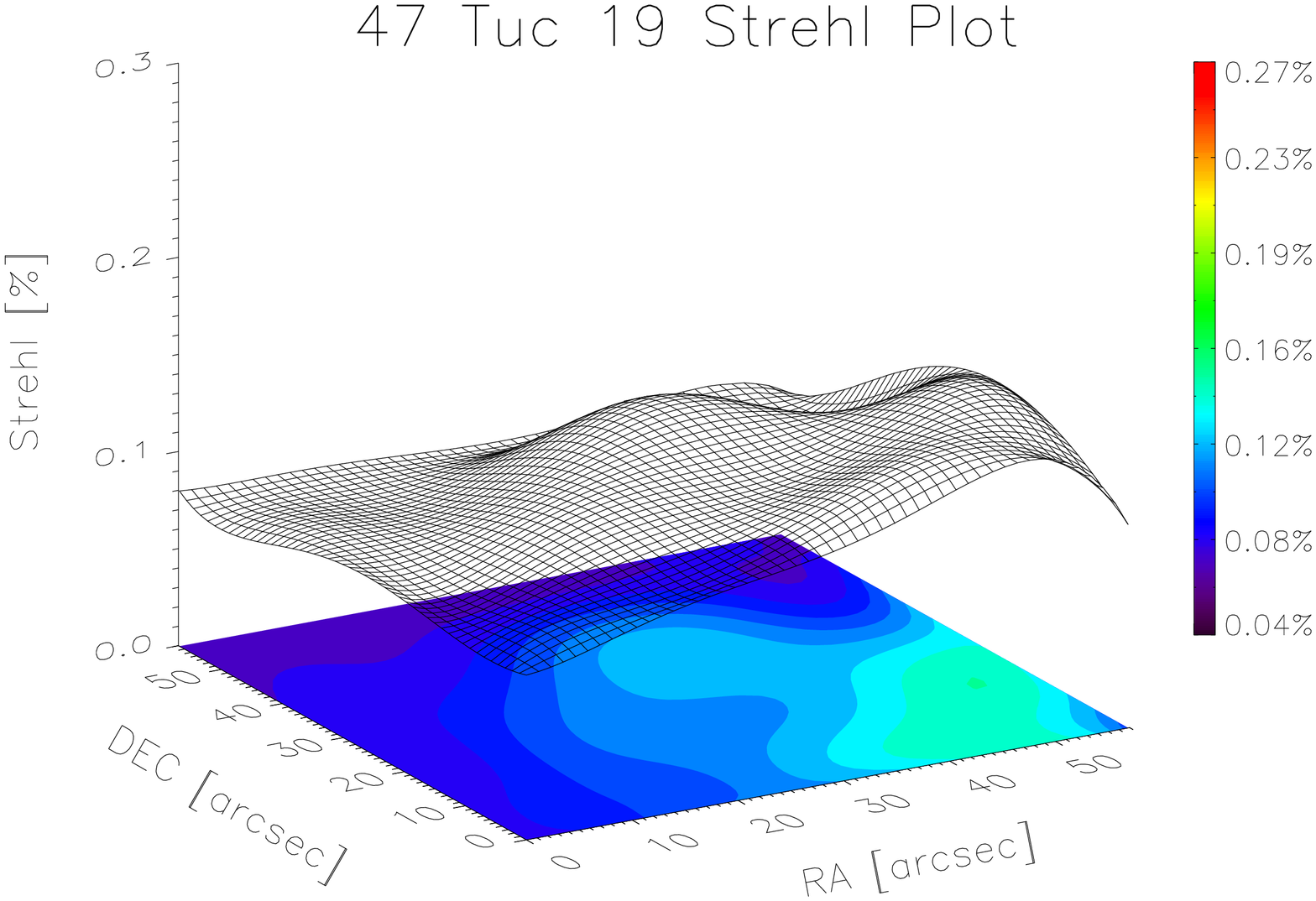}
 \end{center}
 \caption[Strehl-maps for some of the mad frames of the cluster 47~Tuc]
 {Strehl-maps for some of the mad frames of the cluster 47~Tuc obtained with GLAO.
 Shown are the first, middle and last exposure.}
\label{fig:mad_47Tuc_strehl}
\end{figure}

As a check of the AO performance I calculated
Strehl-maps for each frame\footnote{Program provided by Felix
Hormuth}. A theoretical diffraction limited PSF for MAD was
computed and normalized to a flux of one. With the Source
Extractor package \citep{Bertin1996} the stars in the frame were
detected and aperture photometry was used to calculate the flux of
the stars. After normalizing the flux to one, a comparison of the
theoretical peak value and the maximum value of the star yields
the Strehl ratio. After interpolating values for areas where no
stars were found a smooth surface was fitted to the data, leading
to a two dimensional Strehl-map for each frame. In
Fig~\ref{fig:mad_NGC6388_strehl} and \ref{fig:mad_47Tuc_strehl}
some of these Strehl-maps are shown, the first, middle and last
frame.\\

\noindent In the case of NGC~6388 a slight degradation in
performance with time can be seen, following roughly the change in
the seeing. The first five frames obtained with GLAO have the 
highest Strehl ratio and do not change with the changing seeing
conditions. The Strehl is fairly even over the field of view of
the camera with a small drop-off to the edges of the field. This
shows how well and uniformly the layer oriented MCAO approach
corrects wavefront distortions. The drop-off to the edges of the
FoV can partly be explained by the MCAO and the atmospheric
tomography approach. The light coming from the different
directions of the guide stars is optically co-added and a
correction is computed based on this light distribution. But the
footprints of the columns above the telescope in the direction of
the guide stars overlap more in the middle of the field than at
the edges in the higher layer. If the control software is not
optimized to correct over the whole FoV very evenly, the middle of
the field will be corrected better. As the data I am looking at is
the first data of MCAO and layer oriented mode, I am not surprised
to see such an effect. A performance evaluation of these data can
be found in \cite{Arcidiacono2008}.
\noindent In the case of 47~Tuc the Strehl is smaller than in
the case of the NGC~6388 data. The main reason for that is not
that this data is taken with correcting only the ground layer, but
more due to the circumstance that already the initial conditions
for the AO correction was very poor with a seeing of 1.09 - 1.19
arcseconds. An AO system can only enhance the performance
significantly, if the initial atmospheric conditions are fairly
good. Nevertheless, an even Strehl ratio of $\sim 15~\%$ over a
$1\arcmin \times 1\arcmin$ FoV is already an enhancement compared
to the seeing limited case.

\section{PSF Tests}
\label{sec:PSFTests} Before I decided what kind of model or
empiric PSF I should use to fit the positions of the stars in the
frames, I conducted several tests of the PSFs in the frames. I
analyzed the distributions of shape and orientation of the PSF
over the field of view, to check if I can fit the same PSF to all
stars or have to use different PSFs for different parts of the
FoV, which is mostly the case in classic single guide star
adaptive optics imaging. As I first only had the data of the
cluster NGC~6388, I conducted all these PSF tests with this data
set and based the following decisions and steps on the
consequential results. Later, after the analysis of the NGC~6388
data set was finished, I got the opportunity to also analyze GLAO
data of the cluster 47~Tuc for which I conducted the same tests.
Though they show certain differences, I nevertheless analyzed this
data set in the same way as the NGC~6388 data, to conserve the
possibility of a direct comparison of the results of the two data
sets.\\
\begin{figure}
 \begin{center}\hspace{0.3cm}
  \includegraphics[height=6.5cm, angle=90]{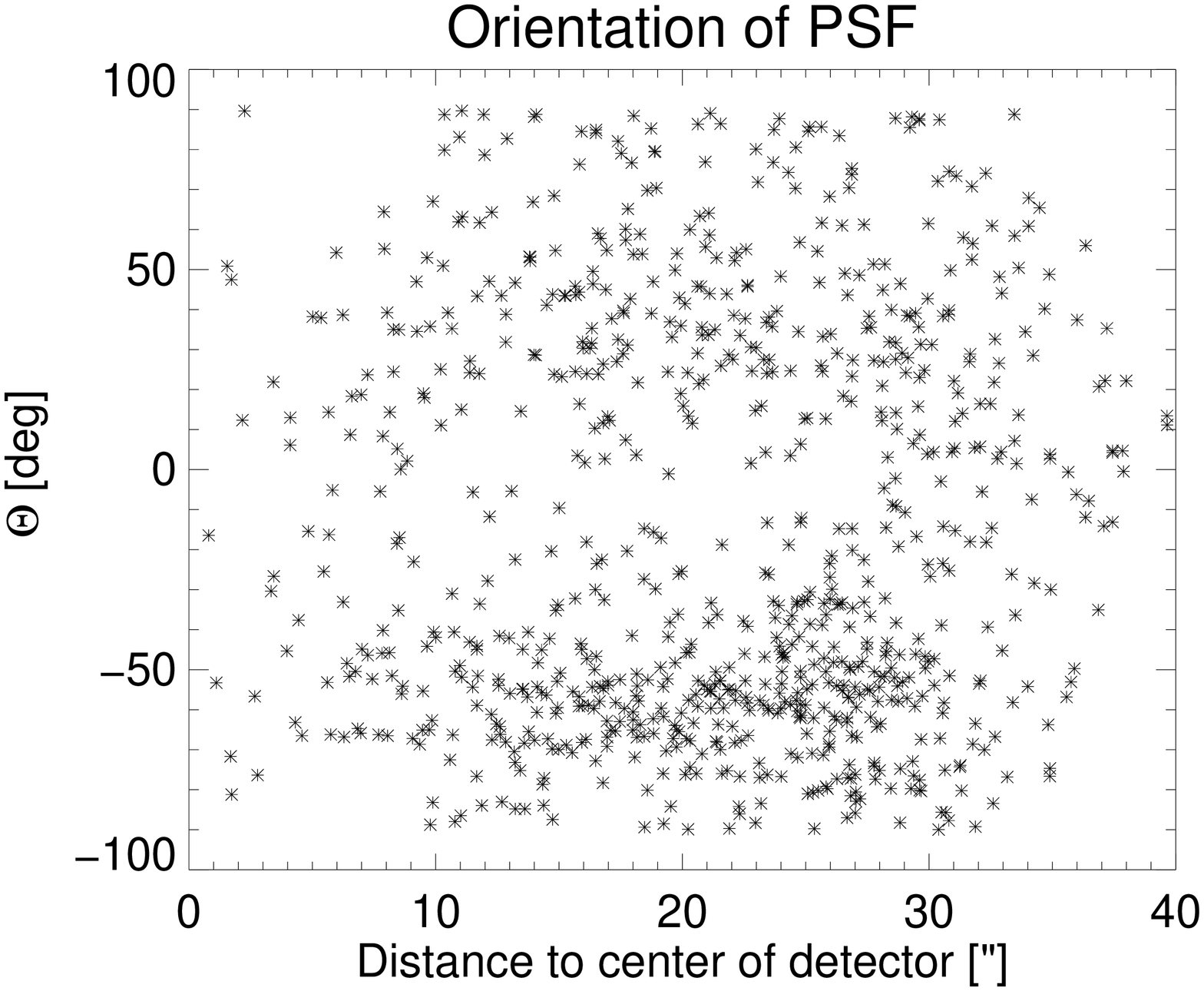}   \vspace{0.5cm} \hspace{0.8cm}
  \includegraphics[height=6.5cm, angle=90]{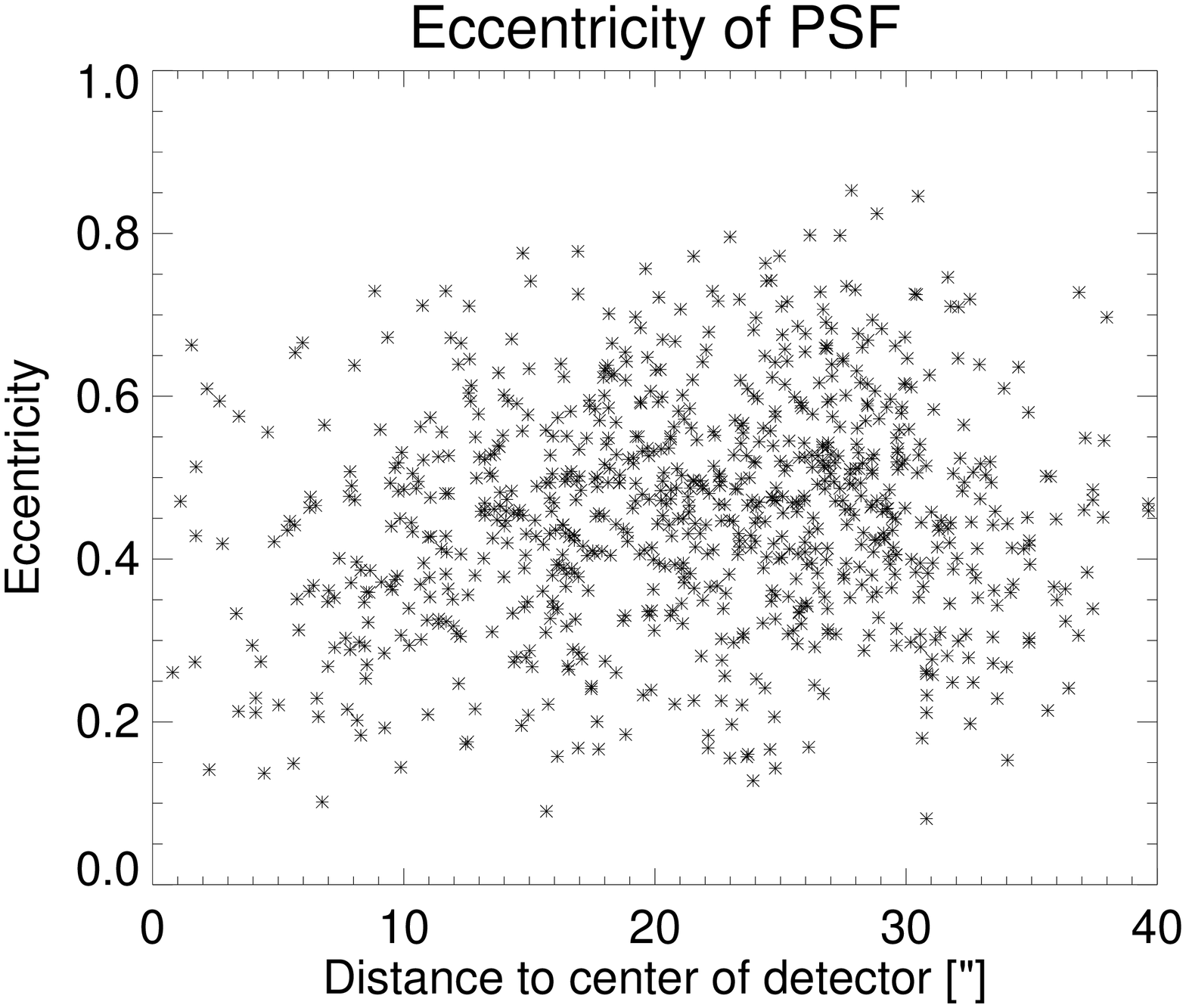} \\ \vspace{0.5cm}
\hspace{0.3cm}
  \includegraphics[height=6.5cm, angle=90]{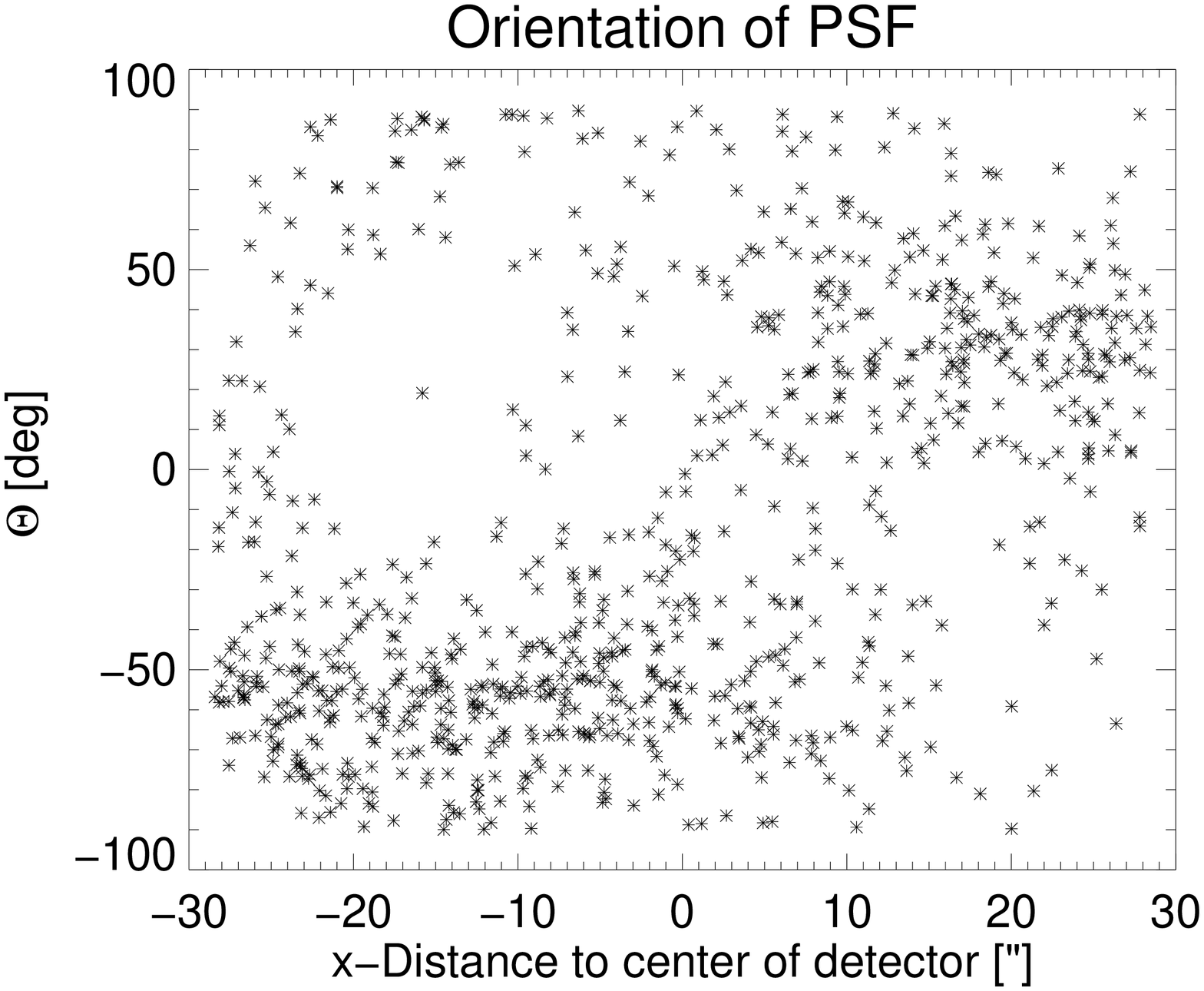}   \vspace{0.5cm} \hspace{0.8cm}
  \includegraphics[height=6.5cm, angle=90]{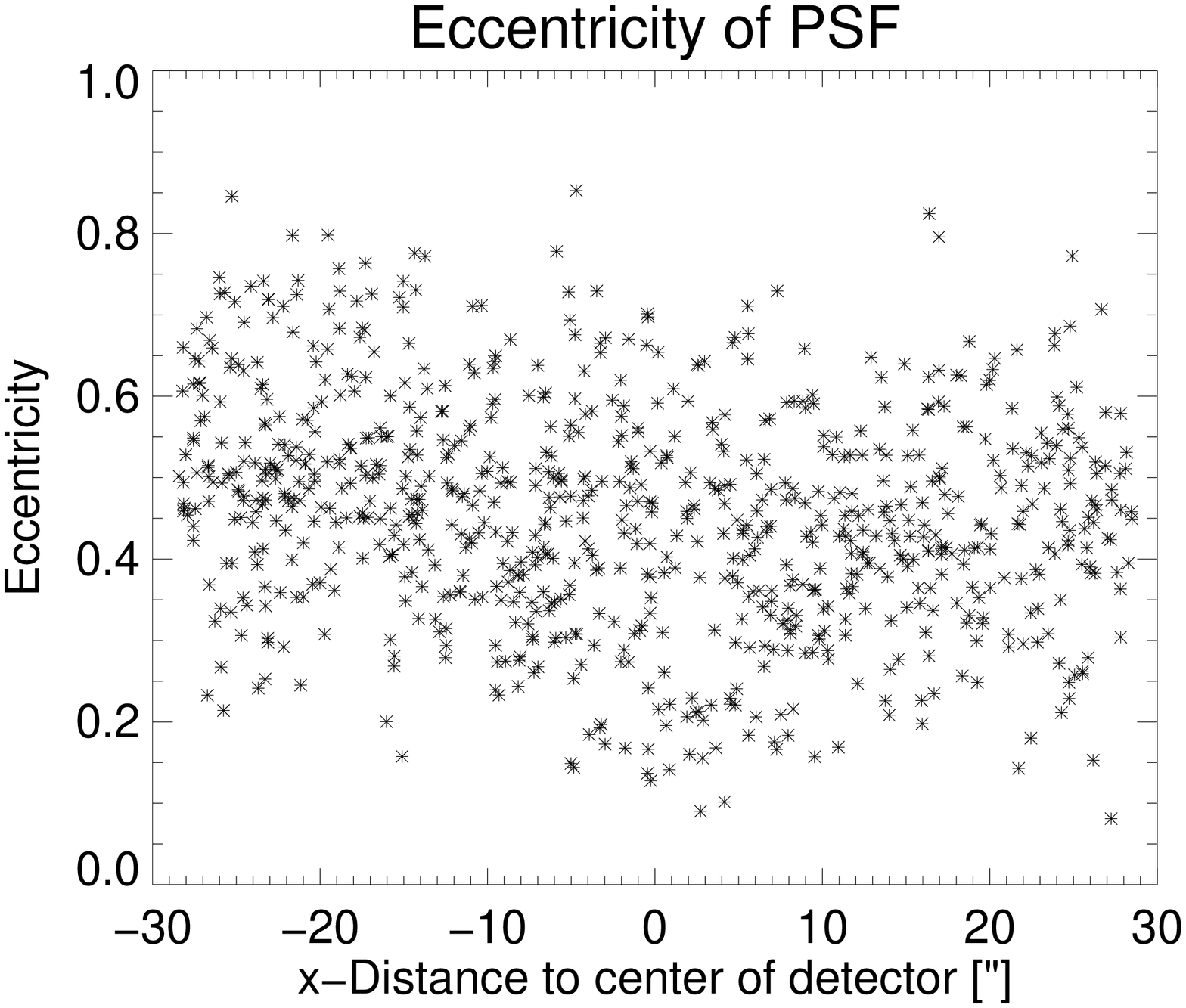} \\ \vspace{0.5cm}
\hspace{0.3cm}
  \includegraphics[height=6.5cm, angle=90]{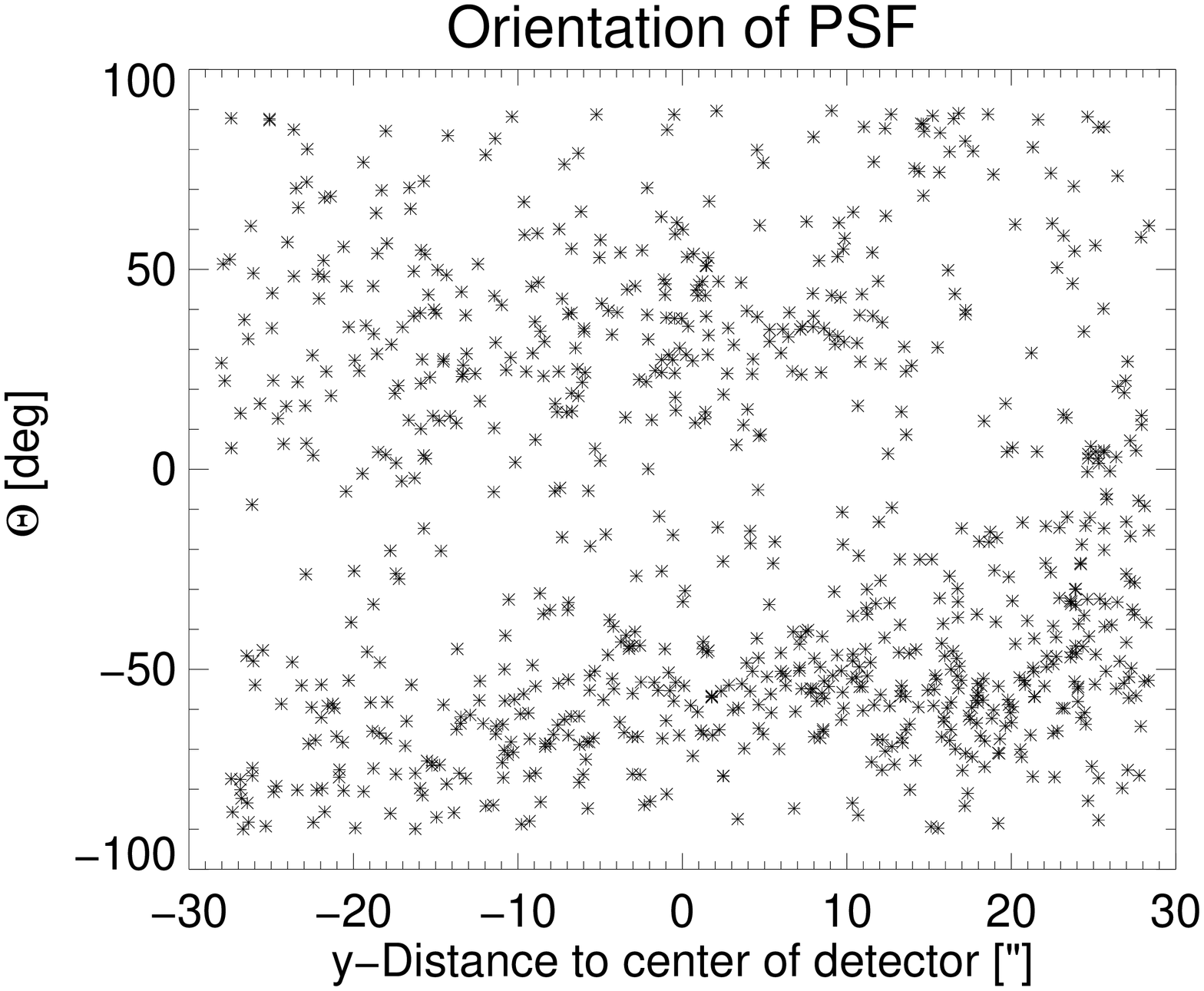} \hspace{0.8cm}
  \includegraphics[height=6.5cm, angle=90]{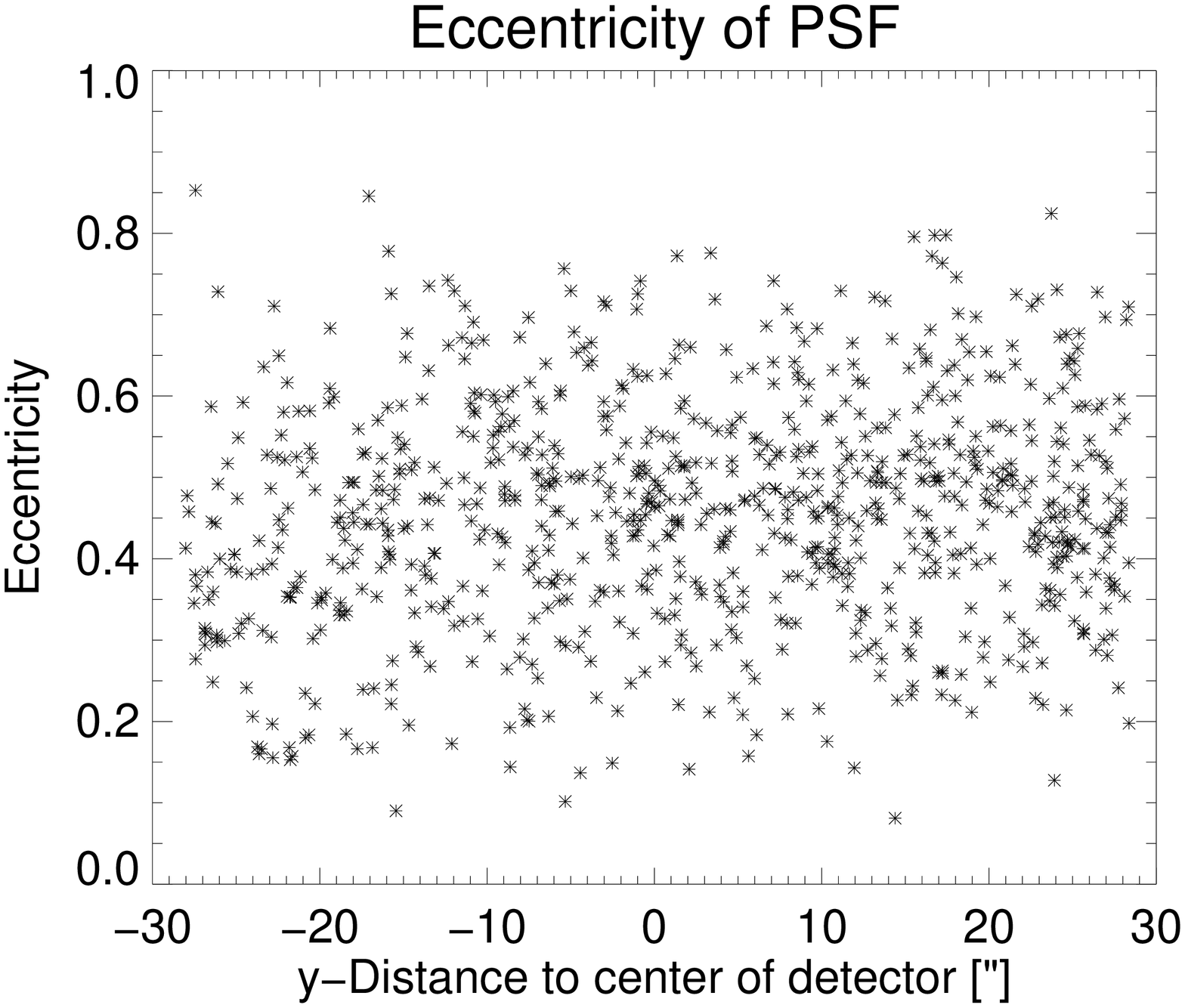}
 \end{center}
 \caption[Distribution of the orientation and eccentricity of the PSF in one of the MCAO frame
  of the NGC~6388 data]{Distribution of the orientation and eccentricity of the PSF
  in one of the MCAO frames of the NGC~6388 data as an example. The upper panels show the
  distribution as function of the distance of the stars from the
  center of the detector. The middle panels the distribution
  as function of the distance in x-direction and the lower
  panels as function of the y-distance.}
\label{fig:angle_dist}
\end{figure}

\noindent For the PSF tests I first used the \textit{StarFinder}
program (see next section for details) to measure the positions of
the stars in the field. I then cut a quadratic box around each
star and fitted a two dimensional, rotatable Moffat function (see
Equ.~\ref{equ:Moffat}) to each star, using the
non-linear least square fitting package MPFIT2DPEAK, provided by
Craig Markwardt
 \citep{Markwardt2009}. After flagging those stars
which were too close to the frame edges or other stars, as well as
those where the fit with the Moffat function obviously gave wrong
parameters, I worked with the remaining 780 - 800 stars in both
data sets. The fitting routine has some problems if more than one
star is present in the box cut around the star which one wants to
fit. If the two stars have similar brightness, the valley between
them is fitted and if the second star is brighter than the one
gained for, always the brighter one is fitted. I therefore checked
for large changes in the position and brightness of the fitted
Moffat function to the values given by \textit{StarFinder} and
flagged those stars where the fit did not went well.

\noindent Looking at the distribution of the rotation angle
$\Theta$ of the fitted Moffat function, a small dependency of the
orientation of the PSFs with the position of the stars can be seen
in the case of the NGC~6388 data. The angle is measured clockwise
from the detector x-axis to the major axis and is in the range
$[-\frac{\pi}{2}, \frac{\pi}{2}]$. In Fig.~\ref{fig:angle_dist},
left panels, an example of the distribution of the orientation of
the PSF is shown for one frame of the MCAO frames of the NGC~6388
data. The upper panel shows the distribution as function of the
distance of the stars from the center of the detector. The middle
panel the distribution as function of the distance in
$x$-direction and the lower panel as function of the $y$-distance.
Although it seems, that fewer orientations near $\Theta = 0\degr$
are fitted, this effect is not dependent on the distance of the
stars from the center. I calculated correlation coefficients for
these plot, yielding values of 0.019, 0.438 and -0.093 for the
distance from the center, the $x$-direction and the $y$-direction
respectively. Whereas the coefficients for the center distance and
the $y$-direction are not significant, the one for the $x$
direction shows a moderate correlation. I plotted the fitted
Moffat functions as enlarged ellipses overimposed on the positions
of the stars. Fig.~\ref{fig:47Tuc_ellips} shows the distribution
in the right panel. One can see the different orientation of the
stars in the lower right corner compared to the ones in the upper
left corner. This behavior changes only slightly over the
different frames. In the case of plotting angles, one has to be
careful with the interpretation of the correlation coefficient,
because of the periodicity of the angle. Plotting the distribution
with different cut values, as for example $[0,\pi]$ or
$[\frac{\pi}{2}, \frac{3}{2}\pi]$, can yield different correlation
coefficients. I therefor calculated the coefficient for different
cut values. The correlation coefficients changes slightly, but
yielded all to the same interpretation as above.\\

\begin{figure}
 \begin{center}
   \includegraphics[width=7cm, angle=90]{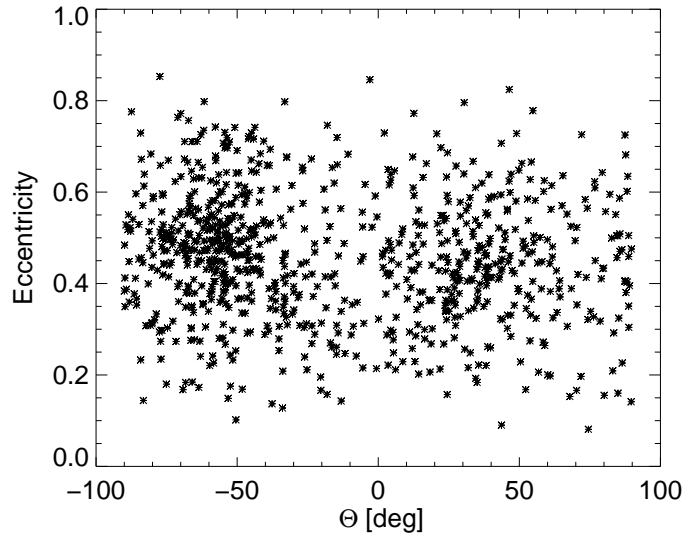}
 \end{center}
 \caption[Correlation of the eccentricity and orientation of
 the PSF in one of the MCAO frames of the NGC~6388 data]{Correlation of the
  eccentricity and orientation of the PSF in one of the MCAO frames of the NGC~6388 data.}
\label{fig:excen_angle}
\end{figure}

\noindent I also looked at the shape, i.e. the eccentricity, of
the fitted Moffat functions. In Fig.~\ref{fig:angle_dist}, right
panels, the distribution of the eccentricity, $e = \frac{\sqrt{a^2
-b^2}}{a}$ as a function of distance to the detector center is
shown. Here $a$ is the half width at half maximum (HWHM) of the
larger axis and $b$ the HWHM of the smaller axis. Although the
mean eccentricity is $\sim 0.4$, no dependency of the shape of the
PSF with the position of the stars on the detector is observable.
The calculated correlation coefficients are 0.125, -0.216 and
0.115 for the distance from the center, the $x$ direction and the
$y$ direction respectively. All these values show a very weak
correlation. I also looked at the distribution of the eccentricity
in the four quadrants of the detector, to check if there is any
correlation, as for example one of the guide stars is located
close to the upper left quadrant for the NGC~6388 observations
(see Fig.~\ref{fig:NGC6388guidestars}). Each quadrant shows
a similar distribution of eccentricities.\\

\noindent As a last check I looked for a correlation between the
orientation of the PSF and its eccentricity. As one can see in
Fig.~\ref{fig:excen_angle} there is no preferred eccentricity for
a specific angle.\\

\noindent Compared to the single guide star AO correction, where
the stars are elongated in the direction to the guide star and the
effect is stronger the further away the star is from the guide
star, these results already show the advantage and improvement of
observing with MCAO.\\
\begin{figure}
 \begin{center}\hspace{2.5cm}
  \includegraphics[width=6.8cm, angle=90]{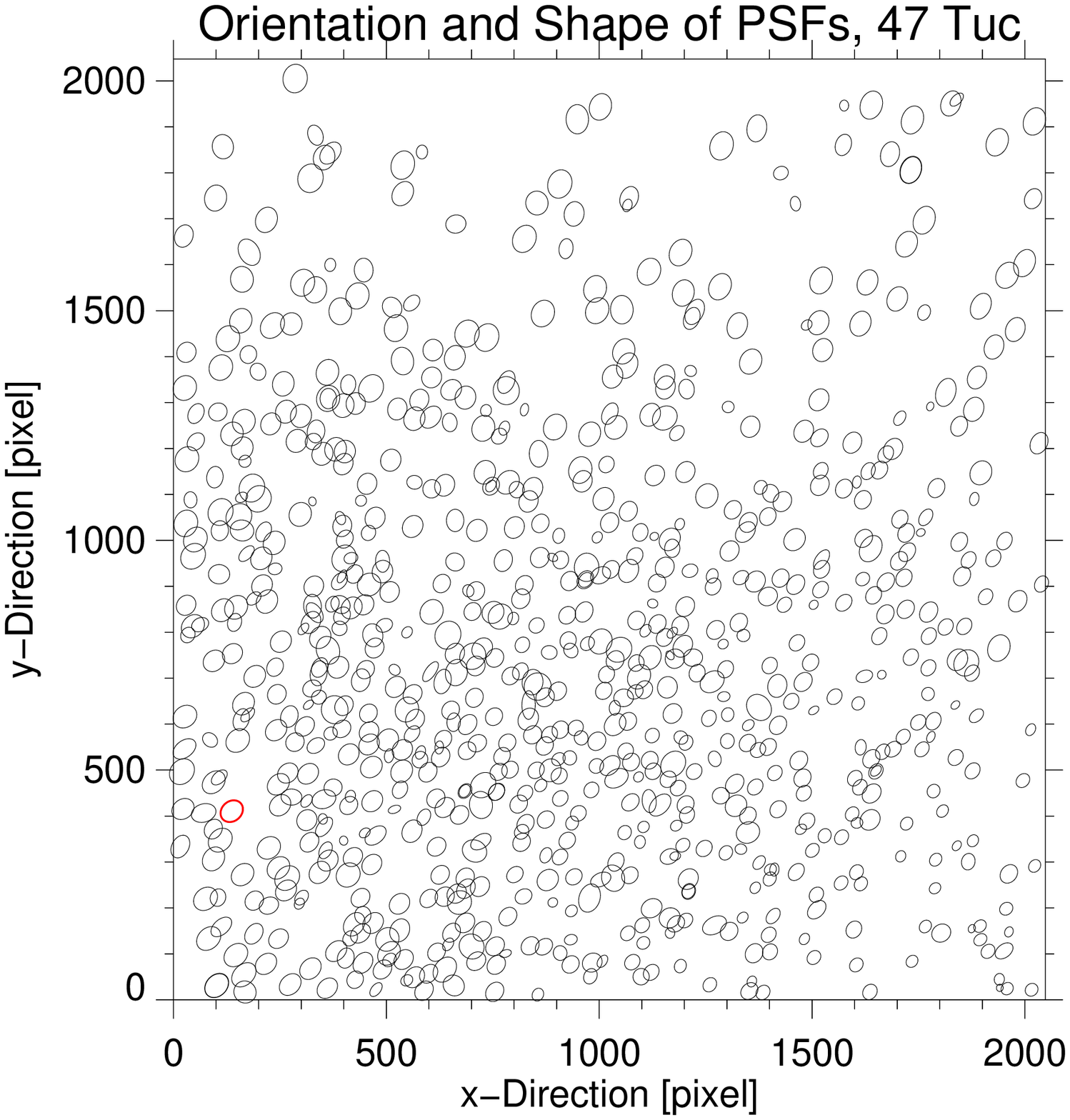} \hspace{3cm}
  \includegraphics[width=6.8cm, angle=90]{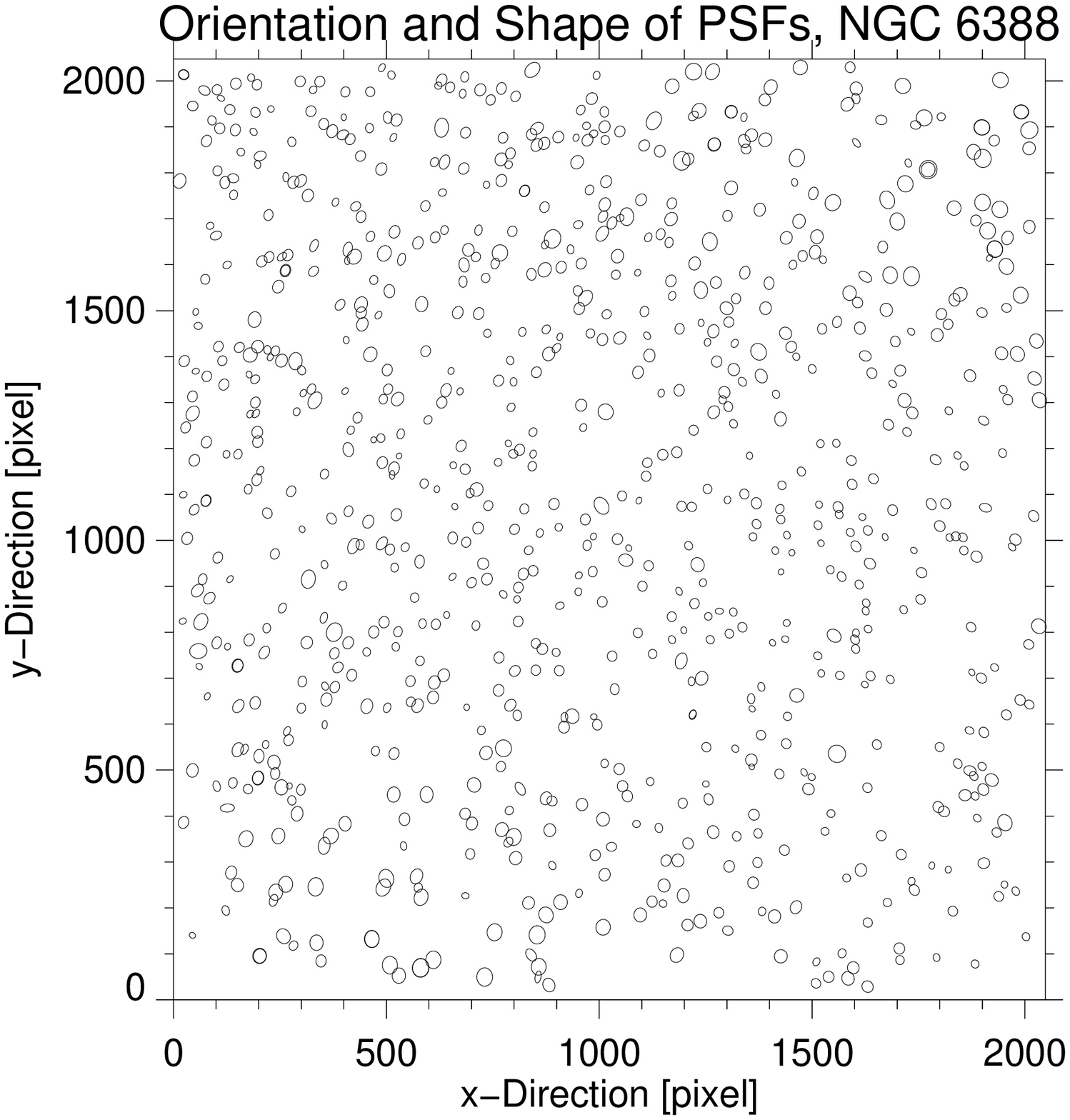}
 \end{center}
 \caption[Fitted Moffat functions, displayed as ellipses (enlarged)
 at the positions of the stars, showing the orientation and shape of
 the PSFs]{Fitted Moffat functions, displayed as enlarged ellipses at the
 positions of the stars, showing the orientation and shape of the
 PSFs in the 47~Tuc data in the left panel. The different sizes of
 the ellipses reflect the different fitted FWHM of the Moffat functions.
 The red ellipse marks the guide star contained in the detector FoV.
 For comparison the same plot is shown for a MCAO frame of the
 NGC~6388 data in the right panel with the same enlargement factor.}
\label{fig:47Tuc_ellips}
\end{figure}

\noindent And although I see some small spatial correlations, I
decided to use one PSF model to fit all stars in the field in the
way as described in the next section (\ref{sec:mad_positions}). Using
the Moffat fit in all frames was also not an option, because of
the above mentioned fitting problems with close stars pairs. \\

\begin{figure}
 \begin{center}\hspace{0.3cm}
  \includegraphics[height=6.5cm, angle=90]{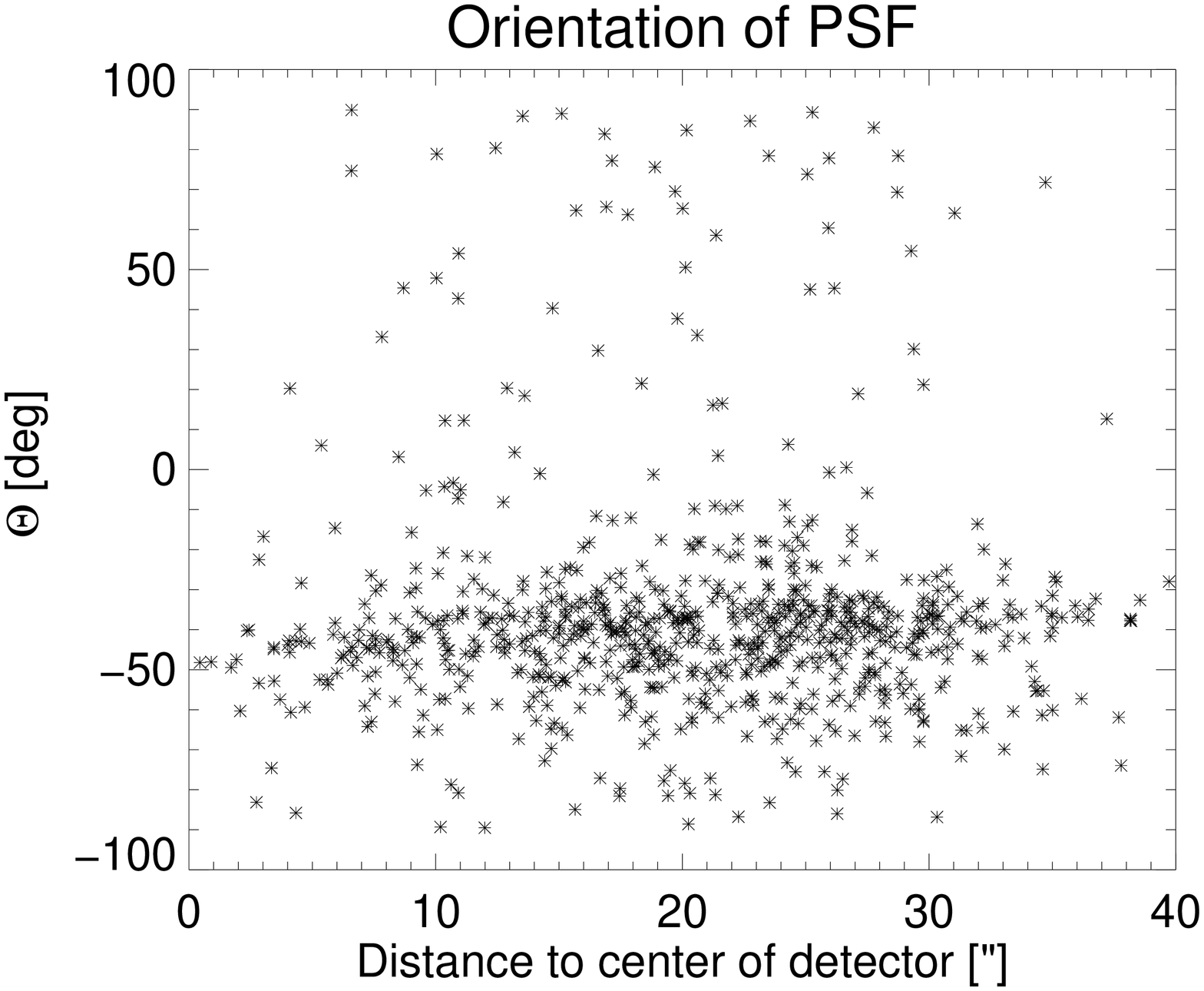}   \vspace{0.5cm} \hspace{0.8cm}
  \includegraphics[height=6.5cm, angle=90]{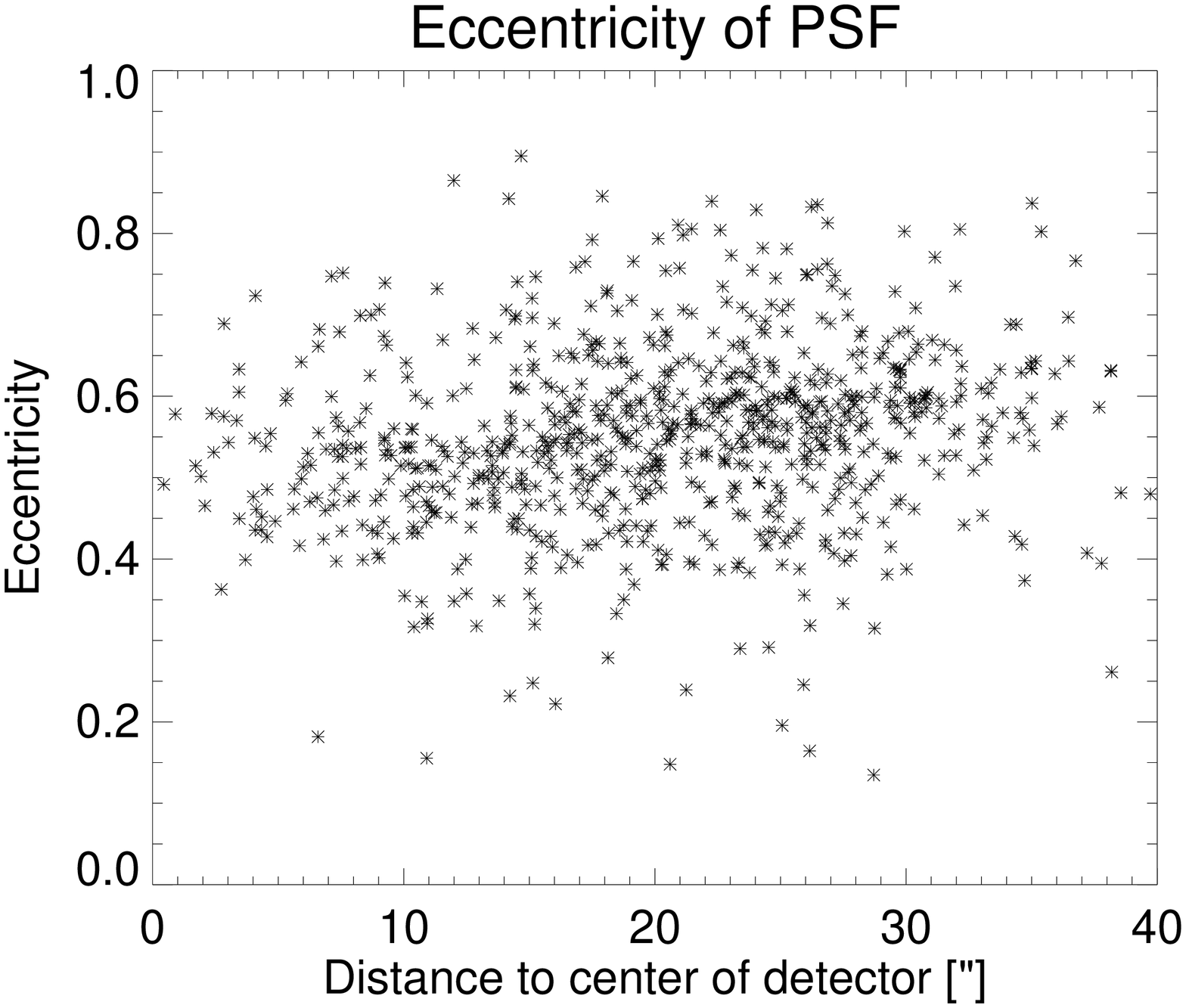}\\  \vspace{0.5cm}
\hspace{0.3cm}
  \includegraphics[height=6.5cm, angle=90]{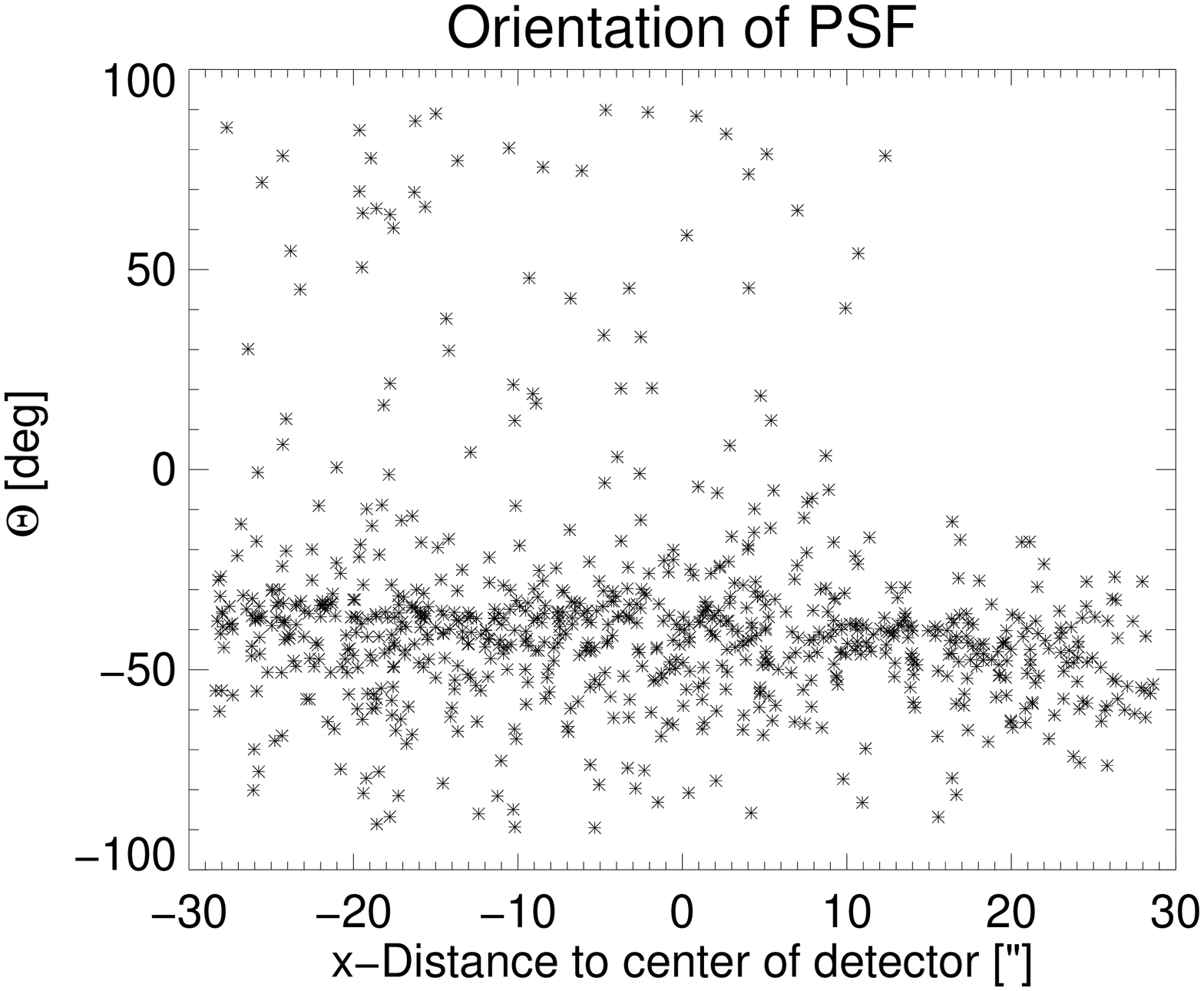}   \vspace{0.5cm} \hspace{0.8cm}
  \includegraphics[height=6.5cm, angle=90]{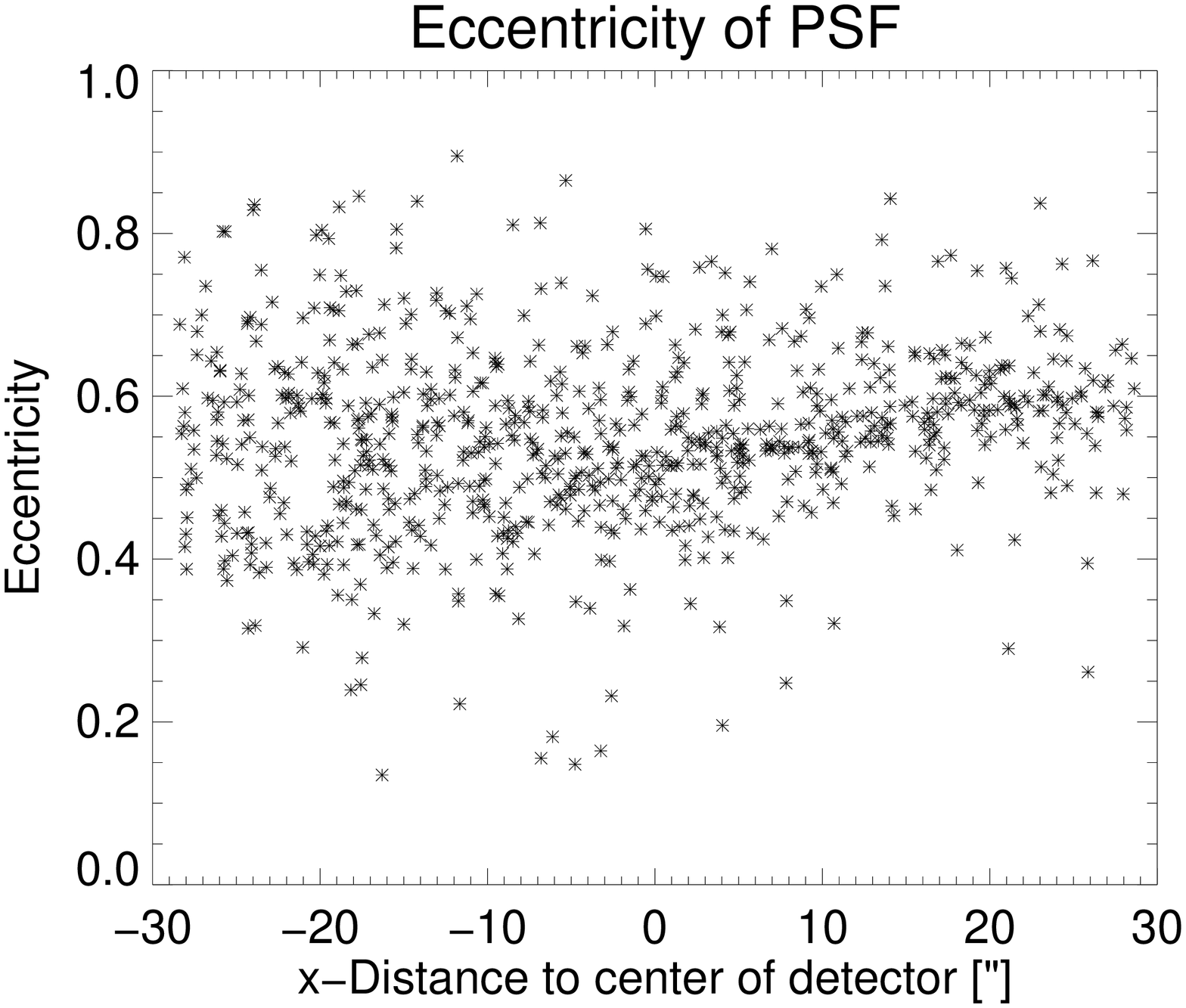} \\ \vspace{0.5cm}
\hspace{0.3cm}
  \includegraphics[height=6.5cm, angle=90]{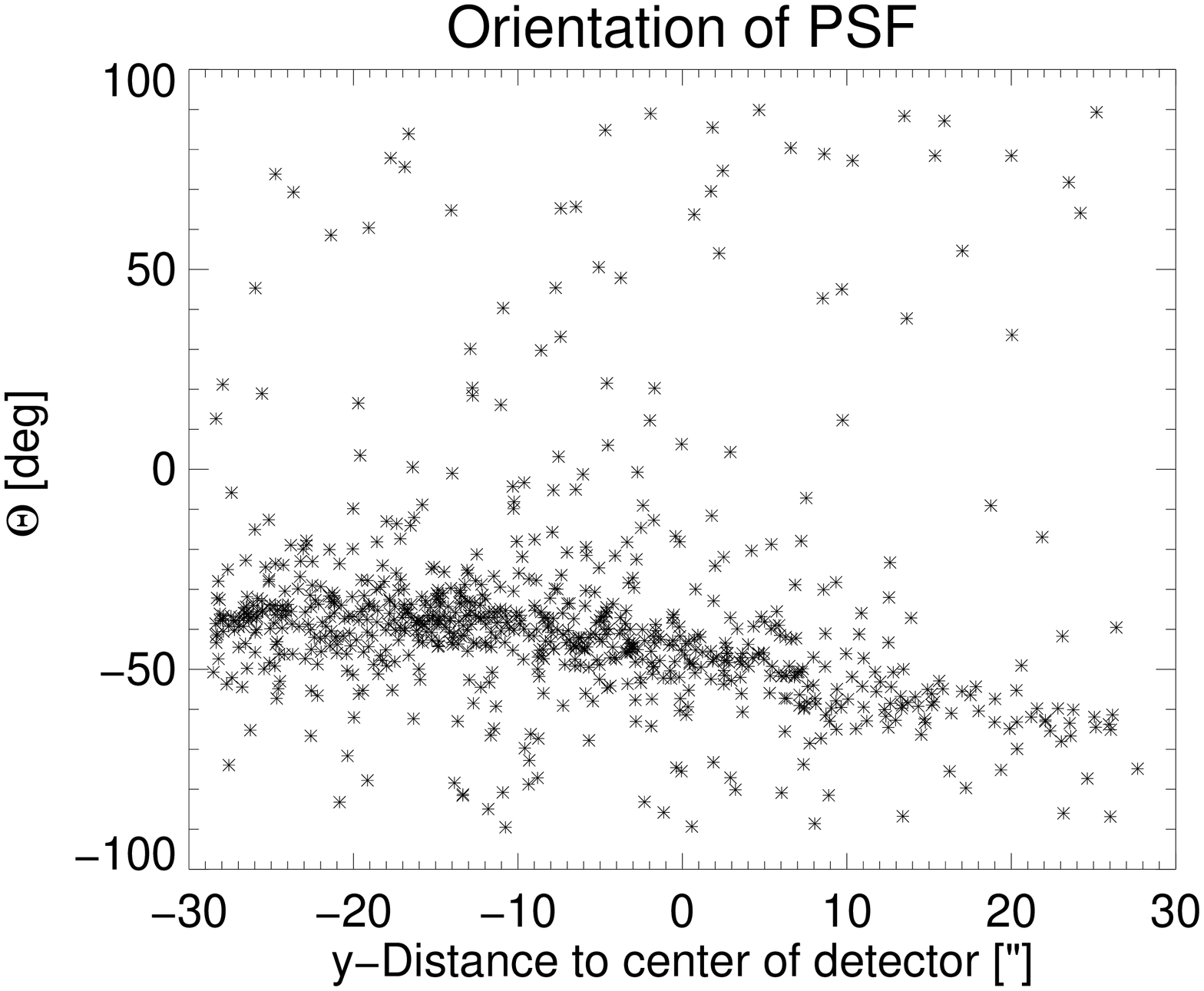} \hspace{0.8cm}
  \includegraphics[height=6.5cm, angle=90]{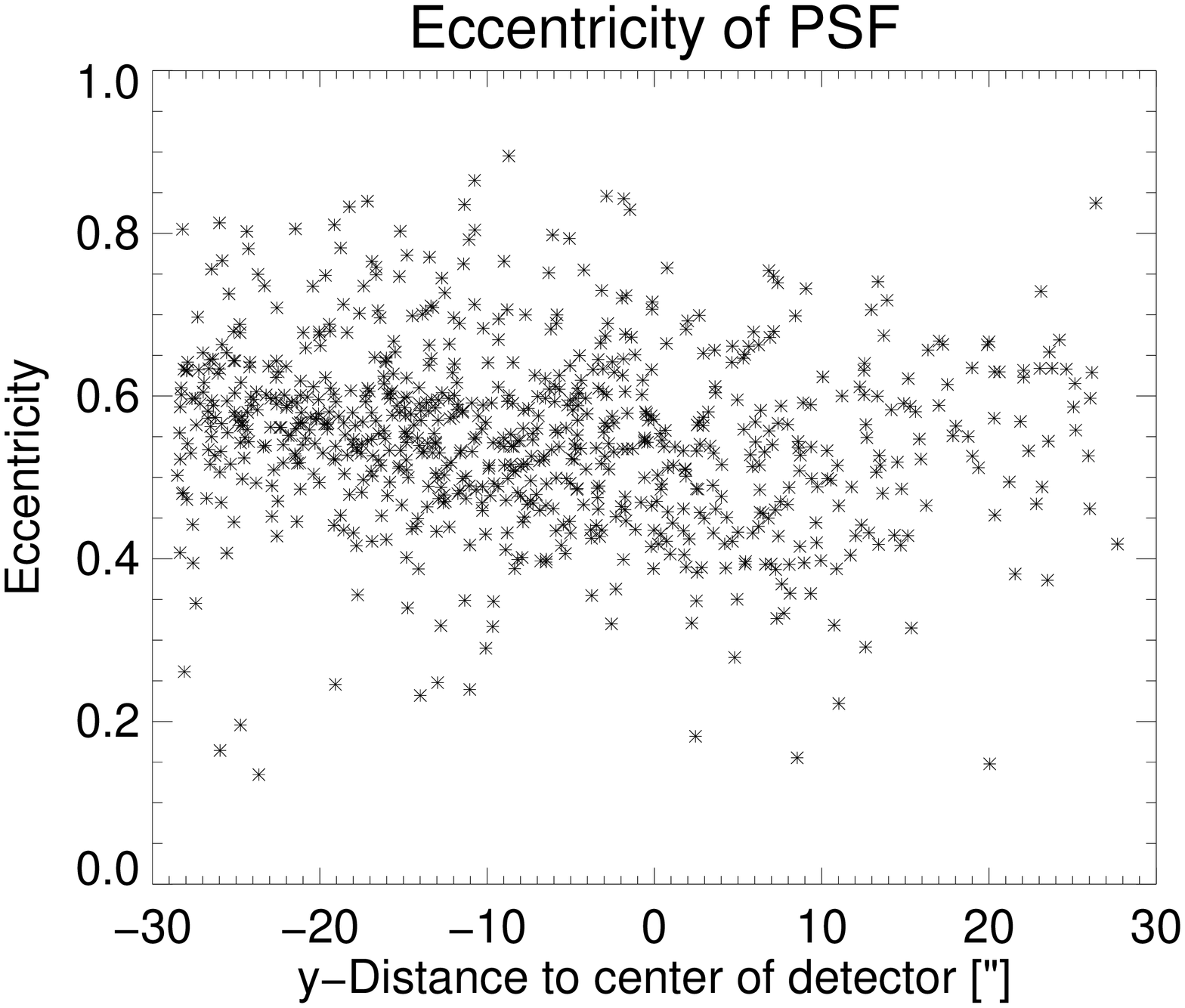}
 \end{center}
 \caption[Distribution of the orientation and eccentricity of the PSF in one frame
  of the 47~Tuc data]{Distribution of the orientation and eccentricity of the PSF
  in one frame of the 47~Tuc data as an example. The upper panel shows the
  distribution as function of the distance of the star from the
  center of the detector. The middle panel the distribution
  as function of the distance in x-direction and the lower
  panel as function of the y-distance.}
\label{fig:47Tuc_angles}
\end{figure}
\noindent When I later analyzed the data of the cluster 47~Tuc, I
performed the same tests. As the initial observing conditions were
worse than in the case of the NGC~6388 data, also the overall
performance was a bit worse (compare
Tables~\ref{tab:observingconditions1} and
\ref{tab:observingconditions2} and
Figs.~\ref{fig:mad_NGC6388_strehl} and
\ref{fig:mad_47Tuc_strehl}). Already in the image one can see an
elongation of the stars in one direction. The analysis of the
orientation of the fitted Moffat function confirms this
impression, see Fig.~\ref{fig:47Tuc_angles}. A clear preference of
orientation angles between $\Theta = -30\degr$ and $\Theta =
-50\degr$, with an average of $\overline{\Theta} = -43.5\degr$, is
recognizable, with a small dependency on the
$y$-positions of the stars in the field. Stars in the upper part
of the detector seem to be oriented a bit more to the y-direction
than the stars in the lower part of the field, see also
Fig.~\ref{fig:47Tuc_ellips}. The eccentricity distribution, with a
mean eccentricity of 0.55, also does only show a very weak
dependency on the position on the detector. But the mean
eccentricity is higher than in the case of the NGC~6388 data
obtained under good initial conditions. One of the used guide
stars is in the lower left corner of the FoV. If the shape of the
stellar PSFs depended on the position of this guide star, $\Theta$
should change over the field depending on the position angle
between the stars and the guide star. Instead the guide star
itself is elongated and oriented in the same way as the stars
around it ($\Theta = 30.25\degr, e = 0.59$). In the left panel of
Fig.~\ref{fig:47Tuc_ellips} the fitted Moffat functions are
displayed as ellipses at the positions of the stars, showing the
orientation and shape of the PSFs. This shows nicely the principle
of the layer oriented correction approach, where the light of all
guide stars is added and measured for the correction of a certain
layer and the correction itself is not so much dependent on the
single position of the guide stars, as in the classical, single
guide star approach. On the other hand it also shows, that the
correction was not yet perfect in this data set, leaving this
residual elongation
and orientation of the stars.\\

\noindent Nevertheless I continued measuring the positions of the
stars and their uncertainties in the same way as in the case for
the data of the cluster NGC~6388 to compare the two sets later
under the aspect of GLAO versus MCAO correction. Most of the stars
in the 47~Tuc data show the same orientation and eccentricity,
making it again feasible to work with one PSF model for the full
FoV.

\section{Position Measurements}
\label{sec:mad_positions}
To measure the positions of the stars in the single images of both
clusters I used the program \textit{StarFinder}
\citep{Diolaiti2000, Diolaiti20002}, which is an IDL based code to
analyze AO images of stellar fields. The following description is
for 47~Tuc, but it is the same for the
NGC~6388 data.\\

\begin{figure}
 \begin{center}
  \includegraphics[height=6.8cm]{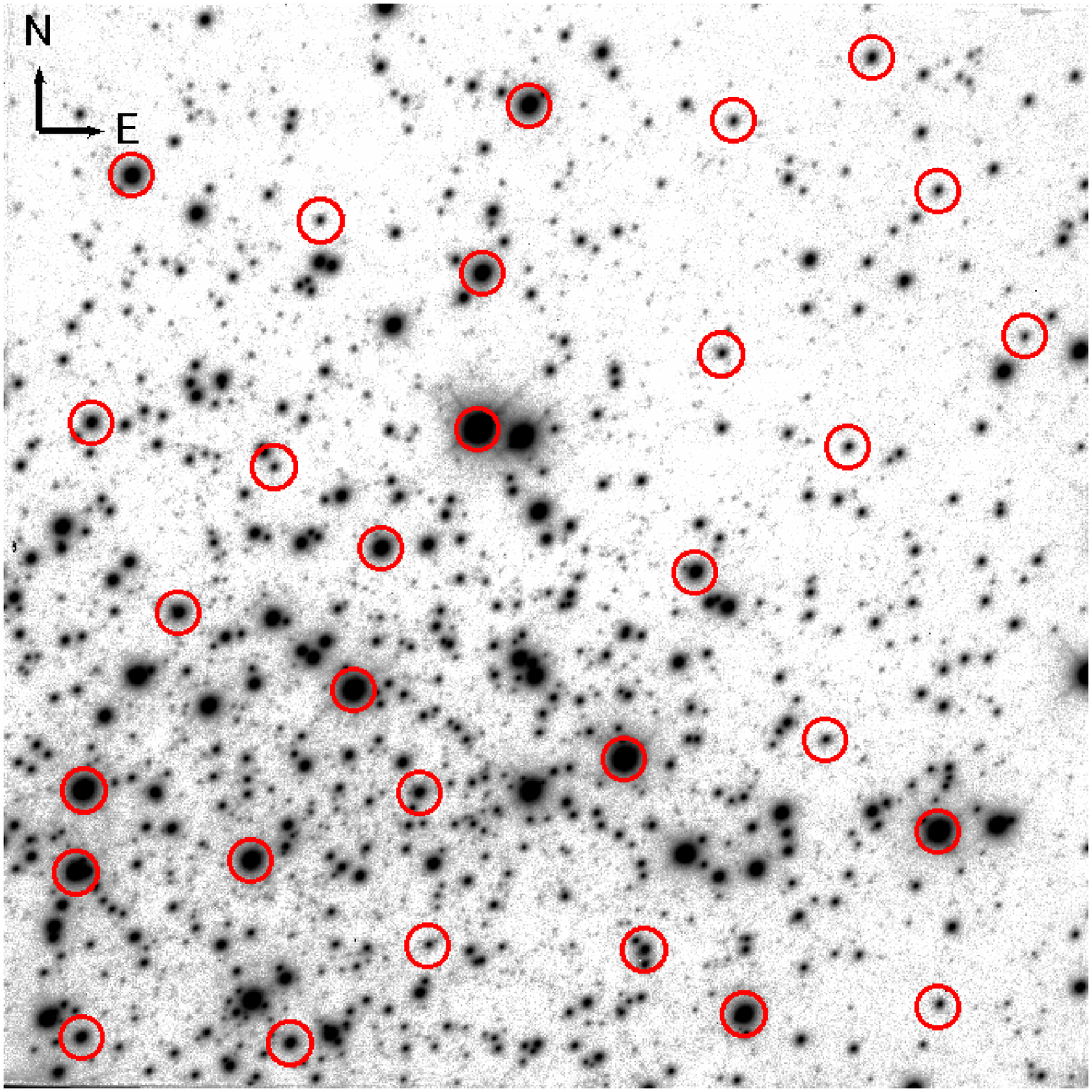} \hspace{0.8cm}
  \includegraphics[height=6.8cm]{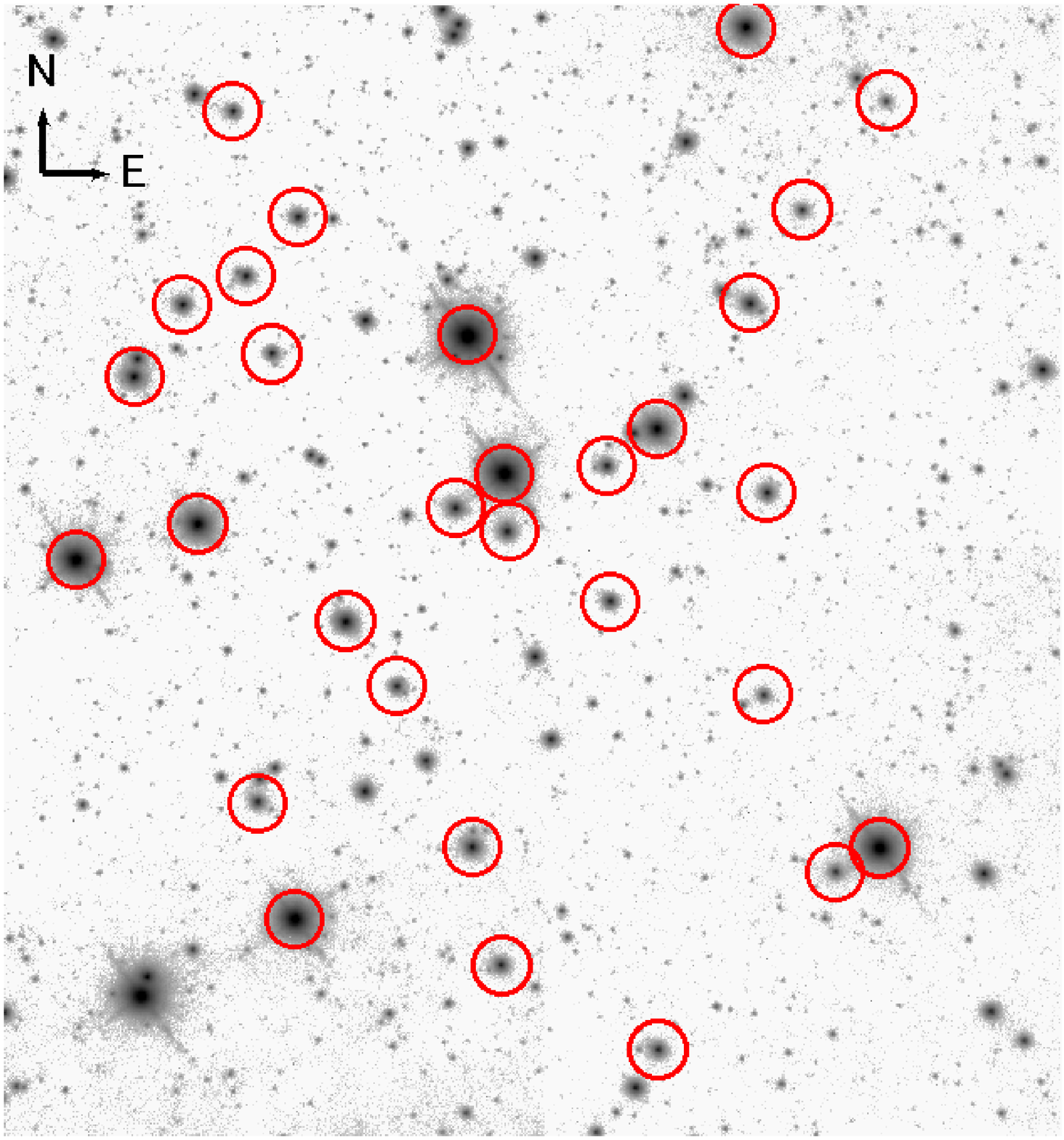}
 \end{center}
 \caption[Stars used for PSF extraction in the fields 47~Tuc and NGC~6388]
 {Stars used for PSF extraction in the fields 47~Tuc (left) and NGC~6388
 (right). Note the different scales: The 47~Tuc field is
 $57\arcsec \times 57\arcsec$ and the NGC~6388 field is cut to
 $42.5\arcsec \times 45.4\arcsec$.}
\label{fig:PSFstars}
\end{figure}

\noindent I directly extracted the Point Spread Functions (PSF)
for the star fits from the images, by using in each frame the same
30 stars. In Fig.~\ref{fig:PSFstars} the stars used for the PSF
extraction are marked for both fields.\\

\begin{figure}
 \begin{center}
  \includegraphics[width=5.2cm, angle=90]{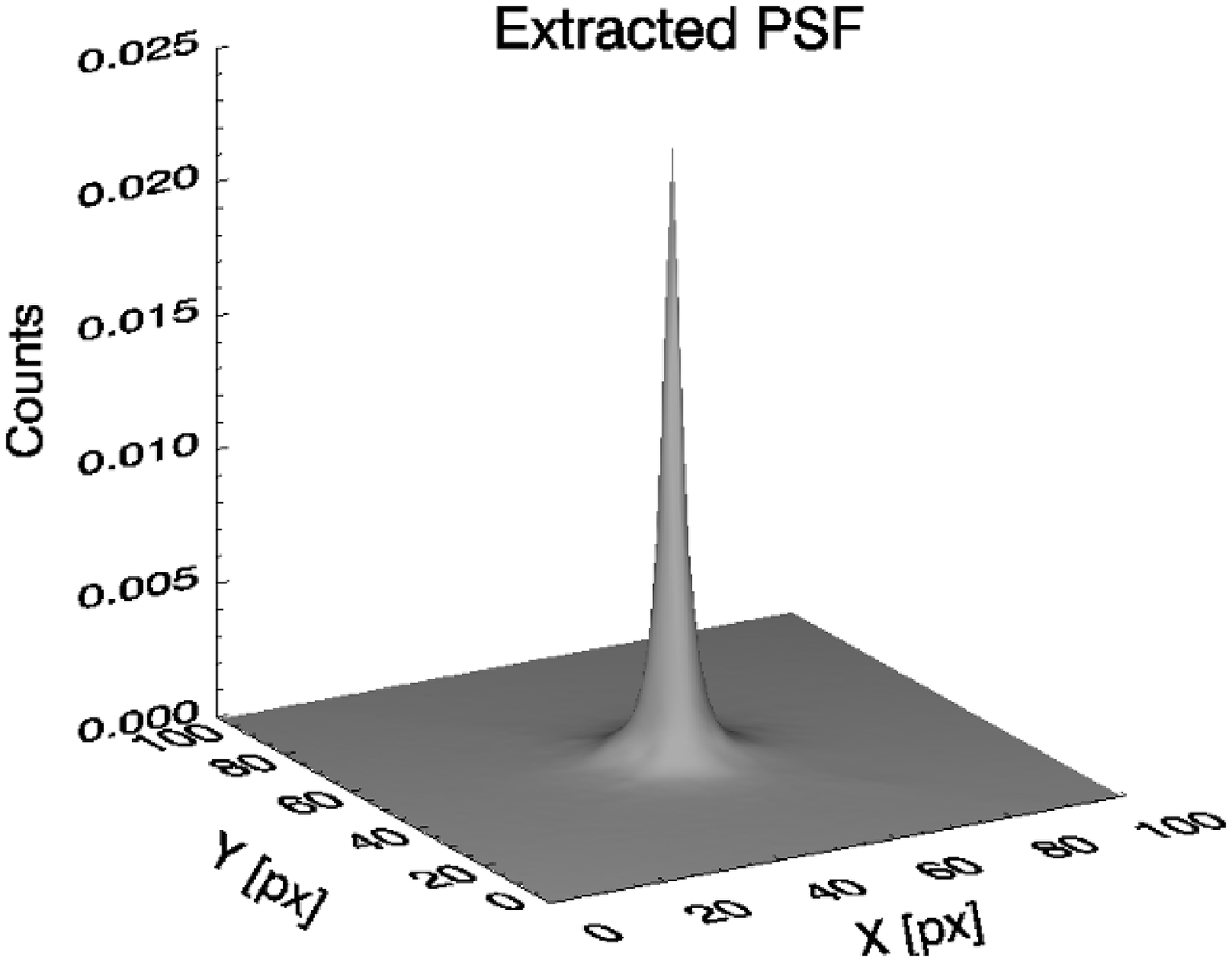}
  \includegraphics[width=5.2cm, angle=90]{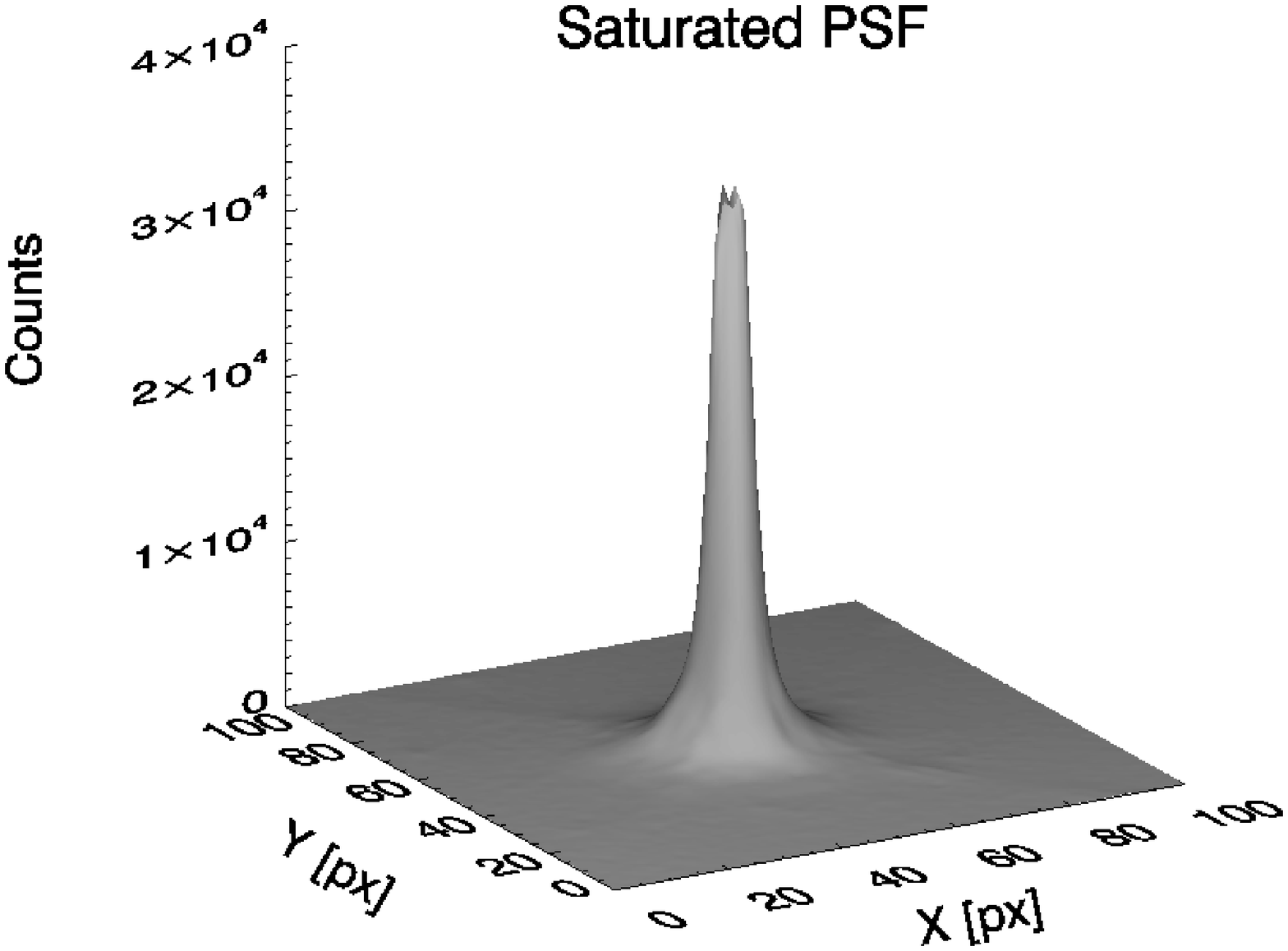}
 \end{center}
 \caption[Example of a with \textit{StarFinder} extracted PSF and a saturated PSF]
 {\textit{Left:} Example of a with \textit{StarFinder} extracted PSF. The flux
 is normalized to one. \textit{Right:} Example of a saturated PSF. The PSF has
 a hole instead of a tip. The different peaks at the 'top-ring' can clearly be
 seen, causing \textit{StarFinder} to fit several stars instead of one.}
\label{fig:PSF}
\end{figure}

\noindent I assumed here that the PSF does not vary strongly over
the FoV, as would be the case in single guide star adaptive
optics. But as seen in the analysis before
(Sect.~\ref{sec:PSFTests}), the PSF did not show a huge variation
over the field and as I do not intend to do photometry, I assume
that the center of the PSF can still be determined accurately
enough from a fit with an averaged PSF. In Fig.~\ref{fig:PSF},
(left), an example of such an extracted PSF is shown. The flux is
normalized to one. The 30 stars were chosen to be equally
distributed over the entire field and to give a good mixture of
brighter and fainter stars. After marking the stars,
\textit{StarFinder} provides the opportunity to remove secondary
sources such as close stars, which may be in the box cut around
the marked stars. The final PSF is then generated by a median
combination of the selected stars. At the end I cut the square box
containing the PSF into a round one of 80 pixel diameter, which
corresponds
more to the base of a PSF.\\
\noindent Some of the brighter stars are saturated. In the case of
MAD with its IR detector these saturated stars do not have a flat
plateau in place of the PSF tip, but instead have a hole in the
middle (Fig.~\ref{fig:PSF}, right).
\noindent \textit{StarFinder} had some problems, fitting those
stars, as the remaining ring of the PSF is not constant in flux.
Several \textit{stars} were fitted along the ring. In principle it
should be possible to also fit the extracted PSF to the saturated
stars by first filling the holes and then using a function of
\textit{StarFinder}, which repairs the PSF during the fit. But as
it is difficult to evaluate how the position measurement is
affected by this procedure and I had only a few saturated stars in
the field, I decided to not use these stars in the following
analysis. After deleting the false detections in all frames, $\sim
1450$ stars remained in the 47~Tuc data and $\sim 1500$ stars in
the first 10 frames of the NGC~6388 data and $\sim 600$ in the
last 20 frames. The huge difference in the numbers for the
NGC~6388 data is due to the change in integration time after the
first 10 frames (see Tab.~\ref{tab:observingconditions2}), that
led to a lower signal to noise ratio in the averaged frame and
therefore to fewer detections of the faint stars.

\noindent The position uncertainties for the stars were computed
with photon statistics as an estimate. The extracted PSF from each
frame was cut at the value of half maximum. Then a rotatable
ellipse was fitted to the slice-plane, giving values for the
semi-major and -minor axes of this ellipse, corresponding to the
HWHM of the two axes. After parametrization the ellipse equation
 I calculated
the maximum projection values, $x_{max}, y_{max}$, of the
semi-major axis onto the $x$- and $y$-axes. The positional errors,
$\Delta x, ~\Delta y$, for the single stars of each frame were
then computed following photon statistics by:
\begin{align}
\Delta x &=\frac{x_{max}}{\sqrt{n}}\\ \notag \Delta y &=
\frac{y_{max}}{\sqrt{n}} \notag
\end{align}
where $n$ is the total number of photons. To derive $n$ one has to
take into account the gain factor of the detector, $g = 2.9 \rm
e^{-}/ADU$ for the CAMCAO camera, and the number of exposures,
NDIT, averaged to create this frame. In
Tab.~\ref{tab:observingconditions1} and
\ref{tab:observingconditions2}, last column, the mean FWHM value,
calculated as the mean of the FWHM of the minor- and major-axis,
is noted. This approach is somewhat conservative compared to the
classical approach, calculating the error by $\sigma / \sqrt{n}$,
 but I did not want to underestimate the errors. The positional
errors from the fit derived like this range from 0.010~mas
(0.00034~px) to 0.555~mas (0.020~px) for the brightest and
faintest star, respectively, in the 47~Tuc data set and from
0.004~mas to 0.141~mas (0.0001 - 0.005~px) in the NGC~6388 data
set. The difference in these values for the two data sets comes
mainly from the smaller FWHM of the PSFs in the NGC~6388 data,
which itself is mainly due to the better initial observing
conditions.

\section{Ensquared Energy}
\begin{figure}
 \begin{center}
  \includegraphics[width=6cm, angle=90]{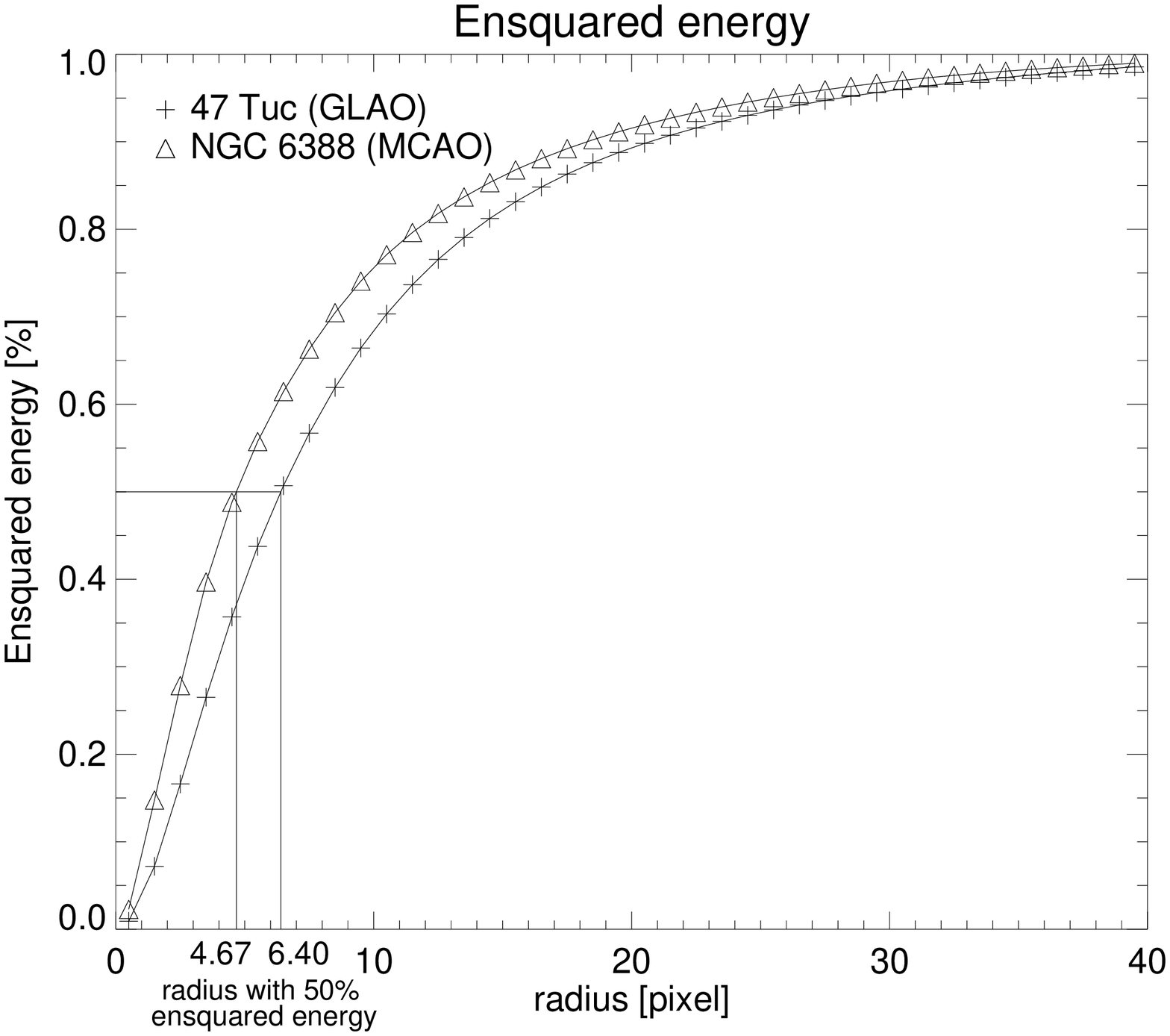} \hspace{3cm}
  \includegraphics[width=6cm, angle=90]{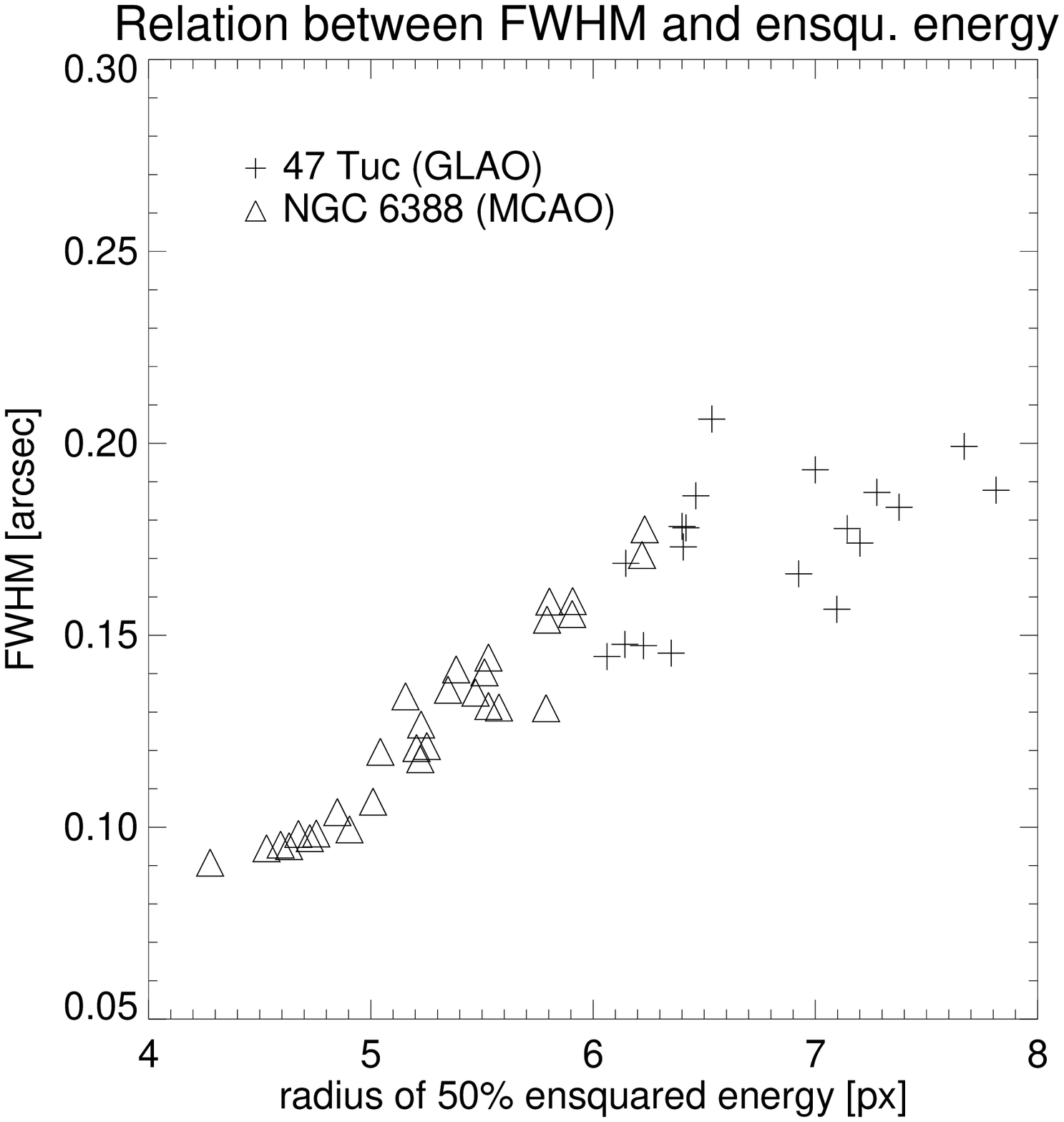}
 \end{center}
\label{fig:e_energy}
\caption[Example of the development of the ensquared energy
 with box size and relation to the FWHM]{\textit{Left:}
 Example of the development of the ensquared energy with box
 size for the MCAO and the GLAO case. The radius containing 50\%
 of the energy is marked. \textit{Right:} Relation between measured
 FWHM and radius containing 50\% of the energy for the MCAO and GLAO correction.}
\end{figure}

In adaptive optics observations often the encircled or ensquared
energy is taken as a measure of performance besides the FWHM. The
encircled energy in percent is defined as the flux of a PSF
contained in a certain radius from the middle point divided by the
total flux. In the case of an image where the flux is given as
flux per pixel, like any detector image, the ensquared energy is
used, which is the flux within a certain quadratic box with the
size of $n$ pixel times $n$ pixel. To correctly calculate the
comprised flux one needs to know how the program used to calculate
the ensquared energy, defines the pixel center. I used IDL in my
analysis, where a pixel is defined from -0.5 to 0.5. For instance,
a peak of a PSF given by the coordinates 50, 50 is therefore
defined in the middle of the pixel and not for example at its
lower left corner. Knowing this, I measured the flux in square
boxes with $1 \times 1, ~3 \times 3, ~5 \times 5 ...79 \times 79$
pixel in diameter, covering the full size of the extracted PSFs.
In Fig.~\ref{fig:e_energy} (left), an example of the development
of the ensquared energy with pixel distance from the center pixel
is shown. Here \textit{radius} denotes the half diameter of the
box. Finally I calculated the radius in pixels within which 50\%
of the energy of the PSF is contained and used this radius as a
performance indicator. The smaller the radius the better the AO
correction, moving flux from the seeing halo of the PSF into the
central peak. As comparison I show in Fig.~\ref{fig:e_energy},
right panel, the relation between the measured FWHM of the PSF and
the radius of 50\% ensquared energy, $r_{50}$. Clearly visible is
a relation, the larger the FWHM, the larger $r_{50}$. For the MCAO
data this relation is sharper and better defined as for the ground
layer data, reflecting the better initial observing conditions and
therefore the better correction in the single frames.

\section{Distortion Mapping}
\label{sec:Distortion} To investigate the stability of the MCAO
and GLAO performance in terms of astrometric precision over time,
I first had to correct for distortions of the field. During the
observations problems with the de-rotator occurred due to a
software problem, leading to a bigger rotational error in several
frames. If the AO correction is very stable over time, the
relative positions of the stars should be the same after
correcting for
effects such as the de-rotator problem.\\
\noindent The first thing I did, was to set up a coordinate
reference frame to which I mapped the single frame coordinates.
For this I set up some criteria for the stars chosen to create
this reference frame: After the measurement of the positions of
the stars in each field following the procedure described in
Sect.~\ref{sec:mad_positions}, I compared the lists of stars found in
the individual frames and identified the subset common to all
frames of the NGC~6388 and 47~Tuc data, respectively. The next
step was to exclude faint stars from the list, namely those that
had less than 10000 counts, i.e. a peak value around 85 ADU in the
case of 47~Tuc and those with less than 2800 counts, peak $\sim
50$ ADU, in the case of NGC~6388. These values were chosen,
because in the NGC~6388 data there are fewer bright stars than in
the 47~Tuc data. Additionally, all stars that have a close
neighbor at less than 2 HWHM separation were excluded. This left
$\sim$ 280 stars for the 47~Tuc field and $\sim$ 130 stars for the
NGC~6388 field, still absolutely
enough to calculate the transformations.\\
In order to create the reference frame, I started to use the best
frame I had, chosen according to the highest mean Strehl in the
images, as a first reference frame and mapped all the positions
from each individual frame onto this reference frame. I then
calculated the shift and scale in $x$ and $y$ and the rotation
between these frames using the
MIDAS\footnote{http://www.eso.org/sci/data-processing/software/esomidas/}
data reduction software and simple affine transformations. I was
not applying any interpolation directly to the images, but instead
just worked with the measured coordinates. After correcting for
the derived rotation, as well as for the shift and scale in $x$
and $y$ of each stellar position I created a
master-coordinate-frame by averaging the position of each star
over all frames. This was my new masterframe for the coming
analysis. I then mapped all coordinates from each frame to this
masterframe, leading to a better transformation. One could think
that one can get even better transformations between the frames by
applying this method iteratively, creating once more a
master-coordinate-frame by calculating  mean positions and
averaging those positions. If the distortions in the images, left
over from the AO or systematic, were homogeneous over the FoV the
transformations should not change or enhance a lot. But if the
distortions are not homogeneous, but depend, for example, on the
camera position in the FoV, I would introduce warpings in the
master-coordinate-frame which I cannot map with a simple shift,
scale and rotation anymore. I therefore stopped after one
iteration.\\
\noindent I then calculated the residual separations between the positions of
the stars in the master-coordinate-frame to the new positions in
every frame calculated with the obtained transformation
parameters and analyzed them as a measure of astrometric precision.

     \cleardoublepage
    \chapter{Results}
       \label{chap:MADResults} 
\begin{figure}
 \begin{center} \hspace{2.2cm}
  \includegraphics[width=6.8cm, angle=90]{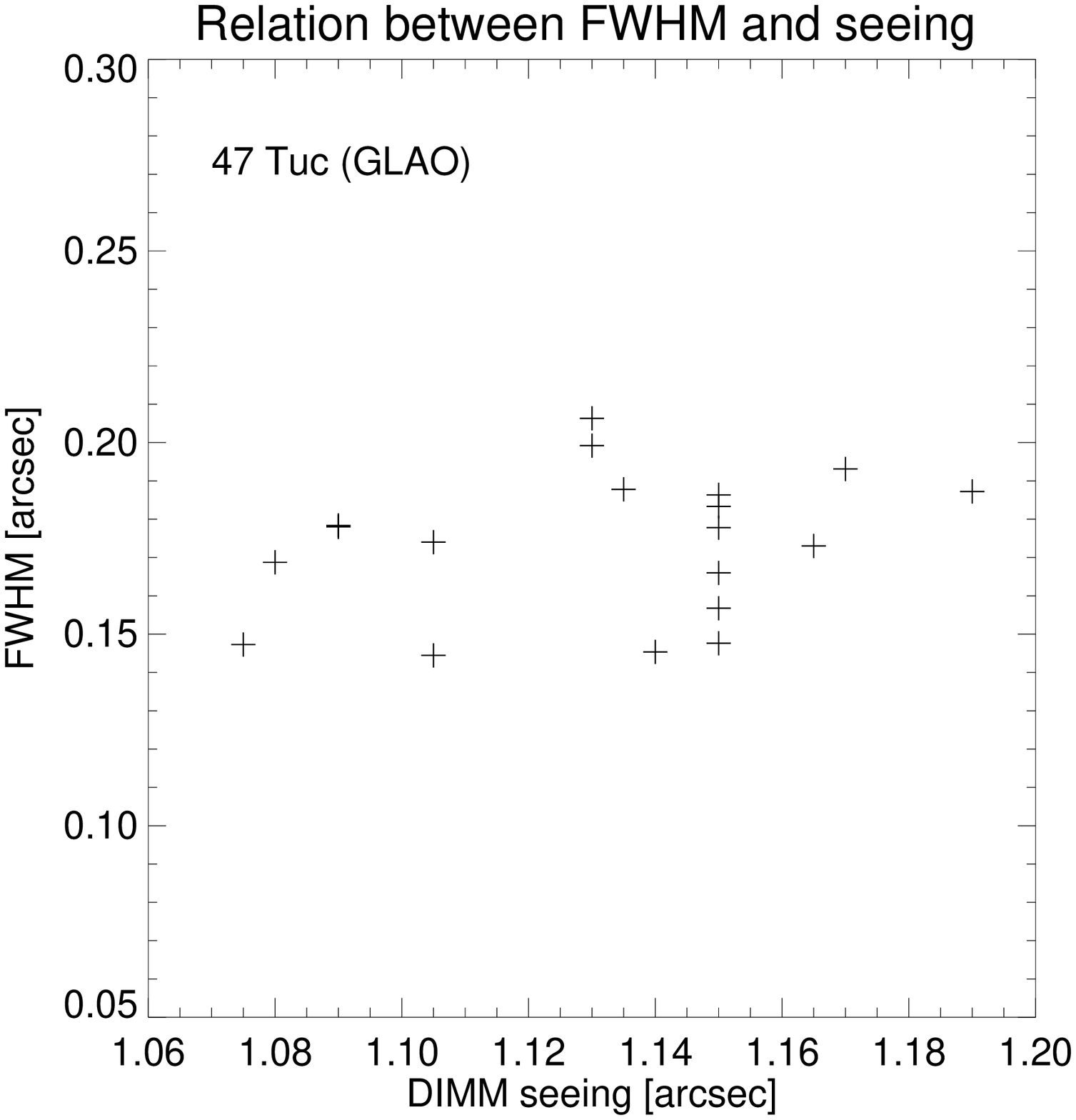} \hspace{2.8cm}
  \includegraphics[width=6.8cm, angle=90]{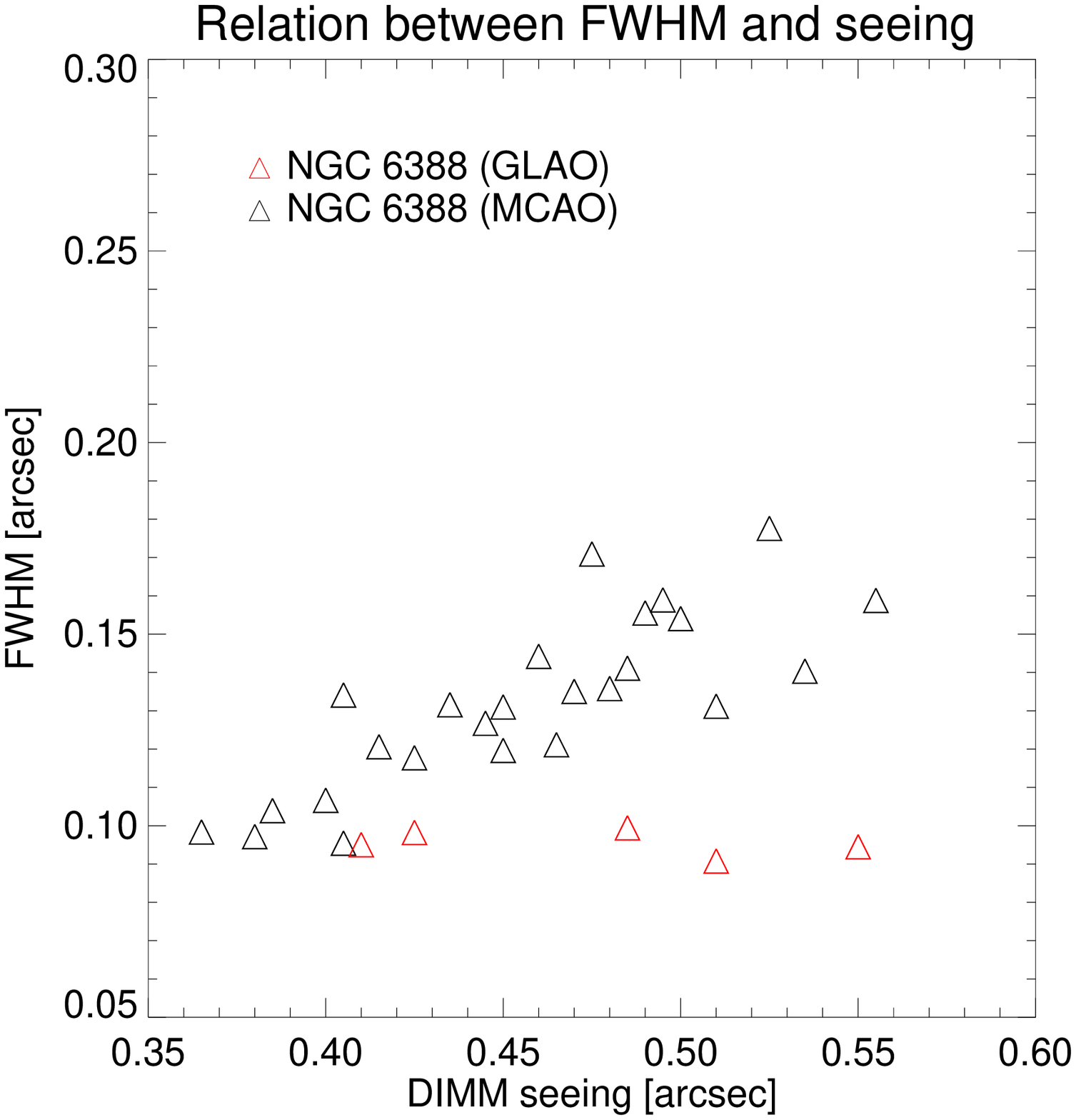}
 \end{center}
 \caption[FWHM over seeing for the 47~Tuc and NGC~6388 data]
 {FWHM over seeing plotted for the 47~Tuc (left) and NGC~6388 (right)
 data. The five ground layer frames are marked in red in the right
 panel with the NGC~6388 data.}
\label{fig:FWHM_Seeing}
\end{figure}

To set my results concerning the
astrometric precision achievable with MCAO into a context, I
looked at overall performance indicators, such as the FWHM of the
fitted PSF and seeing. This led to several interesting results. In
Fig.~\ref{fig:FWHM_Seeing} I have plotted the FWHM calculated from
the extracted PSF over the measured mean DIMM seeing for each
frame. The seeing is measured in the $V$ band, while the
observations were made in $Br_{\gamma}$ (47~Tuc) and $K_{s}$
(NGC~6388). Hence the difference between the seeing and the
measured FWHM cannot be directly taken as the correction factor
achieved. Nevertheless, the principal relation between these two
values is preserved as Fig.~\ref{fig:FWHM_Seeing} demonstrates. In
the case of the NGC~6388 data (right panel), where the first 5
frames were obtained with GLAO correction and the last 25 with
full MCAO correction, I marked the GLAO frames in red.
Interestingly, the FWHM does not change significantly in the GLAO
frames, although the seeing does, but in the MCAO case, I see a
correlation between seeing and FWHM, as expected. A similar
behavior of the FWHM can be seen in the 47~Tuc data, which is also
in ground layer mode. The FWHM gets only slightly larger, the
larger the seeing does. On the one hand the different FWHM values
for nearly the same seeing may be explained by slightly different
eccentricities of the PSFs in the different frames, leading to
different mean FWHM values when calculated as the mean of the FWHM
values of the minor- and major-axes. On the other hand, this can
be interpreted as a sign of the non-correlation of these two
parameters during GLAO correction. But this has to be seen with
caution, as I have no other data set confirming or disproving this
interpretation. A possible reason for the behavior in the NGC~6388
data is that the AO system was optimized for the ground layer
correction in the nights before. The GLAO correction alone worked
very well and even better than expected\footnote{C.Arcidiacono
(private communication)}. Switching on the full MCAO correction
can in a first attempt lead to a small degradation in the
performance, if the controller of the system has not yet been
optimized for the correction of two layers. Also in the case of
MAD, the subsystems for the correction of the two layers are
nested, meaning that one sensor \textquotesingle sees
\textquotesingle the correction applied to the distortions
of the other layer. This could lead to a degradation of the
overall performance, if the system is not yet fully optimized. As
this is the first time the full MCAO approach was tested, this
resulting behavior of the performance is interesting but
comprehensible and should not lead to wrong conclusions about the
full capacity of the MCAO correction. To analyze this behavior
further, one would need more data and more detailed information
about the applied correction parameters of the system, both of
which we do not have. Also our goal is not a performance analysis,
this is done by the group that built the layer oriented part of
MAD \citep{Arcidiacono2008}, but to see how precise astrometric
measurements can be conducted under the given observing
circumstances and corrections. Altogether the seeing does not
change a lot during the observing time in both cases, but rather
varies between $1.08\arcsec - 1.19\arcsec$ in the case of the
47~Tuc data and $0.37\arcsec - 0.56\arcsec$ during the NGC~6388
observations (see also Tables~\ref{tab:observingconditions1} and
\ref{tab:observingconditions2}).\\

\section{Separation Measurements}
\label{subsec:distMeas} One test I performed, was to measure the
relative separation between various pairs of stars all over the
FoV. I wanted to derive a time sequence of the separation, to see
how stable it is. If only a steady distortion were present in the
single frames, then the separations should be stable over time or
only scatter within a certain range given by the accuracy of the
determination of the position of the stars. If differential
distortions between the single frames are present, but these
distortions are random, the scatter of the separations should
increase. A not perfectly corrected defocus for example would
change the absolute separation between two stars, but, to first
order, not the relative one measured in the frames, if this
defocus is stable. An uncorrected rotation between the frames
would change the separation of two stars in the x
and y direction, but not their full separation, $r$.\\
\begin{figure}
 \begin{center}
  \includegraphics[width=13.5cm, angle=180]{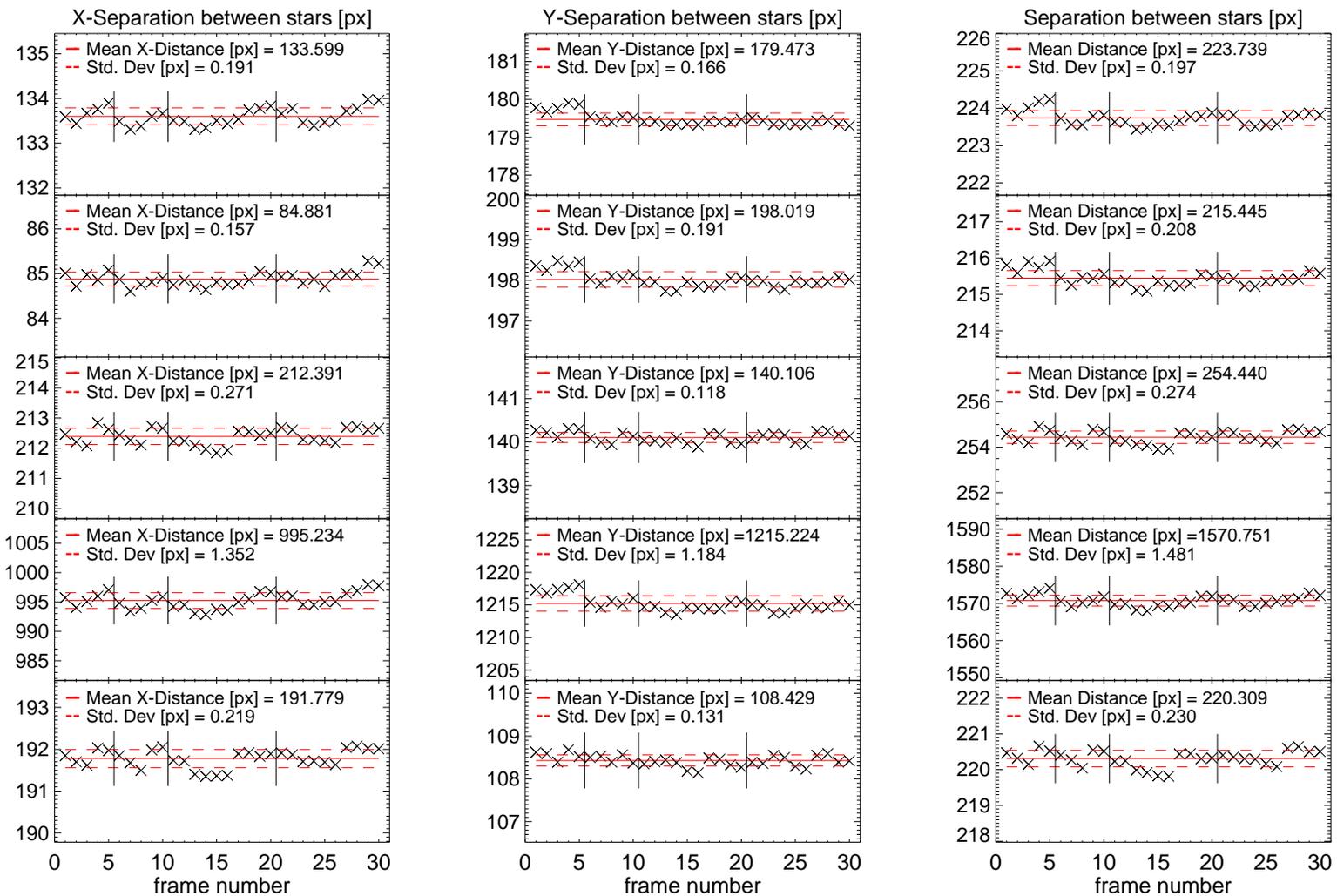}
 \end{center}
 \caption[Separation between pairs of stars over frame number]
 {Separation between pairs of stars plotted against frame number. The left panel
 shows the separation in x-direction, the middle panel in
 y-direction and the right panel the full separation
 $r = \sqrt{\Delta x^2 + \Delta y2}$. The small straight
 lines mark the frames after which a new five points
 sequence of jitter movements was started.}
\label{fig:separation}
\end{figure}

\noindent Performing this test for several stars with short and
large separations and with different position angles between the
stars showed in the case of the NGC~6388 data a recurring pattern
in the separation in $x, y, r$, which is not observable in the
47~Tuc data. Fig.~\ref{fig:separation} shows the separation in $x,
y$ and $r$ over the frame number for five representative pairs of
stars in the NGC~6388 data. Looking at the pattern, which repeats
after five frames for the first 10 frames and after 10 frames for
the following frames (where always two images were taken at the
same jitter position  before moving to the next position), this
change  in separation seems to be correlated with the jitter
movement during the observations which also has a five points
pattern with an additional change of center position (see
Fig.~\ref{fig:NGC6388jitterpattern}). In the case of the MAD
instrument the camera itself is moved in the focal plane to
execute the jitter pattern. This can lead to vignetting effects,
if the jitter offset is too large and it seems to introduce
distortions dependent on the position of the camera in the field
of view. It is unlikely that this pattern is due to the problems
with the de-rotator, because of the uniform repetition of the
pattern. Also this pattern is not seen in the 47~Tuc data, which
was obtained without jitter movements, but experienced the same
de-rotator problems.\\
\noindent I performed this test with the same star pairs after
applying the calculated distortion correction for shift, scale and
rotation (see Sect.~\ref{sec:Distortion}). The strong pattern is
gone, leaving a more random variation of the separation. Also the
calculated standard deviation is much smaller, ranging from a
factor of $\sim 3$ up to a factor of $\sim 19$ times smaller!
Comparing the single standard deviations shows a smaller scatter
among their values than before the distortion correction. All this
yields to the conclusion, that the calculated and applied
distortions remove a large amount of the separation scatter, but
not all of it. The remaining scatter of the separations between
the stars in the single frames still ranges from $\sim 1.2 - 2.8$
mas, well above the scatter expected from photon statistics,
pointing to uncorrected higher order distortions. But the values
here are only calculated for a small fraction of the stars in the
frames. I therefore also had a look on a more global scale of the
positional residuals.

\section{Residual Mapping} 
\begin{figure}[t!]
 \begin{center}
  \includegraphics[width=6.2cm, angle=90]{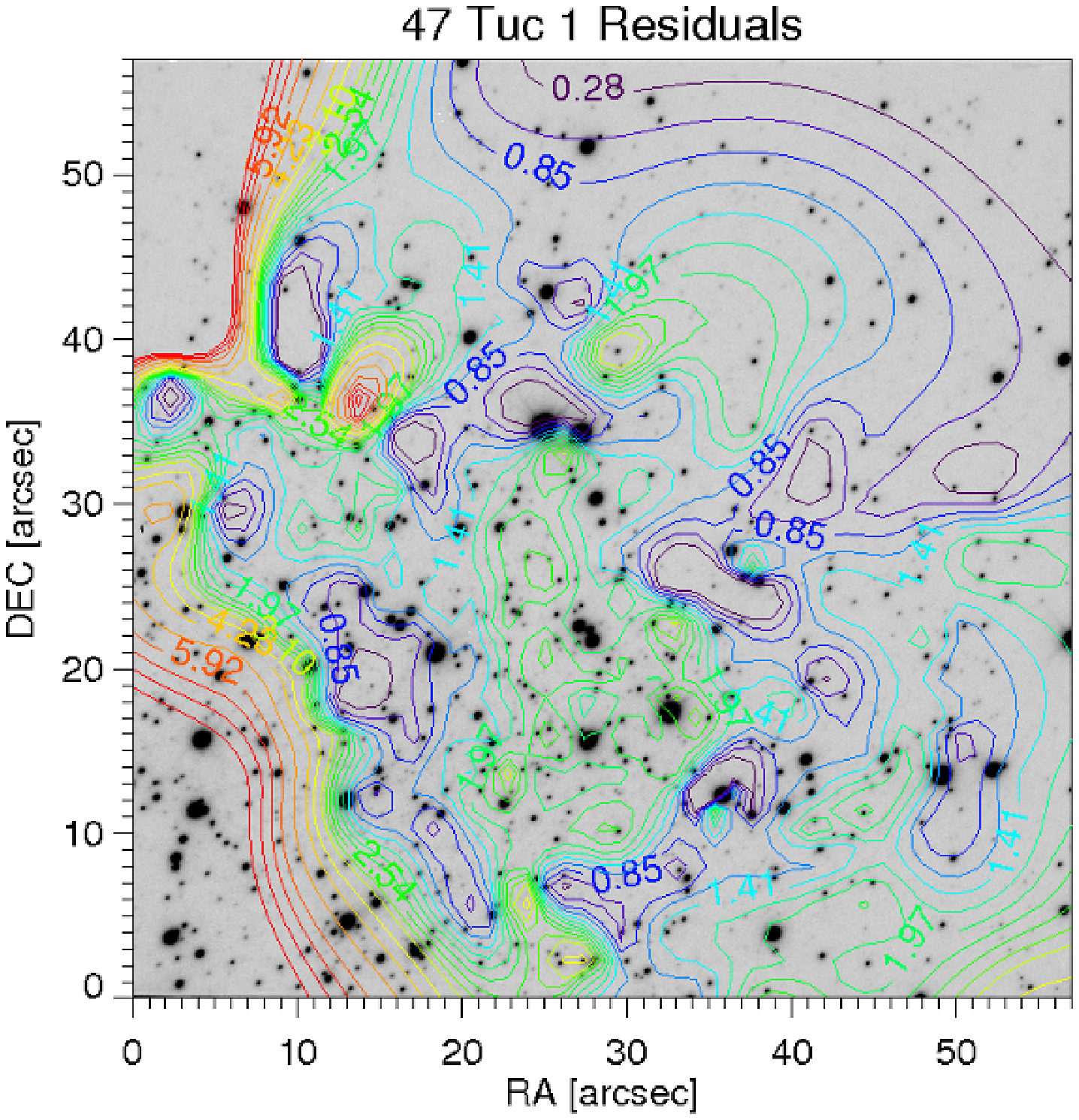}
  \includegraphics[width=6.2cm, angle=90]{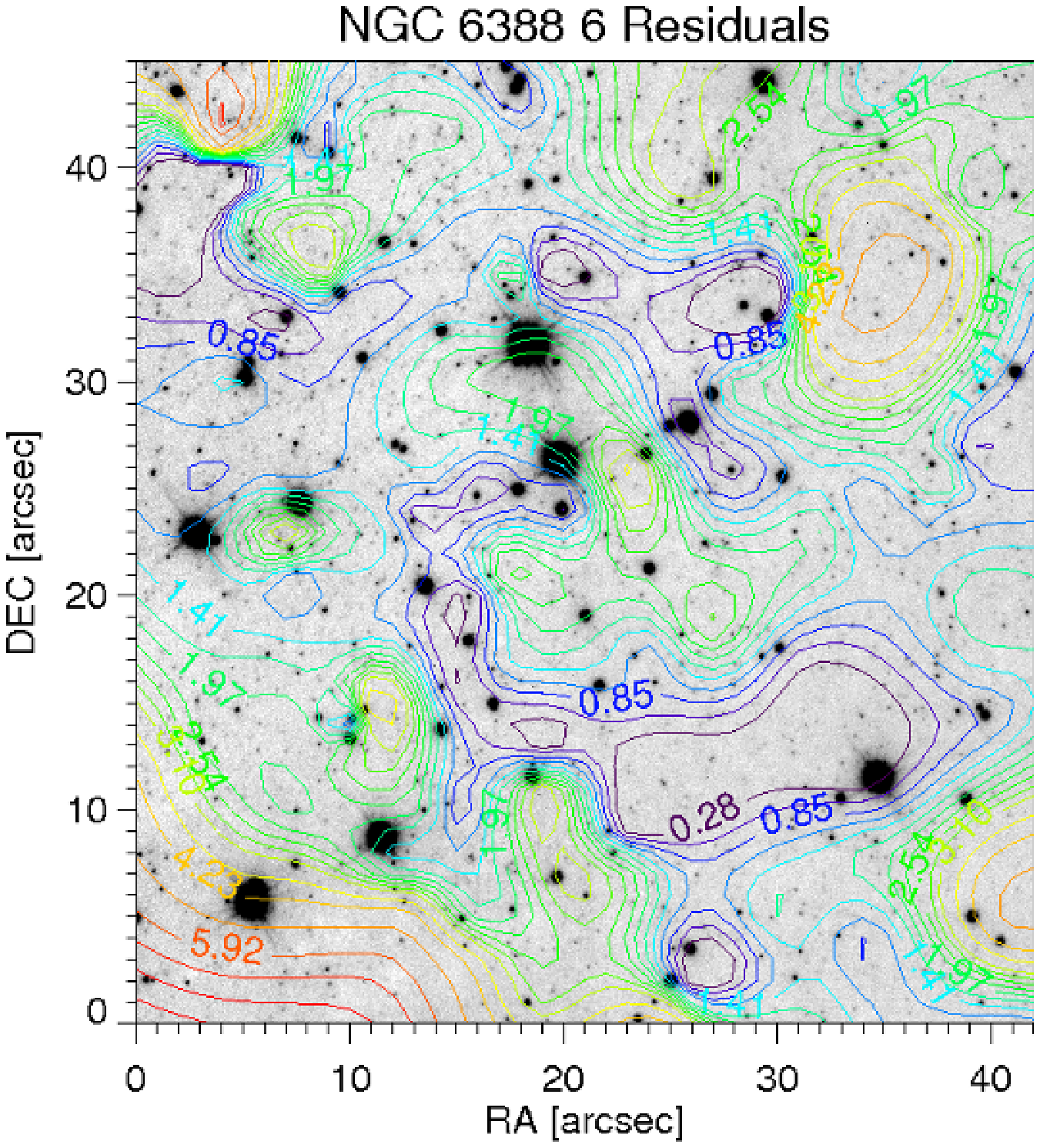}\\ \vspace{0.5cm}
  \includegraphics[height=7.1cm, angle=90]{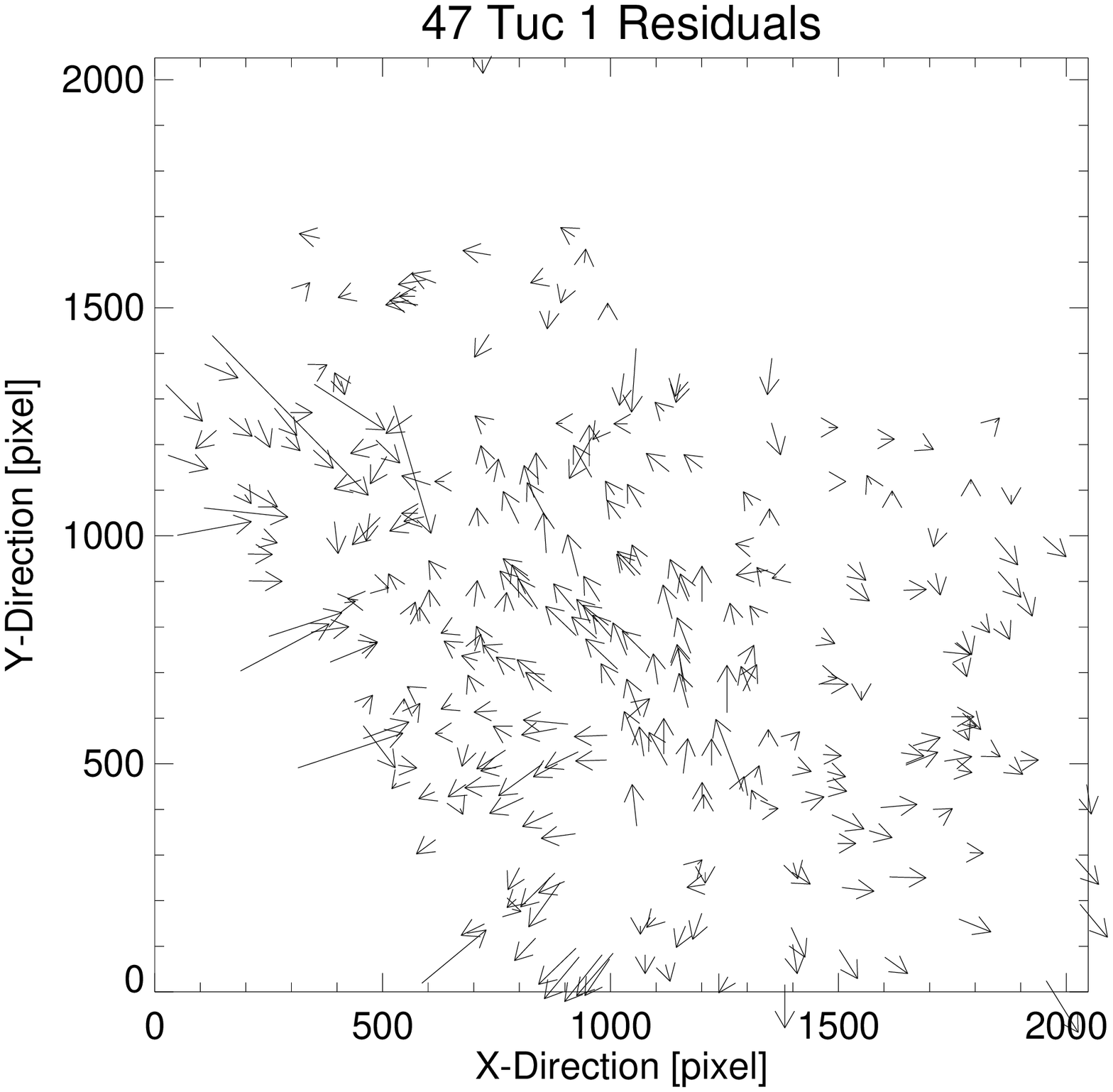}
  \includegraphics[height=7.1cm, angle=90]{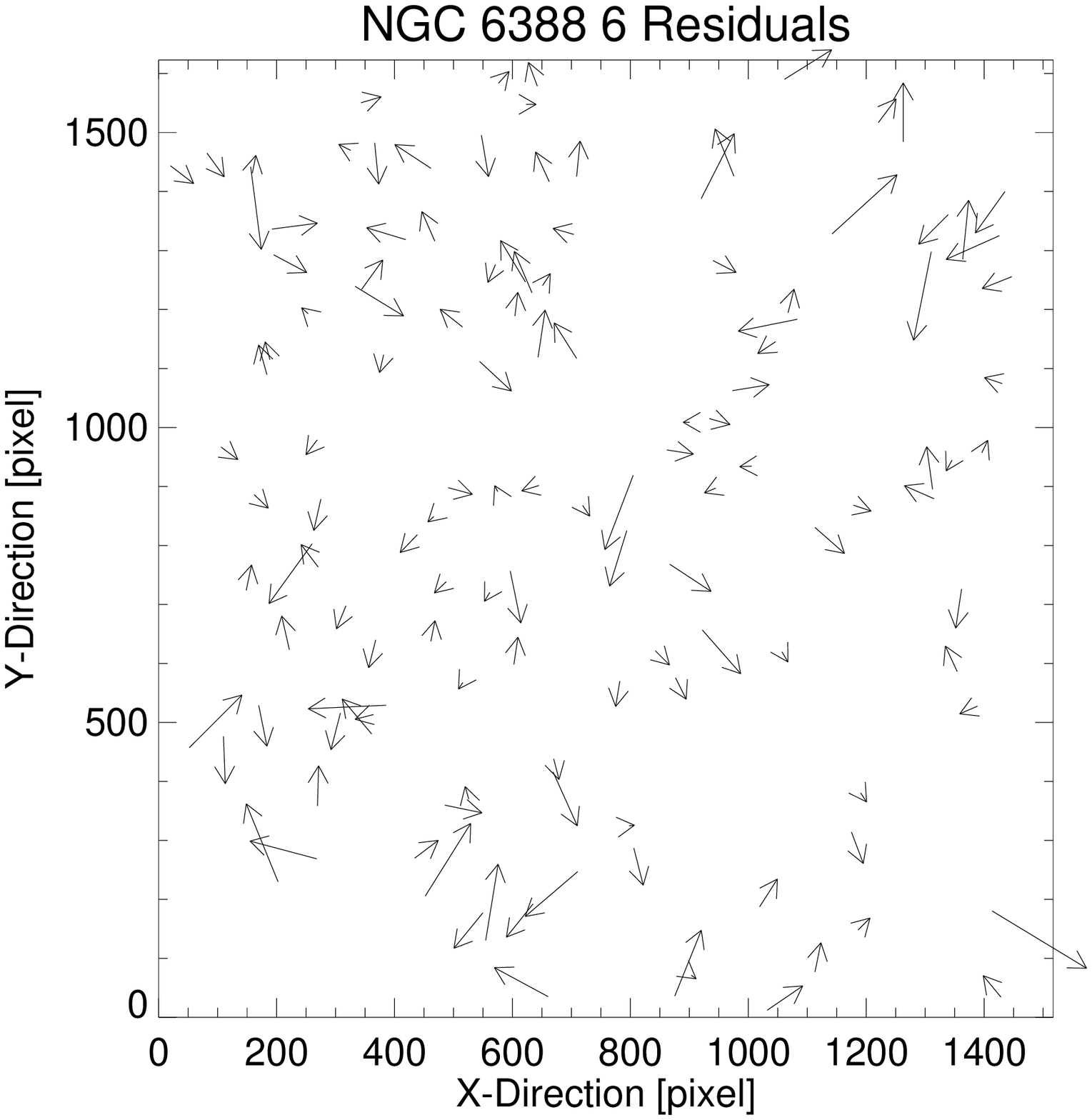}
 \end{center}
 \caption[Example contour and arrow plots of the residuals in the positions of the stars after
 the distortion correction]
 {Example contour plots of the residuals (in mas) of the positions of the stars after
 the distortion correction in the upper panels. In the lower panels, the corresponding arrow maps are shown. The arrows are extended by a factor of 1000. The left panels show the data of the
 cluster 47~Tuc and the right panels the data of the cluster NGC~6388. The empty areas in the 47~Tuc data are due to the applied selection criterias for the stars.}
\label{fig:contours}
\end{figure}
To look at the spatial distribution
of the residuals after the distortion correction, where I
corrected for $x \rm and ~y$-shift, $x \rm and ~y$-scale and
rotation relative to the masterframe, I created contour plots of
the residuals by fitting a minimum curvature surface to the data
of each frame. In Fig~\ref{fig:contours}, upper panels, an example of such a
contour plot is shown for the two data sets. The maps for the
other frames look pretty much the same with small variations in
distribution and size of the residuals. But the main goal of this
test was to check for any strong spatial variation of the
residuals over the FoV. Similar to the Strehl maps and the PSF tests I made in
Chap.~\ref{sec:PSFTests}, no strong spatial variation can be seen,
such as for example a strong gradient in one direction. The high
residuals in the two left corners of the 47~Tuc data are an
artifact of the surface fit, as there were not enough data points
in these areas for a good surface fit. Additionally I created
arrow diagrams showing not only the strength, but also the
direction of the residuals for each star used to calculate the
transformation. Looking at these maps, Fig.~\ref{fig:contours} lower panels, the orientation of the
arrows is random. I corrected for scale and rotation, so I do not
expect any prominent residual due to these parameters. A residual
scale would lead to a pattern, where the arrows all point radially
away from one area and a residual rotation would leave arrows
arranged on circles, all facing in the direction of the rotation.
No pattern of this kind can be seen.

\begin{figure}
 \begin{center}
   \includegraphics[height=7cm, angle=90]{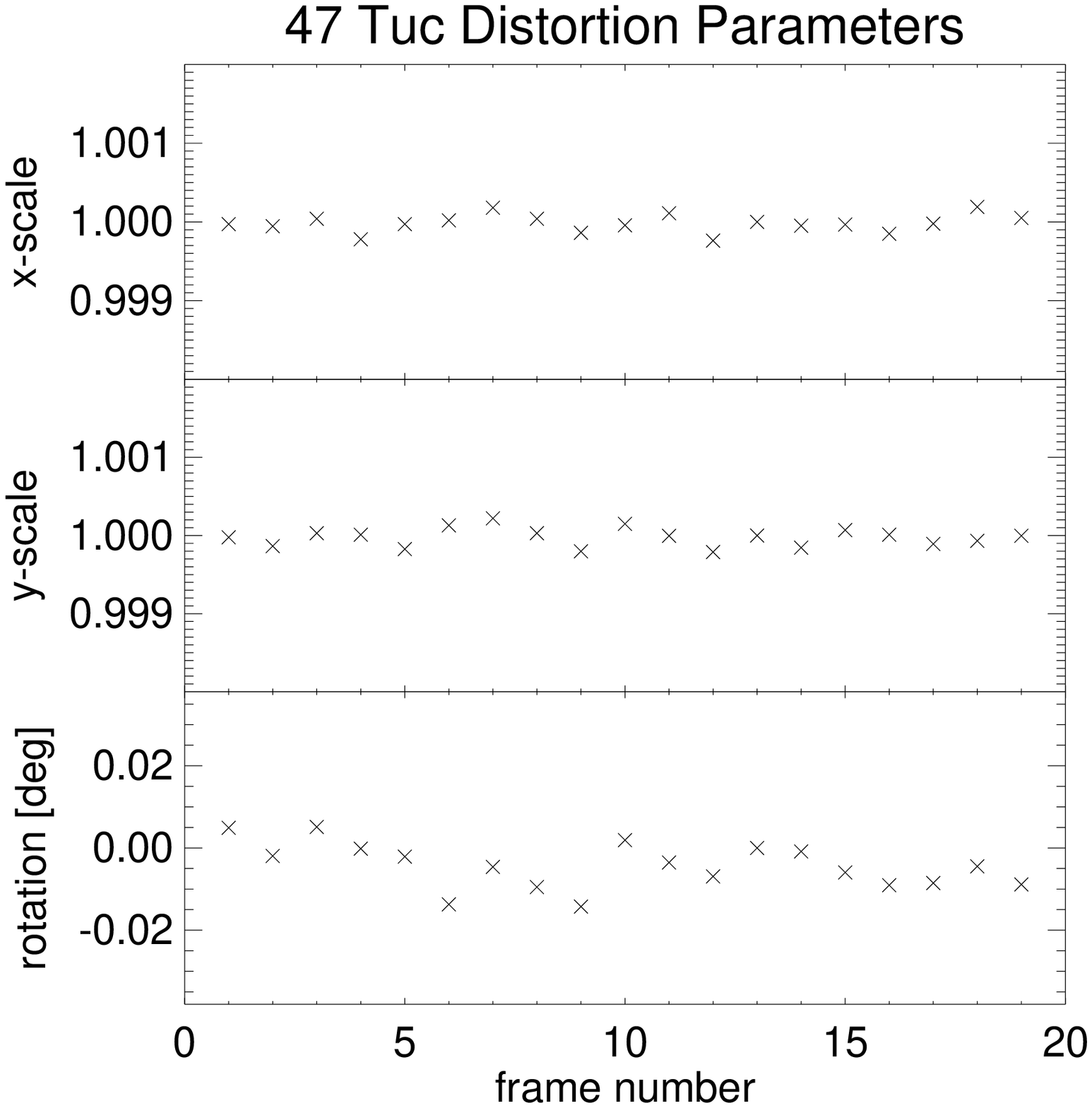} \hspace{0.2cm}
   \includegraphics[height=7cm, angle=90]{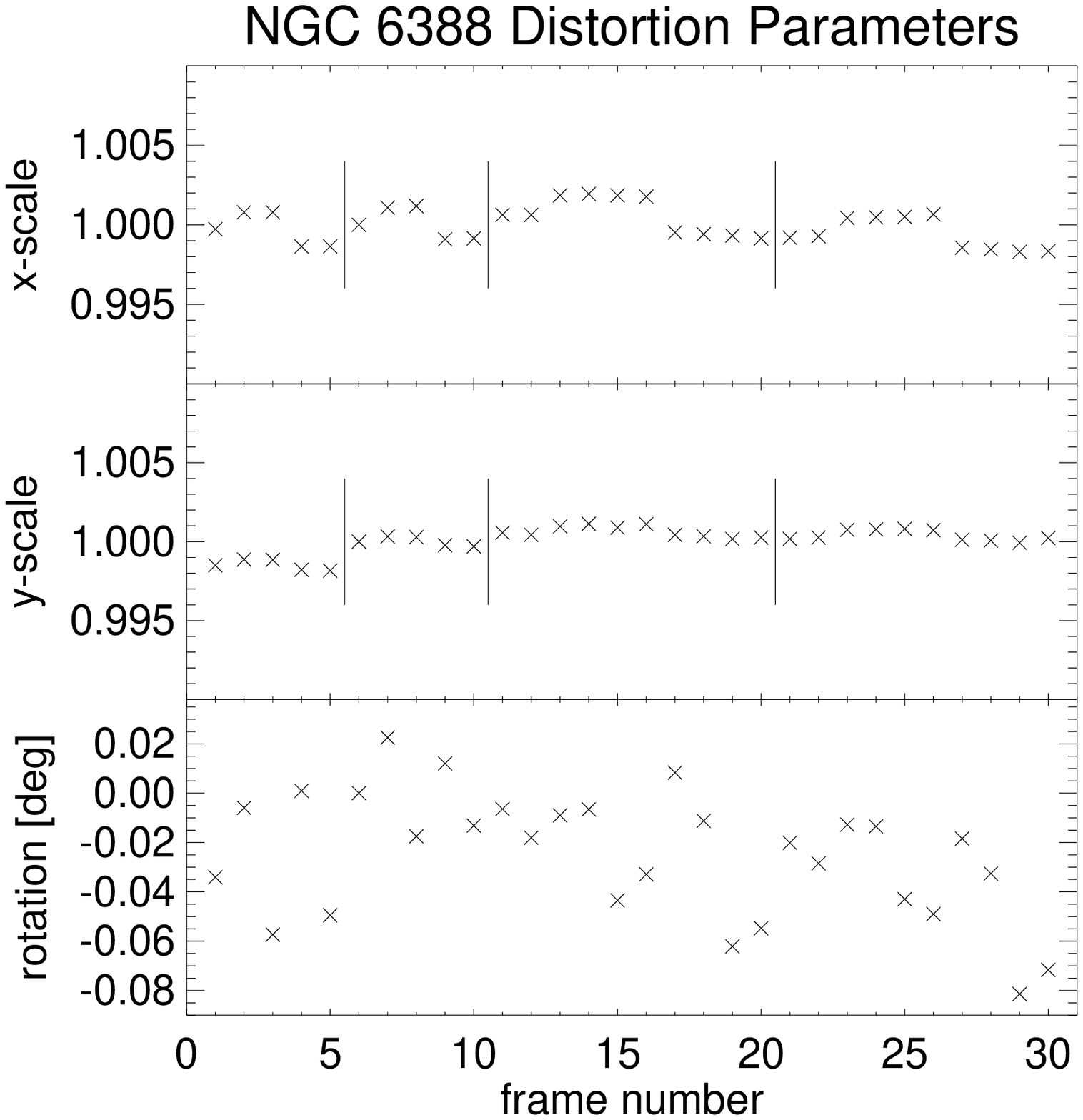}
 \end{center}
\caption[Applied distortion parameters over frame number]{Applied
distortion parameters over frame number for the 47~Tuc (left) and
NGC~6388 data (right). The panels show from top to bottom the
calculated distortion parameters for $x$ scale, $y$ scale and
rotation for each frame. While the rotation parameter is random,
the scale parameter of the NGC~6388 data show a pattern which is
not visible in the 47~Tuc data.} \label{fig:distortions}
\end{figure}

\noindent I plotted the calculated distortion parameters for
$x$-scale,  $y$-scale and rotation over the frame number, which
can be seen as a time-series, in Fig.~\ref{fig:distortions} for
both data sets.
Whereas the parameter for the rotation correction looks random,
but  with some high values, indicating the de-rotator problem, the
correction parameters for the scale in $x$ and $y$ show a pattern
in the case of the NGC~6388 data set (right). This pattern repeats
after five (10) frames, as does the pattern for the separation
measurement. As these are the applied correction parameters, they
show nicely the existence of the pattern and the ability of
correcting for these major distortions. In the 47~Tuc data there
is also some scatter, which is expectable, but no repeating
pattern can be seen. Additionally, the values are smaller in the
case of the ground layer data which
was obtained without jitter (note the different scaling of the two plots).\\

\begin{figure}[t]
 \begin{center}
  \includegraphics[height=6.9cm,angle=90]{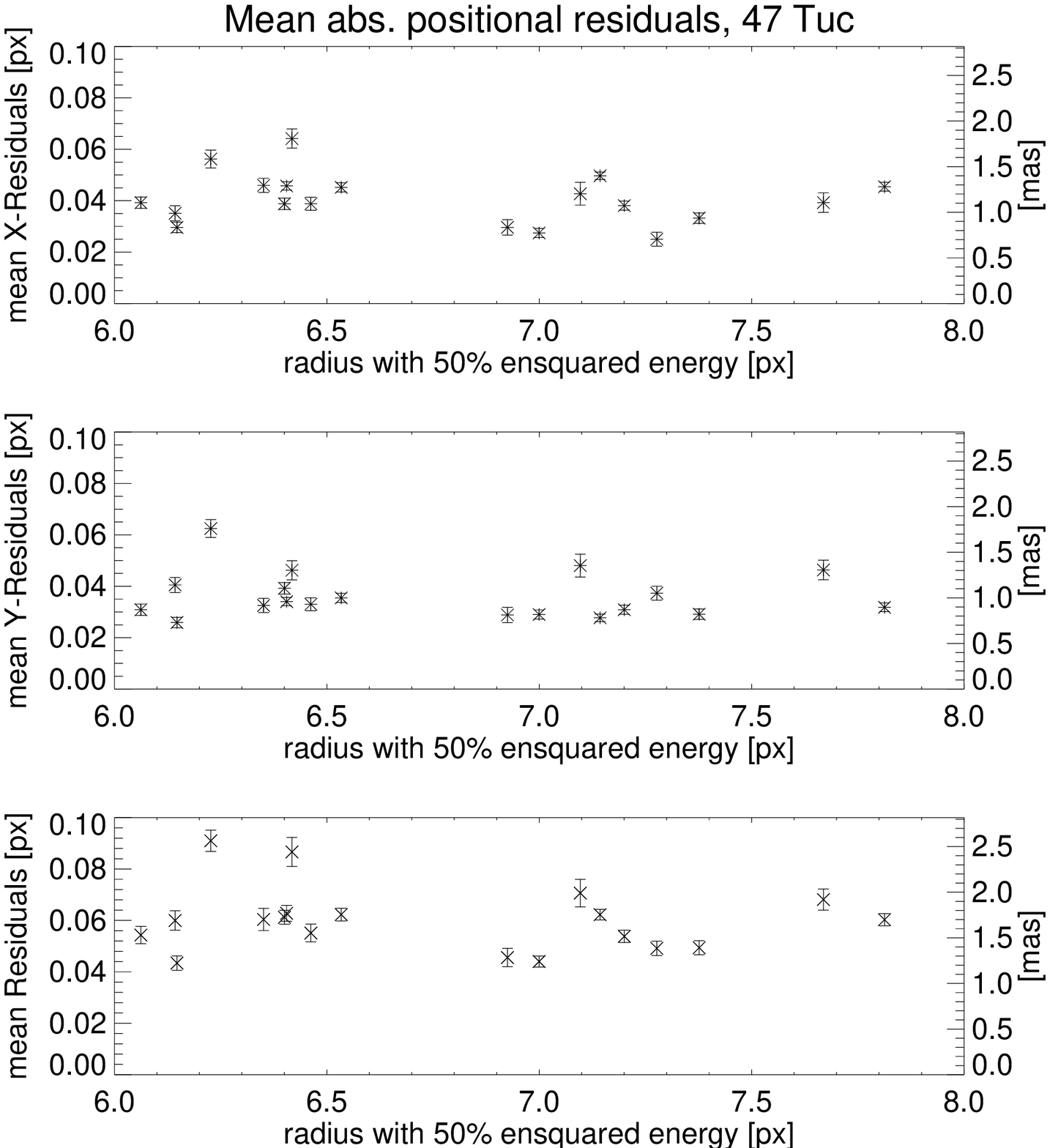} \hspace{0.45cm}
  \includegraphics[height=6.9cm, angle=90]{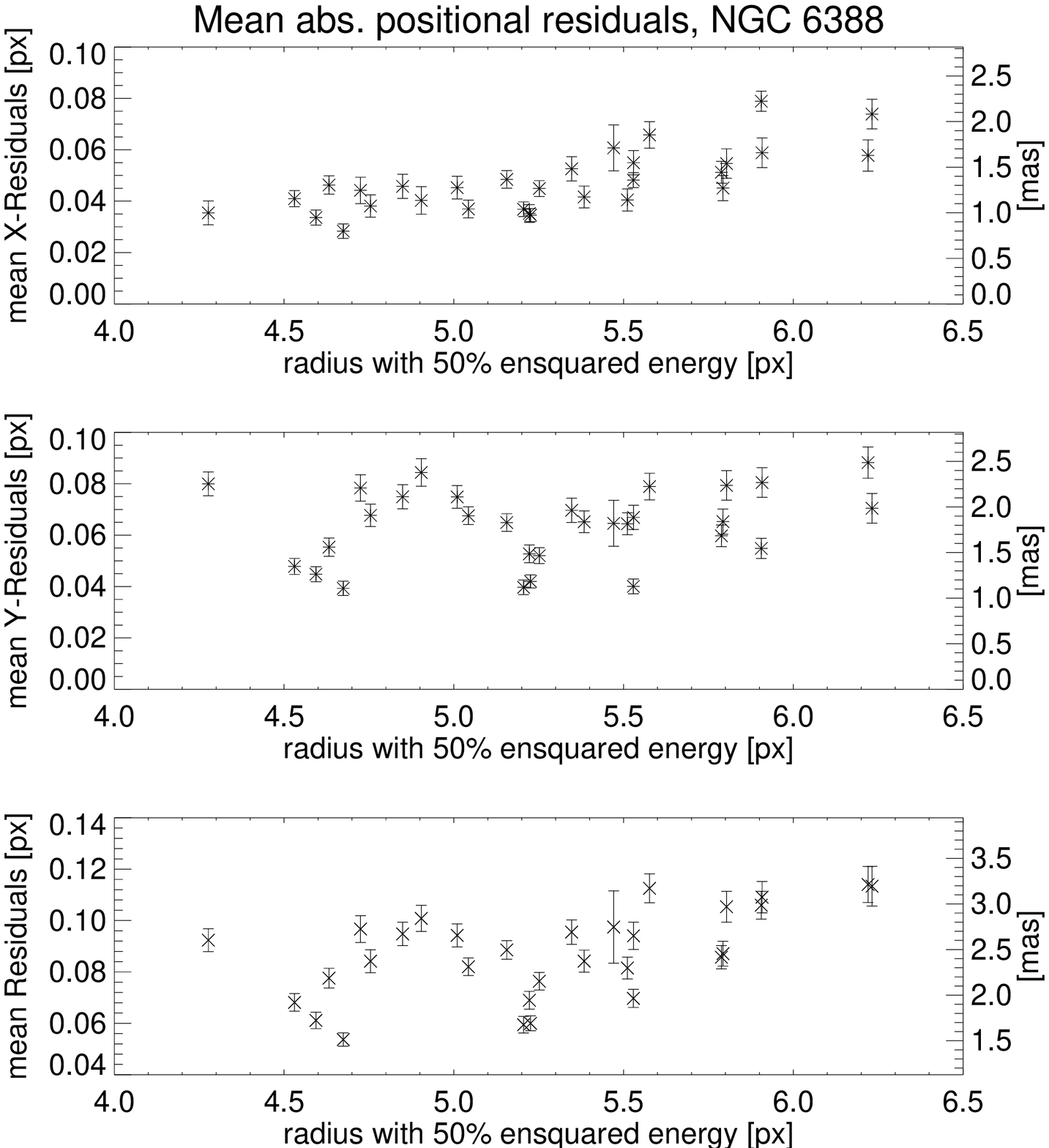}
 \end{center}
 \caption[Mean absolute positional residuals over the radius
 of 50\% ensquared energy]{Mean absolute positional residuals
 over the radius of 50\% ensquared energy. In the left panel for the 47~Tuc
 data set and in the right panel for the NGC~6388 data set. The plot shows
 from top to bottom the mean of the absolute values of the residuals to the masterframe in the
 $x$- and $y$-direction and the separation
 after the correction of $x$ and $y$-shift, $x$ and $y$-scale and rotation. The overplotted error
 bars correspond to the error of the mean value $\left (\sigma /\sqrt{n} \right )$,
 with n equal to the number of stars used to calculate the mean value and
 $\sigma$ being the standard deviation. The left y-axis shows the
 residuals in units of pixel and the right one in units of mas.}
 \label{fig:resid_frame}
\end{figure}

\noindent Finally, I calculated the mean residuals over the full
FoV for both data sets separately for the $x, y$ and $r$ direction
for each frame. The mean values are very close to zero ($\sim
10^{-5} - 10^{-6}$), supporting the results from the arrow plots
of random orientation, but looking at the mean of the absolute
values of the residuals shows how large they still are. In
Fig.~\ref{fig:resid_frame} the mean of the absolute residuals over
the full FoV in the $x$ and $y$ direction and in the separation
are plotted over the radius of 50\% ensquared energy of the
corresponding extracted PSF of each frame and each data set.
No correlation of the size of the residuals with the ensquared
energy can be seen in the 47~Tuc data set but a small
correlation in the $x$-direction in the NGC~6388 data. What is
visible, is that the absolute values of the residuals and their
scatter are larger in the case of the NGC~6388 data set compared
to the 47~Tuc data set, even though the initial observing
conditions were better and the measured FWHM values are smaller.
This gives a first impression on how precise the astrometry is in
these MAD data. The mean absolute residuals are between 0.025~px
and 0.092~px (0.7 - 2.6 mas) in the case for the ground layer
corrected 47~Tuc data set and between 0.028~px and 0.114~px (0.8 -
3.2 mas) in the MCAO corrected NGC~6388 data set. With photon
statistics alone, the positions should vary in a way smaller
range. Taking the positional accuracy calculated from photon
statistics for the faintest stars used in this set, the residuals
should be within 0.005~px (0.14 mas) in the 47~Tuc case and 0.012
(0.33 mas) in the NGC~6388 case. The calculated values show a
residual positional scatter that cannot be explained by simple
statistical uncertainties. It rather shows that even after a basic
distortion correction, the remaining positional scatter is fairly
large for the purpose of high precision astrometry. This scatter
seems to have its origin in higher order distortions present in
the images, as it seems largely independent from the size of the
PSF. Additionally, the residuals and scatter are larger in the
case where the camera was jittering to scan a bigger field of
view. This jitter movement introduced distortion, which I already
saw in the separation measurements and in the distortion
correction parameters calculated for scale and rotation. But also
the AO correction can introduce higher order distortions as it
dynamically adapts to atmospheric turbulence changes.

\section{Mean Positions}
\begin{figure}
 \begin{center}
  \includegraphics[width=5cm, angle=90]{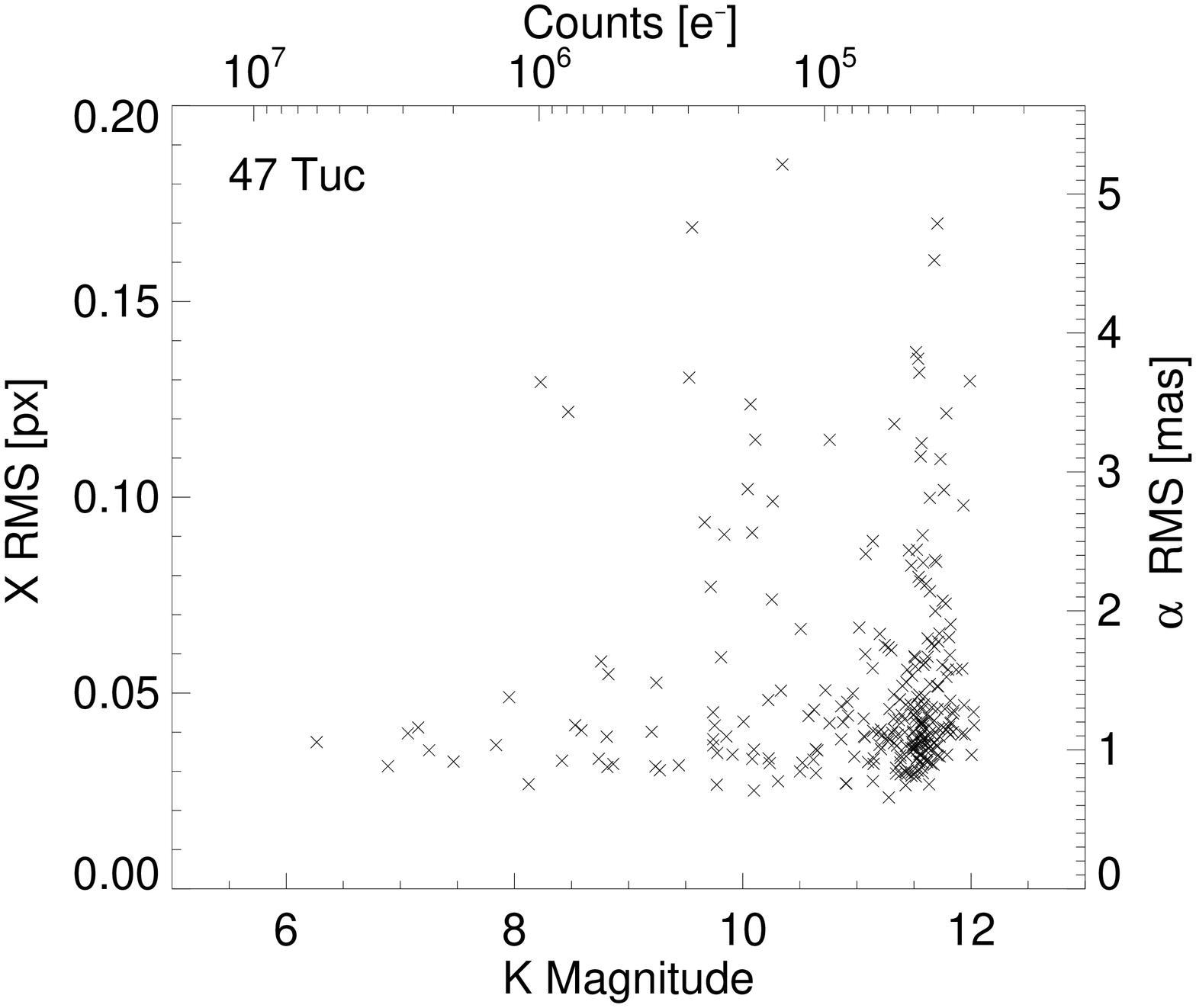} \hspace{1.5cm}
  \includegraphics[width=5cm, angle=90]{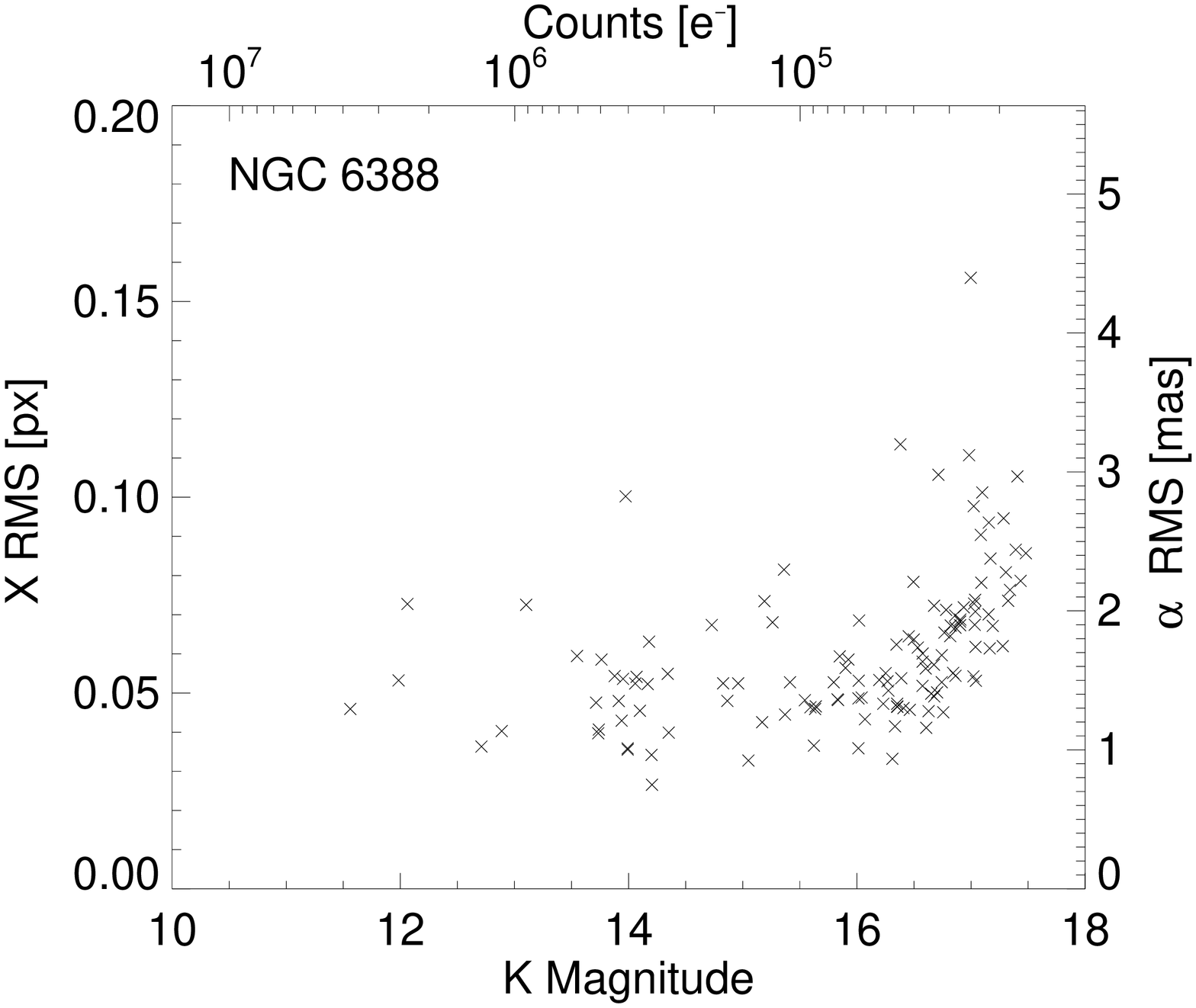} \\ \vspace{1.5cm}
  \includegraphics[width=5cm, angle=90]{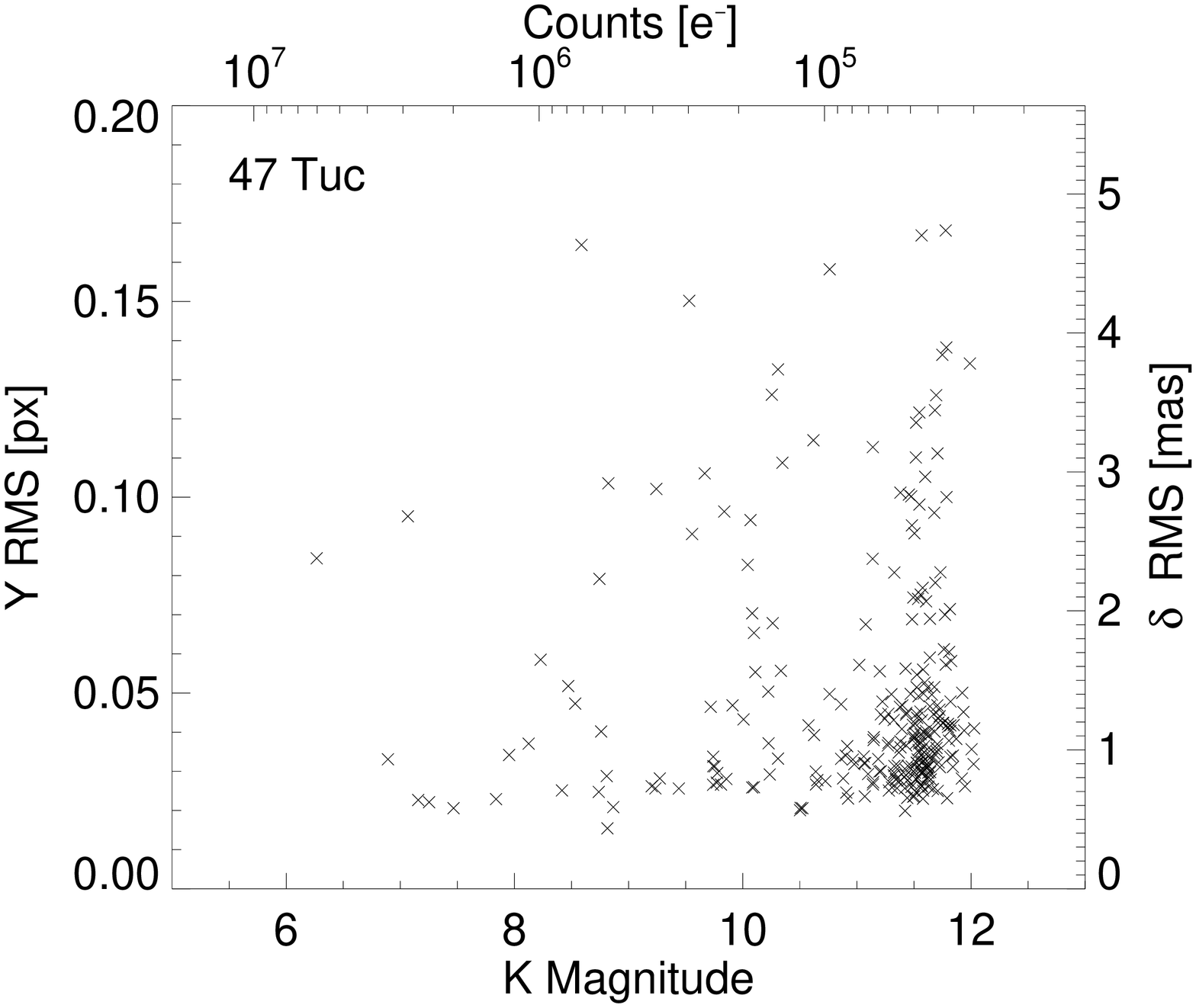} \hspace{1.5cm}
  \includegraphics[width=5cm, angle=90]{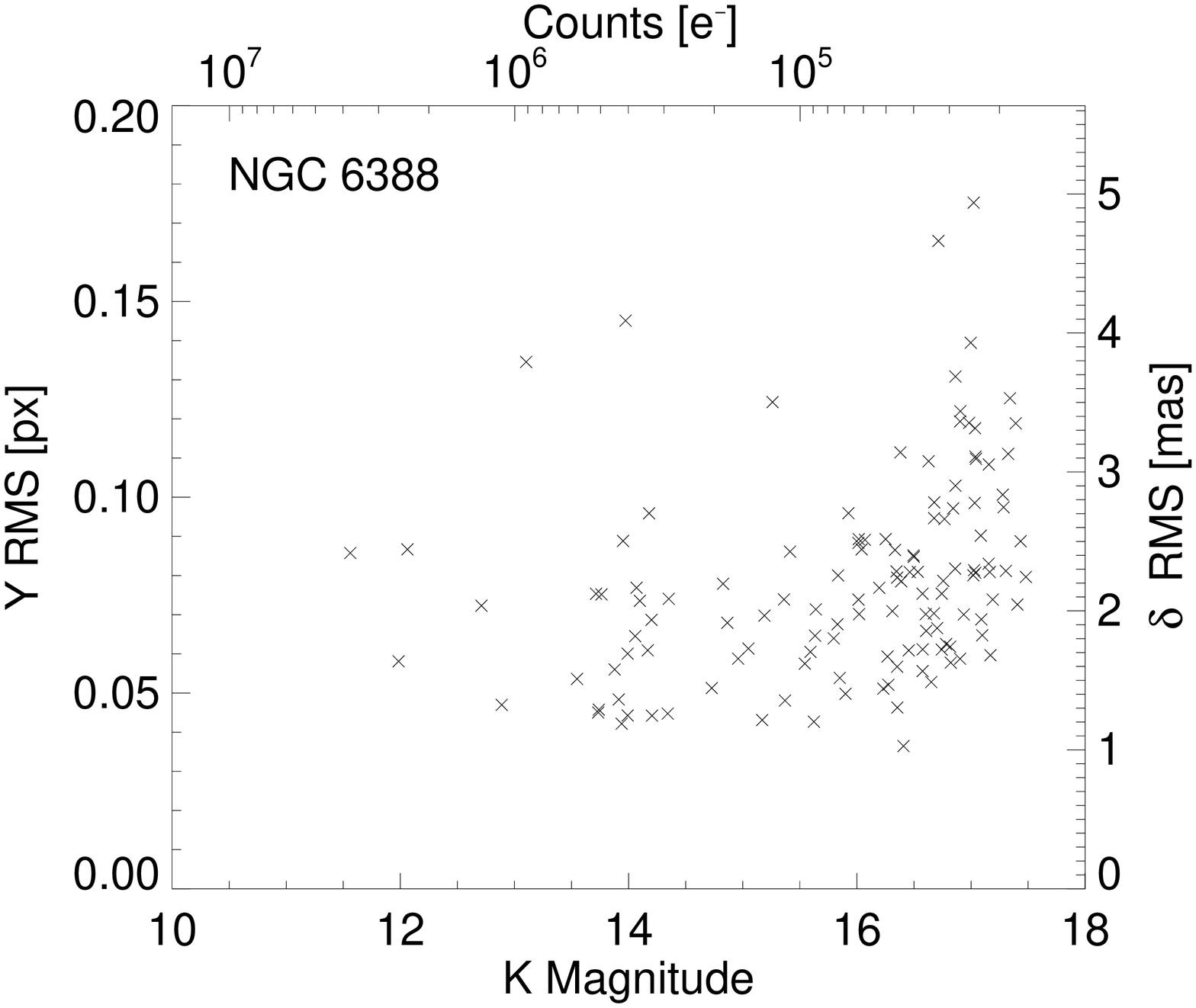}
 \end{center}
\caption[MAD positional RMS over Magnitude]{MAD positional RMS
over Magnitude} \label{fig:astr_position}
\end{figure}

As a last step I calculated the mean position for each star over
all frames. This is one way to measure the astrometric positions
and their uncertainties.
In Fig.~\ref{fig:astr_position} the achieved astrometric precision
is plotted over the K magnitude for each star in the final lists
of both data sets. The given magnitude is not an accurate value,
calculated with careful photometry, but rather represents the from
the measured flux estimated 2MASS K magnitude of the stars. For
the conversion from counts to magnitudes I used the counts of the
stars calculated by starfinder. For the brighter isolated stars I
took the corresponding K magnitudes from the 2MASS catalog and calculated with
theses values the zero point of the conversion for each of these
bright stars. After taking the mean of these zero points I could
convert the measured counts for all stars into their corresponding K
magnitude ($\rm mag = \rm zero-point - 2.5~\log(\rm flux)$). These
values are not meant to be understood as being exact, but will be
good enough to see the principal relation between precision and
intensity. For completeness I indicated the exact counts at the
upper x-axis of the plots. I made this conversion also for the
data set of the globular cluster 47~Tuc even though it was
observed in $Br_{\gamma}$. Therefore note that the given numbers
for the K magnitudes in the plots are not measured, but are
related
to a certain measured flux.\\

\noindent As one can see in the plots for the NGC~6388 data set
(right panels), the fainter stars have less precision in their
position than the brighter ones. The mean positional precision of
the stars between 14 and 18 mag is $\pm 0.073$ pixel or $\pm
2.057$ mas, calculated as a mean value of the $x$- and
$y$-directions. This is the achievable astrometric precision with
the available MAD data in full MCAO mode. Theoretically, as stated
above, the faintest stars in this regime should have a precision
of about $\pm 0.012$~px ($\pm 0.33$ mas) assuming only photon
statistics. The measured precision is a factor of 6.2 worse. This
is a quite large discrepancy, although one has to take more than
photon statistics into account for calculating a correct error
budget. On the other hand, aiming at a in astrometry reasonable
measurement accuracy of 1/25 - 1/50 pixel (1.127 - 0.564 mas) the
achieved mean astrometric precision is still a factor of 1.8 - 3.6
larger
than the expected possible precision.\\
\noindent In the GLAO data set of 47~Tuc the mean astrometric
precision for stars between 9 and 12 mag is $\pm 0.051$ pixel
($\pm 1.437$ mas). Although the fainter stars seem to have
slightly larger uncertainties, this correlation is less
distinctive than in the MCAO case (NGC~6388). And again, comparing
these values with the theoretically achievable values of $\sim
1/25$ pixel shows, by a factor of 1.3, a less precise position
measurement
than expected.\\
\noindent In Tab.~\ref{tab:MAD_results} the expected and achieved
results for the astrometric precision in the two data sets are summarized.\\
\noindent Comparing the results for the ground layer correction
with those of the MCAO correction shows a higher precision in the
GLAO data. One could expect it the other way round as the initial
observing conditions and the average Strehl are better in the MCAO
data. Also the FWHM and the radius of 50\% ensquared energy are
smaller in the MCAO data. Additionally, the exposure time was
longer for the NGC~6388 cluster. But this is no advantage in the
sense of better signal, as NGC~6388 is further away, 9.5 kpc, than
47~Tuc, 4.0 kpc, and is fainter. In Fig.~\ref{fig:astr_position}
one can see by comparing the upper $x$-axes, that the final counts
distribution is the same for the two sets of stars. One of the
main differences in the two data sets is the jitter movement. As
could already be seen, this movement introduces distortions. I
corrected for shift, scale and rotation, but still the afterwards
achieved precision is smaller, indicating distortions of higher
order, quadratic or even higher.
\begin{table}
\begin{center}
\begin{tabular}{|l|l|l|l|l|}
\hline
GJ~1046 & Photon & $\frac{1}{25}$ pixel & 47~Tuc & Photon  \\
 & statistics & & & statistics \\ \hline
$\pm 0.073$~px & $\pm 0012$~px & $\pm 0.040$~px & $\pm 0.051$~px &
$\pm 0.005$~px \\
$\pm 2.057$~mas & $\pm 0.326$~mas & $\pm 1.127$~mas & $\pm
1.437$~mas & $\pm 0.137$~mas \\
\hline
\end{tabular}
\end{center}
\caption[Summary of the expected and achieved astrometric
precisions]{Summary of the expected and achieved astrometric
precisions. The first and fourth columns list the measured mean
astrometric precision for the NGC~6388 and 47~Tuc data,
respectively. The second and last columns list the expected
precision from photon statistics for the faintest star in the
respective magnitude ranges for the two data sets, second column
for the NGC~6388 data and the last column for the 47~Tuc data. The
middle column lists the corresponding pixel and milli-arcsecond
values for a precision of 1/25 pixel.} \label{tab:MAD_results}
\end{table}

\section{Discussion and Conclusion}
I have analyzed the first multi conjugated adaptive optics data
available in the layer oriented approach with respect to
astrometric performance. The data were taken with the MCAO
demonstrator MAD at the VLT. Two sets of data of globular
clusters, observed in two different approaches were analyzed: The
globular cluster 47~Tucanae with only ground layer correction and
the globular cluster NGC~6388 in full MCAO correction.\\

\noindent The first data analyzed was those of the globular
cluster NGC~6388 taken with full MCAO correction. I calculated
Strehl maps for each frame. The Strehl is fairly uniform over the
FoV with a small degradation to the edges of the FoV and average
values between 11\% and 23\%. The performance was slightly
degrading over the time of the observation which can be seen in
the lower Strehl values and larger FWHM values in the later
frames. The first five frames were obtained with only the ground
layer corrected. The FWHM and Strehl are stable in these frames,
even though the seeing measured by the DIMM seeing monitor,
changed slightly.\\
\noindent I created a master frame with positions of bright
isolated stars in the field and calculated distortion parameters
for a shift and scale in $x$ and $y$ direction and a rotation for
each frame to this master frame. Separation measurements between
stars before and after the distortion correction showed that this
correction is indeed reducing the scatter in the separations
measured over all frames. But it also shows a residual scatter,
which is probably due to higher order distortions. A pattern
visible in the separation measurements as well as in the applied
distortion parameters is thought to be due to the jitter movement
of the camera during the observations. This movement introduced
additional distortions which could partly be corrected with the
distortion correction.\\
\noindent The precision of the positions of the stars, calculated
by the scatter of the mean position of the stars over all frames,
is $\pm 0.073$ pixel or $\pm 2.057$ mas in the corresponding 2MASS
K magnitude range from 14 to 18 mag. The positions are calculated
in the detector coordinate system.\\
\noindent The positional precision, as well as the scatter in the
separation measurements and the mean residuals of the positions of
the stars after the distortion correction to the positions in the
master frame, all lie in the same range, showing the astrometric
precision achievable with these data.\\
\noindent I compared my results with the unpublished ones
calculated by Alessia Moretti\footnote{Private communication}. She
analyzed the same data set photometrically with the DAOPHOT
reduction package \citep{Daophot1987}. DAOPHOT uses a variable
Penny function to fit the PSFs of the stars. The main goal of her
work is the photometric analysis of the cluster data, but she also
performed astrometric measurements. Her calculated mean
astrometric precision of stars with a K magnitude between 14 and
18 mag is 0.061 pixel. The precision was calculated in a final
image which was created by combining the four best frames,
allowing for translation, scale and rotation of the images before
addition. Our results are conforming, showing that the use of a
mean PSF for fitting the positions of the stars is not the reason
for the uncertainty in the astrometric precision, and the scatter
seen in my results is really in the data and not an artefact of
the fitting procedure.

\noindent The second data set I analyzed was conducted with only
correcting the ground layer turbulences. The calculated Strehl
maps yielded smaller but still fairly uniform values between 9\%
and 14\%. The lower Strehl in the 47~Tuc data set can partly be
explained by the fact that only the distortions due to the ground
layer were corrected, but also the initial atmospheric conditions
were worse, which leads to a degradation in the possible
performance of an AO system. To disentangle these two possible
causes, one needs to analyze more data sets.\\
\noindent I also created a master frame and calculated distortion
parameters for each frame to this master frame. The separation
measurements and distortion parameters did not show a pattern as
was the case in the NGC~6388 data set.\\
\noindent The precision of the positions of stars in the
corresponding 2MASS K magnitude range between 9 and 12 mag is $\pm
0.051$ pixel ($\pm 1.437$ mas) in this data set.\\
\noindent Astrometric analysis of the core of 47~Tuc were also
performed by \cite{McLaughlin47Tuc}, who used several epochs of
Data from the Hubble Space Telescope (HST). They derived
positional precisions in the single epoch data, taken with the
High Resolution Camera (HRC) of the Advanced Camera for Surveys
(ACS) for most stars in the range of 0.01-0.05 pixel. With a
plate-scale of 0.027 arcsec/pixel this corresponds to 0.27-1.35
mas. The errors were calculated in the same way as in this work,
taking the standard deviation of the positions in all frames as
uncertainties. Detailed distortion corrections were computed for
ACS by \cite{Anderson2002}, which were applied to the data in the
work of McLaughlin et al.. This shows that the precision derived
with MAD is already comparable to HST astrometry and with a good
distortion characterization, future instruments will yield even
higher astrometric precision.

\noindent Although the Strehl is smaller and the FWHM is larger in
the GLAO data of the cluster 47~Tuc, the achieved astrometric
precision is higher. Also the observing conditions were worse
during the GLAO observations compared to the MCAO observations
with a mean seeing of $1.13\arcsec$ and $0.46\arcsec$,
respectively. All this leads to the conclusion that the
degradation of the astrometric precision in the MCAO data set is
mainly due to the jitter movement during the observations, which
introduced additional distortions. But also the more complex
correction of two layers could have introduced higher order
distortions, which we could not correct for. To verify this one
needs to compare all the results with the applied correction
parameters. But going more into detail would be a distortion
characterization, which is indeed a very interesting task, but not
goal of this work. To fully characterize the remaining
distortions, one would need more data, taken under various seeing
conditions and observation configurations. As MAD will not be
offered again, a fully satisfactory analysis is not possible.\\

\noindent For future MCAO observations one should try to either
build an instrument, where the camera is not moved to execute the
jitter movement or completely avoid jittering. As the latter one
is often not possible in IR observations, one should take great
care of the distortions present in the frames and fully
characterize those for high precision astrometric observations.\\
\noindent All the results presented here are still given in
detector coordinates, as I analyzed the data in matters of the
adaptive optics correction and instrumentation stability over the
time of the full length of the observation. Going to celestial
coordinates would involve the correction for effects such as
differential aberration and differential refraction to derive the
true positions of the stars. As the observed FoV is large, these
effects can reach several milliseconds of arc of displacement
between stars at different points on the detector. These
transformations introduce additional position uncertainties,
degrading the astrometric precision further, but need to be done
when comparing data from different epochs (as seen in
Chap.~\ref{chap:Astrometry}).

     \cleardoublepage
    %
    %
    \chapter{Acronyms}
        \begin{tabbing}
    \textbf{AO} \qquad\qquad\qquad \=Adaptive Optics \\ [0.2cm]
    \textbf{AU} \> Astronomical Unit \\ [0.2cm]
    \textbf{DIMM} \> Differential Image Motion Monitor \\ [0.2cm]
    \textbf{DIT} \> Detector Integration Time \\ [0.2cm]
    \textbf{DM} \> Deformable Mirror \\ [0.2cm]
    \textbf{ESO} \> European Southern Observatory \\ [0.2cm]
    \textbf{FCUL} \> Faculdade de Ci\^{e}ncias da Universidade de Lisboa \\ [0.2cm]
    \textbf{FoV} \> Field of View \\ [0.2cm]
    \textbf{GLAO} \> Ground Layer Adaptive Optics \\ [0.2cm]
    \textbf{HLAO} \> High Layer Adaptive Optics \\ [0.2cm]
    \textbf{HST} \> Hubble Space Telescope \\ [0.2cm]
    \textbf{LGS} \> Laser Guide Star \\ [0.2cm]
    \textbf{LO} \> Layer Oriented \\ [0.2cm]
    \textbf{MAD} \> Multi conjugated Adaptive optics Demonstrator \\ [0.2cm]
    \textbf{MCAO} \> Multi Conjugative Adaptive Optics \\ [0.2cm]
    \textbf{NACO} \> NAOS-CONICA \\ [0.2cm]
    \textbf{NDIT} \> Number of Detector Integration Time \\ [0.2cm]
    \textbf{NGC} \> Natural Guide Star \\ [0.2cm]
    \textbf{PSF} \> Point Spread Function  \\ [0.2cm]
    \textbf{PWS} \> Pyramid Wavefront Sensor \\ [0.2cm]
    \textbf{RV} \> Radial Velocity \\ [0.2cm]
    \textbf{SDI} \> Spectral Differential Imaging \\ [0.2cm]
    \textbf{SHS} \> Shack-Hartmann Sensor \\ [0.2cm]
    \textbf{SNR} \> Signal to Noise Ratio \\ [0.2cm]
    \textbf{SO} \> Star Oriented \\ [0.2cm]
    \textbf{SR} \> Strehl Ratio \\ [0.2cm]
    \textbf{TNG} \> Telescopio Nationale Galileo \\ [0.2cm]
    \textbf{VLT} \> Very Large Telescope \\ [0.2cm]
    \textbf{WCS} \> World Coordinate System \\ [0.2cm]
    \textbf{WF} \> Wavefront \\ [0.2cm]
    \textbf{WFS} \> Wavefront Sensor \\ [0.2cm]
    \end{tabbing}

    %
    \cleardoublepage
    %
    \bibliographystyle{my_apa}
    \bibliography{Literatur}
    \addcontentsline{toc}{chapter}{Bibliography}
    %
    %
    \cleardoublepage
    \pagestyle{plain}
    \chapter*{Acknowledgment}
    At this point I would like to thank all those people who helped me on my way to finally hand in this thesis.

\noindent First of all I would like to thank my supervisor Martin K\"urster for his patience and guidance over the last three years. A big thank you also for proofreading this thesis and all the helpful comments.

\noindent Thank you to Prof.\ Hans-Walter Rix and Prof.\ Joachim Wambsgan{\ss} agreeing to referee the thesis.

\noindent PD Henrik Beuther and Prof.\ Werner Aeschbach-Hertig for being jury members at the defense.

\noindent Rainer K\"ohler for fitting and fitting and fitting... and for explaining so much and always having time for me and never get hacked off :)

\noindent Emiliano Arcidiacono for helping me so much with all my MAD questions. And for squeezing me in in his and Jacopo's office for two weeks to help define the goals of the MAD analysis.

\noindent Roberto Ragazzoni and the full MAD team for providing me with the MAD data.

\noindent My former and present office mates for all the conversations in- and outside daily science life. And the great MPIA studends coffee break members for having everyday 30 min to relax and chat.

\noindent Markus for giving me so much support during all my smaller and bigger problems during the last years and for running from printer to printer to find the \textit{best printed version} of the thesis.\\
\noindent Thank you for loving me and asking me to marry you *

\noindent A big thank you to my family, who is always supporting me. My brother for reading part of the thesis. My mother for always believing in me. My father, who cannot celebrate this day with me, for being the best Dad in the world!

\noindent Finally a heartily thank you to all the people I have not mentioned specifically here, but who helped me on my long way.\\[2cm]

\begin{center}
\large{\textit{Thank you!}}
\end{center}

\end{document}